%

\texsis
\elevenpoint
\offparens
\singlespaced
\widenspacing
\voffset=5mm

\showsectIDfalse
\def\Tbf{\twentypoint\bf}

\def\Chapter#1{\chapter{#1}}
\def\Section#1{\section{\the\chapternum.\the\sectionnum\ #1}}
\def\SubSection#1{\smallskip\nosechead{\twelvepoint\it#1}%
                  \addTOC{2}{{\it#1}}{\folio}}
\def\MiniSection#1{\smallskip\nosechead{\twelvepoint\it#1}}
\sectionminspace=0.1\vsize

\Contentstrue

\input epsf.tex
\def\DrawFig#1{\centerline{\hbox{\vbox{#1}}}}
\def\DrawHiLoRes#1#2{#2} 

\font\SetFont=msbm10 at 11pt

\font\TbfBoldItalic=cmbxti10 at 20pt
\font\tbfBoldItalic=cmbxti10 at 12pt

\def\RealSet{\hbox{\SetFont R}}

\def\IntegSet{\hbox{\SetFont Z}}
\def\Lie{\hbox{\it \$}}

\def\BoldBeta{\hbox{\mib$\beta$}}
\def\Boldh{\hbox{\bf h}}
\def\BoldM{\hbox{\bf M}}
\def\BoldP{\hbox{\bf P}}
\def\BoldF{\hbox{\bf F}}
\def\BoldS{\hbox{\bf S}}
\def\BoldLam{\hbox{\mib$\Lambda$}}
\def\Boldm{\hbox{\bf m}}
\def\Boldf{\hbox{\bf f\/}}
\def\Bolds{\hbox{\bf s}}
\def\Boldmu{\hbox{\mib$\mu$}}
\def\Boldpi{\hbox{\mib$\pi$}}

\font\SmallFont=cmr7
\def\SmallWord#1{\hbox{\SmallFont #1}}

\def\Cplusplus{\hbox{C$++$}}
\def\Clawplusplus{\hbox{Claw$++$}}
\def\Clawpack{\hbox{CLAWPACK}}
\def\AmrClaw{\hbox{AMRCLAW}}

\def\InvA{w_{\hskip0.2mm\scriptstyle\rm I}}
\def\InvB{w_{\hskip0.2mm\scriptstyle\rm II}}
\def\InvC{w_{\hskip0.2mm\scriptstyle\rm III}}
\def\InvD{w_{\hskip0.2mm\scriptstyle\rm IV}}

\def\EqNum#1#2{(\use{Eq.{#1}}{\it{#2\/}})}

\def\BoxTop{\hrule\line{\vrule height10pt\hfil\vrule height10pt}\vskip2.5mm}
\def\BoxBottom{\vskip5mm\hbox to \hsize{%
  \hbox{\vrule height10.4pt width0.4pt}\hbox{\vrule height0.4pt width8mm}%
  \hfil%
  \hbox{\vrule height0.4pt width8mm}\hbox{\vrule height10.4pt width0.4pt}%
  }
  \vskip-9.5mm
}
\def\Tab{\hskip9mm}
\def\OpenBrace{{\tt \char123}}
\def\CloseBrace{{\tt \char125}}
\def\Under{{\tt \char95\allowbreak}}
\def\Tilde{{\tt \char126}}
\def\Hash{{\tt \char35}}
\def\Code#1{\leftline{\strut\tt\Tab #1}\nointerlineskip}



\pageno=-1


\nopagenumbers

\hbox{}
\vskip35mm

{\Tbf
\centerline{Numerical Relativity}
\centerline{and}
\centerline{Inhomogeneous Cosmologies}
}

\vskip 25mm

{\fourteenpoint
\centerline{Simon David Hern}
\vskip 2mm
\centerline{Wolfson College}

\vskip 60mm
\centerline{A dissertation submitted to the}
\centerline{University of Cambridge}
\centerline{for the degree of}
\centerline{Doctor of Philosophy}
\vskip 4mm
\centerline{August 1999}
}

\vfill\supereject
\pagenumbers


\hbox{}
\vskip35mm
{\parindent=0.1\hsize
\narrower

\noindent
The work in sections~6.2 and~6.3 of this dissertation was done
in collaboration with my supervisor, J.~M.~Stewart, and has been
published as Hern and Stewart~1998.
In addition, parts of this work have made use of a computer code (the
adaptive mesh refinement algorithm described in chapter~3) which is an
improved version of a code originally written by my supervisor.
All other work described in this dissertation is my own.

\vskip6mm

\noindent
This work was supported by a studentship from the
Engineering and Physical Sciences Research Council.

}
\vfill\supereject


\hbox{}
\vskip35mm
{\parindent=0.1\hsize
\narrower

\noindent
I would like to thank

\vskip6mm

\noindent
my supervisor, John Stewart, for initiating this project and nursing
it through to completion with a wealth of helpful suggestions;

\vskip6mm

\noindent
my parents and my sister, Trudy Lianne, for never being more than a 
phone call away;

\vskip6mm

\noindent
and all the other people who have provided help or encouragement over
the last few years, with individual name-calls being deserved by the
following proofreaders and (coffee) drinking buddies:
Roland Higginson, Madelaine Crush,
Margarida Cunha, David Halls,
Sean Rooney, Myriam Champenois,
Tiarn\'an \'OCl\'eirigh, Clara Ovenston
and Ian Hawke.

}
\vfill\supereject


\leftline{\Tbf{Contents}}
\nobreak\bigskip
 
\font\TbfBoldItalic=cmmi10 at 11pt
\font\tbfBoldItalic=cmmi10 at 11pt

\Contents

\font\TbfBoldItalic=cmbxti10 at 20pt
\font\tbfBoldItalic=cmbxti10 at 12pt



\Chapter{Introductory Comments}
\pageno=1
Some brief comments are made here as way of introduction to highlight
themes that recur throughout this dissertation. 

The field of numerical relativity
is concerned with the development and
application of methods for the generation of approximate numerical
solutions for relativistic theories of gravity (possibly coupled to
non-gravitational fields).
While the subject is of relatively recent origin, the substantial
increase in availability of computing power over recent years has
led to it becoming very broad in scope.
A history of the development of numerical relativity as a field of
research can be reconstructed from the many review articles that have
been written over the years, a sample of which are
Piran~1981, Stewart~1984, Hobill and Smarr~1989, 
Bona and Mass\'o~1993, Seidel~1996 and~1998, and 
Cook and Teukolsky~1999.

Much of the recent research in numerical
relativity has concentrated on the modelling of collisions between
black holes, the motivation for this being the possibility of
observing such events through gravitational wave astronomy.
As a contrast to this, the present work uses numerical methods to
simulate inhomogeneous cosmological spacetimes, and while these
spacetimes are not
claimed to be realistic models of the observed universe (being ruled
out by their lack of matter content and their high degrees of
symmetry) they do provide a good arena in which new numerical and
mathematical approaches can be tested.

This dissertation can be divided into three main parts.
The first part comprises chapters~2 and~3 and discusses numerical
techniques (specifically finite difference and high-resolution
methods, and adaptive mesh refinement) that can be used to produce
approximate solutions to Einstein's field equations.
The mathematical representation of these equations is the subject of the
second part, with chapter~4 examining hyperbolic formulations 
and chapter~5 examining the properties of different gauge
conditions.
The third part of this work (chapters~6 and~7) presents results from
numerical simulations of inhomogeneous vacuum cosmologies with planar
and $U(1)$ symmetries.
This dissertation concludes with a summary of the main results in
chapter~8. 

What follows is a summary of the contents of this dissertation, with
emphasis being placed on ideas that connect the different chapters.

\MiniSection{Development of Numerical Software}
In this dissertation results are presented from numerical simulations
of a range of evolution systems which are introduced in chapters~4,
5, 6 and~7 (more details can be found in particular in
sections~\use{sec.HypNumerical}, \use{sec.GowCollapse},
\use{sec.GowExpand} and~\use{sec.MonCollapse}). 
All of these simulations have been performed using a computer code
developed for the purpose by the author.
The code is capable of generating numerical solutions to first-order
systems of hyperbolic partial differential equations in one or two
spatial dimensions.
It implements two advanced numerical methods
initially developed within the context of computational fluid dynamics.

The code numerically integrates systems of evolution equations using a
multi-dimensional high-resolution algorithm which is described in
section~\use{sec.PdeVolume}. 
This integration method operates by solving a sequence of Riemann
problems for the equations under consideration, and provides
significant improvements in performance over standard finite
difference methods (discussed in section~\use{sec.PdeDifference}) when
used to evolve solutions containing discontinuities. 
Riemann solvers for the systems of equations investigated here are
developed in sections~\use{sec.PdeRiemann} and~\use{sec.HypNumerical}.
While evolution schemes based on Riemann solvers have been used
previously in numerical studies of relativistic hydrodynamics, this
work represents the first use of such methods for the evolution of
gravitational fields. 

The high-resolution integration scheme is complemented by the
inclusion in the code of adaptive mesh refinement capabilities,
described in detail in chapter~3, which allow the local resolution of
a simulation to vary dynamically in response to changing features of
the evolved solution. 
Substantial improvements in performance have been achieved for the
numerical simulations of chapters~5, 6 and~7 through the use of this
method. 
(In addition, the code takes advantage of the availability of multiple
processors, as discussed in section~\use{sec.AmrParallel}.)
Although adaptive mesh refinement has previously been used in a number
of one-dimensional simulations in numerical relativity (see the
introduction to chapter~3), the present work
represents one of only a few applications of the technique to
relativistic simulations in more than one spatial dimension.

\MiniSection{Formulations of the Einstein Field Equations}
One of the main topics investigated in this dissertation is the use of
different formulations of the Einstein equations as bases for numerical
simulations. 
In sections~\use{sec.GowGowdy} and~\use{sec.MonIntro} reductions of
the Einstein equations are introduced for two cosmological models
having particular spatial symmetries, while in section~\use{sec.HypFR}
a family of formulations of the field equations suitable for evolving
fully general spacetimes is discussed.
Numerical results for these evolution systems are compared and
contrasted in section~\use{sec.MonFR}, and some reasons are suggested
for the differences in performance that are observed.

Formulations of the Einstein equations can be categorized according to
various mathematical properties that they may possess, and for
numerical simulations it may be judged desirable
(section~\use{sec.HypIntro}) for an evolution system to be strongly
hyperbolic and to have only physically relevant characteristic speeds. 
Section~\use{sec.HypFR} describes the derivation of a family of hyperbolic
formulations of the Einstein equations which generalize an evolution
system proposed by Frittelli and Reula such that the characteristic
speeds can all be made to be physically relevant.
Numerical simulations are described in this dissertation which are the
first to employ the Frittelli-Reula hyperbolic formulation, and which
are among only a relatively small number of numerical tests of hyperbolic
formulations that have been carried out in more than one spatial
dimension. 
Section~\use{sec.HypNumerical} discusses some of the issues involved
in implementing the Frittelli-Reula system for numerical work, and in
particular explains how it can be used in conjuction with the
advanced numerical methods described in chapter~2.

All of the evolution systems studied in this dissertation resolve
ambiguities in the choice of spacetime coordinates by employing the
harmonic time slicing condition. 
Section~\use{sec.HomHarmonic} discusses this and alternative
gauge conditions that may be used in numerical simulations.
Analytic results are derived which describe the reasonably general
circumstances under which coordinate singularities form in harmonic
slicings of some example spacetimes.
These results are verified through computer simulations, and
different slicings of the same spacetime are compared using an
numerical algorithm presented in section~\use{sec.HomCompare}.

\MiniSection{Simulations of Cosmological Spacetimes}
The numerical methods of chapters~2 and~3 are used in this
work to evolve solutions to cosmological models having varying
degrees of inhomogeneity.
In chapter~5 homogeneous Kasner spacetimes are evolved as part of an
investigation into gauge effects in numerical simulations.
Chapter~6 considers the Gowdy cosmological models which have two
spacelike Killing vectors, while in chapter~7 more general spacetimes
with $U(1)$~isometry groups are studied using Moncrief's reduction of 
the field equations. 
These cosmological models are all vacuum and all have the spatial
topology of a three-torus.

Although the cosmological spacetimes studied in this dissertation
are not realistic models of the observed universe, they nevertheless
provide a setting within which a range of physical phenomena can
be investigated. 
In sections~\use{sec.GowCollapse} and~\use{sec.MonCollapse} results
are presented from numerical simulations of inhomogeneous spacetimes
as they collapse towards cosmological `big bang' singularities;
particular attention is paid to the spikes and other fine-scale
features that are found to develop in the components of the evolved
metrics.
This behaviour appears to be a generic feature of gravitational
collapse.  

Extrapolation of the results for simulations of
collapsing cosmologies allows a picture to be constructed of the
spacetime behaviour at the `big bang' singularity.
In section~\use{sec.GowSing} such numerical evidence is seen to suggest
that the singularities in the Gowdy cosmological models are
asymptotically velocity-term dominated, but that spatial derivatives
in the Einstein equations do not become negligible as a singularity
is approached. 

Section~\use{sec.GowExpand} investigates the behaviour of
gravitational waves in expanding cosmologies.
Large amplitude waves close to the cosmological singularity are
found to lead to the formation of fine-scale spatial structure 
(in the sense that modes of high frequency are excited); however 
numerical evidence suggests that at late times the waves behave much
like linear perturbations of homogeneous spacetimes, regardless of
their initial amplitudes.
Differences between linear and nonlinear gravitational waves
are investigated by comparing numerical simulations to results from
linearized theory, and it is argued that some of the behaviour that
has been attributed to nonlinear gravitational waves in the literature
is better thought of as being a gauge effect.

\MiniSection{Fine-Scale Features in Cosmological Simulations}
For numerical simulations of collapsing inhomogeneous cosmological
models, it is found (sections~\use{sec.GowCollapse}
and~\use{sec.MonCollapse}) that features can develop in 
the evolved solutions on spatial scales which become increasingly small
as the cosmological singularity is approached.
These fine-scale features are very difficult to accurately
capture using traditional numerical approaches, and the adaptive mesh
refinement technique described in chapter~3 proves to be a very useful
tool in this regard: applied to the simulations of planar cosmologies in
chapter~6 it allows accurate results to be produced in very much
shorter times than would otherwise be required, 
and for the simulations of $U(1)$-symmetric cosmologies in
chapter~7 it allows much higher resolutions to be
attained than the available computer resources would otherwise permit.
(The question of whether fine-scale structure can also form in expanding
cosmologies is addressed in section~\use{sec.GowExpand}.)

Making use of some ideas on the identification of coordinate-independent
behaviour discussed in section~\use{sec.HomCompare}, it is argued that
the fine-scale features found in the collapsing cosmological
simulations of chapters~6 and~7 are physically relevant: they are
observed in scalar quantities such as the invariants of the curvature
tensor as well as in the metric components, and they are small with
respect to proper measures of distance.
The importance of this result is made clear by its contrast to the
behaviour found in the numerical simulations of
section~\use{sec.HomHarmonic}.  
There, homogeneous Kasner spacetimes are evolved using harmonic
time slicing, and for foliations in which coordinate singularities
appear the metric components develop fine-scale features which
are superficially similar to those found in the inhomogeneous models,
but which are clearly without any physical significance.


\Chapter{High-Resolution Methods}
The present work is concerned with the development and testing of
methods for generating approximate numerical solutions to the vacuum
field equations of general relativity.
In later chapters (in particular chapter~4) the field equations are
presented as a variety of first-order systems of hyperbolic partial
differential equations, numerical solutions to which can be obtained
by integrating in time sets of data specified on appropriate initial
hypersurfaces. 
The present chapter describes the basic numerical methods used here to
perform these integrations.
There exists a vast literature discussing the theoretical and practical
aspects of such numerical methods, and a formal introduction to the
subject is beyond the scope of the present discussion; rather, this
chapter is intended as a `user's guide' to the methods employed for
reference in later chapters.
Very much more extensive discussions can be found in
the books by LeVeque~(1992), Godlewski and Raviart~(1996),
Gustafsson, Kreiss and Oliger~(1995), Hirsch~(1988, 1990), Potter~(1973),
Press et al.~(1992, chapter~19), Sod~(1985), and Toro~(1997), among
many others.

The most commonly used methods for generating numerical solutions to
systems of partial differential equations are based on finite
differences, and these are discussed in
section~\use{sec.PdeDifference}.
The strengths and weaknesses of the standard finite difference methods
are well understood, and they are used in the present work as a
reference against which the performance of more sophisticated
numerical methods may be judged.
The methods discussed in this chapter are described in forms that are
applicable to problems in one and two spatial dimensions.

A significant limitation of standard finite difference methods is that
they usually only produce acceptable results when the evolved solution
is sufficiently smooth. 
This is important because in many applications the models being
studied are known to admit solutions in which discontinuities form
even for smooth initial data. 
Much research has been done in the field of computational fluid
dynamics into the development of advanced numerical methods which
are capable of accurately evolving solutions irrespective of their
smoothness. 
Finite volume methods achieve this by analysing a series of Riemann
problems for the evolution system, and section~\use{sec.PdeRiemann}
discusses how solutions to Riemann problems can be determined in some
basic cases. 
The numerical simulations carried out in the present work use a
particular finite volume method: the multi-dimensional high-resolution
wave-propagation algorithm of LeVeque~(1997).
The appeal of this method is that it has been demonstrated to work on
a range of different problems, and an implementation of it is freely
available in the form of the \Clawpack\ software (LeVeque~1998).
A description of the method is given in section~\use{sec.PdeVolume}.

The main motivation for the use of high-resolution methods
for work in numerical relativity is the need to correctly evolve the
shock waves that occur in solutions to the Einstein equations when
they are coupled to hydrodynamical sources.
Even when the evolved spacetimes are vacuum (as is the case in the
present work) it may be hoped that the use of a high-resolution method
will improve the accuracy of numerical results when steep gradients
are present in the gravitational fields (a situation which appears
to occur under a variety of circumstances).
Furthermore, the `causal' nature of such methods, which evolve
solutions based on the propagation properties of different
characteristic quantities, may simplify the imposition of physically
realistic boundary conditions, for example at the horizon of a black
hole.

\Section{Finite Difference Methods
         \label{sec.PdeDifference}}
This section summarizes some basic properties of standard one- and
two-dimensional finite difference methods for the numerical evolution
of solutions to first-order systems of hyperbolic partial differential
equations. 
The evolution systems are assumed to be in flux-conservative form, and
the finite difference methods considered here are numerically
conservative. 
The use of operator splitting to extend the applicability of the
methods to evolution systems with source terms is discussed.

\SubSection{Systems of Conservation Laws}
The systems of time-evolution equations considered in the
present work are assumed to be expressible in the standard first-order
form
$$
  \partial_t u(t,x,y) + \partial_x f(u) + \partial_y g(u) = s(u)
  \, , \EQN PdeConsLaws
$$
where~$u(t,x,y)$ is a vector of~$n$ unknown variables,
$f(u)$~and~$g(u)$ are {\it flux vectors\/} in the $x$-~and $y$-directions
respectively, $s(u)$~is a vector of {\it source terms\/}, and
$\partial_t$,~etc.\ represent partial derivatives with respect to the
independent variables.
For the purposes of illustration, the system~\EqNum{PdeConsLaws}{}
is shown as depending on two spatial dimensions (coordinates~$x$
and~$y$) although the main points of the following discussion apply
to systems in any number of dimensions.

The {\it Cauchy problem\/} for the system~\EqNum{PdeConsLaws}{}
comprises the determination of the value of~$u$ (at least within an
interval of the coordinate~$t$) given that its value is known at an
initial time~$t=t_0$.
It is assumed here that the evolution system is {\it hyperbolic\/} in
the sense that the $n \times n$~Jacobian matrices
$$
  A(u) = { \partial f(u) \over \partial u }
  \qquad \hbox{and} \qquad
  B(u) = { \partial g(u) \over \partial u }
  \EQN PdeJacobian
$$
have real eigenvalues and complete sets of eigenvectors, and
consequently that the Cauchy problem is well posed.
(A more complete discussion of hyperbolicity of systems of equations
is given in section~\use{sec.HypIntro}.)

In evolving solutions to systems of the form~\EqNum{PdeConsLaws}{}
numerically, it is usually possible to treat the influence of the
source terms separately from the influence of the {\it transport part\/}
(which is to say, the flux vectors~$f$ and~$g$).
A splitting approach for doing this is described later in this
section, and for much of this discussion it suffices to consider only
evolution systems for which no source terms are present:
$$
  \partial_t u(t,x,y) + \partial_x f(u) + \partial_y g(u) = 0
  \, . \EQN PdeConserve
$$

Systems of equations having the form of equation~\EqNum{PdeConserve}{}
are described as {\it flux conservative\/}, and they are
frequently encountered as models for physical systems.
The reason why systems of this form are so named is made clear if they
are re-expressed in the integral form
$$\eqalign{
  \int_{y_A}^{y_B} \!\! \int_{x_A}^{x_B} u(t_1,x,y) \, dx \, dy
  & {} = 
  \int_{y_A}^{y_B} \!\! \int_{x_A}^{x_B} u(t_0,x,y) \, dx \, dy
  \cr
  & {} - \left[
  \int_{t_0}^{t_1} \!\! \int_{y_A}^{y_B} f(u(t,x_B,y)) \, dy \, dt
  - 
  \int_{t_0}^{t_1} \!\! \int_{y_A}^{y_B} f(u(t,x_A,y)) \, dy \, dt
  \right] \cr
  & {} - \left[
  \int_{t_0}^{t_1} \!\! \int_{x_A}^{x_B} g(u(t,x,y_B)) \, dx \, dt
  - 
  \int_{t_0}^{t_1} \!\! \int_{x_A}^{x_B} g(u(t,x,y_A)) \, dx \, dt
  \right] \, . \cr
  } \EQN PdeIntegral
$$
If the variables~$u$ are thought of as representing densities of
unknown quantities, then equation~\EqNum{PdeIntegral}{} shows that the
total amount of a quantity in a spatial region
$[x_A,x_B]\times[y_A,y_B]$ changes between times $t=t_0$ and $t=t_1$
only because of flow through the boundaries of the region, as
represented by the fluxes~$f$ and~$g$; the variables~$u$ are 
conserved quantities.  

As a first step towards generating numerical solutions to the
evolution system~\EqNum{PdeConserve}{}, the temporal and spatial domain
of the problem is discretized by defining a mesh of points
$$
  \{ \, (t^n,x_i,y_j) = ( t^0 + n \, \Delta t, \,
                          x_0 + i \, \Delta x, \,
                          y_0 + j \, \Delta y ) 
                      : n,i,j \in \IntegSet 
  \, \} \, . \EQN PdeMesh
$$
The time step~$\Delta t$ and the mesh spacings~$\Delta x$ 
and~$\Delta y$ are assumed in the present chapter to be constant so
that the discrete mesh is uniform in both time and space. 
For many of the numerical simulations described in later chapters,
however, the size of the time step~$\Delta t$ is allowed to vary
during the course of the simulation, with the appropriate
modifications to the evolution procedures described here being
straightforward. 

There are two basic ways in which a numerical solution defined on the
mesh~\EqNum{PdeMesh}{} can be considered as approximating an exact
solution~$u(t,x,y)$.
Most typically a numerical solution $\hat{u}^{n}_{i,j}$ will
approximate the values of the exact solution at the 
mesh points:
$$
  \hat{u}^n_{i,j} \simeq u(t^n,x_i,y_j)
  \, . \EQN PdePoint
$$
Alternatively, in consideration of the integral
form~\EqNum{PdeIntegral}{} of the evolution system, a numerical
solution~$\bar{u}^n_{i,j}$ can approximate the average value of the
exact solution taken over cells centred on the mesh points:
$$
  \bar{u}^n_{i,j} \simeq {1 \over \Delta x \, \Delta y}
    \int_{ y_j - {1\over2} \Delta y }^{ y_j + {1\over2} \Delta y }
    \!\!
    \int_{ x_i - {1\over2} \Delta x }^{ x_i + {1\over2} \Delta x }
    u(t^n,x,y)
    \, dx \, dy
  \, . \EQN PdeCellAv
$$
In the present work this cell-based interpretation of a numerical
solution is the one usually employed, and the bar over the
symbol for the conserved variables is often omitted from the notation.

The hyperbolicity of the system~\EqNum{PdeConserve}{} means that
information propagates within it at finite speeds, and a value for the
numerical solution~$\bar{u}^{n+1}$ in a cell~$(x_i,y_j)$ can be
determined as part of a numerical integration scheme based on the
values~$\bar{u}^n$ within a region of cells, described as the {\it
stencil\/} of the method, surrounding the cell~$(x_i,y_j)$.
(Such a `time-marching' procedure for constructing a numerical
solution may more generally make use of known values of the solution at
earlier time levels, $\bar{u}^{n-1}$,~etc., but in the present discussion
only {\it two-level\/} integration methods are considered.)
Since meshes used in numerical simulations cannot be of infinite
extent, for cells at the edges of a mesh the integration stencil
will extend beyond the limits of the computational domain.
To avoid having to use different integration procedures in different
parts of the domain, the present work adopts an approach of enlarging
the mesh to include {\it boundary cells\/} (typically a border region,
one or two cells wide) in which the solution values are not 
evolved but instead are calculated at the start of each time step
based on some specified boundary conditions.

Variables in numerical solutions may be conserved in the same way as in
exact solutions if a discrete version of
equation~\EqNum{PdeIntegral}{} is assumed. 
A cell-based numerical evolution method is {\it conservative\/} if it
can be expressed in the form
$$\eqalign{
  \bar{u}^{n+1}_{i,j} = \bar{u}^n_{i,j} &
    {} - {\Delta t \over \Delta x}
         \left[ F(\bar{u}^n_{i,j},\bar{u}^n_{i+1,j})
              - F(\bar{u}^n_{i-1,j},\bar{u}^n_{i,j}) \right]
  \cr &
    {} - {\Delta t \over \Delta y}
         \left[ G(\bar{u}^n_{i,j},\bar{u}^n_{i,j+1})
              - G(\bar{u}^n_{i,j-1},\bar{u}^n_{i,j}) \right]
  \, , \cr
  } \EQN PdeConsMeth
$$
where~$F$ and~$G$ are called the {\it numerical flux functions\/}.
As an abbreviation of the notation, the numerical flux functions are
shown in equation~\EqNum{PdeConsMeth}{} as depending on the values of
the numerical solution in two grid cells; in fact the functions may be
based on the values of the numerical solution within a stencil of
cells of arbitrary shape and size.
In addition to their usefulness in conserving values of unknown
variables which may have physical meanings as densities of some kind,
conservative numerical methods are of particular importance when
evolving solutions in which discontinuities may develop in some of the
variables, as will be discussed in section~\use{sec.PdeRiemann}.

\SubSection{The Lax-Friedrichs and Lax-Wendroff Methods}
In this subsection one- and two-dimensional versions of two standard
numerical integration schemes are briefly described, together with
some of their basic properties.
These schemes are constructed by replacing the derivatives in the
evolution equations~\EqNum{PdeConserve}{} by 
finite difference approximations which are based on Taylor
expansions of the functions.
Both schemes are numerically conservative in the sense
defined in the previous subsection.

The {\it Lax-Friedrichs\/} scheme is the simplest practical method for
evolving numerical solutions to systems of the
form~\EqNum{PdeConserve}{}.
In one spatial dimension it is written as
$$
  u^{n+1}_i = {1\over2} (u^n_{i-1} + u^n_{i+1}) 
    - {\Delta t \over 2 \Delta x} 
      \left[ f(u^n_{i+1}) - f(u^n_{i-1}) \right]
  \, , \EQN PdeLFOne
$$
which generalizes in an obvious way to problems in two (or more)
spatial dimensions:
$$\eqalign{
  u^{n+1}_{i,j} = {} & {1\over4} (u^n_{i-1,j} + u^n_{i+1,j} 
                                + u^n_{i,j-1} + u^n_{i,j+1}) 
  \cr & {}
    - {\Delta t \over 2 \Delta x} 
      \left[ f(u^n_{i+1,j}) - f(u^n_{i-1,j}) \right]
    - {\Delta t \over 2 \Delta y} 
      \left[ g(u^n_{i,j+1}) - g(u^n_{i,j-1}) \right]
  \, . \cr
  } \EQN PdeLFTwo
$$

The {\it two-step Lax-Wendroff\/} scheme produces more accurate
results (in a sense made clear below) than the Lax-Friedrichs scheme
by in effect using Lax-Friedrichs steps to construct solution values
at intermediate points.
In one dimension it takes the form
$$
  u^{n+1}_i = 
    u^n_i - {\Delta t \over \Delta x} 
            \left[ f(u^{n+1/2}_{i+1/2}) - f(u^{n+1/2}_{i-1/2}) \right]
  \, , \EQN PdeLWOne
$$
where
$$
  u^{n+1/2}_{i+1/2} =
    {1\over2} (u^n_{i} + u^n_{i+1}) 
    - {\Delta t \over 2 \Delta x} 
      \left[ f(u^n_{i+1}) - f(u^n_{i}) \right]
  \, .
$$
There are various ways in which the two-step Lax-Wendroff
method can be extended to higher spatial dimensions, and a
version of the two-step Richtmyer scheme is used here: 
$$\eqalign{
  u^{n+1}_i = u^n_i & {}
    - {\Delta t \over 2 \Delta x} 
      \left( \left[ f^{n+1/2}_{i+1/2,j-1/2} 
                  - f^{n+1/2}_{i-1/2,j-1/2} \right] 
           + \left[ f^{n+1/2}_{i+1/2,j+1/2} 
                  - f^{n+1/2}_{i-1/2,j+1/2} \right] \right)
    \cr & {}
    - {\Delta t \over 2 \Delta y} 
      \left( \left[ g^{n+1/2}_{i-1/2,j+1/2} 
                  - g^{n+1/2}_{i-1/2,j-1/2} \right] 
           + \left[ g^{n+1/2}_{i+1/2,j+1/2} 
                  - g^{n+1/2}_{i+1/2,j-1/2} \right] \right)
  \, , \cr
  } \EQN PdeLWTwo
$$
where
$$\eqalign{
  u^{n+1/2}_{i+1/2,j+1/2} = {} &
    {1\over4} (u^n_{i,j} + u^n_{i+1,j} + u^n_{i,j+1} + u^n_{i+1,j+1})
    \cr 
    & {} - {\Delta t \over 4 \Delta x} 
         \left( \left[ f^n_{i+1,j} - f^n_{i,j} \right] 
              + \left[ f^n_{i+1,j+1} - f^n_{i,j+1} \right] \right)
    \cr
    & {} - {\Delta t \over 4 \Delta y} 
         \left( \left[ g^n_{i,j+1} - g^n_{i,j} \right] 
              + \left[ g^n_{i+1,j+1} - g^n_{i+1,j} \right] \right)
  \, , \cr }
$$
and, as a shorthand, $f^n_{i,j}$~is used to mean~$f(u^n_{i,j})$, and
so on.

Local stability of the above integration schemes requires conditions
to be imposed on the size of the time step~$\Delta t$ relative to the
mesh spacings~$\Delta x$ and~$\Delta y$.
The {\it Courant-Friedrichs-Lewy condition\/} is the requirement that
the domain of dependence of the integration method include the
domain of dependence of the system of evolution equations.
For integration schemes like the ones described here which have
stencils three cells wide, this condition translates as
$$
  \Delta t \le
    { \Delta x \over \max \vert \lambda_k \vert }
  \qquad \hbox{or} \qquad
  \Delta t \le \min 
    \left\{ { \Delta x \over \max \vert \lambda_k \vert } ,
            { \Delta y \over \max \vert \mu_k \vert } \right\}
  \, , \EQN PdeCFL
$$
depending on whether the scheme is in one or two spatial dimensions,
where~$\lambda_1$, $\lambda_2$,~\dots\ and~$\mu_1$, $\mu_2$,~\dots\ are
the eigenvalues of the Jacobian matrices defined
in equation~\EqNum{PdeJacobian}{}.
For the one-dimensional Lax-Friedrichs and Lax-Wendroff methods,
satisfaction of the Courant-Friedrichs-Lewy condition is sufficient
for the scheme to be stable.
Generally, for finite difference methods in two dimensions the
stability restriction on the time step is more severe, with~$\Delta t$
required to be smaller by a factor of~$\sqrt{2}$ than the value given
in equation~\EqNum{PdeCFL}{}; however, the two-dimensional
Lax-Wendroff scheme of equation~\EqNum{PdeLWTwo}{} is an exception to
this, being restricted only by the standard Courant-Friedrichs-Lewy
condition. 

The main difference between the Lax-Friedrichs and Lax-Wendroff
methods is that the former is accurate to first order whereas the
latter is accurate to second order.
Informally, the {\it order\/} of an integration scheme is the rate at which
a numerical solution converges to an exact solution as the discrete
mesh is refined; if the mesh spacings~$\Delta x$ and~$\Delta y$ are
kept in fixed proportions to the time step~$\Delta t$, and a numerical
solution is evolved from a time $t=t_0$ to a time $t=T$ for different
time step sizes, then in general a measure of the error in the
numerical solution at the final time will obey
$$
  \hbox{error} \sim O(\Delta t^{\,p})
  \qquad \hbox{as \ } \Delta t \rightarrow 0
  \, , \EQN PdeConverge
$$
where~$p$ is the order of the integration scheme.
(Mathematically rigorous definitions of the order of convergence can
be found in the text books listed in the introduction to this chapter.) 
In practice, the relationship~\EqNum{PdeConverge}{} is found to
apply even to quantities only indirectly related to the numerical
error, such as deviations in the Einstein constraint equations, and
verification that numerical results converge at a rate appropriate to
the integration method used is an important test of a numerical code.
It should be noted, however, that equation~\EqNum{PdeConverge}{} is
only valid for exact solutions which are everywhere smooth; the
accurate numerical evolution of non-smooth solutions is a topic
returned to later in this chapter.

In the present work, finite difference methods (in particular the
two-step Lax-Wendroff method) are used as a reference against which
more sophisticated methods (section~\use{sec.PdeVolume}) can be
compared. 
Since finite difference methods, unlike some more advanced methods, do
not make explicit use of the eigenstructure of the Jacobian matrices
of the evolved system, equation~\EqNum{PdeJacobian}{}, they can in
fact be applied to systems in which the hyperbolicity condition is
weakened---the Jacobian matrices need not have complete sets of
eigenvectors---and in chapter~7 the Lax-Wendroff method is used to
evolve solutions to such systems.

\figure{PdeWave}
\DrawFig{
  \epsfxsize=\hsize
  \epsfbox{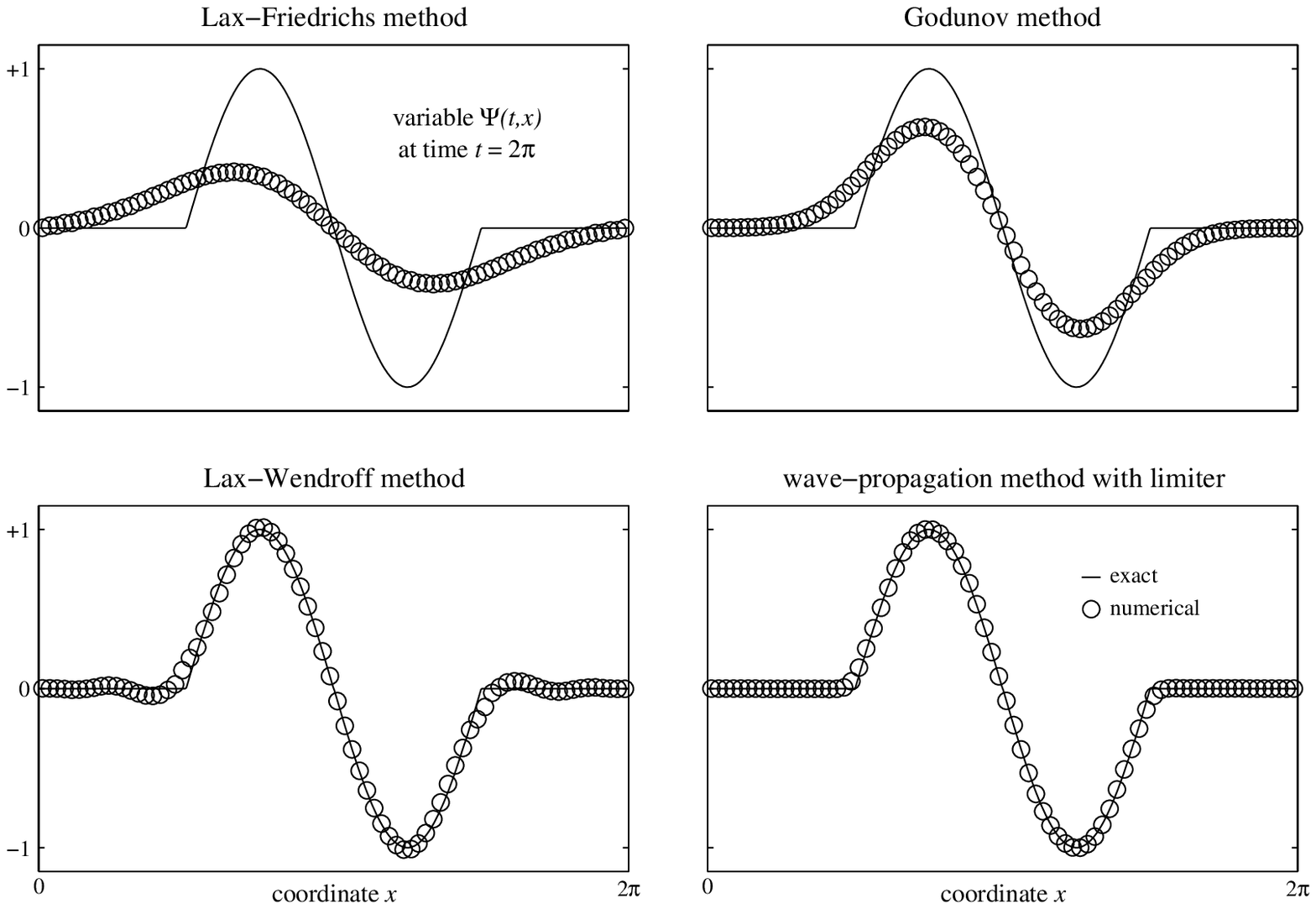}
}
\Caption
Comparison of results for four numerical integration schemes.
The one-dimensional wave equation~\EqNum{PdeWaveEq}{} with speed~$c=1$
is evolved numerically based on the initial
data~\EqNum{PdeWaveInit}{}, and exact and numerical solutions for the
variable~$\Psi$ are shown. 
A mesh of 80~cells is evolved from time $t=0$ to time $t=2\pi$ in 
160~steps. 
The finite difference methods used are discussed in
section~\use{sec.PdeDifference} while the finite volume methods are
discussed in section~\use{sec.PdeVolume}.
The results for the wave-propagation method are produced using the
`superbee' limiter. 
\bigskip
\endCaption
\endfigure

Figure~\use{Fg.PdeWave} shows example results produced using the
Lax-Friedrichs and Lax-Wendroff schemes.
As a test problem, the one-dimensional wave equation for the
variable~$\Psi(t,x)$ is used,
$$
  { \partial^2 \Psi \over \partial t^2 }
  = c^2 { \partial^2 \Psi \over \partial x^2 }
  \, , \EQN PdeWaveEq
$$
since this shares some basic properties with the full Einstein field
equations. 
(In fact equation~\EqNum{PdeWaveEq}{} can be seen to be very similar
in form to the evolution equations for the Gowdy cosmological models
considered in chapter~6.)
The wave speed~$c$ is taken here to be a constant.
The wave equation can be put into the first-order form of
equation~\EqNum{PdeConsLaws}{} by defining as new variables $\phi =
\partial_t \Psi$ and $\chi = \partial_x \Psi$ so that
$$
  \partial_t \!
  \left[ \matrix{ \Psi \cr \phi \cr \chi \cr } \right]
  - \partial_x \!
  \left[ \matrix{ 0 \cr c^2 \chi \cr \phi \cr } \right]
  =
  \left[ \matrix{ \phi \cr 0 \cr 0 \cr } \right]
  \, . \EQN PdeWaveSys
$$
(The source terms that appear in this evolution system are dealt with
using the approach described in the next subsection.)
The wave variable~$\Psi(t,x)$ is taken to be spatially periodic with
period~$2\pi$, and the following choice of initial data is made for
time~$t=0$: 
$$\eqalign{
  \Psi(0,x) = {} & 
    \cases{ -\sin 2x
          & if ${1\over2}\pi < x < {3\over2}\pi$ \cr
            0 & otherwise \cr} \cr
  \partial_t \Psi(0,x) = {} & 0 \, . \cr
  } \EQN PdeWaveInit
$$
This set of initial data is smooth except at two points, and there the
variable $\chi=\partial_x \Psi$ is discontinuous.
The exact solution to the wave equation~\EqNum{PdeWaveEq}{} is
straightforward to derive, and it can be shown that for the initial
data~\EqNum{PdeWaveInit}{} the solution is periodic in time with
period~$2\pi/c$. 

The numerical results presented in figure~\use{Fg.PdeWave} illustrate
some basic traits of the finite difference methods used.
The second-order convergence of the Lax-Wendroff method is responsible
for the overall accuracy of its results in comparison with the
first-order Lax-Friedrichs method.
At the points where the solution is not smooth, however, the
Lax-Wendroff method has introduced oscillations, and this behaviour is
typical: the evolution method is {\it dispersive\/}.
In contrast, the Lax-Friedrichs method has a tendency to smear out
solutions in regions where they are discontinuous or non-smooth: it
is a {\it diffusive\/} method. 
The oscillations introduced by the Lax-Wendroff method during the
evolution of discontinuous solutions are a major drawback to its use
in many applications, and, in response to this, much research has been
done into the development of numerical schemes which are second-order
accurate on smooth solutions, but at the same time give good
resolution of discontinuities.
Section~\use{sec.PdeVolume} discusses one such scheme.

\SubSection{Operator Splitting and Source Terms}
Supposing that a numerical method is known by which a solution can be
evolved to the system of equations~\EqNum{PdeConserve}{}, this
subsection discusses a splitting approach that can be used to generate
solutions to similar systems in which source terms are included,
equation~\EqNum{PdeConsLaws}{}. 

Denote by~${\cal T}_{(\Delta t)}$ the operator for evolving a numerical
solution~$u$ to the source-less system~\EqNum{PdeConserve}{} through a
time step~$\Delta t$ according to any appropriate integration scheme.
A similar operator~${\cal S}_{(\Delta t)}$ can be used to represent
the advance by one time step of a numerical solution to the system of
ordinary differential equations created by the removal of transport
terms from the system~\EqNum{PdeConsLaws}{},
$$
  \partial_t u = s(u)
  \, , \EQN PdeSource
$$
and appropriate numerical methods for evolving such a solution are
discussed below. 
A numerical solution to the full system~\EqNum{PdeConsLaws}{} can then
be advanced one step in time by using a combination of update
operations for the component parts of the system:
$$
  u^{n+1} = {\cal S}_{(\Delta t/2)}
            {\cal T}_{(\Delta t)}
            {\cal S}_{(\Delta t/2)} u^n
  \, . \EQN PdeSplit
$$
This approach is called {\it Strang splitting\/} (Strang~1968, Chapter
19 of Press et al.~1992, LeVeque~1998), and is in general second-order
accurate provided that the two numerical methods represented by
the~${\cal S}$ and~${\cal T}$ operators are individually second-order
accurate. 

Numerical methods for solving systems of the form~\EqNum{PdeSource}{}
are well established, and a particularly good discussion of their
practical application can be found in chapter~16 of Press et al.~1992.  
In the present work the source terms in the
system~\EqNum{PdeConsLaws}{} are evolved by using steps of the
{\it second-order Runge-Kutta\/} method: the 
operator~${\cal S}_{(\Delta t)}$ advances a numerical solution~$u$
through 
$$
  u^{n+1}_{i,j} = u^n_{i,j} + \Delta t \, s(u^{n+1/2}_{i,j})
 \, , \EQN PdeRungeKutta
$$
where an intermediate state is defined by
$$
  u^{n+1/2}_{i,j} = u^n_{i,j} + {\Delta t \over 2} s(u^n_{i,j})
 \, ,
$$
and this procedure is applied to each computational cell individually.

Although the second-order Runge-Kutta method is unsophisticated in
comparison to many other approaches that have been developed for the
solution of systems of ordinary differential equations, it has been
found to perform well for the problems it has been applied to in the
present work.
In particular, only very slight gains in overall accuracy have been
observed when solutions to the system~\EqNum{PdeConsLaws}{}
have been evolved using $N$~second-order Runge-Kutta steps of
size~$\Delta t/N$ to advance the source terms at each stage rather
than a single step of size~$\Delta t$.

\Section{The Riemann Problem
         \label{sec.PdeRiemann}}
In many problems, particularly when hydrodynamical effects are
being modelled, discontinuities are a generic feature of solutions to
the governing equations.
As discussed in the previous section, standard finite difference
methods are typically not well suited to the task of numerically
evolving discontinuous solutions, and in section~\use{sec.PdeVolume} 
a multi-dimensional high-resolution scheme is described which
accurately resolves discontinuities while at the same time giving
second-order accuracy in smooth regions of a solution.
This evolution method operates based on information from the solution
to the Riemann problem for the equations being evolved, and
the nature of this problem and its solution in some simple cases are
the topics of the present section.
The scope of this discussion is limited to concepts that are referred
to later in this work, and more complete accounts can be found in, for
example, LeVeque~1992, and Godlewski and Raviart~1996.

\SubSection{Weak Solutions and Discontinuities}
It is a well-known property of nonlinear hyperbolic evolution systems
that an initially smooth solution may become discontinuous after a
finite amount of time, and such behaviour may often be physically
reasonable, with shock waves in fluids being an obvious example. 
While solutions to systems of partial differential equations
(specifically equation~\EqNum{PdeConserve}{}) are only defined
classically when they are smooth, it is possible to follow the
behaviour of discontinuities using the conservative
properties of a system:
a {\it weak solution\/} to the evolution system~\EqNum{PdeConserve}{} 
is a function~$u(t,x,y)$ which satisfies the integral
form~\EqNum{PdeIntegral}{} of the system in all regions
$[t_0,t_1]\times[x_A,x_B]\times[y_A,y_B]$. 
A consistent numerical method will always produce results which
converge (if they converge at all) to weak solutions if the method is
in the conservative form~\EqNum{PdeConsMeth}{} introduced in the
previous section. 

The {\it Riemann problem\/} for a system of the
form~\EqNum{PdeConserve}{} demonstrates the behaviour of discontinuous
solutions by evolving a set of initial data comprising two constant
states~$u_L$ and~$u_R$ separated by a discontinuity:
$$
  u(0,x) = \cases{ u_L & if $x<0 \, ,$ \cr
                   u_R & if $x\ge0 \, ,$ \cr}
  \qquad \hbox{at time } t = 0 
  \, , \EQN PdeRiemInit
$$
where only one spatial dimension is relevant here.
It can be shown that the solution to the Riemann problem
is always of the form 
$$
  u(t,x) = w(x/t)
  \qquad \hbox{for \ } t > 0
  \, , \EQN PdeRiemSimilar
$$
where the function~$w(\xi)$ comprises a number of regions of constant
state separated by {\it waves\/} spreading out from the
initial discontinuity:
$$
  w(\xi) = \cases{ 
    u_L & if \hskip11.5ex 
          $\xi < s_1-\epsilon_1 \, ,$ \cr
    u_\alpha & if \hskip1.7ex 
          $s_1+\epsilon_1 \le \xi < s_2-\epsilon_2 \, ,$ \cr
    u_\beta & if \hskip1.7ex 
          $s_2+\epsilon_2 \le \xi < s_3-\epsilon_3 \, ,$ \cr
    \ldots &\cr
    u_R & if \hskip0.5ex $s_m+\epsilon_m \le \xi \, .$ \cr }
$$
These waves are of two basic types:
{\it shock waves\/} are simple discontinuities between neighbouring
constant states such that $\epsilon_k = 0$, whereas {\it rarefaction
waves\/} are extended regions (having $\epsilon_k>0$) across
which~$w(\xi)$ is continuous and inside of which it is smooth.
For the evolution systems considered in the present work only shock
waves (more accurately, {\it contact discontinuities\/}) are present
in solutions to the Riemann problem, and this 
situation is illustrated in figure~\use{Fg.PdeShocks}.
Rarefaction waves are not considered further in this work.

\figure{PdeShocks}
\DrawFig{
  \epsfxsize=11cm
  \epsfbox{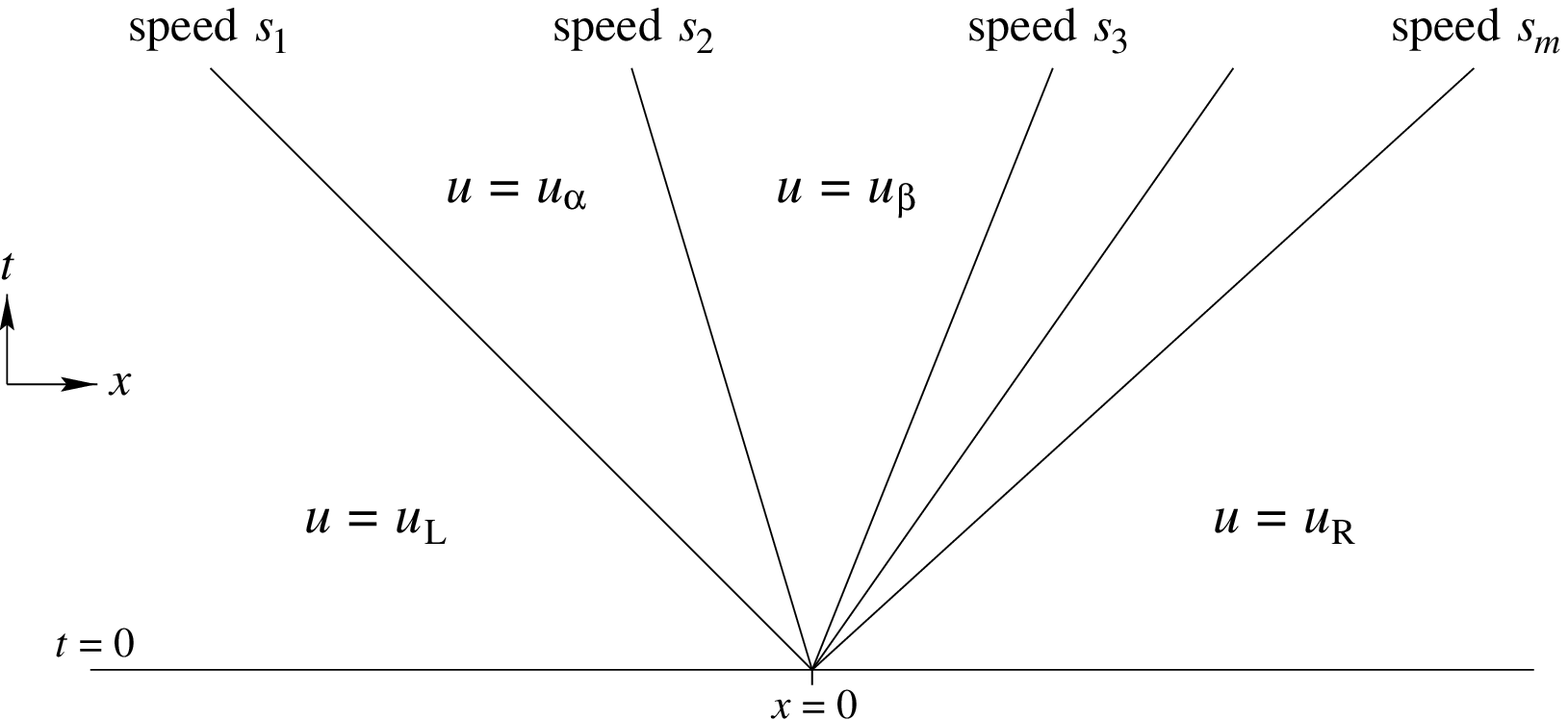}
}
\Caption
Solution to the Riemann problem containing only shock waves.
A number of discontinuities propagating with constant speeds~$s_1$,
$s_2$,~\dots,~$s_m$ separate regions where~$u$ is of constant value.
\bigskip
\endCaption
\endfigure

A solution to the Riemann problem can thus be described by a list
of~$m$ shock waves, each of which has a speed~$s_k$ (as in
figure~\use{Fg.PdeShocks}) and a jump value~${\cal W}_k$, where the
jumps are the differences between neighbouring constant values of the
state vector~$u$:
$$
  {\cal W}_1 = u_\alpha - u_L
  \, , \qquad
  {\cal W}_2 = u_\beta - u_\alpha
  \, , \qquad
  \hbox{and so on,}
$$
so that
$$
  u_R - u_L = \sum_{k=1}^m {\cal W}_k
  \, .
$$
The waves are ordered such that $s_1<s_2<\ldots<s_m$, and the
number~$m$ of waves is not greater than the number~$n$ of unknown
variables in the system. 
It can be shown from the integral form~\EqNum{PdeIntegral}{} of the
evolution system that a shock wave with speed~$s$ separating two
constant states~$u_A$ and~$u_B$ must satisfy
$$
  f(u_B) - f(u_A) = s \, (u_B - u_A)
  \, , \EQN PdeRHJump
$$
and this is called the {\it Rankine-Hugoniot jump condition\/}.
Provided~$u_L$ is sufficiently close to~$u_R$, a unique set of shock
waves $(s_1,{\cal W}_1)$,~\dots,~$(s_m,{\cal W}_m)$ for the solution
to the Riemann problem~\EqNum{PdeRiemInit}{} can always be
determined by requiring that the condition~\EqNum{PdeRHJump}{} be
satisfied across each wave:
$$
  f \left( u_L + \sum_{l=1}^k {\cal W}_l \right)
  - f \left( u_R - \sum_{l=k}^m {\cal W}_l \right)
  = s_k {\cal W}_k
  \, , 
$$
for $k=1,\ldots,m$.

\SubSection{Linear Systems and Approximate Riemann Solvers}
For a general nonlinear system~\EqNum{PdeConserve}{}, the Riemann
problem~\EqNum{PdeRiemInit}{} is typically very difficult to solve.
However for linear systems the solution is straightforward to derive,
and for many applications an approximate solution to the Riemann
problem produced by consideration of a related linear system is
an acceptable substitute for the exact nonlinear solution.

The system~\EqNum{PdeConserve}{} is {\it linear (with constant
coefficients)\/} if the components of the flux vectors are linear
combinations of the variables~$u$, and in this case the system can be
rewritten as 
$$
  \partial_t u + A \, \partial_x u = 0
  \, , \EQN PdeLinear
$$
where only one spatial dimension is relevant here, and the Jacobian
matrix~$A$ has constant coefficients.
(If the matrix~$A$ has an explicit dependence on the independent
variables~$t$ and~$x$, then the system is described as {\it linear
with variable coefficients}; such systems are considered further below.)
It follows from the assumption that the constant-coefficient system is
hyperbolic that the eigenvalues~$\lambda_1$,~\dots,~$\lambda_m$ of~$A$
are all real, and that the matrix has a complete set of eigenvectors
such that any vector~$v$ can be decomposed as
$$
  v = \sum_{k=1}^m r_k
  \qquad \hbox{where \ } A \, r_k = \lambda_k r_k
  \, . \EQN PdeEigenvect
$$
(The number~$m$ of distinct eigenvalues of~$A$ is not necessarily the
same as the number~$n$ of variables in the evolution system.)
For a linear system the Rankine-Hugoniot jump
condition~\EqNum{PdeRHJump}{} reduces to a simple eigenvalue equation
$$
  A \, (u_B - u_A) = s \, (u_B - u_A)
  \, ,
$$
and the solution to the Riemann problem~\EqNum{PdeRiemInit}{} follows
directly from this.
Assuming the eigenvalues of~$A$ to be ordered in magnitude
$\lambda_1 < \lambda_2 < \ldots < \lambda_m$, then the solution
comprises~$m$ shock waves with speeds
$$
  s_k=\lambda_k \qquad \hbox{for \ } k=1,\ldots,m
  \, , \EQN PdeLinSpeeds
$$
and jumps~${\cal W}_1$,~\dots,~${\cal W}_m$ between constant states
which are found by an eigenvector decomposition analogous to
equation~\EqNum{PdeEigenvect}{}: 
$$
  \Delta u = \sum_{k=1}^m {\cal W}_k
  \qquad \hbox{where \ } A \, {\cal W}_k = \lambda_k {\cal W}_k
  \, , \EQN PdeLinJumps
$$
for~$\Delta u = u_R - u_L$.

As a specific example, the solution to the Riemann problem for the
transport part of the first-order form~\EqNum{PdeWaveSys}{} of the
wave equation is presented here.
The system is linear, and has the Jacobian
matrix 
$$
  A = \left[ \matrix{ \,0 & 0  & 0 \cr
                      \,0 & 0  & -c^2 \cr
                      \,0 & -1 & 0 \cr } \right]
  \, ,
$$
where~$c$ is a constant.
Writing the difference in initial states as 
$\Delta u = [ \Delta\Psi, \Delta\phi, \Delta\chi ]^T = u_R-u_L$, the
solution to the Riemann problem comprises one stationary shock wave,
$$
  s_0 = 0 
  \, , \qquad
  {\cal W}_0 = \left[ \matrix{ \Delta\Psi \cr 0 \cr 0 \cr } \right]
  \, ,
$$
and two shock waves propagating in opposite directions at the 
wave speed~$c$, 
$$
  s_{\pm} = \pm c
  \, , \qquad
  {\cal W}_{\pm} = {1\over2} 
      \left[ \matrix{ 0 \cr 
                      \Delta\phi \mp c \, \Delta\chi \cr 
                      \Delta\chi \mp c^{-1} \Delta\phi \cr } \right]
  \, .
$$
(This situation is quite similar to the one illustrated in
figure~\use{Fg.HypRiemann} for the transport part of the full Einstein
equations.)  

Analysis of the Riemann problem can be difficult even for
linear systems if the flux vectors in equation~\EqNum{PdeConserve}{}
have an explicit dependence on the space or time coordinates.
The simplest approach to dealing with this problem is to consider only 
the local behaviour of the system by solving the Riemann problem for a
fixed value of the coordinates, and in section~\use{sec.GowCollapse}
this approach is applied to a linear system with variable coefficients
which is effectively the
wave equation~\EqNum{PdeWaveEq}{} with a time-dependent speed~$c$.
In that case, the solution to the Riemann problem posed at a
time~$t=t_0$ is needed at a time~$t = t_0 + \Delta t$, and the pattern
of shock waves produced by solving the problem for a fixed-speed
system with~$c$ evaluated at the time $t=t_0 + \Delta t/2$ is found to
give second-order accuracy (as the time step~$\Delta t$ is refined)
when used in place of the exact variable-speed solution.
Section~\use{sec.HypNumerical} discusses a more complicated example in
which second-order accuracy is attained only through a more elaborate
treatment of the coordinate-dependent flux terms.

Approximate solutions to the Riemann problem are typically just as
effective as exact solutions when they are used as part of numerical
integration schemes, and in many cases the former may actually be
preferable, either because an approximate solution can be calculated
very efficiently or because no simple form for the exact solution is
known. 
Roe~(1981) describes an approach for approximating the solution of
the Riemann problem for a nonlinear system,
$$
  \partial_t u + \partial_x f(u) = 0
  \, ,
$$
by the exact solution for a related linear system, 
$$
  \partial_t u + \tilde{A} \, \partial_x u = 0
  \, ,
$$
where the matrix~$\tilde{A}$ has constant coefficients which depend on
the initial states~$u_L$ and~$u_R$ of the Riemann problem, and
the linear system is hyperbolic in the sense defined in
section~\use{sec.PdeDifference}. 
Roe suggests two conditions that should be satisfied by
such a matrix~$\tilde{A}$.
Firstly, as~$u_L$ and~$u_R$ tend to a fixed value~$u_0$, the
matrix~$\tilde{A}(u_L,u_R)$ should tend smoothly to the
value~$A(u_0)$, where~$A(u)$ is the Jacobian matrix defined in
equation~\EqNum{PdeJacobian}{}.
A simple way to satisfy this condition is to choose the
matrix~$\tilde{A}$ as equalling~$A(u_{\ast})$ for some
state~$u_{\ast}$ representing an average of~$u_L$ and~$u_R$.
Roe's second condition is that the conservation property
$$
  \tilde{A} \, ( u_R - u_L ) = f(u_R) - f(u_L)
  \EQN PdeRoe
$$
should hold.
A choice of~$\tilde{A}$ which satisfies this condition may be
difficult to find in general; however it can be shown (Roe~1981) that
simply taking~$\tilde{A}=A(u_{\ast})$ 
where~$u_{\ast}={1\over2}(u_L + u_R)$ will work for evolution 
systems in which the components of the flux vector~$f(u)$ are no more
than quadratic in the components of~$u$.

Roe's approach to the construction of approximate Riemann solutions is
used in section~\use{sec.HypNumerical}.
There the Riemann problem for the full Einstein equations (written in
a hyperbolic form) is considered, and exact and approximate solutions
are derived for use with the numerical integration schemes described in
the next section.

\Section{Finite Volume Methods
         \label{sec.PdeVolume}}
This section describes a multi-dimensional high-resolution method for the
numerical integration of hyperbolic systems of partial differential
equations which have the flux-conservative form described in
section~\use{sec.PdeDifference}.
An integration method is described as being {\it high resolution\/} if
it gives second-order (or better) accuracy when evolving solutions
which are smooth, and in addition resolves discontinuities in solutions
without producing oscillations.

The high-resolution method described here is the wave-propagation
algorithm developed by LeVeque~(1997).
This method can be used for problems in one and two spatial dimensions
(for its extension to three dimensions see Langseth and LeVeque~1999),
and operates by considering sequences of Riemann problems
(section~\use{sec.PdeRiemann}) for the evolution equations under
consideration. 
An implementation of the wave-propagation algorithm in Fortran~77,
called \Clawpack\ (Conservation LAWs PACKage), is freely available
(LeVeque~1998).
For the present work, in order to gain familiarity with
the method, the \Clawpack\ routines have been re-implemented in the
\Cplusplus\ language, with the resulting software being named
\Clawplusplus.
(The adaptive mesh refinement code described in chapter~3 is written
in \Cplusplus, and while it would be possible in principle to link
integration routines written in a different language to that code,
there are many advantages in practice to both pieces of software being
implemented in the same programming language.)
A number of example problems related to fluid dynamics are included
in the \Clawpack\ software, and these have also been converted to
\Cplusplus\ and have been used to verify that the \Clawplusplus\
software performs correctly.

The wave-propagation algorithm is a second-order extension of
Godunov's well-known integration method, and this is described here as a
prelude to discussing the high-resolution scheme.
The account given here is concerned only with aspects of
the wave-propagation algorithm which are directly relevant to its use in
numerical simulations; full details about the method and its
theoretical background can be found in the references listed above
together with LeVeque~1992, 1996.

The \Clawplusplus\ software (which incorporates routines for evolving
solutions according to the Lax-Friedrichs and Lax-Wendroff methods of
section~\use{sec.PdeDifference}) is used as the basis for
almost all of the numerical simulations described in this work.
In particular, section~\use{sec.HypNumerical} discusses how the
wave-propagation algorithm can be used to evolve solutions to the
vacuum Einstein equations when the latter are written in a
flux-conservative hyperbolic form.

\SubSection{Godunov's Method}
The starting point for the construction of the high-resolution
wave-propagation algorithm is Godunov's method, and this is briefly
described in the present subsection.
Godunov's method is applicable to evolution systems having the
flux-conservative form of equation~\EqNum{PdeConserve}{}, and it can
also be used to evolve solutions to more general systems having the
form of equation~\EqNum{PdeConsLaws}{} by employing the operator
splitting approach described at the end of
section~\use{sec.PdeDifference} to treat the source terms.
Where Godunov's method differs from the finite difference methods
described in section~\use{sec.PdeDifference} is in its use of
Riemann problems (see section~\use{sec.PdeRiemann}) to
advance numerical solutions in time.
As might be expected, Godunov's method gives better results for
discontinuous solutions than standard finite difference methods.

Godunov's method in the one-dimensional case is considered first.
Given a numerical solution~$u^n_i$ at a time~$t=t^n$, Godunov's
method for advancing the solution to a later time~$t=t^{n+1}$
considers a piecewise constant function~$U^n(x)$ defined according to
$$
  U^n(x) = u^n_i 
  \textstyle
  \qquad \hbox{for\ \ } x \in [ x_i - {1\over2} \Delta x , \,
                                x_i + {1\over2} \Delta x )
  \, . \EQN PdePiecewise
$$
If the numerical solution~$u^n$ is a cell-based
approximation to the exact solution in the sense of
equation~\EqNum{PdeCellAv}{}, then the function~$U^n(x)$ extends 
that discrete data to the real line in a way which preserves
the original cell averages.
If the function~$U^n(x)$ is taken as initial data for the
evolution system~\EqNum{PdeConserve}{}, then the exact
solution~$U(t,x)$ can be determined at least within a short time
interval by piecing together the solutions to a sequence of Riemann
problems centred at the cell interfaces.
Values~$u^{n+1}_i$ for the numerical solution at the next time
level are then determined by
taking cell averages of the solution~$U(t^{n+1},x)$:
$$
  u^{n+1}_i = {1 \over \Delta x}
    \int_{ x_i - {1\over2} \Delta x }^{ x_i + {1\over2} \Delta x }
    U(t^{n+1},x)
    \, dx 
  \, . \EQN PdeGodInterm
$$
In fact, it follows from the integral form~\EqNum{PdeIntegral}{}
of the evolution system that the function~$U(t,x)$ need not be
calculated in full in order to find~$u^{n+1}$.
Equation~\EqNum{PdeGodInterm}{} can be rewritten as
$$
  u^{n+1}_i = u^n_i - {\Delta t \over \Delta x} 
    \left[ f( U^{\ast}_{i+1/2} ) - f( U^{\ast}_{i-1/2} ) \right]
  \, , \EQN PdeGodunovOne
$$
where $U^{\ast}_{i+1/2}$ is the value of~$U(t,x)$ evaluated
at the cell interface~$x=x_{i+1/2}$, and this value is
known (from the general form~\EqNum{PdeRiemSimilar}{} of the solution
to the Riemann problem) to be independent of the time~$t$.

Equation~\EqNum{PdeGodunovOne}{} is Godunov's method in one spatial
dimension.
It is explicitly conservative (as can be seen by comparison with
equation~\EqNum{PdeConsMeth}{}) and it is stable provided that the
Courant-Friedrichs-Lewy condition~\EqNum{PdeCFL}{}
is satisfied. 
Godunov's method is first-order accurate, and it can be seen from
figure~\use{Fg.PdeWave} that numerical results produced using the
method are not typically as good as results produced using a
higher-order method such as the Lax-Wendroff scheme.
However, no oscillations are produced by Godunov's method at points
where the evolved solution is not smooth, and there is less smearing
of the data than in results produced using the first-order
Lax-Friedrichs finite difference scheme.

A slightly different form of equation~\EqNum{PdeGodunovOne}{} is
useful when considering the extension of Godunov's method to a
second-order scheme. 
The value of~$U^{\ast}_{i+1/2}$ is determined by solving a Riemann
problem with initial states~$u^n_{i}$ and~$u^n_{i+1}$, and from
the discussion of section~\use{sec.PdeRiemann} it can be seen that
$$
  U^{\ast}_{i+1/2}  = u_i + \sum_{k=1}^{p} {\cal W}_k 
                    = u_{i+1} - \! \sum_{k=p+1}^{m} \! {\cal W}_k 
  \, ,
$$
where~${\cal W}_1$,~\dots,~${\cal W}_m$ are the shock jumps into which
the vector $\Delta u = u^n_{i+1} - u^n_i$ is decomposed, and the
value~$p$ is such that the wave speeds satisfy
$$
  s_1 < \ldots < s_p \le 0 < s_{p+1} < \ldots < s_m 
  \, .
$$
It is then convenient to define from the solution to this Riemann
problem a {\it left-going fluctuation\/}, denoted
${\cal A}^{-} \!\Delta u$, and a {\it right-going fluctuation\/}, 
denoted ${\cal A}^{+} \!\Delta u$, according to
$$\eqalign{
  {\cal A}^{-} \!\Delta u & {} = 
    \sum_{k=1}^m \min(s_k,0) {\cal W}_k 
    \, , \cr
  {\cal A}^{+} \!\Delta u & {} = 
    \sum_{k=1}^m \max(s_k,0) {\cal W}_k 
    \, . \cr
  } \EQN PdeFluxDiff
$$
From the Rankine-Hugoniot jump condition~\EqNum{PdeRHJump}{} these
fluctuations can be seen to satisfy 
$$
  {\cal A}^{-} \!\Delta u +
  {\cal A}^{+} \!\Delta u = f(u_{i+1}) - f(u_i)
  \, , \EQN PdeFluxSplit
$$
and with respect to them Godunov's method~\EqNum{PdeGodunovOne}{} can
be written in the wave-propagation form
$$
  u^{n+1}_i = u^n_i - {\Delta t \over \Delta x} 
    \left[ {\cal A}^{+} \!\Delta u_{i-1/2} 
         + {\cal A}^{-} \!\Delta u_{i+1/2} \right]
  \, . \EQN PdeGodFluxOne
$$

The evaluation of the fluctuations~${\cal A}^{-} \!\Delta u$
and~${\cal A}^{+} \!\Delta u$ can be done in a particularly neat way if
the evolution system is linear or if a Roe solver is used.
The constant-coefficient matrix~$A$ which represents the flux of the
system can then be diagonalized as
$$
  A = R \, \hbox{diag}[\lambda_1,\ldots,\lambda_n] \, R^{-1}
  \, ,
$$
where the non-singular matrix~$R$ has columns which are eigenvectors
of~$A$, and~$\lambda_1$,~\dots,~$\lambda_n$ are the corresponding 
eigenvalues (which are not necessarily distinct).
If two new matrices are defined as
$$\eqalign{
  A^- & {} = R \, \hbox{diag}
             [ \, \min(\lambda_1,0),\,\ldots,\,\min(\lambda_n,0) \, ]
             \, R^{-1}
  \, , \cr
  A^+ & {} = R \, \hbox{diag}
             [ \, \max(\lambda_1,0),\,\ldots,\,\max(\lambda_n,0) \, ]
             \, R^{-1}
  \, , \cr
  } \EQN PdeLinMatrices
$$
then $A = A^+ + A^-$, and the fluctuations for the Riemann problem
can be calculated by performing matrix multiplications on the difference
vector~$\Delta u$: 
$$
  {\cal A}^{-} \!\Delta u = A^{-} \Delta u
  \qquad \hbox{and} \qquad
  {\cal A}^{+} \!\Delta u = A^{+} \Delta u
  \, . \EQN PdeLinearFlux
$$
(Fluctuations and the wave-propagation form of Godunov's method
are discussed in much greater detail in the references listed in the
introduction to this section.)

In two dimensions, Godunov's method (in wave-propagation form) is a
straightforward generalization of equation~\EqNum{PdeGodFluxOne}{}:
$$\eqalign{
  u^{n+1}_{i,j} = u^n_{i,j} & {}
    - {\Delta t \over \Delta x} 
    \left[ {\cal A}^{+} \!\Delta u_{i-1/2,j} 
         + {\cal A}^{-} \!\Delta u_{i+1/2,j} \right] \cr
    & {}
    - {\Delta t \over \Delta y} 
    \left[ {\cal B}^{+} \!\Delta u_{i,j-1/2} 
         + {\cal B}^{-} \!\Delta u_{i,j+1/2} \right] \, ,\cr
  } \EQN PdeGodFluxTwo
$$
where the fluctuations~${\cal B}^+ \!\Delta u$ 
and~${\cal B}^- \!\Delta u$ are analogous to~${\cal A}^+ \!\Delta u$ 
and~${\cal A}^- \!\Delta u$, and are defined by solving Riemann problems
in the $y$-direction rather than the $x$-direction.
The method~\EqNum{PdeGodFluxTwo}{} is first-order accurate, and is
stable only for time steps~$\Delta t$ which are smaller by a
factor~$1/2$ at least than the maximum value permitted by the 
Courant-Friedrichs-Lewy condition~\EqNum{PdeCFL}{}.
(A better stability bound than this is one of the features of the
two-dimensional wave-propagation method described below.)

\SubSection{Wave Propagation in One Dimension}
A high-resolution numerical algorithm is typically produced by
combining two different integration schemes: a first-order scheme
which produces good results for discontinuous solutions, and a
higher-order scheme which is accurate for smooth solutions.
The hybrid scheme then varies its order with position
depending on whether the evolved solution is judged as being
discontinuous there or not.
(There exists an extensive literature on the theoretical aspects of
high-resolution methods; for an overview relevant to the present
discussion see chapter~16 of LeVeque~1992.) 

The high-resolution wave-propagation method of LeVeque~1997 is based
on Godunov's first-order scheme with correction terms being added to
increase the accuracy of the method to second order when the solution
is smooth.
In one dimension the method can be written as
$$
  u^{n+1}_i = u^n_i 
  - {\Delta t \over \Delta x} 
    \left[ {\cal A}^{+} \!\Delta u_{i-1/2} 
         + {\cal A}^{-} \!\Delta u_{i+1/2} \right]
  - {\Delta t \over \Delta x} 
    \left[ \tilde{F}_{i+1/2} - \tilde{F}_{i-1/2} \right]
  \, , \EQN PdeWavePropOne
$$
where~$\tilde{F}_{i-1/2}$ and~$\tilde{F}_{i+1/2}$ are
`high-resolution' correction terms to equation~\EqNum{PdeGodFluxOne}{}.
Whereas Godunov's method only makes use of the 
fluctuations~${\cal A}^{\pm} \!\Delta u$ calculated at each cell
interface, the correction terms in the wave-propagation algorithm
incorporate information about each of the waves in the solution to the
Riemann problem:
$$
  \tilde{F}_{i+1/2} = {1\over2} \sum_{k=1}^m
    \vert s_k \vert 
    \left( 1 - {\Delta t \over \Delta x} \vert s_k \vert \right)
    \tilde{\cal W}_k
  \, , \EQN PdeFluxCorrect
$$
where the wave speeds $s_1$,~\dots,~$s_m$ and the (limited) wave
jumps $\tilde{\cal W}_1$,~\dots,~$\tilde{\cal W}_m$ are computed for
the Riemann problem between the states~$u^n_{i}$ and~$u^n_{i+1}$.

If the unmodified wave jumps~${\cal W}_1$,~\dots,~${\cal W}_m$ are
used in equation~\EqNum{PdeFluxCorrect}{} by setting
$\tilde{\cal W}_k = {\cal W}_k$, then the
method~\EqNum{PdeWavePropOne}{} is second-order accurate.
It is not, however, high resolution since it still produces
spurious oscillations at discontinuities in evolved solutions.

To enable the wave-propagation method to accurately evolve
discontinuous solutions, a {\it limiter function\/},~$\phi(\theta)$,
is used to define modified wave jumps 
$\tilde{\cal W}_1$,~\dots,~$\tilde{\cal W}_m$ which are equal to the
original jumps only when the solution is sufficiently smooth. 
The limiter function is applied independently to each family
of waves that appears in the solution to the Riemann problem.
The limited version~$\tilde{\cal W}_k$ of the wave jump~${\cal W}_k$
at the cell interface~$x=x_{i+1/2}$ depends on the speed~$s_k$ of that
wave together with the jumps labelled by the same index~$k$ for the cell
interfaces to the left and right, denoted here as~${\cal W}^L_k$
and~${\cal W}^R_k$. 
A value~$\theta_k$ is defined as a measure of the smoothness of the
solution (with respect to the $k$-th family of waves) according to
$$
  \theta_k = \cases{ ( {\cal W}^L_k \cdot {\cal W}_k )
                   / ( {\cal W}_k \cdot {\cal W}_k ) &
                   if $s_k > 0 \, ,$ \cr
                     ( {\cal W}^R_k \cdot {\cal W}_k )
                   / ( {\cal W}_k \cdot {\cal W}_k ) &
                   if $s_k \le 0 \, ,$ \cr}
$$
where the dot represents inner product.
A value of~$\theta_k$ close to unity is taken as an indication that
the solution is smooth.
Modified wave jumps are then defined by rescaling the original jumps
by a factor determined through a limiter function~$\phi(\theta)$:
$$
  \tilde{\cal W}_k = \phi(\theta_k) {\cal W}_k
  \, . \EQN PdeLimiter
$$
A range of different limiter functions have been proposed in the
literature, and several of these (the {\it minmod\/} limiter, the 
{\it superbee\/} limiter, the {\it van Leer\/} limiter and the 
{\it monotonized centred\/} limiter; see LeVeque~1996 for details)
have been incorporated in the implementation of the wave-propagation
algorithm used in the present work.

Figure~\use{Fg.PdeWave} presents some sample numerical results
produced using the wave-propagation method with a limiter function.
Where the exact solution is smooth it can be seen that the
method produces accurate results which are comparable to those
produced using the second-order Lax-Wendroff scheme; however, unlike
the Lax-Wendroff scheme, the wave-propagation method accurately
resolves the exact solution in regions where it is not smooth, and no
spurious oscillations are introduced.
The wave-propagation method~\EqNum{PdeWavePropOne}{} is stable for any
time step~$\Delta t$ satisfying the Courant-Friedrichs-Lewy
condition~\EqNum{PdeCFL}{}. 
Because of the way wave jumps are compared when applying the limiter
function, the wave-propagation method differs from the other
integration schemes that have been described so far in this chapter in
that it has a stencil that is, strictly speaking, five (rather than
three) cells wide.

\SubSection{Wave Propagation in Two Dimensions}
The extension of the wave-propagation method~\EqNum{PdeWavePropOne}{}
to problems in two spatial dimensions is not straightforward.
In particular, transverse propagation terms need to be included in the
method, and these add complexity not only to the cell-update formula
but also to the implementation of Riemann solver routines for the
evolution systems to which the method is applied.

The basic form of the method generalizes the two-dimensional version
of Godunov's scheme, equation~\EqNum{PdeGodFluxTwo}{}, to
$$\eqalign{
  u^{n+1}_{i,j} = u^n_{i,j} & {}
  - {\Delta t \over \Delta x} 
    \left[ {\cal A}^{+} \!\Delta u_{i-1/2,j} 
         + {\cal A}^{-} \!\Delta u_{i+1/2,j} \right]
  - {\Delta t \over \Delta x} 
    \left[ \tilde{F}_{i+1/2,j} - \tilde{F}_{i-1/2,j} \right] \cr
  & {}
  - {\Delta t \over \Delta y} 
    \left[ {\cal B}^{+} \!\Delta u_{i,j-1/2} 
         + {\cal B}^{-} \!\Delta u_{i,j+1/2} \right]
  - {\Delta t \over \Delta y} 
    \left[ \tilde{G}_{i,j+1/2} - \tilde{G}_{i,j-1/2} \right] 
  \, , \cr
  } \EQN PdeWavePropTwo
$$
where~$\tilde{G}_{i,j-1/2}$ and~$\tilde{G}_{i,j+1/2}$ are second-order
correction fluxes in the $y$-direction, analogous to the correction
terms~$\tilde{F}$ for the $x$-direction in
equation~\EqNum{PdeWavePropOne}{}. 
While the flux corrections~$\tilde{F}$ and~$\tilde{G}$ could simply be
defined according to equation~\EqNum{PdeFluxCorrect}{} (with Riemann
problems being solved independently in the $x$-~and $y$-directions),
such an approach is not by itself capable of making the
scheme~\EqNum{PdeWavePropTwo}{} second-order accurate. 

\figure{PdeTransverse}
\DrawFig{
  \epsfxsize=0.95\hsize
  \epsfbox{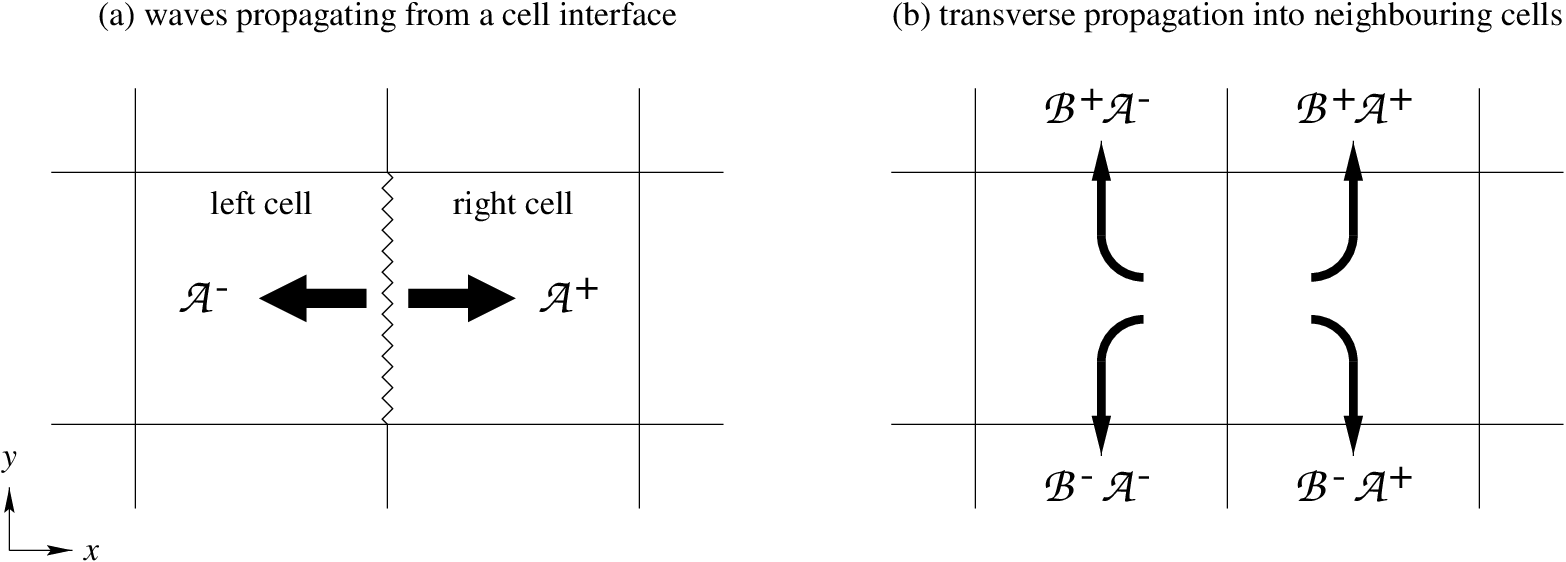}
}
\Caption
Transverse wave propagation.
The left-going and right-going fluctuations~${\cal A}^- \!\Delta u$
and~${\cal A}^+ \!\Delta u$ that emerge from the cell interface (first
plot) are split into four transverse 
fluctuations~${\cal B}^\pm \!{\cal A}^\pm \!\Delta u$ which affect the
solution in the cells above and below the interface (second plot).
\bigskip
\endCaption
\endfigure

One way to make the two-dimensional form of the wave-propagation
algorithm second-order accurate follows from the realization that the
left-going waves produced by the Riemann problem at the interface
between the cells~$u^n_{i,j}$ and~$u^n_{i+1,j}$ should affect not only the
solution value~$u^{n+1}_{i,j}$ but also the solution in the cells
above and below: $u^{n+1}_{i,j+1}$ and~$u^{n+1}_{i,j-1}$. 
The wave-propagation algorithm accounts for transverse wave motion of
this kind by splitting the left-going 
fluctuation~${\cal A}^- \!\Delta u$ into two 
{\it transverse fluctuations\/}, one  
up-going,~${\cal B}^+ \!{\cal A}^- \!\Delta u$, and one
down-going,~${\cal B}^- \!{\cal A}^- \!\Delta u$.
The right-going waves are treated in a similar manner, and for each
Riemann problem that is solved, four transverse 
fluctuations~${\cal B}^\pm \!{\cal A}^\pm \!\Delta u$
are calculated in total. 
(Analogous quantities~${\cal A}^\pm {\cal B}^\pm \!\Delta u$ account
for transverse wave propagation for Riemann problems in the
$y$-direction.) 
This situation is illustrated in figure~\use{Fg.PdeTransverse}.

Transverse fluctuation terms are incorporated in the update
step~\EqNum{PdeWavePropTwo}{} through modifications made to the flux
corrections~$\tilde{F}$ and~$\tilde{G}$.
The exact details of this can
be found in LeVeque~1997,~1998. 
Limiters are applied to the waves produced at each cell interface
according to equation~\EqNum{PdeLimiter}{}, and the
effect of the limiters can also be propagated transversely together
with the second-order corrections of
equation~\EqNum{PdeFluxCorrect}{}. 
A geometrical argument motivating the use of transverse fluctuations
in the case of advection of a scalar quantity can be found in
LeVeque~1996. 

The exact definition of the transverse 
fluctuations~${\cal B}^\pm \!{\cal A}^\pm \!\Delta u$ 
and~${\cal A}^\pm {\cal B}^\pm \!\Delta u$ depends on the nature of
the evolution system being studied.
For a linear system, the Jacobian matrices~$A$ and~$B$ of
equation~\EqNum{PdeJacobian}{} have constant coefficients, and the
transverse fluctuations can be defined through a generalization of
equation~\EqNum{PdeLinearFlux}{}: 
$$
  {\cal B}^\pm \!{\cal A}^\pm \!\Delta u
  = B^\pm A^\pm \Delta u
  \qquad \hbox{and} \qquad
  {\cal A}^\pm {\cal B}^\pm \!\Delta u
  = A^\pm B^\pm \Delta u
  \, , \EQN PdeLinTransverse
$$
where the four matrices~$A^\pm$ and~$B^\pm$ are constructed as in 
equation~\EqNum{PdeLinMatrices}{}.
A similar approach can be employed for nonlinear systems if Roe's
approach (section~\use{sec.PdeRiemann}) is used to produce approximate
solutions to the Riemann problem.
Considering the waves produced between two cells~$u^n_{i,j}$
and~$u^n_{i,j+1}$, a constant matrix~$\tilde{A}$ is used in Roe's
approach to define left-going and right-going 
fluctuations~${\cal A}^\pm \!\Delta u$.
This matrix~$\tilde{A}$ is typically determined by evaluating the
Jacobian matrix~$A(u)$ at a fixed state~$u=u_{\ast}$.
A constant matrix~$\tilde{B}$ can be defined by
evaluating the other Jacobian matrix~$B(u)$ at the same
state~$u_\ast$, and the transverse 
fluctuations~${\cal B}^\pm \!{\cal A}^\pm \!\Delta u$ can then be 
evaluated using matrix multiplication as in
equation~\EqNum{PdeLinTransverse}{}.
(An analogous argument determines values for the transverse
fluctuations~${\cal A}^\pm {\cal B}^\pm \!\Delta u$ in the
$y$-direction.) 

With transverse fluctuations appropriately determined, the
two-dimensional wave-propagation method~\EqNum{PdeWavePropTwo}{} is a
second-order integration scheme for smooth solutions.
Through the use of wave limiters, it can evolve solutions with
discontinuities without introducing spurious oscillations.
The method is stable for any time step~$\Delta t$ satisfying the 
Courant-Friedrichs-Lewy condition~\EqNum{PdeCFL}{}. 
As in the one-dimensional case, the way in which the limiters are
applied means that the wave-propagation method uses an integration
stencil of width five.

Example numerical results produced using the two-dimensional
wave-propagation algorithm can be found in LeVeque~1997,~1998. 
In the present work, a range of numerical simulations of vacuum
cosmological spacetimes have been carried out using the method.
In particular, section~\use{sec.HypNumerical} describes how exact and
approximate Riemann solvers can be used in conjunction with the
wave-propagation method to evolve solutions to a hyperbolic
formulation of the Einstein equations.
When spacetimes described by metrics with smooth components are
evolved, it has been demonstrated that the wave-propagation method
(without limiters) produces results which converge at second
order and which are comparable in accuracy to results produced using
the second-order Lax-Wendroff finite difference scheme; given the
complexity of the evolution systems involved and the experimental
nature of the wave-propagation method, this is a non-trivial result. 
None of the spacetimes that have been evolved in this work have
developed discontinuities or steep gradients of the kind that
benefit from the application of wave limiters (although in some 
simulations this does not at first sight appear to be the case; see
section~\use{sec.GowCollapse}). 
The use of a high-resolution integration method in the present work
may be thought of as preliminary research towards the use of the method
to evolve Einstein's equations coupled to hydrodynamical matter
fields, a situation in which the sophisticated behaviour of a
high-resolution numerical scheme is sure to be needed.


\Chapter{Adaptive Mesh Refinement}
The local accuracy of a numerical solution to a system of
hyperbolic partial differential equations is typically a
function of the mesh spacing of the discrete grid used, as
explained in the previous chapter (section~\use{sec.PdeDifference}).
Evolving a numerical solution on a single uniform grid
can be very inefficient if a fine mesh spacing is needed in
part of the spatial and temporal domain while a
coarse mesh spacing would suffice elsewhere.
Adaptive mesh refinement (henceforth AMR) is a method which
allows the resolution of a numerical simulation to vary in space
and time so that the accuracy of the evolved solution is maintained
above a specified level across the entire solution.

The key reference to the AMR method is the~1984 paper by
Berger and Oliger, and a number of other significant references
are cited at relevant points in this chapter.
An overview of some of the computational work performed using AMR
is given by Plewa~(1999).
Worthy of note is
the \AmrClaw\ software package (Berger and LeVeque~1997), a freely
available two-dimensional implementation of the AMR algorithm 
combined with the \Clawpack\ wave-propagation methods discussed 
in chapter~2.
(Though the standard of the \AmrClaw\ software is very high, its
implementation in Fortran~77 makes it complicated and too inflexible
to be used in the present work; section~\use{sec.AmrStructure} argues
that a modern computer language such as Fortran~90 or \Cplusplus\ is
much better suited to the development of an AMR code.)

Most applications of the AMR method have been to problems
in fluid dynamics.
The complexity of the algorithm, combined with the complexity
of Einstein's equations, has meant that AMR has not to date been
extensively used in numerical relativity.
Its first use was by Choptuik~(1989, 1992,~1993) to study the 
gravitational collapse of a massless scalar field in spherical
symmetry, work which would have been impractical without the use of
AMR since very fine mesh spacings are needed to resolve the
small-scale structure that develops in the model.
A different implementation of the AMR algorithm was used by 
Hamad\'e and Stewart~(1996) to look at the same problem expressed 
in double-null coordinates.
Several other spherically symmetric collapse scenarios have been
examined using these two AMR codes: 
collapse of an~$SU(2)$ gauge field 
by Choptuik, Chmaj and Bizo\'n~(1996),
collapse of an axion/dilaton system by
Hamad\'e, Horne and Stewart~(1996), and
collapse of a perfect fluid by Hamad\'e~(1997).
The development of fully three-dimensional AMR codes for use in
relativity has been motivated by the massive computational
requirements needed to simulate black hole collisions (as outlined in
Seidel~1998; the first work in this direction, Mass\'o, Seidel and
Walker~1995, used a one-dimensional AMR code to evolve a Schwarzschild
black hole).
Br\"ugmann~(1996, 1999) reports preliminary results of the 
BAM (bifunctional adaptive mesh) code in evolving single and double
black hole spacetimes.
Papadopoulos, Seidel and Wild~(1998) use the HLL (hierarchical linked
list) AMR code of Wild~(1996) to simulate gravitational radiation
produced by a perturbed black hole.
Mitra, Parashar and Browne~(1998) developed the DAGH (distributed
adapted grid hierarchies) code for use by the Binary Black Hole NSF
Grand Challenge project. 
Diener et al.~(1999) have constructed initial data sets for the
Einstein equations by using an FTT (fully threaded tree) AMR code
to solve elliptic equations.
The Cactus code (Bona et al.~1998, Mass\'o and Walker 1999), which is
being developed by a number of researchers, is expected to incorporate
AMR capabilities in the future.

This chapter discusses the AMR code developed for the present work.
Section~\use{sec.AmrOutline} summarizes the AMR algorithm
and section~\use{sec.AmrDetails} describes some of its details
(in particular ideas which are new to this work, or which have
received little attention in the standard literature).
Section~\use{sec.AmrStructure} describes the modular structure of
the code and comments on some of the \Cplusplus\ object-oriented
language features used to implement it.
Section~\use{sec.AmrParallel} explains how the code has been
implemented to make use of multiple processors, and finally
section~\use{sec.AmrPractice} assesses how well the code performs
in practice.
The prototype of the code was written by J.~M.~Stewart.
A brief description of it is given in Hern and Stewart~1998.

\Section{Outline of the AMR Algorithm
         \label{sec.AmrOutline}}
The AMR code developed for the present work functions on
the same basic principles as do the majority of AMR codes
described in the literature, of which Berger and Oliger~1984,
Berger and Colella~1989, and Bell et al.~1994
form a good sample.
This section summarizes the main features of the basic
algorithm, which is essentially the same for all AMR codes,
and about which more can be found in the above references.
In section~\use{sec.AmrDetails} some esoteric aspects
of the algorithm---in particular some ideas new to this
work---are discussed in greater depth.

In this work, AMR is used for numerical simulations in one and two
spatial dimensions.
As is the general case with AMR codes, the algorithm in one dimension
is significantly simpler than the two-dimensional version, both
in terms of the concepts involved and the programming required to
implement them.
By contrast, the algorithm in three or more
dimensions is conceptually no different from the two-dimensional case,
although the amount of work required to implement it is significantly
greater.
With this in mind, the discussion in this chapter concentrates on the
two-dimensional version of the algorithm, with its application to
higher dimensions, and its reduction to a single dimension,
being straightforward.

The main features of the AMR algorithm are illustrated in
figure~\use{Fg.AmrPicture}.

\fullfigure{AmrPicture}
\vss
\DrawFig{
  \epsfxsize=\hsize
  \epsfbox{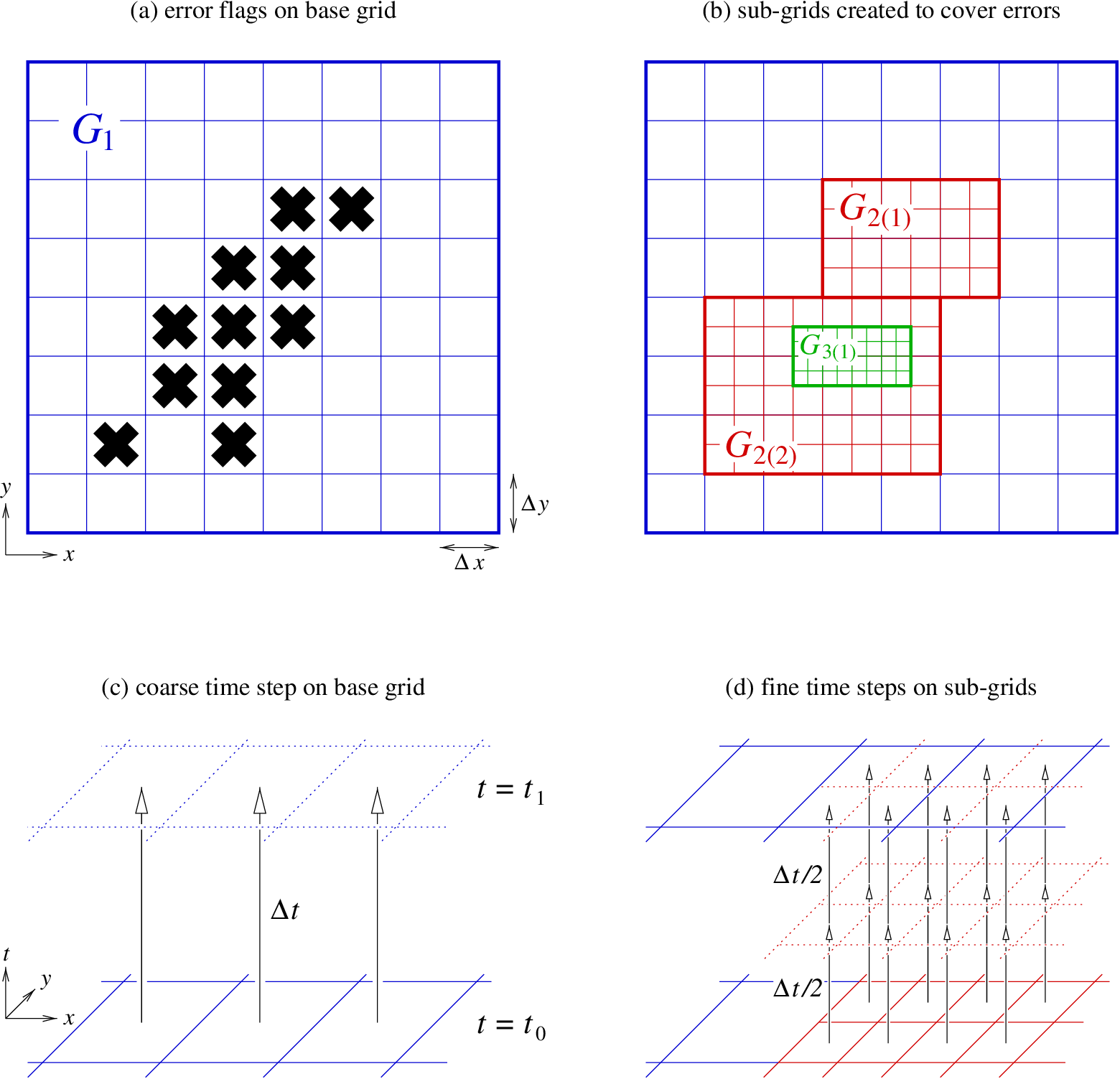}
}
\vskip1.1cm
\Caption
Basic features of an adaptive mesh refinement algorithm.
\vskip2cm
\endCaption
\endfigure

\SubSection{The Grid Hierarchy}
Chapter~2 introduced the uniform grid as a representation
of data on a spatial domain in a discrete form.
Uniform grids are used as building blocks by the AMR algorithm,
with fine-meshed grids set in place to cover regions of
coarse-meshed grids which would otherwise poorly
represent the data in the underlying spatial domain.
For the purposes of this discussion, a grid is taken to be
a mesh of cells rather than points (see equations~\EqNum{PdeCellAv}{}
and~\EqNum{PdePoint}{} of chapter~2), although the ideas in this
section apply just as readily to grids of the latter type.
The extent to which the AMR algorithm is affected by the type
of grid used is a topic returned to later in this chapter.

A hierarchical structure of grids is built up by the AMR
algorithm in the following manner.
Take as a starting point a grid~$G_{1}$ which has a mesh
size $(\Delta x, \Delta y)$ which is too coarse to
accurately describe certain regions of the data it represents.
The accuracy of the grid could be improved by decreasing
its mesh size, but to do this everywhere on the
grid would be a waste of computational resources, and to
give the grid a non-uniform spacing would be, in all but
simple cases, prohibitively complex.
The AMR approach is to create a number of finely meshed sub-grids
$G_{2(1)}$,~\dots,~$G_{2(n)}$ that substitute for the
original grid in the regions where it is lacking.
The mesh size of a child grid~$G_{2(i)}$ will be related
to that of its parent~$G_{1}$ by
$$
\Delta x_{\SmallWord{child}} = \Delta x / \hbox{\tt Refine}
\, , \qquad
\Delta y_{\SmallWord{child}} = \Delta y / \hbox{\tt Refine}
\, ,
$$
where {\tt Refine} is an integer parameter in the code,
a typical value for which is four.

If the newly formed sub-grids are still not sufficiently fine
to resolve the data in all regions then the process of
creating child grids continues recursively.
The result is a family tree of grids in which each generation
has a mesh spacing smaller than the last by a factor {\tt Refine}.
The original grid is said to form level one of the AMR grid 
hierarchy, with level two comprising the child grids, level three
the grandchildren, and so on for an arbitrary number of
levels.
The construction of a grid hierarchy is illustrated in
figure~\use{Fg.AmrPicture}(a,b) for a value of {\tt Refine} 
equal to two.

In principle a lot of freedom is allowed as to how grids can be
arranged to form a grid hierarchy, with the early work of 
Berger~(1982) testing out much of this freedom. 
However the AMR algorithm used for the present work applies some 
straightforward restrictions that lead to a significantly simpler
code which is nonetheless (it is hoped) just as effective.
First of all, children are required to be aligned with their parents,
which is to say that the edges of cells in a child grid must be
parallel to edges of the parent's cells, and the
silhouette of the child must completely cover a rectangular block
of cells in the parent.
(This situation is shown in figure~\use{Fg.AmrPicture}(b).
More generally a child grid could use a completely different
coordinate system to that of its parent.)
Furthermore no grids at the same level may overlap, and every child
grid must be completely contained within its parent.
(The latter condition is not usually part of AMR algorithms,
and complications caused by multiple parents result.
Section~\use{sec.AmrDetails} returns to this point.)

According to the description above, a grid hierarchy has only one
grid in its first level.
In fact this restriction is not enforced by the AMR algorithm,
and section~\use{sec.AmrParallel} suggests one reason for using
multiple base grids.
However, in practice a single base grid is more convenient to work
with, and none of the simulations performed for this work have used
multiple base grids.

\SubSection{Time-Stepping the Grid Hierarchy}
The discussion so far has concentrated on the structure of a
grid hierarchy at one instant,~$t_0$, in time.
Taking the hierarchy itself to be fixed, this subsection outlines
the way in which the AMR algorithm advances the family of grids
to a later time, $t_1 = t_0 + \Delta t_0$.

The time-stepping operates by the following recursive
procedure: all of the grids at level~$N$ are advanced one time step
to time~$t + \Delta t$, then all of the grids at level~$N+1$ are
advanced to the same time.
This is not as simple as it sounds since to be consistent
with the Courant-Friedrichs-Lewy condition
(equation~\EqNum{PdeCFL}{}) the grids at level~$N+1$ must be advanced by
taking {\tt Refine} steps of size $\Delta t / {\tt Refine}$, rather
than one step of size~$\Delta t$ (refinement takes place in time as
well as in space); and since the time-stepping is
recursive, each of the {\tt Refine} steps at level~$N+1$ is immediately
followed by {\tt Refine} correspondingly smaller steps at level~$N+2$,
and so on up the grid tree.
Advancing level one to time~$t_1 = t_0 + \Delta t_0$ in this way leads
to the entire grid hierarchy being advanced to that time.
This is illustrated in figure~\use{Fg.AmrPicture}(c,d).

A grid can only be advanced in time if it has up-to-date
boundary data (as discussed in section~\use{sec.PdeDifference}), and
the AMR algorithm renews these data each time a step is taken by the
grid. 
External and internal grid boundaries must be dealt with.
External boundaries are the kind familiar from numerical simulations
based on single grids, where the grid edges coincide with
the edges of the spatial domain, and the boundary data
are determined by the problem being solved.
Grids at any level of the grid hierarchy can have external boundaries.
The remaining boundaries are internal, and interpolation is used to
supply the required boundary data; the construction of the grid
hierarchy is such that a grid can always
generate internal boundary data by interpolation of data
in its parent grid.
Since the time steps taken by a child grid are smaller than those taken
by its parent, the times of child and parent often differ at the
point when internal boundary data are needed by the child, and
interpolation in time as well as in space becomes necessary; to allow
for this each grid retains a copy of the data it held prior to its most
recent time step.
The use of interpolation is discussed further in
section~\use{sec.AmrDetails}.

A child grid contains more accurate data than the region of its parent
grid that it covers.
In order for this difference in accuracy not to lead to a divergence
in the approximations on the two grids as they are advanced in time,
the data in the child grid must be used to update the data in the
parent grid.
This is done as part of the time-stepping procedure: whenever a child
grid is advanced to the same time as its parent, the child's data are
used to replace the parent's data in the region covered by the child.

\SubSection{Adaption of the Grid Hierarchy}
So far in this account the grid hierarchy has remained fixed while
the grids in it have been advanced in time.
The real power of the AMR algorithm comes from its ability to
automatically adapt the hierarchy so as to maintain good resolution of
data at all times during the evolution.
A regridding phase in the algorithm creates, moves and destroys
sub-grids in the hierarchy so that refined regions occur
wherever and whenever they are needed.

A parameter {\tt Regrid} (typically given the value four) determines 
the number of time steps taken by each level before it 
undergoes regridding, the result of which is rearrangement of grids
on the levels above.
This is worked into the recursive time-stepping procedure such that 
regridding at level~$N+1$ takes place {\tt Refine} times more often than
at level~$N$.
The regridding phase includes two notable sub-problems: how to decide
where refinement is needed, and how to arrange sub-grids to achieve 
this refinement.

Regions are marked for further refinement at the start of the 
regridding phase.
Each grid cell has an error flag which is set if
the data there fail to meet some kind of accuracy criterion.
The nature of the problem itself may suggest appropriate 
criteria (such as close proximity to a shock front in
a hydrodynamics simulation), but while such approaches are useful
if they are available a more general method is to flag points at which
an estimate of the local truncation error is higher than a pre-set
tolerance.
Berger and Oliger~(1984) explain in detail how the local truncation
error can be estimated using Richardson extrapolation;
for the present discussion a quick sketch of the method will suffice.
As a grid is advanced in time, the data on it are stored for the
current time,~$t_n$, and also for the previous two time steps,
$t_{n-1} = t_n - \Delta t$ and $t_{n-2} = t_n - 2 \Delta t$.
If the data at time~$t_{n-2}$ are coarsened such that their resolution
is changed from $(\Delta x, \Delta y)$ to $(2 \Delta x, 2 \Delta y)$,
and then advanced by a time step~$2 \Delta t$, then the result is a
second approximation to the exact solution at time~$t_n$.
The difference between the two approximations is an estimate of the 
local truncation error.

Given a set of error flags the next task is to find an arrangement of
sub-grids which covers all of the flags but which has as small an
area as possible.
This process is described as clustering.
While the original work by Berger~(1982, 1986) used quite elaborate
clustering heuristics, more straightforward approaches based on
computer vision algorithms have since been developed
(Bell et al.~1994), and these are adopted for the clustering
algorithm used in the present work.

Regridding at level~$N$ means that all of the grids at level~$N+1$
are replaced by a new arrangement of grids based on the results of the
clustering of level~$N$'s error flags; if level~$N+1$ does not
exist it is created at this stage.
Typically some of the new level~$N$ grids will overlap the old ones,
and where this happens the old data are copied across to the new grids.
The remaining data needed to complete the new grids are provided by
interpolation of the data in the level~$N$ grids.
Interpolation methods and the regridding algorithm are topics
considered in greater depth in the next section.

\Section{Details of the AMR Algorithm
         \label{sec.AmrDetails}}
The previous section summarized the AMR algorithm used in the 
present work.
Most of the features discussed there are common to all AMR algorithms
based on Berger and Oliger's original design~(1984).
In this section several extensions to the standard algorithm are
considered.
Flux conservation is achieved by implementing the ideas of Berger
and Colella~(1989), and some conservative interpolation methods
are developed.
The standard regridding algorithm (Berger and Oliger~1984) is replaced
by a new approach which avoids some of the complexities usually 
associated with AMR in two (or more) dimensions.
Provisions are made for variable time steps, and the importance of
communication between neighbouring grids is acknowledged.

\SubSection{Flux Conservation}
The AMR algorithm described in section~\use{sec.AmrOutline} uses grids
which are made up of cells rather than points.
Such grids can be time-stepped using the flux-conservative methods of
chapter~2.
However, for the AMR algorithm itself to be conservative, operations 
involving the transfer of data between grids
must be appropriately designed.
Berger and Colella~(1989) and Berger and LeVeque~(1998) describe the
modifications needed to make the standard AMR algorithm 
fully conservative, and this subsection considers these modifications 
in the context of the present work.
(For an alternative approach see Berger~1987.)

The main idea used to make the AMR algorithm conservative is 
straightforward: recalling that the unknown variables represent 
`densities' of some kind (section~\use{sec.PdeDifference}), whenever
grid cells are coarsened or refined it should be done in such a way as
to conserve the  total `mass' of data contained within each of the
coarse cells. 
This is illustrated in figure~\use{Fg.AmrConserve}.
Grid cells are coarsened whenever a parent grid is updated
using data on a child grid, and as part of the local truncation error
estimation procedure, and the conservative method for coarsening data 
is unambiguous: a coarse grid cell is produced by averaging the data
in the fine grid cells.
Sub-grids require data to be refined when they are created, and also 
to provide them with boundary data, and in both cases interpolation 
is used.
Conservative interpolation methods are the subject of the next 
two subsections.

\figure{AmrConserve}
\DrawFig{
  \epsfxsize=9cm
  \epsfbox{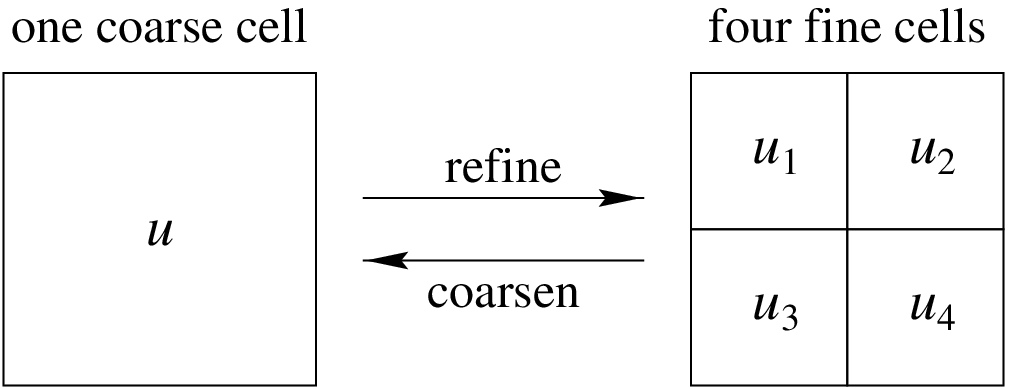}
}
\vskip1mm
\centerline{\twelvepoint 
            $u = {1 \over 4}( u_1 + u_2 + u_3 + u_4 )$}
\Caption
Conservation of data `mass' when converting between coarse and fine 
grid cells.  (Refinement here is by a factor of two.)
\bigskip
\endCaption
\endfigure

By using a flux-conservative integration method, conservation is
guaranteed for individual grids as they are advanced in time.
However problems still arise at the interfaces between grids. 
The fluxes at the edges of a child grid affect both the outermost
cells of the child and the cells in the parent which surround that child,
but their values depend on whether they are calculated as part of the
parent's time step or the child's time step.
For the AMR algorithm to be fully conservative, corrections must be
made to all of the cells in a parent grid which border onto child
grids to account for this discrepancy, as explained by
Berger and LeVeque~(1998).
These corrections have not however been implemented in the AMR
code used for the present work.
Though the reasons for this are mainly practical, the omission
can be partially justified.
None of the numerical simulations performed for this work actually
required global conservation since (because of the presence of source
terms) none of the systems of equations were themselves strictly 
conservative.
Also, while numerical conservation is important for evolving
solutions with discontinuities, it is only required locally;
as long as the grids covering a discontinuity are at the same level of
refinement there should be no need for boundary corrections.
One final (mainly aesthetic) argument against the implementation of
boundary corrections is that, because they introduce a lot of 
additional communication between grids on different levels, they
make the scheme for separating grids from their data---described in 
section~\use{sec.AmrStructure}---significantly more convoluted.

This subsection ends on a brief technical aside.
The flux-conservative wave-propagation methods described in chapter~2
have symmetrical integration stencils of width five, and 
consequently the grids used in the AMR algorithm have boundary
regions two cells wide (see section~\use{sec.PdeDifference}).
In many ways this is a disadvantage; in particular it increases the
amount of work needed to advance each grid.
However it turns out to be advantageous when it comes to
interpolating boundary data for child grids: because a child's 
boundary cells never extend to the limits of its parent, there is
always sufficient data on the parent grid to allow interpolation.
(This is not the case for some other AMR algorithms which occasionally
need to interpolate boundary data from grandparent grids.)

\SubSection{Interpolation I: Piecewise Polynomial Methods}
An AMR algorithm frequently needs to refine regions of coarse data,
and to do this some form of interpolation is used.
The choice of interpolation method depends on whether the data to
be interpolated are stored at grid points or (as in the present work) 
are averaged over grid cells. 
(The distinction is made clear in section~\use{sec.PdeDifference}.)
This subsection and the one that follows describe conservative
interpolation methods for grids of cells in two dimensions (though
the arguments are straightforward to apply to any number of dimensions).

Consider first a coarse grid made up of a two-dimensional mesh of 
points,
$$
(x_i, y_j) = (x_0 + i \, \Delta x, \, y_0 + j \, \Delta y) 
\, , \qquad
i,j \in \IntegSet
\, ,
$$ 
together with data~$f_{i,j}$ defined at these points.
In order to generate fine data from this coarse grid an interpolating
function~$F(x,y)$ which satisfies $F(x_i,y_j) = f_{i,j}$ is needed.
Given such a function, fine data can be defined at points which lie 
in between the points of the coarse mesh; for example,
$\hat f_{i+1/2,j+1/2} := F(x_i + \Delta x / 2, y_j + \Delta y / 2)$.
Chapter~3 of Press et al.~1992 describes a range of interpolation
methods suitable for generating a function~$F(x,y)$.
First- or second-order polynomial interpolation is typically used in
AMR algorithms.

For a flux-conservative AMR algorithm based on grid cells rather than
points the situation is a little different.
The data~$u_{i,j}$ in a coarse grid cell are defined by an average 
over the area of the cell, and an interpolating function~$U(x,y)$ for 
this coarse data is required to satisfy
$$
{1 \over {\Delta x \, \Delta y}}
\int^{x_i + \Delta x / 2}_{x_i - \Delta x / 2}
\int^{y_j + \Delta y / 2}_{y_j - \Delta y / 2}
U\!(x,y) \, dy \, dx = u_{i,j}
\, .
$$
Fine grid data are constructed by integrating an interpolating
function over sub-intervals of the coarse grid cells.
This is illustrated in one dimension in 
figure~\use{Fg.AmrInterpSketch}.

\figure{AmrInterpSketch}
\DrawFig{
  \epsfxsize=15cm
  \epsfbox{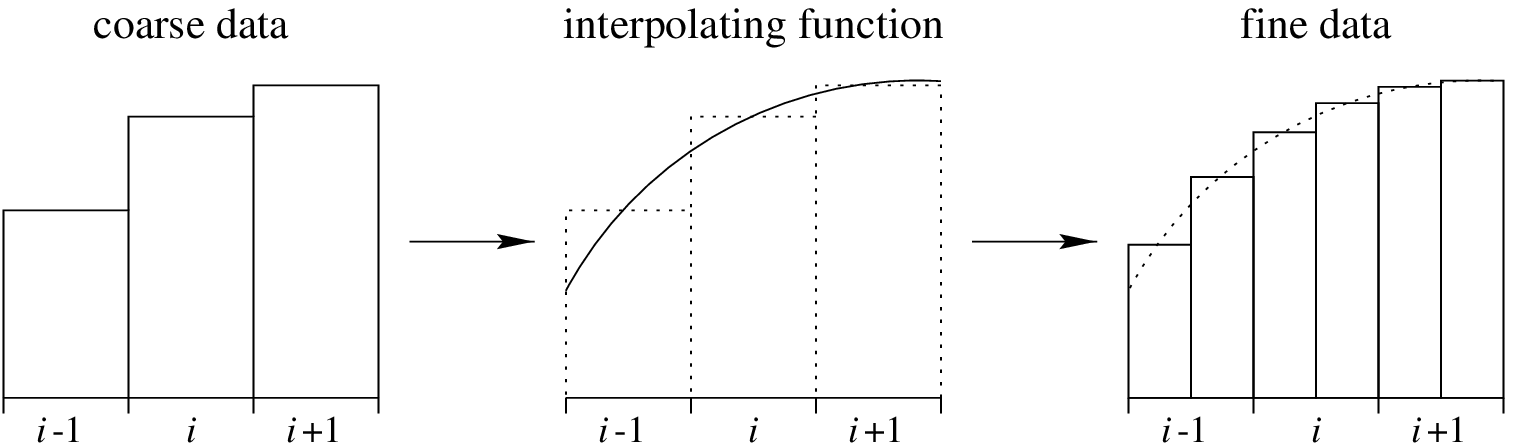}
}
\Caption
Conservative interpolation in one dimension. 
A set of fine data (third plot) is generated to replace the available
coarse data (first plot) using an interpolating function 
as an intermediate step (second plot).
The `mass' of data in each coarse cell (represented by the area under
the graph) is conserved.
(Refinement here is by a factor of two.)
\bigskip
\endCaption
\endfigure

The AMR code used for the present work allows the user to
choose from three different types of conservative interpolation:
piecewise linear, piecewise quadratic and spline.
Sample results are shown in figures~\use{Fg.AmrInterpSmooth}
and~\use{Fg.AmrInterpDiscon}.
Other flux-conservative AMR codes described in the literature
invariably use some form of linear interpolation, and the reasons for
this are discussed below.

\figure{AmrInterpSmooth}
\DrawFig{
  \epsfxsize=\hsize
  \DrawHiLoRes{\epsfbox{intsamp1.eps}}
              {\epsfbox{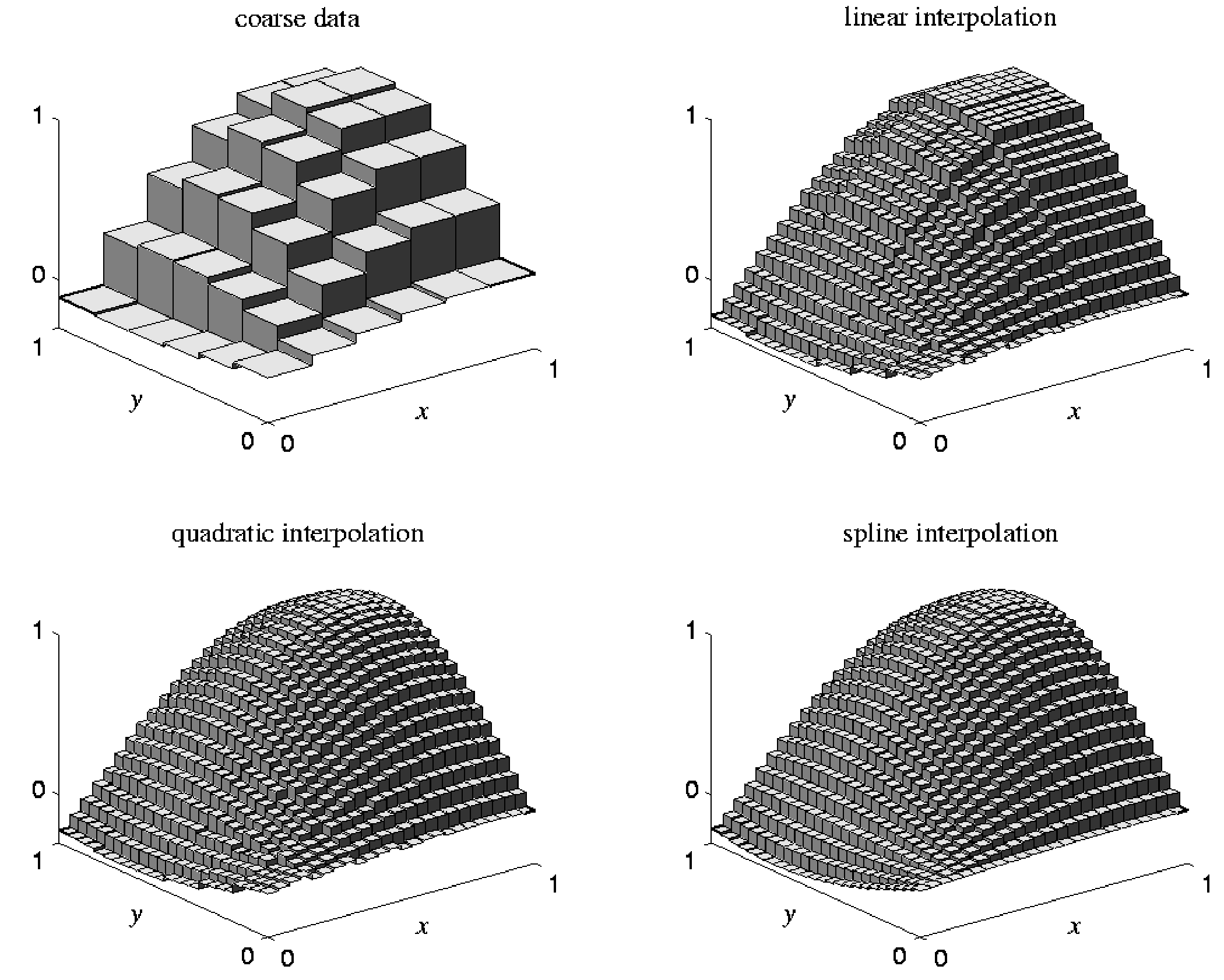}}
}
\Caption
Sample interpolation results for smooth initial data.
Three different conservative interpolation methods are used to
refine a coarse grid of data by a factor of four.
The coarse grid consists of $8 \times 8$~cells (the outermost cells
provide boundary data for the interpolation and only the~$6 \times 6$
innermost cells are shown) and this is refined to 
a grid of size~$24 \times 24$.
The coarse data are produced by taking cell averages of the 
function $\sin(2.25 x - 0.0125) \sin(2.25 y - 0.3125)$.
\bigskip
\endCaption
\endfigure

\figure{AmrInterpDiscon}
\DrawFig{
  \epsfxsize=\hsize
  \DrawHiLoRes{\epsfbox{intsamp2.eps}}
              {\epsfbox{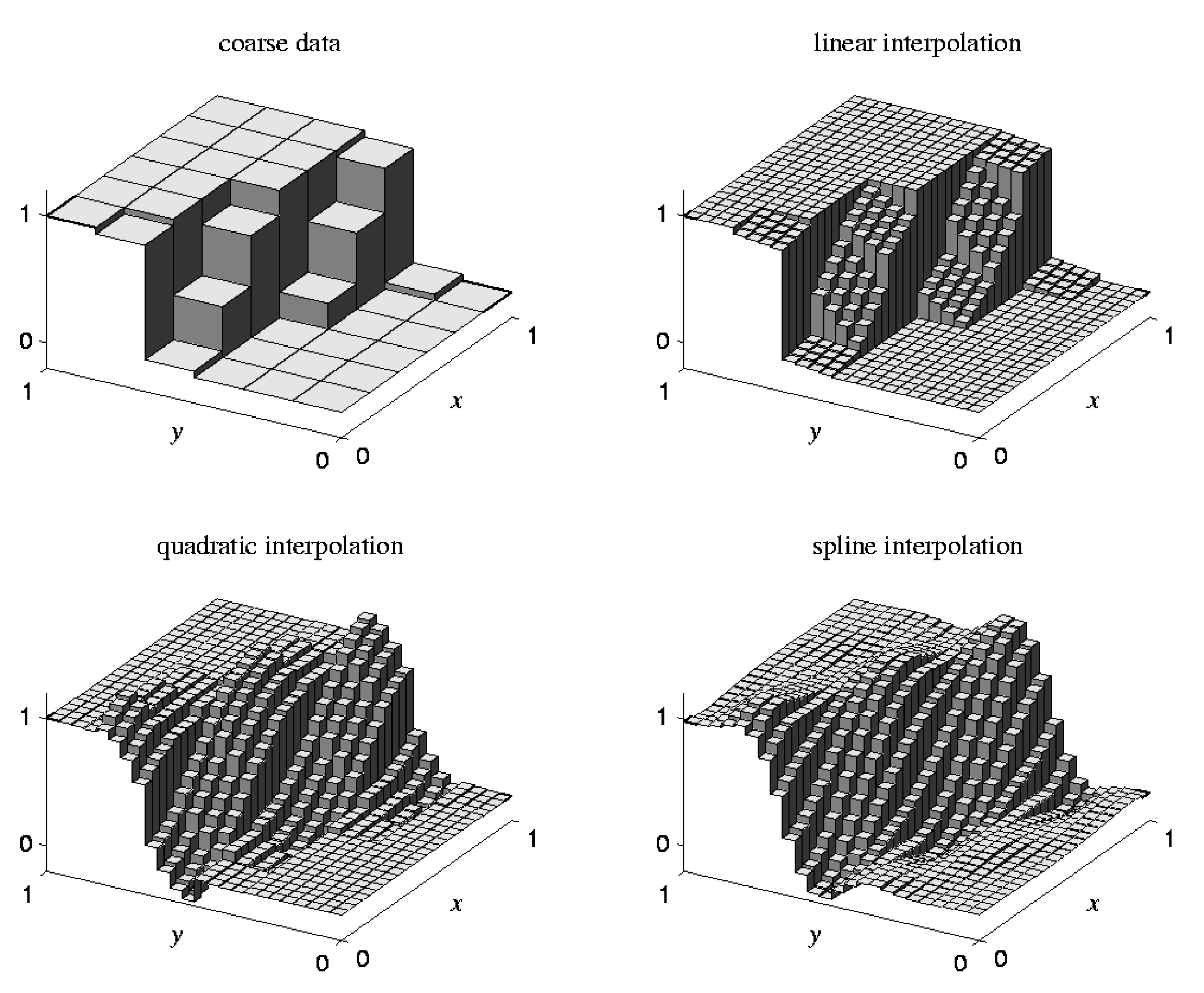}}
}
\Caption
Sample interpolation results for discontinuous initial data.
The setup is the same as in figure~\use{Fg.AmrInterpSmooth} except
that here the coarse data are cell averages of a piecewise constant
function which takes the value one for $x + 2.5 y \geq 1.75$ and
is zero elsewhere.
\bigskip
\endCaption
\endfigure

The idea underlying piecewise polynomial interpolation methods is
to construct an interpolating function~$U(x,y)$ for the entire grid
by piecing together polynomial functions~$U_{i,j}(x,y)$ which are
defined only on individual grid cells.
For linear interpolation a grid cell centred on~$(x_i,y_j)$
will have a local interpolating function of the form
$$
U_{i,j}(x,y) = a + b_x (x-x_i) + b_y (y-y_j)
\, ,
$$
where $a$, $b_x$ and~$b_y$ are constants to be determined in some way
from the cell averages $u_{i,j}$, $u_{i+1,j}$, $u_{i-1,j}$,
$u_{i,j+1}$ and~$u_{i,j-1}$; in fact, the requirement that the
integral of~$U_{i,j}$ over the grid cell be equal to the cell average
automatically fixes $a = u_{i,j}$.
Berger and LeVeque~(1997) choose~$b_x$ and~$b_y$ to be
$$
b_x = \hbox{minmod} 
      ( u_{i+1,j} - u_{i,j} , \, u_{i,j} - u_{i-1,j} )
      / \Delta x
\, ,
$$
$$
b_y = \hbox{minmod} 
      ( u_{i,j+1} - u_{i,j} , \, u_{i,j} - u_{i,j-1} )
      / \Delta y
\, ,
$$
and this choice is adopted by the present work.
Slopes in both the~$x$ and~$y$ directions are selected according to
the min-mod function,
$$
\hbox{minmod} (\alpha,\beta) = 
\left\{ \eqalign{ \alpha & \quad \hbox{if } 
                           \vert \alpha \vert < \vert \beta \vert
                           \hbox{ and } \alpha \beta > 0
                           \, , \cr
                  \beta  & \quad \hbox{if } 
                           \vert \beta \vert < \vert \alpha \vert
                           \hbox{ and } \alpha \beta > 0
                           \, , \cr
                  0      & \quad \hbox{if } \alpha \beta \leq 0
                           \, , \cr
}\right.
$$
which has its origins as a slope-limiter for high-resolution numerical
methods (see LeVeque~1992, page~186), and which prevents the linear
sections from overshooting their 
neighbours if the data to be interpolated are not particularly smooth:
the interpolating function is monotonic if the coarse data are monotonic.
There are of course other reasonable ways of choosing~$b_x$ and~$b_y$.

For piecewise quadratic interpolation the local interpolating function
takes the form
$$
\eqalign{
U_{i,j}(x,y) = a &+ b_x (x-x_i) + c_x (x-x_i)^2 \cr
                 &+ b_y (y-y_j) + c_y (y-y_j)^2 \, . \cr
}
$$
The five constants~$a$, $b_x$, $b_y$, $c_x$ and~$c_y$ can be fixed in 
a straightforward way by requiring that the integral of $U_{i,j}(x,y)$
over the five centremost cells be equal to the respective coarse cell 
averages $u_{i,j}$, $u_{i+1,j}$, $u_{i-1,j}$, $u_{i,j+1}$ and~$u_{i,j-1}$.

The advantages and disadvantages of linear compared to quadratic 
interpolation are clear from figures~\use{Fg.AmrInterpSmooth}
and~\use{Fg.AmrInterpDiscon}.
When the data to be interpolated are reasonably smooth then, as
figure~\use{Fg.AmrInterpSmooth} shows, quadratic interpolation is
preferable to linear interpolation, with the results of the latter 
being quite uneven and jagged.
On the other hand, if the data on the coarse grid are discontinuous
(figure~\use{Fg.AmrInterpDiscon}), then the results of linear 
interpolation, though by no means perfect, are better than the
results of quadratic interpolation in which the tops and bottoms of
discontinuities may be overshot quite badly.
Linear interpolation is thus the method of choice for numerical 
simulations in which the data have, or are expected to develop, 
discontinuities, and this is the reason for its use as 
part of flux-conservative AMR codes.

However for many of the numerical simulations performed for the
present work no discontinuities develop in the data and
the shortcomings of linear interpolation become apparent.
The jaggedness produced by linear interpolation introduces noise
(with wavelengths on the order of the width of a coarse grid cell) 
into the solution every time a new sub-grid is created or internal
boundary data for a sub-grid are produced.
Over time this noise accumulates on the sub-grids and can eventually 
build up to such a level that the simulation is forced to terminate.
Experimentation suggests that the noise is not significantly amplified
by the evolution in comparison to the background solution, and
its initial amplitude is the main concern.
Increasing the accuracy of the interpolation would reduce the
amplitude of the noise, and this can be done in two ways.
Firstly, if the mesh spacing of a coarse grid is decreased such that
the data on it are very well resolved then the inaccuracies in the
interpolated data will be small.
The effect of this is apparent in the differences between one- and
two-dimensional simulations.
In one dimension the meshes used are typically fine enough that linear
interpolation produces good results, and noise is not found to be a
problem. 
However the meshes used in two-dimensional simulations are much
coarser and the resulting poor resolution of data means that
errors in interpolation are significant.
It is usually impractical to substantially decrease the mesh spacings
used in numerical simulations (after all, this is the reason why
AMR is used in the first place), and the alternative approach to
improving interpolation accuracy is to use higher-order methods.
This is the motivation for implementing quadratic interpolation in the
AMR code used in this work, and the result is a reduction in noise
amplitude with a corresponding increase in simulation long-term 
stability.

As can be seen from figure~\use{Fg.AmrInterpSmooth}, quadratic
interpolation still introduces some unevenness into the refined data
(most obvious close to the $x$-$y$~origin in the figure),
and the resulting level of noise is still unacceptable for some
simulations.
The piecewise nature of the interpolating function~$U(x,y)$ is the
main cause of this problem, and by using conservative spline
interpolation methods to produce globally smooth data it may be
hoped that the effect of noise on sub-grids could be minimized.

\SubSection{Interpolation II: Spline Methods}
Press et al.~(1992, section~3.3) describe how cubic splines can be
used to interpolate data which are defined on a mesh of points.
In this subsection a quadratic spline method is developed for the
conservative interpolation of data defined on meshes of cells.
In the same way as in cubic spline interpolation, one-dimensional
quadratic splines are constructed as part of the interpolation
procedure in two and higher dimensions, and these will be described
first. 

A one-dimensional cubic spline is an interpolating function~$F(x)$
which takes the values~$f_1$, \dots,~$f_n$ at the points~$x_1$,
\dots,~$x_n$, and
is constructed by requiring that it be a cubic function on each
interval $[ x_i, x_{i+1} ]$ (for $1 \leq i < n$) and that it be
continuous and smooth with continuous second derivative at each
point~$x_i$ (for $1 < i < n$).
The specification of a conservative quadratic spline proceeds along
similar lines.
For a uniform mesh, $x_i = x_0 + i \, \Delta x$, with cell 
averages~$u_1$, \dots,~$u_n$, an interpolating function~$U(x)$
is constructed which is quadratic on each interval
$[ x_i - \Delta x/2, \, x_i + \Delta x/2 ]$ (for $1 \leq i \leq n$),
with
$$
{1 \over {\Delta x}}
\int_{x_i - \Delta x/2}^{x_i + \Delta x/2} U\!(x) \, dx 
= u_i
$$
(for $1 \leq i \leq n$), and which is continuous and smooth at 
each point~$x_i + \Delta x/2$  (for $1 \leq i < n$).
From these requirements the relationship
$$
U_{i-1/2} + 4 \, U_{i+1/2} + U_{i+3/2} = 3( u_i + u_{i+1} )
$$
is found for the values $U_{i+1/2} := U(x_i + \Delta x / 2)$ of 
the interpolating function at the cell edges, for 
$2 \leq i \leq n-2$.
Two boundary conditions are needed to uniquely specify the
spline and these are provided by taking the spline to be linear in the
first and last cells, $[x_1 - \Delta x/2, \, x_1 + \Delta x/2]$ and
$[x_n - \Delta x/2, \, x_n + \Delta x/2]$.
The result is a system which completely determines the values of the
interpolating function at the cell edges:
$$
\pmatrix{ 3&1& &             \cr
          1&4&1&             \cr
           &1&4&             \cr
           & & &\ddots       \cr
           & & &      &4&1&  \cr
           & & &      &1&4&1 \cr
           & & &      & &1&3 \cr }
\pmatrix{ U_{1+1/2} \cr U_{2+1/2} \cr U_{3+1/2} \cr \vdots \cr
          U_{n-2-1/2} \cr U_{n-1-1/2} \cr U_{n-1/2} \cr }
=
\pmatrix{ 3u_2 + u_1 \cr 3(u_3+u_2) \cr 3(u_4+u_3) \cr \vdots \cr
          3(u_{n-2}+u_{n-3}) \cr 3(u_{n-1}+u_{n-2}) \cr u_n+3u_{n-1} \cr }
\, .
$$
This tridiagonal system can be solved for the values~$U_{i+1/2}$ by 
a method of forward- and back-substitution
(see Press et al.~1992, section~2.4).
Knowledge of $U_{1+1/2}$, \dots,~$U_{n-1/2}$ and $u_1$, \dots,~$u_n$
is sufficient to evaluate the quadratic spline~$U(x)$ in its entirety,
and from this cell averages on a more refined grid can be defined as
before.

A conservative interpolation method in one dimension can be used to
interpolate data conservatively in any number of dimensions, and this
is the way quadratic spline interpolation is applied to
two-dimensional grids.
Figure~\use{Fg.AmrInterpSweeps} illustrates the procedure.
The coarse grid cells are first interpolated along lines parallel to
the $x$-axis.
The partly-refined cells that result are then refined in the
$y$-direction. 

\figure{AmrInterpSweeps}
\DrawFig{
  \epsfxsize=15cm
  \epsfbox{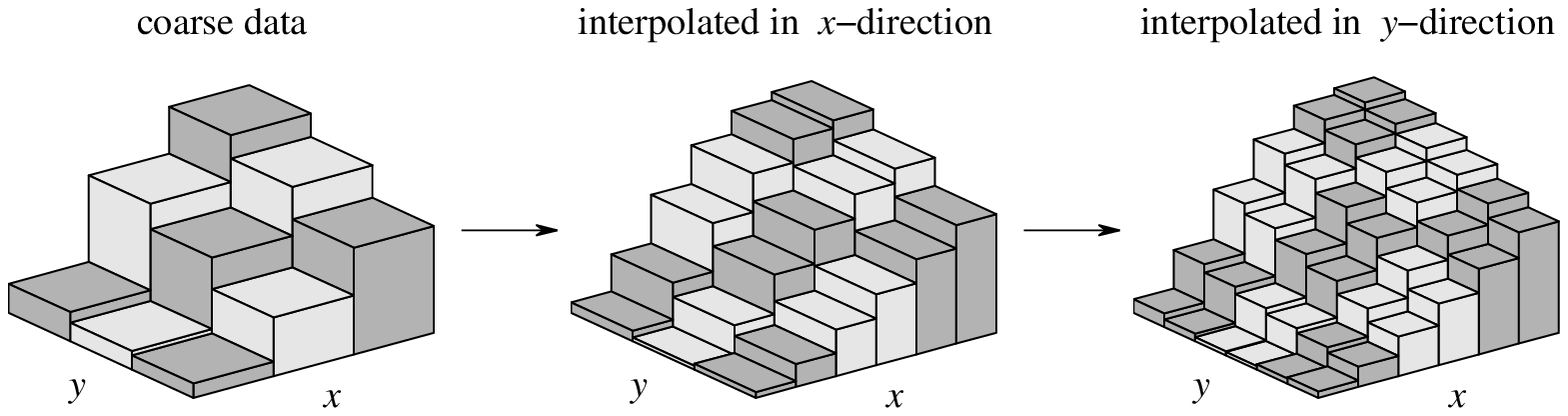}
}
\Caption
Conservative interpolation of two-dimensional data using a
one-dimensional method.
Interpolation sweeps in the $x$-direction are followed by sweeps
in the $y$-direction.
(Refinement here is by a factor of two.)
\bigskip
\endCaption
\endfigure

Figures~\use{Fg.AmrInterpSmooth} and~\use{Fg.AmrInterpDiscon}
show examples of results produced by quadratic spline interpolation.
For smooth initial data (figure~\use{Fg.AmrInterpSmooth}) the results 
are smooth, without the jaggedness at the edges of coarse
cells that occurs with the piecewise polynomial methods.
However for discontinuous initial data (figure~\use{Fg.AmrInterpDiscon})
quadratic spline interpolation suffers from the same problem that plagues
piecewise quadratic interpolation: 
it produces oscillations around discontinuities, and
in contrast to the piecewise method the oscillations affect
(albeit very slightly) the entire grid not just the cells along the 
discontinuities.
This aside, when spline interpolation has been used in simulations in 
which the data do not become discontinuous, it has been found to work 
well and the problem of build-up of noise on sub-grids is much less 
severe.

For data defined at points, spline interpolation might be expected to
be less of an improvement over piecewise methods than it is in the
case of data defined over cells since there is never any difficulty in
constructing a pointwise interpolating function which is continuous.
However in practice the lack of smoothness in piecewise interpolated
data is still found to lead to higher levels of noise in some
simulations than is acceptable.

The conservative interpolation methods described in this and the
previous subsections have been
developed by adapting in a natural way the ideas in
chapter~3 of Press et al.~1992.
In fact the methods in that reference can be used
to derive conservative interpolation methods directly.
For a one-dimensional uniform grid $x_1$,~\dots,~$x_n$ with cell averages
$u_1$,~\dots,~$u_n$, a discrete function $f_{1/2}$,~\dots,~$f_{n+1/2}$
can be defined which takes values at the cell edges
$x_{1/2}$,~\dots,~$x_{n+1/2}$:
$$
f_{1/2} = 0 
\, , \qquad
f_{i+1/2} = \sum_{k=1}^{i} u_k \Delta x 
\quad
\hbox{for } 1 \leq i \leq n 
\, .
$$
Standard methods of interpolation can then be applied to find an
interpolating function~$F(x)$ that passes through the points
$(x_{i+1/2},f_{i+1/2})$, and it can be seen that 
$U(x) := F'(x)$ is a conservative interpolating function for the
original grid cells.
Bypassing the evaluation of~$U(x)$,
cell averages for a refined grid can be calculated directly
from~$F(x)$: for a cell $[x_A,x_B]$, the interpolated 
average~${\hat u}_{AB}$ is
$$
\eqalign{
{\hat u}_{AB} &= { 1 \over x_B - x_A }
                 \int_{x_A}^{x_B} U\!(x) \, dx \cr
              &= { F(x_B) - F(x_A) \over x_B - x_A }
                 \, . \cr}
$$
If an order~$n$ interpolation method is used to determine~$F(x)$,
then~$U(x)$ is constructed by the equivalent of order~$(n-1)$
interpolation, and
this is why quadratic splines are natural for
interpolating grid cells while cubic splines are
used for grid points.
This approach for constructing conservative interpolation methods
extends in an obvious way to higher dimensions by taking
$$
f_{i+1/2, \, j+1/2} = \sum_{k=1}^{i} \sum_{l=1}^{j} 
                      u_{k,l} \Delta x \Delta y
\, , \quad\hbox{etc.}
$$
Essentially the same results are produced by both approaches.
However for some methods (in particular the min-mod
interpolation method described in the previous subsection)
the construction is more intuitive when viewed as a cell interpolation
method rather than the reduction of a point interpolation method.

\SubSection{The Regridding Procedure}
A surprising feature of Berger and Oliger's AMR algorithm~(1984)
is that a child grid is not limited to having only one parent:
a level~$N+1$ grid is allowed to overlap any number of level~$N$
grids as long as it is completely contained within their union.
This makes many of the grid operations performed in an AMR code based
on Berger and Oliger's design significantly more complicated than they 
would otherwise be since multiple parent grids must be accounted for
during any communication of data between grids at different levels.
The reason why multiple parents are usually allowed is that the standard AMR
regridding procedure cannot maintain proper child-parent nesting when
it makes alterations to the grid hierarchy.
(This is actually only the case in two or more 
dimensions, and is part of the reason why one-dimensional AMR codes
are much more common than multi-dimensional ones.)
For the AMR algorithm used in the present work the author has designed
a new regridding procedure which ensures proper child-parent nesting
and which consequently simplifies the implementation of the code as a
whole: each grid need only maintain a pointer to its parent, not a
list of pointers to a number of parents.
Figure~\use{Fg.AmrRegrid} shows the new regridding algorithm.

Regridding at level~$N_{\SmallWord{regrid}}$ results in 
changes being made to the grid layouts on all of the more refined 
levels,~$N_{\SmallWord{regrid}}+1$ 
to~$N_{\SmallWord{finest}}$, as well as the possibility of a new
level,~$N_{\SmallWord{finest}}+1$, being created.
The standard method of regridding, as described by Berger and
Colella~(1989), operates as follows.
Starting at level~$N=N_{\SmallWord{finest}}$ and working down to
level~$N_{\SmallWord{regrid}}$, error flags at level~$N$
are clustered and, based on the results, new sub-grids are created to
replace the grids at level~$N+1$ (that level being created if it
does not already exist).
To ensure that sub-grids are always contained within the union of
grids at the level below, 
if level~$N+2$ exists when level~$N$ is due to be clustered, error flags
are set in level~$N$ in any region covered by a grid in level~$N+2$.
At the time of its creation a child grid is always completely
contained within a single parent grid;
nesting problems arise because the parent's level may be rearranged
immediately after.

\figure{AmrRegrid}
\BoxTop
\Code{void example() \OpenBrace}
\Code{\Tab Level base{\Under}level;}
\Code{\Tab base{\Under}level.regrid{\Under}tree();}
\Code{\CloseBrace}
\Code{}
\Code{void Level::regrid{\Under}tree() \OpenBrace}
\Code{\Tab flag{\Under}errors{\Under}in{\Under}tree();}
\Code{\Tab regrid{\Under}level();}
\Code{\CloseBrace}
\Code{}
\Code{void Level::flag{\Under}errors{\Under}in{\Under}tree() \OpenBrace}
\Code{\Tab {\it flag errors in grids at this level}}
\Code{\Tab if ( next{\Under}level.contains{\Under}grids() ) \OpenBrace}
\Code{\Tab\Tab next{\Under}level.flag{\Under}errors{\Under}in{\Under}tree();}
\Code{\Tab\Tab {\it set flags in this level where there are errors 
                    in\/ {\tt next{\Under}level\/}'s grids}}
\Code{\Tab \CloseBrace}
\Code{\CloseBrace}
\Code{}
\Code{void Level::regrid{\Under}level() \OpenBrace}
\Code{\Tab Level new{\Under}level = ({\it new sub-grids created to cover
                                         error flags at this level\/});}
\Code{\Tab {\it copy data and flags from grids 
                in\/ {\tt next{\Under}level\/} 
                to overlapping grids in\/ {\tt new{\Under}level}}}
\Code{\Tab {\it destroy all grids in\/ {\tt next{\Under}level}}}
\Code{\Tab {\it move grids from\/ {\tt new{\Under}level\/}
                into\/ {\tt next{\Under}level}}}
\Code{\Tab if ( next{\Under}level.contains{\Under}grids() ) \OpenBrace}
\Code{\Tab\Tab next{\Under}level.regrid{\Under}level();}
\Code{\Tab \CloseBrace}
\Code{\CloseBrace}
\BoxBottom
\Caption
Skeleton code for the new regridding algorithm, written in \Cplusplus.
A {\tt Level\/} object contains the grids at one level of refinement.
The function {\tt regrid{\Under}tree\/} regrids the hierarchy starting
at the current level: grids at all finer levels are replaced.
Child grids are always properly nested within their parents and hence
multiple parents are avoided.
\bigskip
\endCaption
\endfigure

The regridding procedure used in the present work avoids nesting
problems and the need for multiple parents by reversing the order in
which the levels are regridded (coarse to fine instead of fine to
coarse) and making a preliminary pass through the grid hierarchy to
prepare the error flags.
To ensure that no useful information is lost at the finest levels,
before any alterations are made to the grid hierarchy the error flags
on the coarser levels are made up-to-date.
Starting at level~$N=N_{\SmallWord{finest}}-1$ and working down
to level~$N_{\SmallWord{regrid}}$, for each grid at level~$N$
error flags are set at all points at which errors are flagged in its 
child grids.
(The function {\tt flag{\Under}errors{\Under}in{\Under}tree} performs
this by recursion in figure~\use{Fg.AmrRegrid}.)
The actual regridding then proceeds from 
level~$N=N_{\SmallWord{regrid}}$ up to level 
$N_{\SmallWord{finest}}$.
For each grid at level~$N$ error flags are clustered and
new child grids are created.
These new grids are replacements for the existing grids at 
level~$N+1$, but before any grids are destroyed data and, importantly,
error flags are copied from old grids to new wherever there is overlap
between them.
(The function {\tt regrid{\Under}level} performs
this by recursion in figure~\use{Fg.AmrRegrid}.)
Since the grid hierarchy is always rebuilt starting from the coarsest
levels, and since a child grid is always properly nested
within a single parent when it is created, multiple parents are not
a problem when this new regridding procedure is used.

\SubSection{Variable Time Steps}
The Courant-Friedrichs-Lewy condition (equation~\EqNum{PdeCFL}{}) puts
an upper limit on the sizes of times steps that can be taken when
evolving numerical solutions to hyperbolic systems.
Since the upper limit usually varies during the course of the
evolution it is common practice for numerical codes to allow the time
steps to adapt, with each new time step being chosen based on an
estimate of the current Courant-Friedrichs-Lewy upper limit.
However, because of the complexity of the hierarchical time-stepping
and the use of double time steps in the error estimation procedure,
variable time steps are not as straightforward to implement in an AMR
code as they are in a traditional single-grid code.
(Examples of AMR codes using variable time steps are Hamad\'e~1997
and Berger and LeVeque~1997.)

The AMR code used for the present work allows variable time steps, but
with limitations.
The time steps can be changed only at the same time that level one of
the grid hierarchy is regridded.
(Recall that all grids on the same level are advanced with the same
time step and that the time step for level~$N$ grids is
$\Delta t / \hbox{\tt Refine}^{(N-1)}$ where $\Delta t$ is the base
level time step; obviously when the time step is changed it is changed
consistently throughout the grid hierarchy.)
This means that the time step size can only be altered after the grids on 
the coarsest level have taken {\tt Regrid} time steps.

A side effect of the implementation of variable time steps concerns
the production of internal boundary data for sub-grids.
The interpolation of fine boundary data from coarse parent grids is
performed temporally as well as spatially.
Since data are stored on each grid for three stages in time (the
current time and the two previous time steps; this is required by the
error estimation routine), time interpolation could be performed at
second order.
However, while this was done in an early version of the AMR code,
the implementation of variable time steps restricts interpolation in 
time to being first order.

The value of variable time steps is exemplified by the Gowdy
cosmological collapse simulations of chapter~6.
The Courant-Friedrichs-Lewy limit on the time steps in that problem 
increases exponentially with time, although in practice a
limit lower than this must be accepted if the source terms in the system 
are to be dealt with accurately at late times.
By allowing time steps to increase during the simulations (but only up
to some preselected maximum), a dramatic decrease in running time is
achieved compared to simulations which use fixed time steps.

\SubSection{Neighbouring Grids}
The AMR algorithm developed for the present work gives special
attention to grids on the same level which abut, that is share a
common edge.
Each grid maintains a list of the grids that abut it, and data from
these neighbouring grids can be used during some operations performed
by the AMR algorithm.
Most significantly, abutting grids provide boundary data for each
other along their common edges, and these are used in preference to data 
interpolated from a parent grid (see Berger and Colella~1989).
Similarly, boundary data from neighbouring grids can be used for 
the `coarsened steps' taken by grids during the error estimation procedure.
In addition, as part of the error clustering process the error flags
on each grid are extended by a buffer region to pre-empt the movement
of features in the evolved solution, and these buffer regions are
allowed to cross into neighbouring grids.

By keeping track of neighbouring grids 
the AMR algorithm ensures that the results produced at grid interfaces
are of no lesser accuracy than the results produced elsewhere.
However, it could be argued that the number of grids which abut is
unlikely to be large enough to justify
the increase in algorithm complexity required to take them into account.
Furthermore, in one dimension there should be no
neighbouring grids at all: two abutting grids can always be combined
to form one single grid.
The motivation for implementing communication between neighbouring
grids in the present work is a consequence of the way in which the 
AMR code has been parallelized.
As section~\use{sec.AmrParallel} explains, large grids can be split 
into smaller abutting grids by the clustering process to improve code
performance, and this can lead to a significant increase in the number of
abutting grids even in the one-dimensional case.

Many of the numerical simulations performed for the present work use
periodic boundary conditions and in such cases the situation can
arise in which two grids abut along external (periodic) boundaries
rather than along internal ones.
In fact when periodic boundaries are used it is possible to have a
grid which abuts itself on all sides.
Periodic boundary conditions are sufficiently important in the present
work that the AMR code takes them into account when constructing lists
of neighbouring grids.

\Section{Structure of the AMR Code
         \label{sec.AmrStructure}}
The complexity of the AMR algorithm calls for more
sophisticated programming techniques than are usually
necessary for traditional numerical codes.
In particular, an AMR code is likely to make extensive use of
dynamic memory allocation, data structures and recursive functions,
and its lack of support for these programming constructs makes
implementing an AMR algorithm in Fortran~77 difficult
(though certainly not
impossible, as Berger and LeVeque~1997 demonstrates).
The AMR code used in the present work is written in
the \Cplusplus\ language (Stroustrup~1997, Lippman and Lajoie~1998) 
and the object-oriented features of that language are used to develop 
data structures within the code that reflect the conceptual structure 
of the algorithm.
In particular the implementation respects the distinction between the 
AMR algorithm itself and the algorithm used to evolve data on grids,
and makes it easy to link different evolution procedures to the same 
AMR code.

In this section the AMR algorithm is decomposed into four modules:
clustering, grid management, grid evolution and grid interaction.
(The term module is used here to refer to a collection
of functions or classes
which share a common purpose, and usually the same source files.)
The interfaces between the modules are simple and well defined, and
in principle each could be implemented by a different programmer
in a different programming language.
Some of the object-oriented capabilities of \Cplusplus\ which
facilitate the implementation of these modules are discussed.
The modules are entirely independent of the problem they are used to
solve.
(For the sake of clarity the names given to classes and functions in
the following explanation often differ from the names used in
the program source code.)

\SubSection{Clustering and Grid Management}
The first point to note is that the clustering phase of the AMR
algorithm (see section~\use{sec.AmrOutline}) is completely
self-contained, and should be implemented in a way that respects this.
The clustering module takes an array of error flags and
returns a list of regions where new sub-grids should be
created.
Its complex operation can be treated as a `black box' by other parts
of the AMR code.
(Illustrating the independence of modules, during extensive writing
and rewriting of the AMR code the clustering module,
written by J.~M.~Stewart, has not needed altering at all.)

At the highest level of abstraction, the AMR algorithm is concerned
solely with the manipulation of
grids in the grid hierarchy: it creates and destroys
grids, advances them in time, orchestrates the exchange of boundary
data between neighbouring grids, and so on
(recall figure~\use{Fg.AmrPicture}).
These aspects of the AMR algorithm form the grid management module.
Although the module is, in terms of source code, relatively small, it
contains the most complicated parts of the AMR algorithm.
To a large extent the module is independent of the dimension of
the grids it manages.

It is natural to create a data structure to represent grid
objects that are manipulated by the grid management module,
and figure~\use{Fg.AmrAmrgrid} shows the outline of a
suitable \Cplusplus\ class, {\tt AmrGrid}.
The contents of the class can be divided into three basic groups.
The first group comprises information used directly by the grid 
management module to fit grids into the grid hierarchy,
for example the size and coordinate origin of a grid and the
details of its sub-grids.
The second group comprises functions through which the grid management
module manipulates the solution data stored on grids, for example a function
{\tt advance{\Under}grid} which integrates the data in time.
The final group comprises additional data structures and functions
used by the grids themselves; in particular it would be useful for each grid
to maintain an array of error flags.

\figure{AmrAmrgrid}
\BoxTop
\Code{class AmrGrid \OpenBrace}
\Code{public:}
\Code{}
\Code{\Tab // information used for grid management}
\Code{\Tab {\it location of grid within spatial domain of the problem}}
\Code{\Tab {\it relative position of grid in hierarchy:
                pointer to parent, list of children, etc.}}
\Code{}
\Code{\Tab // functions for manipulating solution data}
\Code{\Tab virtual void advance{\Under}grid(double dt) = 0;}
\Code{\Tab virtual void flag{\Under}errors(double tolerance) = 0;}
\Code{\Tab virtual AmrGrid* make{\Under}subgrid%
                             ({\it sub-grid details}) = 0;}
\Code{\Tab {\it etc.}}
\Code{}
\Code{\Tab // data and functions mainly for internal use}
\Code{\Tab {\it array of error flags}}
\Code{\Tab {\it functions for manipulating error flags
                (buffering, etc.)}}
\Code{}
\Code{\CloseBrace;}
\BoxBottom
\Caption
Sketch of the {\tt AmrGrid} class.
It is an abstract \Cplusplus\ class which represents the 
parts of a grid used by the grid management module.
It contains no actual solution data.
\bigskip
\endCaption
\endfigure

One thing conspicuously absent from the {\tt AmrGrid} class is the
actual solution data that a grid stores.
The reason for this is that to specify a format for the
solution data is to limit the AMR code to that format;
among other things the code would be tied to either
the grid point or the grid cell framework.
Since the grid management module is not actually affected by the data
format, there is no reason for the {\tt AmrGrid} class to be, and
instead it is declared an abstract class: by itself it is
incomplete and objects of its type cannot be used in the AMR code.
The member functions designed to manipulate the solution data are declared
to be pure virtual functions by the syntax
\par\centerline{
{\tt virtual} {\it function} {\tt = 0;}
}\noindent
in the class declaration (figure~\use{Fg.AmrAmrgrid}), and this
means that they have no definition within the {\tt AmrGrid} class but
must instead be defined in a class derived from it (as demonstrated in the
next subsection).

The clustering and grid management modules by themselves form a basic
AMR software package. 
The next subsection explains how extensions to the package
allow it to work with different data formats and integration methods.

\SubSection{Grid Evolution and Grid Interaction}
A grid evolution module comprises two elements:
solution data, and an algorithm for time-stepping the data.
By itself such a module is the central part of a standard single-grid
numerical code, and rather than describing how a grid evolution module
can be especially written for use in the AMR code, this subsection
explains how a pre-existing grid evolution module can be connected
to the grid management module of the previous subsection.
Figure~\use{Fg.AmrDatagrid} gives an example grid evolution module,
a \Cplusplus\ class called {\tt DataGrid}.
In practice this class could be a `wrapper' for a time-stepping
procedure written in a different programming language with each
instance of the class being allocated its own set of solution data
in the format required by the time-stepper.

\figure{AmrDatagrid}
\BoxTop
\Code{class DataGrid \OpenBrace}
\Code{public:}
\Code{}
\Code{\Tab {\it array of solution data}}
\Code{\Tab void timestep(double dt);}
\Code{}
\Code{\CloseBrace;}
\BoxBottom
\Caption
Outline of a basic grid evolution module, the {\tt DataGrid} class.
Each {\tt DataGrid} object contains solution data which it
is capable of evolving in time using an appropriate approximation
scheme.
\bigskip
\endCaption
\endfigure

The {\tt AmrGrid} class and the {\tt DataGrid} class represent
different aspects of the grids used in the AMR algorithm.
The two classes are unified within the grid interaction module:
in figure~\use{Fg.AmrHygrid} multiple inheritance is used to create
a class {\tt HyGrid} that contains all of the functions and data of
both the {\tt AmrGrid} class and the {\tt DataGrid} class.
Together with a battery of functions for copying, coarsening and 
interpolating data, the {\tt HyGrid} class makes up the grid interaction
module.
This bridges the gap between the grid management module and the grid
evolution module, and completes the AMR package.

\figure{AmrHygrid}
\BoxTop
\Code{class HyGrid : public AmrGrid, public DataGrid \OpenBrace}
\Code{public:}
\Code{}
\Code{\Tab // implementation of AmrGrid's pure virtual functions}
\Code{\Tab virtual void advance{\Under}grid(double dt) 
           \OpenBrace\
           DataGrid::timestep(dt); {\it \dots} 
           \CloseBrace}
\Code{\Tab virtual void flag{\Under}errors(double tolerance);}
\Code{\Tab virtual AmrGrid* 
           make{\Under}subgrid({\it sub-grid details});}
\Code{\Tab {\it etc.}}
\Code{}
\Code{\Tab // functions and data for internal use}
\Code{\Tab {\it solution data from previous two time 
                steps---used for error estimation}}
\Code{\Tab {\it functions for data interpolation}}
\Code{\Tab {\it etc.}}
\Code{}
\Code{\CloseBrace;}
\BoxBottom
\Caption
The {\tt HyGrid} class is crossbred from the {\tt AmrGrid} and 
{\tt DataGrid} classes of figures~\use{Fg.AmrAmrgrid}
and~\use{Fg.AmrDatagrid}. 
Multiple inheritance allows the data and functions of the two base 
classes to be combined in a single class.
The {\tt HyGrid} class completes the definitions of the pure virtual
functions declared in the {\tt AmrGrid} class.
It is the centre-piece of the grid interaction module.
\bigskip
\endCaption
\endfigure

Inheritance in \Cplusplus\ makes it simple to create new classes that
are modified versions of existing classes or that, like the 
{\tt HyGrid} class, combine the features of two or more classes.
More than this though, an instance of a derived class such as the 
{\tt HyGrid} class can be used just as if it were an instance of one
of the base classes: the grid management module will manipulate a
hierarchy of {\tt HyGrid} objects without being aware that they are
not the {\tt AmrGrid} objects it has been programmed to use.
Virtual functions complete the illusion:
whenever the grid management module calls (for example) the 
{\tt advance{\Under}grid} function of the {\tt AmrGrid} class, it is
actually the definition of the corresponding function in the 
{\tt HyGrid} class that is used.
(This is fortunate since {\tt advance{\Under}grid} is a pure virtual
function in the {\tt AmrGrid} class and has no definition there.)

One of the member functions inherited by the {\tt HyGrid} class from
the {\tt AmrGrid} class deserves more detailed consideration.
The function {\tt make{\Under}subgrid} causes a {\tt HyGrid} object
to create a new {\tt HyGrid} object, a sub-grid of the original grid.
The grid management module always uses this function to create grids 
rather than creating them itself, and the reason for
this is quite subtle.
In the grid management module only one type of grid, the 
{\tt AmrGrid} class, is recognized, and if that
module were to create new grids directly they would be instances 
of that class.
However the grids in the grid hierarchy are actually 
{\tt HyGrid} objects; because the {\tt HyGrid} class is derived
from the {\tt AmrGrid} class, instances of the former can be used as
if they were instances of the latter.
The problem is that the grid management module does not know enough
about the grids it is using to create one directly, and so it uses 
{\tt make{\Under}subgrid}, a member function of the {\tt AmrGrid}
class, to do the job for it.
Since {\tt make{\Under}subgrid} is a virtual function it can be
implemented so that it always creates a sub-grid which has the same
class as the parent grid, and this ensures that all the grids in the grid 
hierarchy have the same type.
(A similar problem occurs when the grid management module destroys a
grid, but the solution is slightly more technical and beyond the scope
of this discussion.)

The AMR code developed for the present work actually includes two
different grid evolution modules, each with an associated grid
interaction module.
During the early stages of its development the AMR code stored data at
grids points and time-stepped them using a two-step Lax-Wendroff method.
Independently, a single grid code, \Clawplusplus, was written that uses
an advanced high-resolution method to evolve data stored on a mesh of
cells (chapter~2 discusses this code).
By adopting the modular construction described in this section in the
AMR code, the \Clawplusplus\ code (transformed into a grid evolution 
module and outfitted with a grid interaction module) has become
interchangeable with the original Lax-Wendroff integration
routine and its associated data format.
This approach has reduced the amount of source code needed to compile
several different numerical codes and has been found to be very
flexible; the same source code is compiled to produce both the standalone
\Clawplusplus\ code and the related grid evolution module used in
the AMR code.
In addition, comparing the results produced by the two pairings
of grid evolution and interaction modules has proved to be a useful
way of identifying programming errors.

\SubSection{Problem-Specific Code}
The modules described in the previous two subsections make up the complete
AMR package.
However some additional code must be supplied before the
package can be used to solve specific problems:
the grid evolution module must be provided with the details of the
equations it is to solve, typically in the form of procedures to evaluate
flux and source terms; 
boundary conditions must be specified at the edges of the spatial
domain; and initial data for the evolution must be set up.
In addition, since data output is outside the scope of the AMR
package, suitable routines for saving the results of the simulation
must be written.

For some problems it may be useful for each grid in the hierarchy to
carry with it auxiliary data fields that are not declared as part of
the {\tt HyGrid} class.
For example, in a code for evolving solutions to the Frittelli-Reula
equations (described in chapter~4) there might be some advantage to
storing data describing the background gauge fields on each grid in
addition to the data stored for the unknown variables.
Instead of rewriting the {\tt HyGrid} class to include the extra
data a neater solution is to create a new class, {\tt FRGrid}, which
derives from the {\tt HyGrid} class and which makes additions and
alterations to the latter to account for the additional data.
(In particular, the {\tt make{\Under}subgrid} function must be
updated in the {\tt FRGrid} class for the reasons given in the
previous subsection.)

A driver program is used to initiate the AMR algorithm.
It creates a grid of the appropriate type (either a {\tt HyGrid} 
object or an {\tt FRGrid} object) to form the first level of the
grid hierarchy, and presents it to the AMR package.
It then sets the AMR algorithm in motion.

\Section{Parallelization of the AMR Code
         \label{sec.AmrParallel}}
On a multiprocessor system, program performance can be improved by
sharing the program's workload among a number of processors.
To allow this, the program must have been written with parallel
execution in mind: its workload must be decomposable into separate
tasks that can be performed independently.
Ideally the time taken to execute a program should be inversely
proportional to the number of processors employed, but this is rarely
the case since a program cannot usually be parallelized in its
entirety (and, in addition, hardware limitations are a factor).

The AMR code developed for the present work has been parallelized to 
run on a Silicon Graphics Origin2000 multiprocessor 
(see Fier~1996 and Cortesi~1998 for technical details).
The machine uses a shared memory architecture which allows all of the
processors direct access to the full memory of the computer.
To avoid bottlenecks, memory is physically distributed and as a
consequence memory access times vary depending on which processor is
accessing which memory address.
To compensate for this the Origin2000 makes extensive use of data and
instruction caching.

\SubSection{Approaches to Parallelization}
This subsection weighs up the advantages and disadvantages of some of
the possible approaches to parallelizing an AMR code.

Nearly all of an AMR code's time is spent performing data manipulation
operations such as time-stepping the solution data on a grid or
interpolating data for use in a sub-grid.
One approach to parallelizing the code would be to individually
parallelize each of these operations;
for instance, when the AMR algorithm calls for the data on a grid to
be time-stepped, all of the processors at the code's disposal could be
employed for the operation.
The advantage of this approach is that, if the individual operations
are parallelized well, a very even distribution of work between
processors should be achieved.
However the approach suffers from two disadvantages.
Firstly, since all of the processors operate in parallel on the same
data (for example, all helping to time-step the same grid) performance
is certain to be hampered by memory contention and poor access times
due to memory locality.
The other disadvantage is that a lot of work would be needed to
implement parallelization in this form: each operation to be
parallelized would need special treatment, and for some operations
(such as time-stepping by a high-resolution method) the
parallelization would be particularly complicated.
In addition, since parallelization by this approach affects only the
grid evolution and the grid interaction modules (to use the
terminology of section~\use{sec.AmrStructure}) it would need to be
implemented anew for each time-stepping algorithm included in the AMR
code. 

Another approach to AMR parallelization would be to assign grids to
different processors such that all the data manipulation operations on
one grid are performed by the same processor.  
The majority of operations can be performed for the grids on one level
independently (for example, all of the grids on level~$N$ can take one 
time step independently, provided they have appropriate boundary
data), and so the code can be run in parallel as long as the
processors are synchronized between stages.
This is the approach adopted for the present work, and
Quirk and Hanebutte~(1993) also use it for an AMR code built on
a message passing paradigm.
Since grids are to a large extent independent entities there should
be few memory conflict problems in a code parallelized in this way.
The approach is also simple from the programming point of view,
and all of the parallelization takes place in the grid management
module.
The main drawback with the method is that for good performance it
relies on grids being assigned to processors in such a way that each
processor has the same workload, and this may be difficult to 
achieve in practice; it may be necessary to break big grids into a lot
of much smaller grids.

Wild~(1996) has developed an AMR code that is based on a different
design to the standard algorithm described in
section~\use{sec.AmrOutline}, and it is interesting to consider this
code in the context of AMR parallelization.
Instead of overlaying grids of varying resolutions to adaptively
refine data, Wild's `hierarchical linked list' approach uses a single
base grid into which additional points are inserted to increase the
local resolution.
This method avoids the need for clustering during the regridding phase
of the algorithm, and does not wastefully refine regions that are not
flagged as needing refinement.
However its extensive use of linked lists does add a degree of
complexity to many of the operations that are used to manipulate
solution data, and the code's performance may suffer because of this.
Wild's design can be thought of as a standard AMR code in which
only very small, uniformly sized sub-grids are used, and as such
it seems to lend itself to a parallelization approach midway between the
operation-based and the grid-based approaches of the preceding
paragraphs (see Wild~1996, appendix~A); however it is possible that the
disadvantages described above would affect a parallelization of Wild's
code less severely than they would a standard AMR code.

The final parallelization approach discussed here offers very good
performance, but is probably unusable in practice.
If the spatial domain of the problem is split into separate
regions then each could be evolved by an AMR code executing on a
different processor, with synchronization between processors only
being needed at the end of each coarse time step to allow boundary
data to be exchanged.
This approach would be simple to implement and would almost completely
avoid memory contention problems.
However it suffers from two serious problems:
it only achieves a good workload balance for simulations in which 
refined regions are evenly distributed across the spatial domain,
and it suffers from a loss of accuracy in the refined data at 
the `physical boundaries' between the domains of the different processors.
While both of these problems could in principle be circumvented, it
seems unlikely that the approach could be made to work wholly
satisfactorily in the general case.

\SubSection{Grid-Based Parallelization and Scheduling}
The AMR code used in the present work has been parallelized using a
grid-based approach, briefly described in the previous subsection.
Operations such as integration of solution data and error flagging can
be performed simultaneously for all of the grids at one level of the
grid hierarchy, and the code exploits this by assigning grids to
different processors.
Figure~\use{Fg.AmrParallelize} shows how simple modifications to the
code (all within the grid management module) allow the grid operations
to be performed in parallel. 

\figure{AmrParallelize}
\BoxTop
\Code{// parallel code to replace:}
\Code{// for ( {\it each grid\/ {\tt gr\/} in this level} )
         gr.do{\Under}something();}
\Code{}
\Code{\Hash pragma parallel}
\Code{\OpenBrace\ // begin parallel region}
\Code{}
\Code{\Tab int this{\Under}cpu = mp{\Under}my{\Under}threadnum();}
\Code{\Tab for ( {\it each grid\/ {\tt gr\/} in this level} ) \OpenBrace}
\Code{\Tab\Tab if ( gr.cpu == this{\Under}cpu ) gr.do{\Under}something();}
\Code{\Tab \CloseBrace}
\Code{}
\Code{\CloseBrace\ // end parallel region}
\BoxBottom
\Caption
Code fragment to illustrate parallelization.
The member function {\tt do{\Under}something} is called for each grid
in the current level.
The code is made to run in parallel by assigning grids to different
processors. 
\bigskip
\endCaption
\endfigure

Grids are assigned to processors at the time of their creation
and for the code to give good performance it
is important that the same processor be used for all operations on a
grid, particularly those involving memory allocation.
The assignment of grids to processors needs to be done such that the
total amount of work required to deal with the grids on one level is
distributed among the processors as evenly as possible, and
how well this is done is the main factor affecting the reduction in
execution time of the code when it runs in parallel.
During the regridding of level~$N$ a list of suggested sub-grids
for level~$N+1$ is generated by the clustering module, and for a code
running in parallel this list must be analysed before the
sub-grids are created to decide on a mapping of grids to
processors. 
This involves two stages: grid splitting and grid scheduling.

The number of cells making up a grid is taken to be a measure of 
the amount of work required to perform operations on that grid.
(This is only a first approximation and it does not take into
account factors such as cache usage and number of boundary cells.)
Suppose that~$n$ grids are to be created at the current level by the
regridding procedure, and that these grids are estimated as having
work requirements $w_1$, \dots,~$w_n$.
The total amount of work needing to be performed on this level is then
$$
W_{\SmallWord{total}} = \sum_{i=1}^{n} w_i \, ,
$$
and if that work is distributed evenly between~$m$ processors
the workload of each will be
$$
W_{\SmallWord{ideal}} = {W_{\SmallWord{total}} \over m} \, .
$$
In assigning grids to processors the aim is to minimize the amount by
which the workload of the most overworked processor 
exceeds~$W_{\SmallWord{ideal}}$.
This problem is known as task scheduling in the computer science
literature, and good results can be obtained by using a simple
heuristic method, described below.

When assigning grids to processors in an AMR code an additional element
of freedom is available: large grids can be split into a number
of smaller grids.
For the measure of work used here, the value 
of~$W_{\SmallWord{total}}$ is not affected if grids are split into
smaller grids, although in practice there will be an overhead for each
additional grid created.
(On the other hand, cache behaviour may mean that splitting grids up
sometimes reduces the total workload.
This effect was investigated, but it was found to be too complicated to
make allowances for in the code.)
For the present work, the creation of large numbers of small grids is
avoided and splitting of grids is limited to one frequently encountered
situation: if a single grid has a work requirement that 
exceeds~$W_{\SmallWord{ideal}}$ then that grid is split into smaller
grids, all of them requiring less than~$W_{\SmallWord{ideal}}$ units
of work.
This avoids the inefficient case in which there are less grids on a
level than there are processors available to deal with them.
A side-effect of the grid splitting is that,
since~$W_{\SmallWord{ideal}}$ depends on the number of processors
being used, so does the final arrangement of grids, and consequently
so do the code's results.
While this may not be an ideal situation, the dependence of the results
on the number of processors is in practice only slight, and splitting of
grids allows substantial improvements in performance.

When all of the grids are of an appropriate size, the code
assigns them to processors using the
LPT (longest or largest processing time) scheduling rule.
This simple but effective algorithm is discussed by
Horowitz, Sahni and Rajasekaran (1998, sections~12.3.1 and~12.4.1)
and Cosnard and Trystram~(1995, chapter~10).
It operates by iteration of the following step:
of the remaining grids, assign the one with the highest work
requirement to the processor with the smallest workload.
While this algorithm is not guaranteed to produce an optimal grid
schedule, its results are usually good, and the following bound can be
proven (Horowitz, Sahni and Rajasekaran~1998, theorem~12.5):
$$
W_{\SmallWord{LPT}} \leq
W_{\SmallWord{optimal}} ( {4 \over 3} - {1 \over 3 m} )
\, ,
$$
where~$W_{\SmallWord{LPT}}$
and~$W_{\SmallWord{optimal}}$ are the workloads of the most
overworked processors for an LPT schedule and an optimal schedule,
and~$m$ is the number of processors.
It is usually not practical to try to calculate an optimal schedule 
directly (the problem is NP-complete).

The performance of the parallelized AMR code is analysed in the next
subsection. 
Although the scheduling system described here works well, there is
room for improvement in all aspects of it: the workload estimation
could be made more accurate, a more advanced task scheduling algorithm
could be used and more sophisticated use could be made of grid
splitting. 

\SubSection{Performance of the Parallel AMR Code}
The speedup of a program when it runs using multiple processors is
calculated as
$$
\hbox{speedup}(m)
    = { \hbox{completion time on one processor}
  \over \hbox{completion time on } m \hbox{ processors} } \, .
$$
Ideally the speedup will be equal to the number of
processors the program employs, but in practice hardware limitations and
incomplete program parallelization mean that the speedup is generally
less than this.

Figure~\use{Fg.AmrSpeedup} shows speedup results for a typical
simulation performed by the parallelized AMR code.
The actual (elapsed time) speedup is compared with the speedup that
would be expected based on the processor workload schedules generated
during the regridding process.
(In producing these results, grids were split to fit the number of
available processors, and
consequently each speedup value in figure~\use{Fg.AmrSpeedup} was
calculated based on the times of two slightly different 
evolutions.)

\figure{AmrSpeedup}
\DrawFig{
  \epsfxsize=4in
  \epsfbox{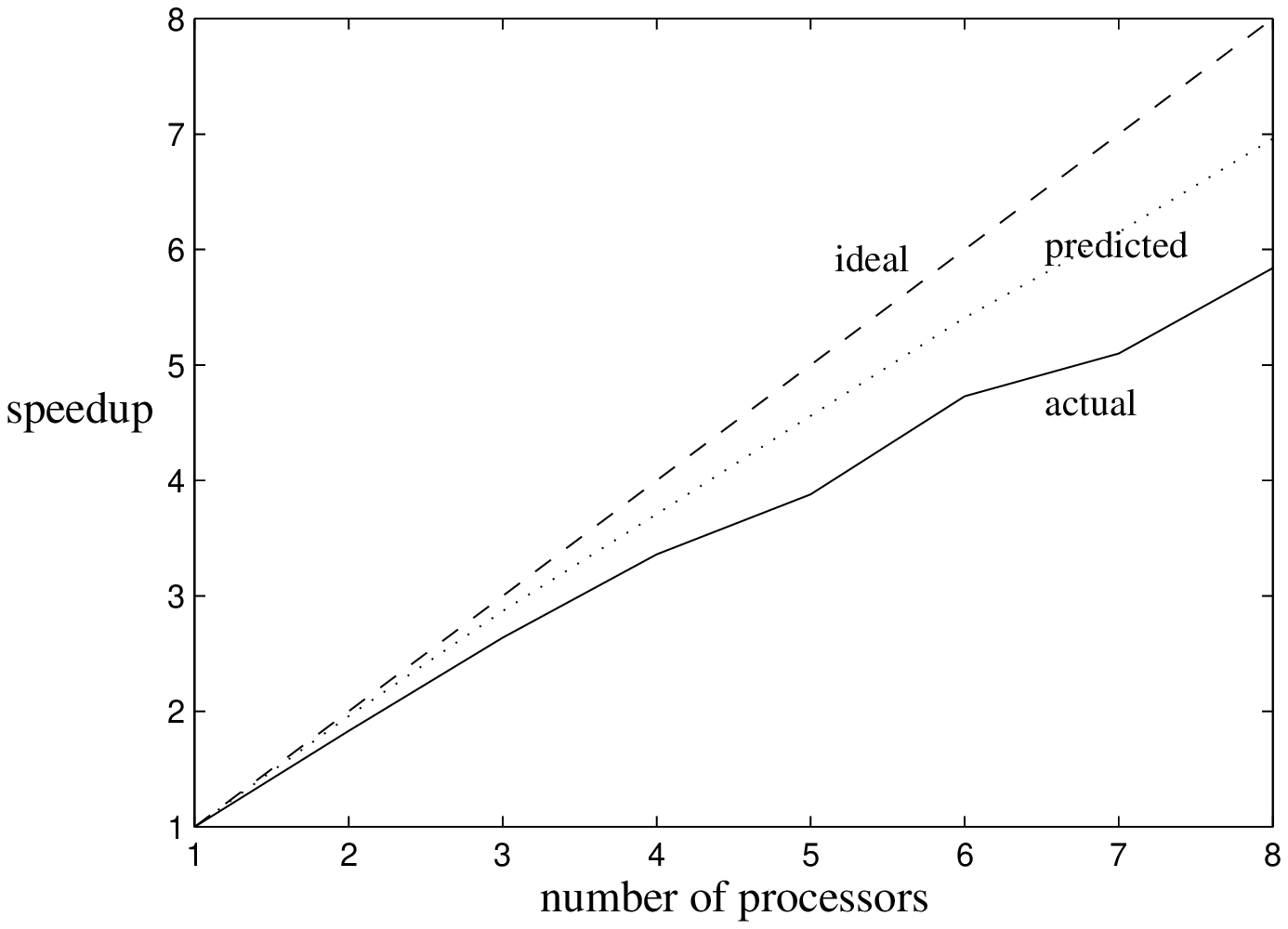}
}
\Caption
Typical speedup results for the parallelized AMR code.
The dotted line is the speedup predicted from the processor workload
schedule. 
The solid line is the actual speedup measured.
The dashed line shows the ideal case of linear speedup.
\bigskip
\endCaption
\endfigure

The speedup predicted from the distribution of work between
processors is the maximum speedup that could be expected from
the code.
It falls short of the ideal case of linear speedup because the 
workload scheduling procedure described in the previous subsection is
not able to produce a perfectly even distribution of work between
processors.
For the simulation used to produce the results in
figure~\use{Fg.AmrSpeedup}, as with all of the simulations performed
for this work, only one grid was used in the first level of the grid
hierarchy, and so operations on that level were never run in parallel. 
Allowing the code to split the base grid would improve the code's
parallel performance, particularly for simulations which use only a
small amount of refinement, although in general the base grid
accounts for only a very small amount of the overall work performed by
the code.
For the results plotted in figure~\use{Fg.AmrSpeedup}, splitting the
base grid would increase the predicted speedup of the code to
about~7.2 (from~7.0) in the eight processor case.

There are several reasons why the actual speedup is less than the
speedup predicted from the workload distribution results.
First of all, the work scheduling procedure is quite unsophisticated,
and the amount of `work' it assigns to a processor is only a first
approximation to the amount of time that the processor will be
occupied for. 
Secondly, the predicted speedup does not account for grid management
operations that the code cannot perform in parallel.
However, excluding the output of solution data to disk, these
operations take up an insignificant amount of the code's time:
execution profiles for the simulation used in
figure~\use{Fg.AmrSpeedup} reveal that only a tiny fraction of a percent of
the code's time is spent performing functions which are not parallelized.
Thirdly, hardware factors (memory contentions, lack of cache reusage
and inefficient data placement) add an overhead to the code's running
time when multiple processors are used.
Profiles of the simulation used in figure~\use{Fg.AmrSpeedup} show
that the total time spent performing data manipulation
operations increases as the number of processors increases.
For eight processors the increase is approximately seven percent, and
this accounts for about half of the discrepancy between the predicted
speedup and the actual speedup.
(An additional factor affecting the code's speedup results is the 
environment in which it runs: in a multi-user system several tasks may
be competing for the available processors, and if one of the
processors used by the code is delayed by another task the remaining
processors must wait for it to catch up.
This skews the timing measurements.)

The parallel performance of a code is usually described using Amdahl's
law, a formula which relates speedup to the fraction~$p$ of the code
that can be run in parallel: 
$$
\hbox{speedup}(m) = { 1 \over (p/m) + (1-p) } \, ,
$$
where~$m$ is the number of processors used (see Fier~1996).
Although this simplistic breakdown into serial parts and
parallel parts is not altogether appropriate for the AMR
code, for comparison purposes approximate values
of~$p$ can be calculated for the
data plotted in figure~\use{Fg.AmrSpeedup}.
The predicted speedup corresponds to a parallel fraction of
$p \simeq 0.97$ while the actual speedup has $p \simeq 0.93$.

The discussion here has been illustrated with performance statistics
from a `typical' AMR simulation. 
In fact the parallel performance can vary significantly from one
simulation to another, and from one set of input parameters to
another, with the best results usually being produced when a
large amount of refinement is used.
The results shown in figure~\use{Fg.AmrSpeedup} are from a
two-dimensional gas dynamics test simulation (an example problem from 
LeVeque~1998) and represent good-to-average performance of the AMR code;
much better code performance has been observed, together with much
which was significantly worse.

\Section{AMR in Practice
         \label{sec.AmrPractice}}
This section concludes the discussion of AMR by looking at how the
algorithm operates on some example problems and by examining the
general advantages and disadvantages of the method.

\fullfigure{AmrDepths}
\vss
\DrawFig{
  \epsfxsize=0.9\hsize
  \epsfbox{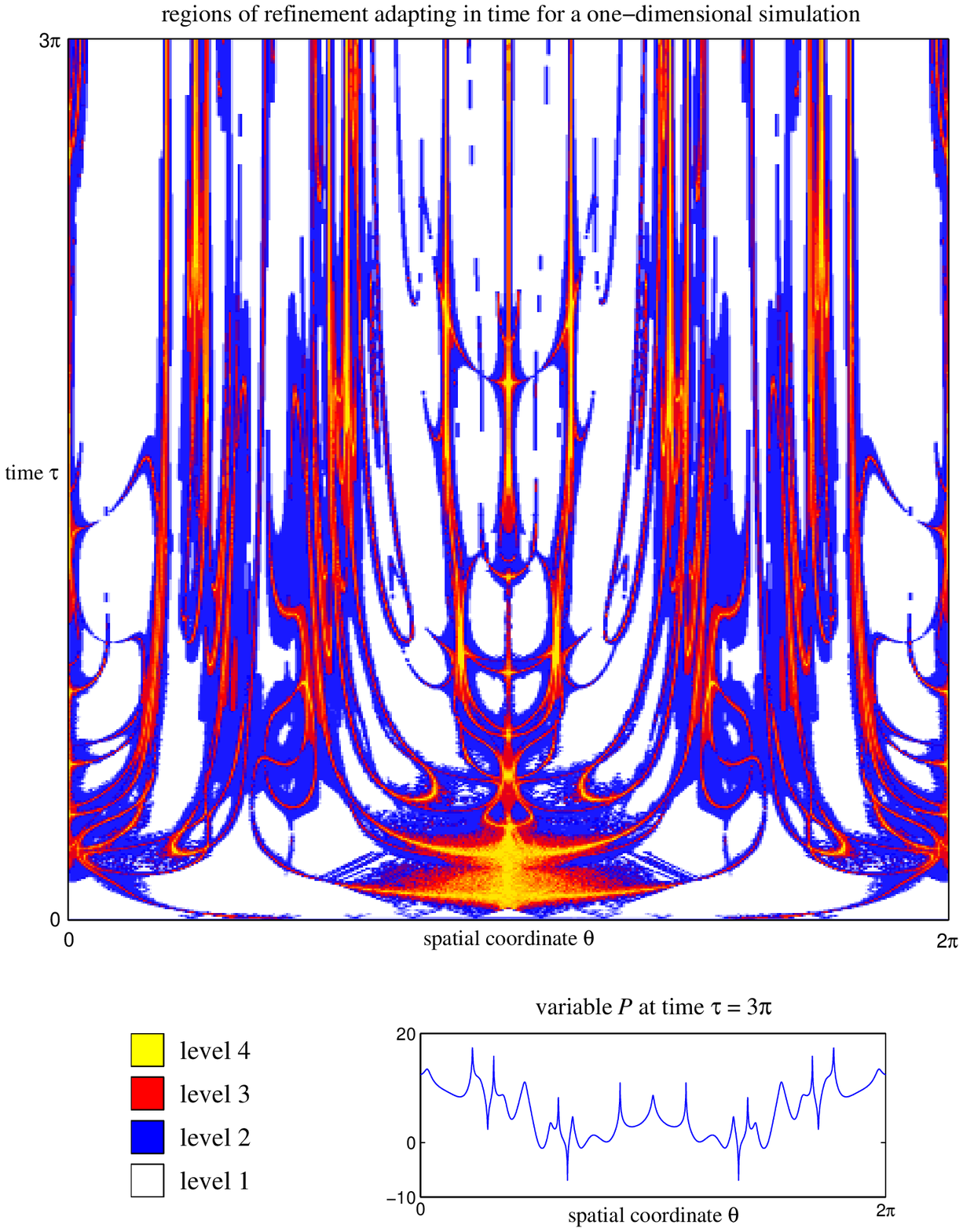}
}
\Caption
The AMR grid hierarchy adapts over time to follow features of the
evolving solution data.
The top frame shows the level of refinement used by the AMR code
varying in space and time for a simulation of a one-dimensional
cosmological model (see chapter~6 for details).
The bottom frame shows one of the variables at the
end of the simulation.
\vskip2mm
\endCaption
\endfigure

\fullfigure{AmrLayout}
\vss
\DrawFig{
  \epsfxsize=0.76\hsize
  \DrawHiLoRes{\epsfbox{layout.eps}}
              {\epsfbox{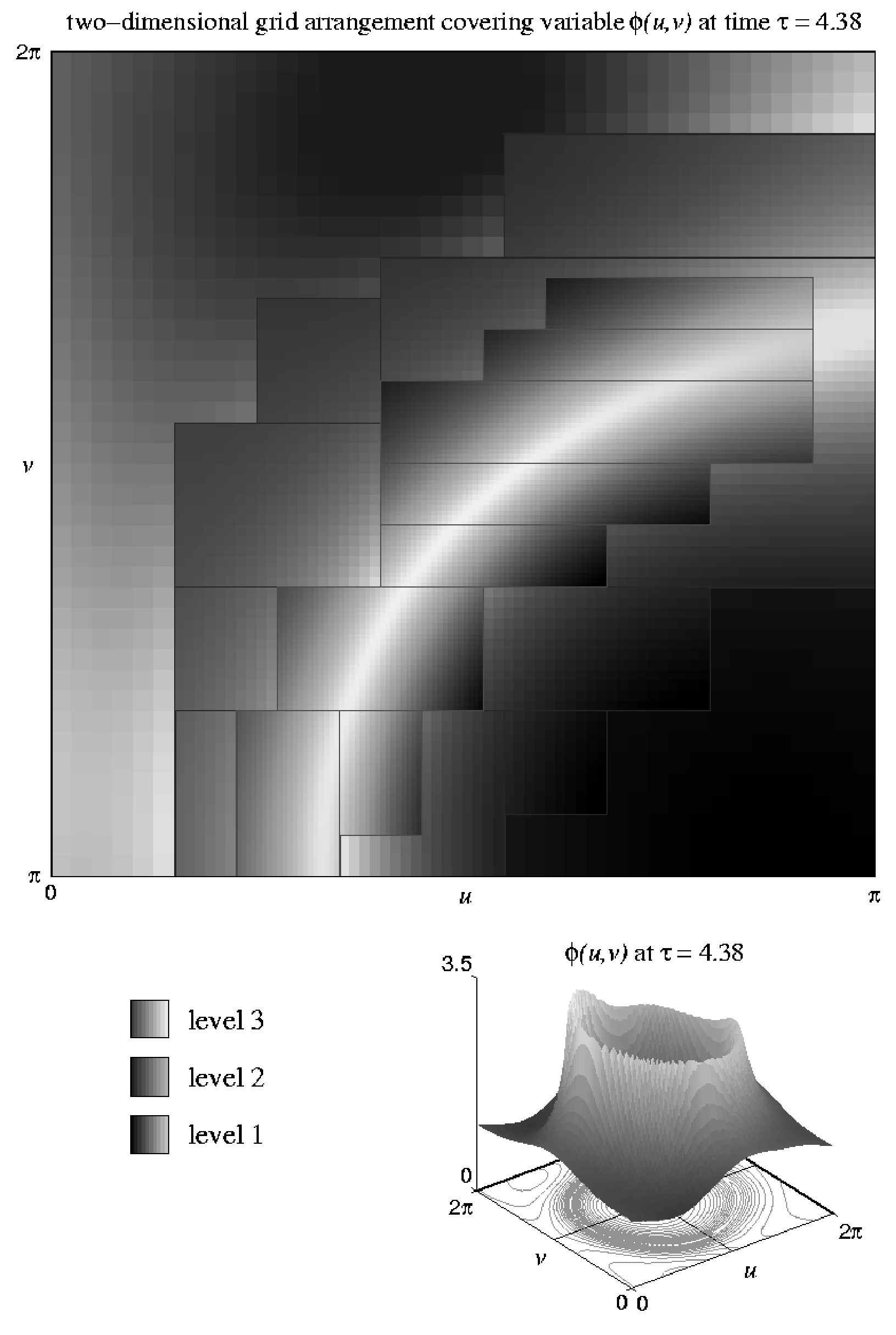}}
}
\Caption
Grids at different resolutions for a two-dimensional simulation.
The top frame shows part of a shaded contour plot of one of the
evolved variables at a fixed time.
Data from all grids are plotted, with different colour schemes
used for grids at different levels.
The $80 \times 80$ base grid is refined by a factor of two
at each new level.
The bottom frame shows a surface plot of the same variable for
the whole spatial domain.
Chapter~7 discusses the cosmological model used.
\endCaption
\endfigure

\figure{AmrTimings}
\centerline{\vbox{
\halign{#\hfil&\hfil#&#\hfil\cr
        Module&Time\%&\hskip8.5mm Breakdown\cr
        \noalign{\medskip\hrule\medskip}
        Grid evolution:&95\%&$\quad\cases{
          \>\hbox to 43mm{Riemann solver: \hfil 50\%}\cr
          \>\hbox to 43mm{Wave propagation: \hfil 24\%}\cr
          \>\hbox to 43mm{Source terms: \hfil 21\%}\cr
        }$\cr
        \noalign{\smallskip}
        Grid interaction:&5\%&$\quad\cases{
          \>\hbox to 43mm{Interpolation: \hfil 2\%}\cr
          \>\hbox to 43mm{Parent update: \hfil 2\%}\cr
          \>\hbox to 43mm{Other: \hfil 1\%}\cr
        }$\cr
        \noalign{\smallskip}
        Grid management:&0\%&\cr
        \noalign{\smallskip}
        Clustering:&0\%&\cr}
}}
\Caption
Timing measurements for the AMR code, broken down by module 
and operation.
Times marked as~$0\%$ are non-zero but negligible.
(The error estimation procedure contributes to the times measured
for the grid evolution module.)
\bigskip
\endCaption
\endfigure

\figure{AmrTimetrial}
\DrawFig{
  \epsfxsize=\hsize
  \epsfbox{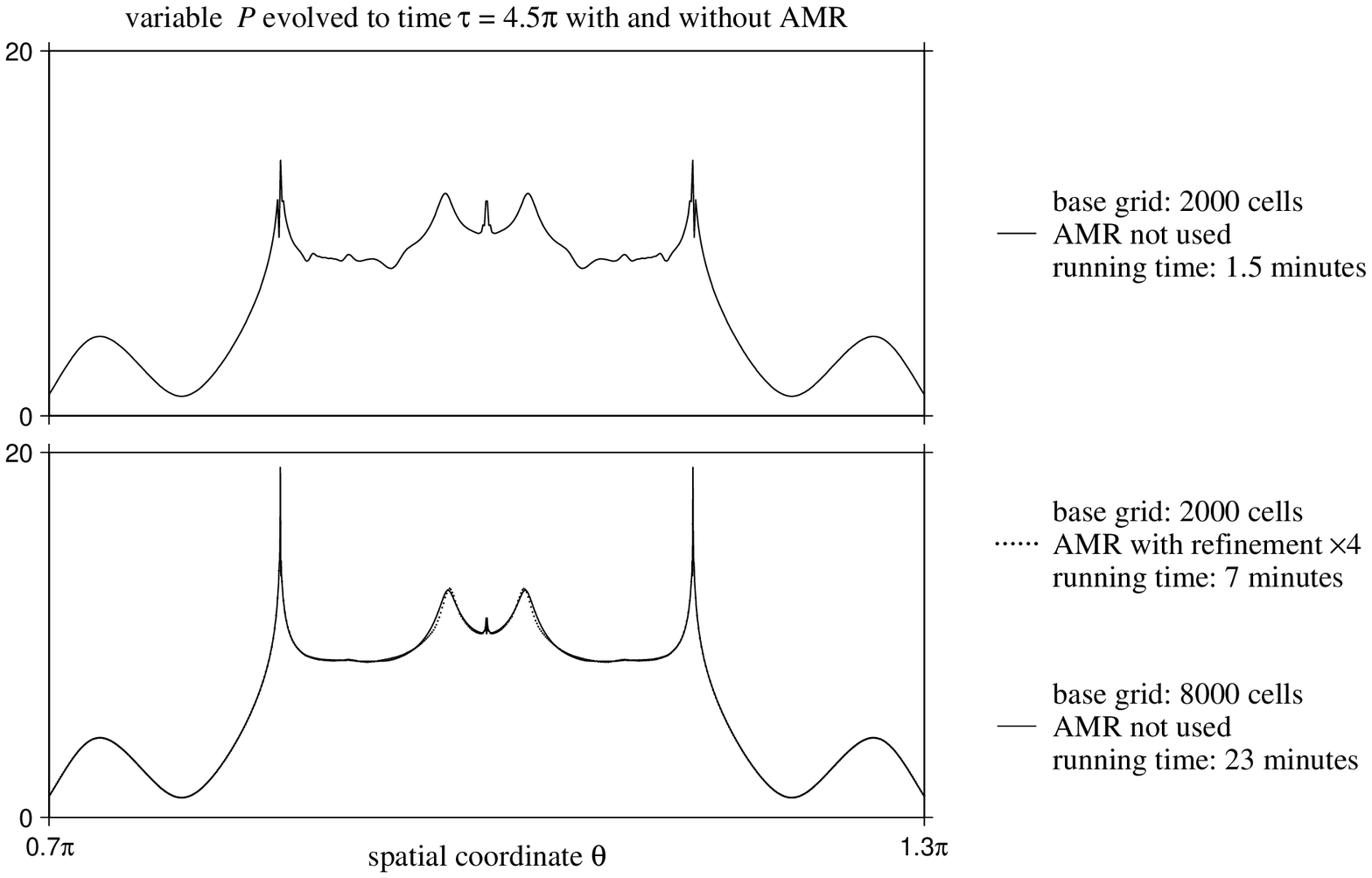}
}
\Caption
Comparison of running times with and without AMR.
The same simulation is performed for three different set-ups.
In two cases AMR is not used: for one the base grid has low
resolution, for the other high resolution.
In the third case AMR is used on a low-resolution base grid such that
it has sub-grids which are as finely meshed as the high-resolution
base grid.
The results with AMR are comparable with the results on the 
high-resolution base grid; however the former are produced in
significantly less time.
(The one-dimensional cosmological model used for the simulations is
described in chapter~6.)
\bigskip
\endCaption
\endfigure

Figure~\use{Fg.AmrDepths} shows the time evolution of a
one-dimensional grid hierarchy produced during the integration of
a collapsing Gowdy~$T^3$ cosmological model, a system described in
detail in chapter~6.
The base grid comprising two thousand grid cells is refined by a
factor of two at each new level, with no sub-grids being created
above level four.
(A refinement factor of four is usually used in
full-scale simulations.)
During the evolution the Gowdy variables show very complicated
behaviour and sharp spikes form in them at various points.
The grid hierarchy adapts to track this behaviour and positions
highly refined grids over the emerging spikes.
(Figures~\use{Fg.GowDominAt} and~\use{Fg.GowDominBt} illustrate 
the complexity of the behaviour in
the Gowdy models; their similarity with figure~\use{Fg.AmrDepths}
is clear.)

The grid arrangement for a two-dimensional simulation (a collapsing
$U(1)$-symmetric cosmology, discussed in chapter~7) is shown at one 
instant in time in figure~\use{Fg.AmrLayout}.
As structure begins to form at small scales in the solution, refined
sub-grids are created to keep it well resolved.
The difference in resolution of data between grids on different
levels can be seen in the shaded contour plot of the figure.
Grid splitting (for three processors running in parallel) was used by
the code as part of the regridding procedure in this example.

An approximate breakdown of the time spent by the AMR code in
performing different operations is given in
figure~\use{Fg.AmrTimings}.
The grid evolution module accounts for the majority of the code's
running time, whereas the amount of time spent performing grid
management and clustering operations is negligible.
(The values in the table were produced by profiling a two-dimensional
evolution based on the Frittelli-Reula system of equations described in
chapter~4.
The code evolved grid data using the wave-propagation method of 
chapter~2 with an exact Riemann solver, and interpolation was by 
quadratic splines.)

It is worth emphasising that the AMR code does not produce numerical 
solutions which are more accurate than those produced by single
grid methods;
rather it produces solutions attaining a pre-specified accuracy in
less time than would be taken using a single high-resolution grid.
Figure~\use{Fg.AmrTimetrial} illustrates this point.
Also, the use of sub-grids has an advantage in addition to reducing
the local error in the data on the base grid: it allows fine-scale
features of the solution to be resolved even though they may be less 
than a coarse grid cell width in size.

The adaptive nature of the AMR algorithm means that a simulation can
run very fast at times at which a low resolution of data is sufficient
(typically during its initial stages).
However this adaptivity makes it very difficult to predict the overall
running time of a simulation:
if a low error tolerance is used such that a lot of grid cells are
flagged for refinement then the time required to run the simulation
increases massively (though, of course, it does produce more accurate
results). 
Successful use of the AMR code requires experimentation for each new
problem being investigated.
In particular, there is no single ideal criterion for deciding which grid
cells should be flagged for refinement:
although the error estimation process (section~\use{sec.AmrOutline})
is unambiguous, how best to weight the errors in the different unknown
variables, and whether to use absolute or relative error magnitudes,
depends on the nature of the simulation.

The production of internal boundary data for sub-grids has long been
recognized as a potential source of problems in the AMR approach.
Berger~(1982, 1985) has proved some basic stability results for
interfaces between coarse and fine grids when dissipative numerical
evolution methods are used.
However, analysis by Wild~(1996, chapter~5) shows that phase and amplitude
errors inevitably occur when waves cross between regions of high and low
resolution, and that there is a tendency for high frequency waves to
become trapped in high-resolution regions.
Wild's results are based on the use of non-dissipative integration
methods, and it may be hoped that the effects of the grid interfaces
will be less severe if dissipative integration methods are used.
In the present work using high-resolution and Lax-Wendroff integration
methods, small amounts of noise are found
at the edges of child grids where boundary data are
produced by interpolation from parent grid data.
The effect is similar to, but less severe than, the production of
noise by insufficiently accurate interpolation, as discussed in
section~\use{sec.AmrDetails}, and a similar remedy might be
applicable: boundary data could be required to match smoothly onto the 
existing data on the sub-grid.
(Indeed, smoothing data at grid interfaces was found by Wild to be a
potential cure for the problem of noise build-up.)
Taking into consideration the points made about interpolation of
continuous and discontinuous data in section~\use{sec.AmrDetails},
an advanced interpolation method could be developed which varies its
order automatically according to the smoothness of the data being
interpolated, and which interpolates coarse data to give a smooth match
onto existing regions of fine data.

The AMR code described in this chapter has played a very important
role in the work presented here.
Without it the numerical simulations performed for chapters~6 and~7
would have taken up significantly more computer time and would have
been greatly restricted by memory limitations, and
the formation of fine-scale structure that occurs in the cosmological
models studied could not have been tracked with such high accuracy.


\Chapter{Hyperbolic Formulations}
The preceeding chapters~2 and~3 describe a range of sophisticated
numerical methods which can be used to produce approximate
solutions to systems of hyperbolic partial differential equations.
The present chapter considers the application of such methods to
general relativity.
Because of their covariant nature the Einstein equations do not in
their standard form constitute a time evolution system, and the ADM
formulation of the equations as a Cauchy problem is usually used as a
basis for numerical simulations.
However, written in the ADM form, the Einstein equations become an
evolution system which is hyperbolic only in a weak sense, and there
are many drawbacks to this lack of hyperbolicity, not least of which
is a restriction on the range of numerical methods that can be used to
evolve the equations.
The needs of numerical relativity have led to a number of new
hyperbolic formulations of the Einstein equations being proposed in
recent years. 
For a review of some of the theoretical aspects of these formulations
see Reula~1998. 

This chapter concentrates on a particular hyperbolic formulation
proposed by Frittelli and Reula~(1996).
The formulation is extended to a family of evolution systems, a subset
of which have properties which make them well-suited for numerical
work; they are the most compact of the hyperbolic formulations of the
Einstein equations that have been constructed to date.
Section~\use{sec.HypFR} discusses the generalized Frittelli-Reula
formulation, with section~\use{sec.HypIntro} reviewing relevant
background material (including the ADM formulation, some basic
definitions of hyperbolicity, and the factors that have motivated the
development of hyperbolic formulations).

In section~\use{sec.HypNumerical} a numerical code that implements the
Frittelli-Reula hyperbolic formulation is discussed.
This is, to the author's knowledge, the first time the Frittelli-Reula
formulation has been used for numerical work.
The code is spatially two-dimensional in the sense that it assumes the
spacetime being investigated to have a single spacelike Killing vector.
By solving the Riemann problem for the Frittelli-Reula evolution
system, the code is able to use the high-resolution wave-propagation
method of chapter~2 to evolve spacetimes at second-order accuracy.
The code has been used to evolve $U(1)$-symmetric vacuum cosmological
models, and results from these simulations are presented much later in
this work in section~\use{sec.MonFR}.

\Section{Hyperbolicity and the Einstein Equations
         \label{sec.HypIntro}}
This section reviews background material relevant to the discussion
of the Frittelli-Reula hyperbolic formulation of the Einstein equations
which follows in section~\use{sec.HypFR}.
The standard ADM formulation upon which the Frittelli-Reula system is
based is discussed and some basic definitions of hyperbolicity for
first-order systems of partial differential equations are given.
The reasoning behind the development of hyperbolic formulations for the
Einstein equations is explained and the distinguishing features of a
number of hyperbolic formulations are summarized.

\SubSection{The ADM Formulation of the Einstein Equations}
The most commonly used approach for producing numerical solutions to
the Einstein equations is to reformulate the equations as a Cauchy problem:
given suitable data describing the geometry of an initial spacelike
hypersurface~$\Sigma_{t_0}$, the Einstein equations are used to describe
the evolution, with respect to some time coordinate~$t$, of the data
into a foliation of spacelike hypersurfaces 
$\{ \Sigma_t : t_0 \le t < T \}$ making up a region of spacetime.
This decomposition of four-dimensional spacetime into space and time
parts is described as a `three-plus-one' splitting, and the standard
approach to splitting the Einstein equations is the Arnowitt-Deser-Misner
(typically abbreviated as~ADM) formulation, which is described in
standard references such as Misner, Thorne and Wheeler~1973
(chapter~21) and Wald~1984 (chapter~10).
The classic~1979 article by York discusses the ADM formulation from the
point of view of numerical relativity,
and in the present subsection the main features of York's interpretation
of the ADM formulation are reviewed.

While the ADM description of a spacetime is fully covariant, for 
simplicity the following account assumes that
a coordinate system $\{t,x^1,x^2,x^3\}$ adapted to the formulation has
already been chosen:
the hypersurfaces of constant time coordinate~$t$ are the spacelike
slices of a {\it foliation\/} of the spacetime, with the spatial
coordinates~$x^1$, $x^2$ and~$x^3$ describing the individual slices.
The geometry of the spacetime is interpreted in terms of time-dependent
three-dimensional scalars, vectors and tensors defined on the spatial
slices. 
The {\it intrinsic metric},~$h_{ij}$, is, for each value of~$t$, a
three-dimensional Riemannian metric defined by the projection of the
four-dimensional spacetime metric,~$g_{\mu\nu}$, onto the spatial
slices. 
The convention adopted in the present work is that the indices~$i$,
$j$,~\dots\ range over the set $\{1,2,3\}$ while the indices~$\mu$,
$\nu$,~\dots\ range over the set $\{0,1,2,3\}$.
Viewed as $3\times3$~matrices, the inverse of the intrinsic
metric~$h_{ij}$ is denoted by~$h^{ij}$, and the indices of spatial
tensors are raised and lowered using this metric.
The {\it extrinsic curvature},~$K_{ij}$, of the foliation is another
symmetric spatial tensor which is, in effect, the time derivative of the
intrinsic metric; it is defined via equation~\EqNum{HypADMEvolve}{a}.

The coordinate freedom present in the Einstein equations appears in
the ADM formulation through the lapse,~$N$, and the shift
vector,~$N^i$. 
The {\it lapse},~$N$, is a positive scalar quantity which determines
how the hypersurfaces of constant~$t$ foliate the spacetime. 
The magnitude of the lapse determines (in a sense made explicit below)
the separation in proper time of two adjacent spatial slices with
coordinate times~$t$ and~$t+dt$.
The {\it shift vector},~$N^i$, accounts for the freedom to redefine
spatial coordinates on the slices of the foliation.
When the shift vector is identically zero the world lines described by
fixed values of the spatial coordinates, 
$x^i=\hbox{{\it constant}}$, run normal to the spatial slices.
In the general case, for a point $(t,x^i)$, the normal to the spatial
slice leads to a point $(t+dt, x^i-N^i dt)$ on a second spatial slice
infinitesimally removed.
The two points are separated by a proper time of
$$
  d\tau = N \, dt 
  \, ,
$$
demonstrating how the structure of the foliation is related to the lapse.

Given values for the intrinsic metric, lapse, and shift on a
slice of the foliation, the four-metric of the spacetime can be
reconstructed. 
In the coordinate system $\{t,x^1,x^2,x^3\}$ it takes the form
$$
  g_{\mu\nu} = \pmatrix{
    g_{00} & g_{0j} \cr
    g_{i0} & g_{ij} \cr
  } = \pmatrix{
    N^k N_k - N^2  & N_j    \cr
    N_i            & h_{ij} \cr
  } \, , \EQN HypADMMetric
$$
where, as stated earlier, the indices of the shift vector,~$N^i$, are
lowered using the three-metric,~$h_{ij}$.

The Einstein equations,
$$
  G_{\mu\nu} = \kappa T_{\mu\nu}
  \, , 
$$
where~$\kappa$ is the matter coupling constant,
can now be decomposed into equations for the spatial quantities.
On the left-hand side of the equations are expressions based on the
four-dimensional Riemann curvature tensor~$R^{\mu}{}_{\nu\eta\sigma}$
associated with the metric~$g_{\mu\nu}$; the Gauss-Codazzi equations
can be used to rewrite these in terms of the three-dimensional
curvature tensor~$\bar{R}^{i}{}_{jkn}$ associated with the intrinsic
metric~$h_{ij}$.
The energy-momentum tensor,~$T^{\mu\nu}$, on the right-hand side of
the equations is decomposed into three spatial quantities according to
$$\EQNalign{
  \rho & {} = 
    N^2 T^{00}
    \, , \cr
  J^i & {} =
    N ( T^{0i} + N^i T^{00} )
    \, , \EQN HypADMMatter \cr
  S^{ij} & {} = 
    T^{ij} + 2 N^{(i} T^{j)0} + N^i N^j T^{00}
    \, , \cr
}$$
which are, respectively, the {\it energy density}, the 
{\it momentum density\/} and the {\it stress\/} relative to the
foliation.
The standard index symmetrization notation is used throughout this work:
$A^{(i} B^{j)} = {1\over2}( A^i B^j + A^j B^i )$, and so on.

The time-time and time-space components of the Einstein equations give
rise to equations which only involve quantities that can be entirely
determined on a single spatial slice, that is, they involve no
explicit derivatives with respect to the time coordinate. 
These are the {\it constraint equations},
$$\EQNalign{
  \bar{R} + K^2 - K_{ij} K^{ij} & {} = 2 \kappa \rho
\, , \EQN HypADMConstr;a \cr
  \bar{\nabla}_{\!k} ( K^{ik} - K h^{ik} ) & {} = \kappa J^i
\, , \EQN HypADMConstr;b \cr
}$$
where~$K=K^{i}{}_{i}$, and~$\bar{R}$ and~$\bar{\nabla}_{\!i}$ are the 
Ricci scalar and the covariant derivative associated with the spatial
metric~$h_{ij}$. 
Equation~\EqNum{HypADMConstr}{a} is the {\it Hamiltonian\/} or 
{\it energy constraint}, and equation~\EqNum{HypADMConstr}{b} is 
the {\it vector\/} or {\it momentum constraint}.
These equations are independent of the lapse and the shift vector.

The space-space components of the Einstein equations relate the time
derivative of the extrinsic curvature,~$K_{ij}$, to quantities which
are fully determined within each spatial slice.
Recalling that the extrinsic curvature is closely related to the time
derivative of the intrinsic metric, it is possible now to write a set
of {\it evolution equations\/} for the two quantities:
$$\EQNalign{
  \partial_t h_{ij} & {} = \textstyle
    - 2 N K_{ij} + 2 \bar{\nabla}_{\!(i} N_{j)}
\, , \EQN HypADMEvolve;a \cr
  \partial_t K_{ij} & {} = \textstyle N (
    K K_{ij} - 2 K_{ik} K^{k}{}_{j} 
    - \kappa S_{ij} + {1\over2} \kappa h_{ij} ( S - \rho )
    + \bar{R}_{ij}
  ) \cr
  & {} \quad 
    - \bar{\nabla}_{\!i} \bar{\nabla}_{\!j} N
    + N^k \bar{\nabla}_{\!k} K_{ij}
    + 2 K_{k(i} \bar{\nabla}_{\!j)} N^k
\, , \EQN HypADMEvolve;b \cr
}$$
where~$S=S^{i}{}_{i}$, and $\bar{R}_{ij}$ is the spatial Ricci tensor,
and~$\partial_t$ denotes differentiation with respect to the time 
coordinate~$t$.
The conservation laws for the energy-momentum tensor,
$$
  \nabla_{\!\mu} T^{\mu\nu} = 0
  \, ,
$$
allow additional evolution equations to be derived for the energy and
momentum densities of the matter fields:
$$\EQNalign{
  \partial_t \rho & {} = 
    N ( S^{ij} K_{ij} + K \rho - \bar{\nabla}_{\!i} J^i )
    - 2 J^i \bar{\nabla}_{\!i} N
    + N^i \bar{\nabla}_{\!i} \rho
\, , \EQN HypADMFluid;a \cr
  \partial_t J^i & {} = 
    N ( 2 K^{ik} J_k + K J^i - \bar{\nabla}_{\!k} S^{ik} )
    - S^{ik} \bar{\nabla}_{\!k} N
    - \rho \bar{\nabla}^i N
    + N^k \bar{\nabla}_{\!k} J^i
    - J^k \bar{\nabla}_{\!k} N^i
\, . \EQN HypADMFluid;b \cr
}$$
These equations are generalizations of, respectively, the continuity
and Euler equations.

For the evolution system~\EqNum{HypADMEvolve}{}
and~\EqNum{HypADMFluid}{} to be complete, some method must be
prescribed for determining the behaviour of the lapse,~$N$, the
shift,~$N^i$, and the stress,~$S^{ij}$, on each spatial slice; 
no information about the evolution of these quantities is supplied by
the Einstein equations.
The behaviour of the stress tensor,~$S^{ij}$, will be determined by
the properties of the matter present in the spacetime, and in simple
cases (for example, when the matter is a perfect fluid) an equation of
state will specify the stress in terms of the energy and momentum of
the matter.
However typically it may be more appropriate to replace the evolution
equations~\EqNum{HypADMFluid}{} with an evolution system which is
specialized to the type of matter being treated, with the
quantities~$\rho$, $J^i$ and~$S^{ij}$ in
equations~\EqNum{HypADMConstr}{} and~\EqNum{HypADMEvolve}{} being
rewritten in terms of an alternative set of matter variables.

The specification of a method for determining the lapse,~$N$, and the
shift,~$N^i$, constitutes a choice of coordinate system for the
spacetime being studied.
The question of how best to determine these quantities is a difficult
one: poor choices can result in foliations which develop coordinate
singularities, which only cover a small region of the spacetime, or
which intersect curvature singularities.
York~(1979) discusses a variety of methods for choosing the lapse and
the shift which are believed to avoid some of these problems.
The role of the lapse in numerical relativity is a topic which is
considered further in chapter~5.

Assuming that a suitable specification of the lapse,~$N$, the shift
vector,~$N^i$, and the stress tensor,~$S^{ij}$, on each
spatial slice is given, then if the intrinsic metric,~$h_{ij}$, the
extrinsic curvature,~$K_{ij}$, the energy density,~$\rho$, and the
momentum density,~$J^i$, are all known at an initial time~$t=t_0$,
the evolution equations~\EqNum{HypADMEvolve}{}
and~\EqNum{HypADMFluid}{} can be used to determine their values
at later times $t_0 \le t < T$ (where~$T$
is not usually known beforehand and may be infinite).
Such a solution will satisfy the Einstein equations if the
constraints~\EqNum{HypADMConstr}{} hold for every value of~$t$, and it
follows from the Bianchi identity,
$$
  \nabla_{\!\mu} G^{\mu\nu} = 0
  \, ,
$$
that this will be the case if they hold at the initial time~$t=t_0$.
One way in which the ADM formulation can be used to generate numerical
solutions to the Einstein equations is now clear:
an initial data set $\{h_{ij},K_{ij},\rho,J^i\}$ is chosen which
satisfies a discretized version of the constraint
equations~\EqNum{HypADMConstr}{} and, with appropriate conditions
determining the auxiliary variables $\{N,N^i,S^{ij}\}$ at
each step, the data are advanced in time based on a discretized
version of the evolution equations~\EqNum{HypADMEvolve}{}
and~\EqNum{HypADMFluid}{}. 
(The failure of numerical solutions of the evolution equations to
exactly satisfy the constraint equations away from the initial time slice
has led to the use of `constrained' evolution methods by some
researchers in which some of the unknown variables are determined at
each time step by solving the constraint equations instead of the
evolution equations; while such methods typically show improved
stability, their use is not without drawbacks, and in the present work
only unconstrained evolution is used.)

Direct numerical implementations of the ADM formulation have been used
to successfully evolve solutions to the Einstein equations in a large
number of cases.
However some properties of the formulation make it
less well suited for numerical work than might be hoped, and this has
led to the development of various alternative `hyperbolic'
formulations, the subject of the present chapter.

\SubSection{Hyperbolic Partial Differential Equations}
In this subsection some basic properties of first-order systems of
partial differential equations relevant to the present work are
briefly reviewed, with attention being given to systems for which
the Cauchy problem is well posed.
Much more detailed discussions of this subject can be found in the
books by Gustafsson, Kreiss and Oliger~(1995),
Courant and Hilbert~(1962), Kreiss and Lorenz~(1989), and John~(1982),
among many others.

Consider a $k$-dimensional first-order system,
$$
  \partial_t u 
  + B^i(u,t,x^j) \partial_i u
  = S(u,t,x^j)
  \, , \EQN HypPDESystem
$$
where~$u(t,x^j)$ is a vector of~$n$ unknowns, the~$B^i$ are $n \times n$
matrices with the index~$i$ taking values $1$,~\dots,~$k$, and~$S$ is a
column vector.
(For simplicity here the components of~$u$, $B^i$ and~$S$ are assumed
to be real.)
It is worth noting that the flux-conservative system of
equation~\EqNum{PdeConsLaws}{} can easily be rewritten so that it
takes the above form. 
The Cauchy problem for the system~\EqNum{HypPDESystem}{} involves
finding a solution~$u(t,x^i)$ which satisfies the equations on a time
interval $t_0 \le t < T$ (for~$T$ as large as possible) and which
coincides with a specified set of initial data: $u(t_0,x^i)=f(x^i)$.
If the system~\EqNum{HypPDESystem}{} represents a physical process
then it may be expected that a unique solution~$u(t,x^i)$ to the Cauchy
problem will always exist given a reasonable set of initial
data~$f(x^i)$, and furthermore that the solution will depend
continuously (in some sense) on the initial data.
A first-order system for which these properties, defined in a
rigorous way, are known to hold is described as having a 
{\it well-posed\/} Cauchy problem.  
(Formal definitions of well-posedness, as given in
Kreiss and Lorenz~1989, usually consider the rate of growth of an `energy
norm' defined on the solution, with well-posedness typically requiring
that the norm grow no faster than exponentially.)

The well-posedness of the system~\EqNum{HypPDESystem}{} can be
determined from its {\it principal part}, by which is meant the spatial
derivative terms~$B^i \partial_i$.
With this in mind the {\it symbol\/} of the system is defined to be
$$
  P(\omega) = \omega_i B^i
  \, , \qquad
  \hbox{where } 
  \omega = (\omega_1,\ldots,\omega_k) \in \RealSet^k
  \, , \EQN HypPDESymbol
$$
a linear combination of the matrices $B^1$,~\dots,~$B^k$.
The symbol~$P$ can be classified according to its
{\it hyperbolicity}, with well-posedness of the Cauchy problem holding
for systems in which~$P$ satisfies sufficiently strong hyperbolicity
conditions.
(The kinds of hyperbolicity discussed here are the same as in
Gustafsson, Kreiss and Oliger~1995.)

The system~\EqNum{HypPDESystem}{} is ({\it strongly\/}) {\it hyperbolic\/} 
if the matrix~$P(\omega)$ has real eigenvalues and a complete set of
eigenvectors for all non-trivial~$\omega$, and in this case it has a
well-posed Cauchy problem.
If the eigenvalues of~$P(\omega)$ are real but its eigenvectors do not
form a complete set then the system is only {\it weakly hyperbolic\/}
and it is not well posed.
(These definitions of hyperbolicity are extensions of the definitions
used for constant-coefficient linear systems in which~$P$ does not
depend on the unknowns~$u$ or the coordinates $\{t,x^i\}$.
For nonlinear systems the hyperbolicity conditions on~$P$ are
required to hold for all physically reasonable values of the unknowns
and the coordinates.)

Two more restrictive classes of hyperbolicity are also defined,
both of them implying (by standard results of linear algebra) strong
hyperbolicity and hence well-posedness of the system.
If for all non-trivial~$\omega$ the eigenvalues of~$P(\omega)$ are
real and distinct, then the system is {\it strictly hyperbolic}.
(This type of hyperbolicity, though powerful, occurs rarely in
physical systems.)
If for all~$\omega$ the matrix~$P(\omega)$ is symmetric,
then the system is {\it symmetric hyperbolic}.
This latter form of hyperbolicity is important for physical systems,
and its definition is often extended:
the system~\EqNum{HypPDESystem}{} may be described as symmetric
hyperbolic (strictly speaking it is {\it symmetrizable}) if there
exists a positive-definite symmetric $n \times n$ matrix~$M$ such that
$MP(\omega)$ is a symmetric matrix for all~$\omega$.
Symmetric (or symmetrizable) hyperbolicity of a system can also be
demonstrated by explicit construction of a suitable `energy norm' for
the principal part.

In hyperbolic systems, information always propagates at finite
speeds: if two solutions to a system are identical outside of a
region of $x^i$-space given by $\vert x^i \vert < \epsilon$ at some
time~$t$, then at a later time~$t + \delta t$ the solutions will
still be identical outside of a region $\vert x^i \vert < \alpha$
for some value~$\alpha < \infty$.
The {\it characteristic surfaces\/} of a hyperbolic system provide
important details about how information propagates within the
system: different combinations of the unknown variables can be
pictured as advecting along the different characteristic surfaces,
with the outermost surfaces defining the maximum propagation speed
within the system.
Different characteristic surfaces may be identified with different
physical processes occurring within the system.

A local picture of the characteristic surfaces that pass through a
point~$O$ can be built up using the symbol~$P(\omega)$ evaluated at
that point.
A spatial direction vector $v=(v_1,\ldots,v_k)$ is chosen at~$O$,
normalized so that $\vert v \vert = 1$ (where the standard $L_2$-norm is
used here, and the speeds referred to below are with respect to the
spatial coordinates; the presence of a metric in general relativity
provides an alternative way by which the vector could be
normalized).
The eigenvalues of the matrix~$P(v)$ are then the 
{\it characteristic speeds\/} of the system in the direction~$v$:
if~$\lambda$ is an eigenvalue of~$P(v)$ then the vector 
$(1,\lambda v_i)$ in $(t,x^i)$-space is tangential to a
characteristic surface at the point~$O$; furthermore the eigenvector
of~$P(v)$ associated with~$\lambda$ indicates which combinations of
the unknowns~$u$ propagate along that characteristic surface.

It is worth noting that the hyperbolic properties and the
characteristic structure of a system are essentially unaffected by
changes in its dependent variables.
It is straightforward to show that for a change of variables
$u \rightarrow \hat{u}$ for which the Jacobian matrix
$J = \partial u / \partial \hat{u}$ is non-singular, the
eigenvalues and the number of associated independent eigenvectors of
the new symbol $\hat{P}(\omega) = J^{-1} P(\omega) J$ do not change.
Consequently the characteristic speeds and the satisfaction of
conditions for weak, strong and strict hyperbolicity are the same for
the new system as for the original one.
While the explicit symmetry of the symbol is in general affected by
a change of variables, a system which is symmetrizable remains so:
if~$M$ is a symmetrizer for the original system then $J^{T} M J$
is a symmetrizer for the new system.

\SubSection{Motivation for the Development of Hyperbolic Formulations}
It is well known that Einstein's equations form a hyperbolic system
in suitable coordinates (see, for example, chapter~10 of
Wald~1984). 
What is surprising, however, is that the ADM formulation of the equations
is not hyperbolic in any strong sense, and does not constitute a
well-posed Cauchy problem for general relativity.
(In section~\use{sec.HypFR} it is demonstrated that a first-order form
of the ADM formulation which uses harmonic time slicing is
only weakly hyperbolic in the sense defined in the previous subsection.)
In recent years there has been much interest in the development of
formulations of Einstein's equations which are explicitly
hyperbolic, and for which the characteristics speeds correspond to the
physical speeds of the system, with the main motivation for this being
the possible use of such systems in numerical relativity.
Hyperbolicity of first-order evolution systems is important for the
construction of proofs of existence and uniqueness of solutions, and
it is perhaps incongruous that numerical solutions to a system should
be generated without these fundamental properties of analytic
solutions having been established.
Hyperbolicity should also facilitate the demonstration of
well-posedness in other aspects of numerical relativity: given that
violations of the Einstein constraint equations in numerical solutions
are unavoidable, a hyperbolic system describing their propagation
may be important when considering the stability of numerical simulations
(see, for example, Frittelli~1997);
also, little is known about the stability of Einstein's equations when
boundary data are specified at the edges of numerical grids, and
hyperbolicity may allow the well-posedness of this `initial boundary
value problem' to be investigated (see, for example, Stewart~1998).

Hyperbolic formulations are expected to be particularly important for
long-term numerical evolutions of black hole spacetimes.
Since these formulations respect the causality of the Einstein
equations, black hole horizons can be considered as natural internal
boundaries for numerical simulations, and the problematic regions
inside the black holes can be excised from the domain of evolution.
Bona, Mass\'o and Stela~(1995) demonstrate the potential of this
approach using a hyperbolic formulation with a causal numerical
evolution method.

Another advantage of hyperbolic formulations is that they can be used
in conjunction with sophisticated numerical evolution methods.
As discussed in chapter~2, many advanced numerical methods developed
for use in computational fluid dynamics require the evolution
equations to be expressed in first-order flux-conservative form (see
equations~\EqNum{PdeConsLaws}{}), and furthermore some make use of the
characteristic structure of the equations, requiring the evolution
systems to be strongly hyperbolic.
Most hyperbolic formulations of the Einstein equations are compatible
with these requirements, and in cases where
the Einstein equations are coupled to hydrodynamical fields, a
hyperbolic formulation for the former would allow the complete system
to be treated in a unified manner using standard methods from
computational fluid dynamics.

An idea that is often quoted in the literature concerning hyperbolic
formulations suggests that if a system of equations has mathematical
properties which closely resemble the properties of the physical
system being modelled (in particular, if a formulation of general
relativity is hyperbolic and has only physically relevant
characteristic speeds) then it may be expected to perform better in
numerical simulations than alternative systems.
While this idea is appealing, it does not appear to be supported by
empirical evidence: in section~\use{sec.MonFR} results from a
direct numerical comparison of a number of formulations of Einstein's
equations are presented, and it can be seen that hyperbolic
formulations (without their hyperbolicity being exploited) do not in
general produce significantly more accurate results than
the standard ADM formulation.

The next subsection briefly reviews some of the hyperbolic
formulations that have been developed for the Einstein equations.
Prior to that a comment is made here concerning gauge conditions
for hyperbolic formulations.
As described earlier in this section, the ADM formulation puts no
restrictions on the possible values taken by the lapse~$N$ (except
that it be positive) or the shift vector~$N^i$.
However several hyperbolic formulations of the Einstein equations
require the lapse to satisfy a relation with the intrinsic metric,
$$
  N = Q(t,x^k) \sqrt{\det h_{ij}}
  \, , \EQN HypHarmFixed
$$
where~$Q(t,x^i)$, called the slicing density, is an arbitrary
(positive) function of the spacetime coordinates which cannot depend
on any of the unknown variables.
Typically the shift vector is also required to be independent of the
unknowns, although it is otherwise arbitrary: $N^i \equiv N^i(t,x^j)$. 
The condition~\EqNum{HypHarmFixed}{} on the lapse is described as
generalized harmonic time slicing, and an alternative way of
expressing it is as an evolution equation,
$$
  ( \partial_t - N^i \partial_i )N = N f(t,x^i) - N^2 K
  \, , \EQN HypHarmEvolve
$$
where~$K$ is the trace of the extrinsic curvature,
and~$f(t,x^i)$ is another arbitrary gauge function related 
to~$Q(t,x^j)$ and~$N^i(t,x^j)$ by
$$
  f = ( \partial_t - N^i \partial_i ) \ln Q
      + \partial_i N^i
  \, .
$$
If the gauge function~$f$ is identically zero then the time slicing is
described as simple harmonic; in the case that the shift~$N^i$ is zero
this corresponds to the slicing density~$Q$ being independent of the
time coordinate.
Equation~\EqNum{HypHarmFixed}{} may be used to remove the lapse
completely from the evolution equations of the formulation;
alternatively, the lapse (possibly in addition to its spatial
derivatives) may be included in the set of dynamical variables of the
formulation using equation~\EqNum{HypHarmEvolve}{}.
Although, as equation~\EqNum{HypHarmFixed}{} makes clear, any
spacetime foliation can be reproduced using generalized harmonic
time slicing, in practice it is not clear how the gauge functions
should be chosen so as to produce well-behaved foliations; for this
reason generalized harmonic time slicing may be considered less useful
than gauges such as maximal slicing for which singularity avoidance
properties have been established. 
Time slicing conditions are considered further in
section~\use{sec.HomHarmonic}.

\SubSection{A Review of Hyperbolic Formulations}
In the present work a family of hyperbolic formulations based on a
system constructed by Frittelli and Reula~(1996) is developed for use
in numerical simulations, and the remainder of this chapter concentrates
on this generalized formulation.
In the next section the literature relating to the
Frittelli-Reula system is reviewed.
The present section ends by briefly reviewing work that has been
done on other hyperbolic formulations, concentrating on those which
have been used or have been proposed to be used in numerical simulations.

The hyperbolic formulation of Bona, Mass\'o and co-workers has been
the one most widely used in numerical simulations to date.
It has developed through several incarnations:
Bona and Mass\'o~1989, 1992; Bona et al.~1995, 1997.
The formulation is based on the ADM equations and comprises a
one-parameter family of physically equivalent evolution systems (with
the parameter controlling the inclusion of terms based on the
Hamiltonian constraint) which are strongly hyperbolic and can be
written in first-order flux-conservative form.
For vacuum spacetimes the formulation uses thirty-seven unknowns of
which thirty are the intrinsic metric and its derivatives (including
the extrinsic curvature), four are the lapse and its spatial derivatives
(treated in this case as dynamical variables), and three are additional
variables related to the momentum constraints.
The formulation is hyperbolic for a variety of algebraic time slicing
conditions (including maximal slicing as a limit) but of these only
simple harmonic slicing leads to the system having physically relevant
characteristic speeds.
The shift vector in the formulation is an arbitrary gauge function
allowed to depend on the spacetime coordinates but not the unknown
variables. 

The range of spacetimes studied using numerical implementations of the
Bona-Mass\'o hyperbolic formulation is extensive.
It has been used to evolve spherically symmetric (Bona, Mass\'o and
Stela~1995) and three-dimensional (Bona et al.~1998) black hole
spacetimes as well as to study the propagation of gravitational waves
in three dimensions (Anninos et al.~1997).
Boson stars have been evolved in spherical symmetry 
(Arbona and Bona~1999), and a three-dimensional code has been
constructed which couples the hyperbolic formulation to a perfect
fluid source (Font et al.~1998).
The formulation has also been used to investigate the behaviour
produced by algebraic slicing conditions (Alcubierre~1997).

Choquet-Bruhat, York and co-workers have constructed several
hyperbolic formulations of the Einstein equations which are reviewed
in Choquet-Bruhat, York and Anderson~1998.
Their `Einstein-Ricci' formulation (Choquet-Bruhat and York~1995, 
Abrahams et al.~1995, 1997)
resembles the ADM formulation but uses evolution equations derived
from a wave equation for the spatial components of the 
Ricci tensor.
It incorporates time derivatives of the metric up to
third order.
The formulation is spatially covariant and uses the generalized
harmonic slicing condition.
Written as a flux-conservative first-order system it is symmetric
hyperbolic and has only physically relevant characteristic speeds.
With the lapse included as a dependent variable, the formulation uses
a total of sixty-seven unknowns to describe vacuum spacetimes.

The Einstein-Ricci hyperbolic formulation has been used in
a one-dimensional numerical code for the evolution of spherically
symmetric vacuum spacetimes by Scheel et al.~(1997, 1998) (see also
Cook and Scheel~1997).
It has also been used in a form linearized about the Schwarzschild
metric to extract gravitational radiation and to provide outer
boundary conditions for three-dimensional numerical evolutions
(Rupright, Abrahams and Rezzolla~1998).
A `fourth-order' form of the Einstein-Ricci system has also been
derived (Abrahams et al.~1996) in which the lapse and the shift can be
specified arbitrarily.
Although this system is probably not suitable for numerical work it
has been used as the basis of a gauge-invariant perturbation theory by
Anderson, Abrahams and Lea~1998.

The `Einstein-Bianchi' system (Anderson, Choquet-Bruhat and York~1997)
is a variation of the Einstein-Ricci system in which the dynamical
variables all have clear interpretations: they are the intrinsic metric,
the extrinsic curvature, the lapse, the spatial Christoffel symbols,
and the Riemann curvature tensor decomposed into separate `electric'
and `magnetic' parts.
This formulation has the same hyperbolic structure and the same number
of variables as the Einstein-Ricci system.

Van Putten and Eardley~(1996) use the orthonormal frame formalism to
derive a fully covariant hyperbolic formulation of the Einstein
equations based on the Bianchi identity for the Riemann tensor.
Using the tetrad elements and their connections as dynamical variables
a wave equation is derived from which a first-order evolution system
may be obtained if the components of the Riemann tensor are introduced
as variables.
As with most other hyperbolic formulations, four arbitrary gauge
functions are available through which the slicing of the spacetime can
be controlled.
In van Putten~1997 a one-dimensional numerical implementation of the
formulation is tested against the Gowdy~$T^3$ cosmological model.

Friedrich~(1996) discusses various approaches for constructing
symmetric hyperbolic formulations of the Einstein equations.
The Bianchi equation for the Riemann tensor is used to construct
hyperbolic systems based on both the ADM formulation and
the orthonormal frame formalism.
(A side effect of using the Bianchi identity in this form is
that unphysical, albeit slower than light, characteristic speeds
appear in all of the systems constructed.)
The hyperbolic ADM systems use variables formed by decomposing the
Weyl tensor into electric and magnetic parts, and they slice
spacetimes according to the generalized harmonic gauge condition.
These systems can be written in flux-conservative form, and for the
evolution of vacuum spacetimes they use either forty or fifty
unknowns, depending on how various gauge quantities are specified.
The hyperbolic formulations based on the orthonormal frame formalism
(which are similar to the formulation of van Putten and Eardley
described above) use around forty unknowns and are based on a variety of
gauge conditions which provide alternatives to the standard time slicing
conditions used in numerical relativity.
It is not clear that these systems can be written in flux-conservative
form. 
(In related work, symmetric hyperbolic evolution systems are used to
investigate the initial boundary value problem for general relativity;
Friedrich and Nagy~1999.)

Although no numerical work directly implementing any of Friedrich's~(1996)
hyperbolic formulations has been published to date, H\"ubner~(1998) and
Frauendiener~(1998) have both developed conformally rescaled versions
of the equations for numerical use.
The conformal approach allows asymptotically flat spacetimes sliced by
hyperboloidal hypersurfaces to be compactified and treated
using finite numerical grids.
Both evolution systems are symmetric hyperbolic and in the vacuum
case employ around fifty unknown variables.

Other symmetric hyperbolic formulations based on the use of
orthonormal frames have been constructed by 
Estabrook, Robinson and Wahlquist~(1997),
Friedrich~(1998), and van~Elst and Ellis~(1999);
the latter two both consider spacetimes in which perfect fluid sources
are present, and attach the orthonormal frames to the matter flow vectors.
There has also been some interest in the construction of symmetric
hyperbolic formulations based on the Ashtekar variables  
(Iriondo, Leguizam\'on and Reula~1997, 1998; 
Yoneda and Shinkai~1999a,b; Shinkai and Yoneda~1999).

Recently, Anderson and York~(1999) have proposed a new symmetric
hyperbolic formulation, called the `Einstein-Christoffel' system,
which is closely related to the generalized Frittelli-Reula
formulation described in section~\use{sec.HypFR}: both formulations
use sets of thirty unknown variables which are essentially the
components of the intrinsic metric and the first derivatives of these
components with respect to space and time.
Apart from the (essentially cosmetic) differences in the sets of
variables used, the Einstein-Christoffel system is distinguished from
the generalized Frittelli-Reula formulation in that it adds slightly
different multiples of the constraint quantities to the ADM evolution
equations.
(If the generalized Frittelli-Reula formulation were to be generalized
further by the addition of a term $-2 N \chi h^{ij} {\cal C}_k$
to the right-hand side of equation~\EqNum{HypFRSchem}{b}, then the
Einstein-Christoffel system could be considered as a special case of
the formulation corresponding to the parameter choice $\chi=\eta=1$,
$\gamma=\Theta=0$.)

In other recent work Alcubierre et al.~(1999) have constructed a
family of first-order strongly hyperbolic evolution systems with
physically relevant characteristic speeds which separate out the
non-conformal degrees of freedom from the conformal ones.
(The determinant of the intrinsic metric and the trace of the
extrinsic curvature are assumed to be known functions of the spacetime
coordinates.)
Although these evolution systems are different from the generalized
Frittelli-Reula systems described in the next section, the method by
which they are derived is similar.
Frittelli and Reula~(1999) have also investigated similar
conformally-decomposed evolution systems.

\Section{The Generalized Frittelli-Reula Hyperbolic Formulation
         \label{sec.HypFR}}
In their~1996 paper Frittelli and Reula (expanding on a result from
their~1994 paper on the Newtonian limit of general relativity)
present a one-parameter family of symmetric hyperbolic formulations
of the Einstein equations, one of which (described here as the
{\it original\/} FR~system) has only physically relevant characteristic
speeds.
Frittelli and Reula derive their result by writing the ADM formulation
of the Einstein equations as a first-order system into which five
parameters have been introduced (two through the definition of new
variables, one through a condition on the lapse, and two controlling
the addition of constraint terms to the evolution equations).
By positing
an energy norm for the system they find a one-dimensional parameter
subspace for which the equations are symmetric hyperbolic, and by
requiring the 
characteristic surfaces to coincide with the light cones they fix the
parameter values uniquely.

Unfortunately a mistake in Frittelli and Reula's proof of symmetric
hyperbolicity results in the original FR~system being, in actual fact,
only weakly hyperbolic (although its characteristic speeds are as
claimed).
It is reasonably straightforward to `correct' the evolution system so
that the energy norm argument given by Frittelli and Reula is valid,
and the resulting system (which is similar to the one originally
considered in Frittelli and Reula~1994) is referred to here as the
{\it modified\/} FR~system.
It admits a number of unphysical characteristic speeds.  
The complete set of evolution equations for the modified FR~system was
derived by the author in collaboration with J.~M.~Stewart.
In Stewart~1998 the equations are used as the basis of an analysis of the
numerical tractability of Einstein's equations formulated as an
unconstrained initial boundary value problem.
In Brodbeck et al.~1999 the modified FR~equations are embedded in a
much larger evolution system of seventy unknowns for which (it is
suggested) constraint violations will be dissipated away during the
course of numerical simulations.

In the present work a `generalized' version of the Frittelli-Reula
formulation is derived in which additional parameters are introduced.
By examining the principal part of the generalized formulation a
one-parameter family of strongly hyperbolic evolution systems each
having only physically relevant characteristic speeds is identified.
These are described as the {\it causal\/} FR~systems.
The generalized formulation is described in detail in the present
section, and in section~\use{sec.HypNumerical} its implementation in a
numerical code is described.
Using $U(1)$-symmetric vacuum cosmological models as a test bed, in
section~\use{sec.MonFR} the relative numerical performances of a
number of special cases of the generalized formulation are assessed.

\SubSection{Derivation of the Generalized Frittelli-Reula Formulation}
In this subsection the derivation of a generalized form of the
Frittelli-Reula (henceforth~FR) hyperbolic formulation of Einstein's
equations is outlined.
In the next subsection the characteristic structure of the formulation
is examined and a one-parameter family of strongly hyperbolic systems
with physical characteristic speeds is identified.

The FR~formulation is based on the ADM evolution
equations~\EqNum{HypADMEvolve}{a} and~\EqNum{HypADMEvolve}{b} 
for the intrinsic metric~$h_{ij}$ and the extrinsic curvature~$K_{ij}$. 
These equations constitute an evolution system which is first order in time
but second order in space, and new variables representing the spatial
derivatives of the intrinsic metric must be introduced to reduce the
equations to first order.
The FR~formulation uses the variables
$$\eqalign{
  h^{ij} = {} & (h_{mn})^{-1}
  \, , \cr
  M^{ij}{}_{k} = {} & \textstyle {1\over2} ( \partial_k h^{ij} 
                             - h^{ij} h_{mn} \partial_k h^{mn} )
  \, , \cr
  P^{ij} = {} & K^{ij} - h^{ij} K
  \, , \cr
} \EQN HypFRVars
$$
where~$h^{ij}$ is the contravariant form of the intrinsic metric, and
as a notational convenience all indices $i$,~$j$,~\dots\ are raised and
lowered using the metric~$h^{ij}$, even for quantities
like~$M^{ij}{}_{k}$ which are not tensors. 
Taking into account the symmetry of the quantities in their
$i$-$j$~indices, equation~\EqNum{HypFRVars}{} represents a set of
thirty unknown variables.

(In Frittelli and Reula~1996 the definitions of the
variables~$M^{ij}{}_{k}$ and~$P^{ij}$ include two parameters,~$\alpha$
and~$\beta$, through which the form of the resulting evolution system
can be controlled.
However, as was noted in section~\use{sec.HypIntro}, the hyperbolicity
of a first-order system is essentially unaffected by non-singular
transformations of its dependent variables, and so there is no loss of
generality in fixing the definitions of the variables from the start.
The definitions of equation~\EqNum{HypFRVars}{} correspond to the
parameter choice $\alpha=\beta=-1$ which is used in the original
FR~system.) 

The gauge quantities~$N$ and~$N^i$ in the ADM formulation are fixed in
the FR~system by imposing the generalized harmonic time slicing
condition described in section~\use{sec.HypIntro}:
the lapse~$N$ is determined from the determinant of the three-metric
according to equation~\EqNum{HypHarmFixed}{} with a slicing
density~$Q$ which is a known (that is, not dynamically determined)
function of the spacetime coordinates, and the shift vector~$N^i$ is
also a known function of the coordinates.
To summarize,
$$\eqalign{
  N = {} & Q(t,x^k) \sqrt{\det h_{ij}}
  \, , \cr
  N^i = {} & N^i(t,x^k)
  \, . \cr
} \EQN HypFRGauge
$$
(In Frittelli and Reula~1996 a parameter~$\epsilon$ is included in the
relationship between the lapse and the intrinsic metric.
Since none of the slicing conditions that result from this appear to
be any more useful than harmonic slicing, in the present work the
parameter is fixed at $\epsilon=1/2$, the same value that is used for
the original FR~system.)

The constraint equations~\EqNum{HypADMConstr}{a}
and~\EqNum{HypADMConstr}{b} of the ADM formulation can be rewritten in
terms of the FR~variables: the Hamiltonian constraint is 
${\cal C}^0 = 0$ and the momentum constraint is ${\cal C}^i = 0$ where
$$\EQNalign{
  {\cal C}^0 = {} & \textstyle
      - \partial_k M^{kn}{}_{n}
      + M^{km}{}_{n} M_{k}{}^{n}{}_{m}
      - M_{k} M^{kn}{}_{n}
      + {1\over4} M_{k} M^{k}
    \cr & {} \textstyle
      - {1\over2} M^{k}{}_{n}{}^{m} M_{k}{}^{n}{}_{m}
      - {1\over2} P_{k}{}^{n} P_{n}{}^{k}
      + {1\over4} P^2
      - \kappa \rho
  \, , \EQN HypFRConstr;a \cr
  {\cal C}^i = {} & \textstyle
        \partial_k P^{ik}
      - 2 M^{in}{}_{k} P_{n}{}^{k}
      - {1\over2} M^{i} P
      + M^{n}{}_{k}{}^{i} P_{n}{}^{k}
      + {3\over2} P^{ik} M_{k}
      - \kappa J^{i}
  \, . \EQN HypFRConstr;b \cr
}$$
The notation $P = P_{k}{}^{k}$ and $M_{i} = M_{k}{}^{k}{}_{i}$
is used throughout this work.
Recalling that the constraint equations are satisfied for any
(analytic) solution of the Einstein equations, it is clear that
evolution systems can differ in terms that are proportional to the
constraint quantities~$\displaystyle {\cal C}^{\mu}$ while still
admitting the same physical solutions.
The FR~formulation makes use of this by introducing constraint terms,
controlled by parameters~$\eta$ and~$\gamma$, into a first-order form
of the ADM evolution system, with factors chosen such that
cancellations can occur in the principal part.
Schematically the FR~evolution equations are
$$\EQNalign{
  \partial_t h^{ij} = {} &
    (\,\ldots\,) 
  \, , \EQN HypFRSchem;a \cr
  \partial_t M^{ij}{}_{k} = {} &
    (\,\ldots\,) - 2 \eta N \delta_{k}{}^{\!(i} {\cal C}^{j)}
  \, , \EQN HypFRSchem;b \cr
  \partial_t P^{ij} = {} &
    (\,\ldots\,) + 2 \gamma N h^{ij} {\cal C}^{0}
  \, , \EQN HypFRSchem;c \cr
}$$
where the equations for~$h^{ij}$ and~$P^{ij}$ follow from the ADM
equations~\EqNum{HypADMEvolve}{a,b}, and the equation
for~$M^{ij}{}_{k}$ is found by differentiating its
definition~\EqNum{HypFRVars}{} with respect to time and commuting
spatial and temporal derivatives of the intrinsic metric.
(The parameter~$\eta$ does not explicitly appear in Frittelli and
Reula~1996; its value there is fixed at $\eta=1$ throughout.)

Since spatial derivatives of the intrinsic metric commute, it follows
from the definition~\EqNum{HypFRVars}{} of~$M^{ij}{}_{k}$ that the
FR~variables must satisfy
$$
  \partial_n M^{ij}{}_{k} + M_{n} M^{ij}{}_{k} =
  \partial_k M^{ij}{}_{n} + M_{k} M^{ij}{}_{n}
  \, . \EQN HypFRCommute
$$
This can be viewed as another constraint equation for the FR~system.
Spatial derivatives of~$M^{ij}{}_{k}$ occur in the principal part of
the evolution equation for~$P^{ij}$, and multiples of
equation~\EqNum{HypFRCommute}{} can be added to the evolution system
to alter its hyperbolicity just as multiples of the Hamiltonian and
momentum constraints are added.
A parameter~$\Theta$ is introduced here to control the extent to which
equation~\EqNum{HypFRCommute}{} is used to modify the system.
(In Frittelli and Reula~1996 the parameter value $\Theta=1$ is
implicitly assumed, while the modified FR~system of Stewart~1998
uses~$\Theta=0$.) 

The energy-momentum quantities~$\rho$, $J^i$ and~$S^{ij}$ from
equation~\EqNum{HypADMMatter}{} are assumed to be determined from the
behaviour of the matter that is present in the spacetime.
(For a vacuum spacetime the quantities are of course zero.)
Typically the matter model considered will provide a hyperbolic
evolution system (coupled to gravity) through which values for the
quantities can be determined.
Alternatively the conservation laws~\EqNum{HypADMFluid}{a,b} can be
written as evolution equations for~$\rho$ and~$J^i$ which are closed
by an equation of state specifying~$S^{ij}$: 
$$\EQNalign{
  \partial_t \rho +
    \partial_k ( N J^k - N^k \rho )
  = {} & \textstyle
    N ( 
          S_{k}{}^{n} P_{n}{}^{k}
        - {1\over2} P ( S + \rho )
        - J^{k} M_{k}
        - Q_{,k} Q^{-1} J^{k}
    ) - \rho N^{k}{}_{\!,k}
  \, , \>\> \EQN HypFRMatter;a \cr
  \partial_t J^i +
    \partial_k ( N S^{ik} - N^k J^i )
  = {} & \textstyle
    N ( 
          2 S_{k}{}^{n} M^{ik}{}_{n}
        - S_{k}{}^{n} M_{n}{}^{ki}
        + {1\over2} S M^{i}
        - {3\over2} S^{ik} M_{k}
        + 2 P^{ik} J_{k}
  \cr & {} \quad \textstyle
        - {3\over2} P J^{i}
        - {1\over2} \rho M^{i}
        - Q^{,i} Q^{-1} \rho
    ) - J^{k} N^{i}{}_{\!,k}
      - J^{i} N^{k}{}_{\!,k}
  \, , \EQN HypFRMatter;b \cr
  S^{ij} 
  = {} & \textstyle
    S^{ij} ( \rho, J^k, h^{nm} )
  \, , \EQN HypFRMatter;c \cr
}$$
where derivatives of the gauge functions~$Q$ and~$N^i$ are denoted by,
for example, $Q^{,ij} = h^{ik} h^{jn} \partial_n \partial_k Q$, and
these terms appear in the source part of the system rather than in its
principal part. 

The complete evolution system for the generalized FR~formulation which
incorporates the parameters~$\eta$, $\gamma$ and~$\Theta$ is given in
figure~\use{Fg.HypFREqns}.
The system is written in the flux-conservative form of
equation~\EqNum{PdeConsLaws}{}. 
The REDUCE computer algebra package (Hearn~1995, MacCallum and
Wright~1991) played an important part in deriving and checking these
equations. 
In the next subsection the hyperbolicity of the generalized system is
considered and several interesting ranges of the parameters are
identified. 

\figure{HypFREqns}
\BoxTop
$$\eqalign{
\partial_t h^{ij}
  & {} \textstyle 
    - \partial_n ( N^{n} h^{ij} )
    = N ( 2 P^{ij} - P h^{ij} )
    - 2 N^{(i,j)} - h^{ij} N^{n}{}_{\!,n}
  \vrule width0pt depth10pt
  \, , \cr
\partial_t M^{ij}{}_{k}
  & {} \textstyle
    + \partial_n ( 
        2 N \eta \, \delta_{k}{}^{(i} P^{j)n}
      - N \delta_{k}{}^{n} P^{ij}
      - N^{n} M^{ij}{}_{k} )
  \cr & {} \textstyle
    = N \Big(
        P^{ij} M_{k} 
      - P M^{ij}{}_{k}
      - 2 \eta \, \delta_{k}{}^{(i} [
          P^{j)n} M_{n}
        - {1\over2} P M^{j)}
        - Q_{,n} Q^{-1} P^{j)n}
  \cr & \hskip13.7em {} \textstyle
        + M_{nm}{}^{j)} P^{nm}
        - 2 M^{j)m}{}_{n} P_{m}{}^{n}
        - \kappa J^{j)} ] \Big)
  \cr & \quad {} \textstyle
      - 2 N^{(i}{}_{\!,n} M^{j)n}{}_{k}
      + N^{m}{}_{\!,k} M^{ij}{}_{m}
      - N^{n}{}_{\!,n} M^{ij}{}_{k}
      + h^{ij} N^{n}{}_{\!,nk}
      - N^{(i,j)}{}_{k}
  \vrule width0pt depth10pt
  \, , \cr
\partial_t P^{ij}
  & {} \textstyle
    + \partial_n \Big( 
        2 N (1-\Theta) h^{n(i} M^{j)k}{}_{k}
      + 2 N \Theta M^{n(ij)}
      - N M^{ijn}
      + 2 (\gamma-1) N h^{ij} M^{nk}{}_{k}
      - N^{n} P^{ij} \Big)
  \cr & {} \textstyle
    = N \Big(
        4 M^{n}{}_{k}{}^{(i} M^{j)}{}_{n}{}^{k}
      - M^{k}{}_{n}{}^{i} M^{n}{}_{k}{}^{j}
      - 2 M^{i}{}_{k}{}^{n} M^{jk}{}_{n}
      + {3\over2} M^{ij}{}_{k} M^{k}
      - 3 (1-\Theta)  M^{(i} M^{j)n}{}_{n}
  \cr & \qquad\>\>\, {} \textstyle
      + {1\over2} M^{i} M^{j}
      + 4 (1-\Theta) M^{in}{}_{n} M^{jk}{}_{k}
      + 2 (2\Theta-1) M^{ik}{}_{n} M^{jn}{}_{k}
      - 3 \Theta M_{k} M^{k(ij)}
  \cr & \qquad\>\>\, {} \textstyle
      + 4 (\gamma-1) M^{ij}{}_{k} M^{kn}{}_{n}
      - Q^{,ij} Q^{-1}
      - 2 (1-\Theta) [ Q_{,k} Q^{-1} M^{k(ij)}
                     - Q^{,(i} Q^{-1} M^{j)n}{}_{n} ]
  \cr & \qquad\>\>\, {} \textstyle
      + h^{ij} [ Q_{,k}{}^{k} Q^{-1}
               + \gamma ( 2 Q_{,k} Q^{-1} M^{kn}{}_{n}
                        + {1\over2} P^2 
                        - P_{n}{}^{k} P_{k}{}^{n} )
               + (1-2\gamma) \kappa \rho
  \cr & \hskip5.2em {} \textstyle
               + 2 (\gamma-1) ( M^{km}{}_{n} M_{k}{}^{n}{}_{m}
                              - {3\over2} M_{k} M^{kn}{}_{n}
                              + {1\over4} M_{k} M^{k}
                              - {1\over2} M^{k}{}_{n}{}^{m} 
                                          M_{k}{}^{n}{}_{m} ) ] 
  \cr & \qquad\>\>\, {} \textstyle
      + 2 P_{k}{}^{i} P^{kj}
      - {3\over2} P P^{ij}
      - \kappa S^{ij}
    \Big)
  \cr & \quad {} \textstyle
    - 2 P^{k(i} N^{j)}{}_{\!,k}
    - P^{ij} N^{k}{}_{\!,k}
  \cr
}$$
\BoxBottom
\Caption
The generalized Frittelli-Reula evolution equations incorporating
parameters~$\eta$, $\gamma$ and~$\Theta$.
The unknown variables~$h^{ij}$, $M^{ij}{}_{k}$ and~$P^{ij}$ are
defined by equation~\EqNum{HypFRVars}{}, and the gauge quantities~$N$
and~$N^i$ are defined by equation~\EqNum{HypFRGauge}{}.
The system satisfies the constraint equations~\EqNum{HypFRConstr}{a,b}
and~\EqNum{HypFRCommute}{}.
The matter variables~$\rho$, $J^i$ and~$S^{ij}$ are assumed to satisfy
a hyperbolic evolution system of their own, possibly derived from the
conservation equations~\EqNum{HypFRMatter}{}, and~$\kappa$ is the
constant coupling the Einstein tensor to the energy-momentum tensor.
\bigskip
\endCaption
\endfigure

\SubSection{Characteristic Structure of the Generalized 
            Frittelli-Reula Formulation}
In this subsection the principal part of the FR~evolution system of
figure~\use{Fg.HypFREqns} is investigated for general values of the
parameters~$\eta$, $\gamma$ and~$\Theta$.
A one-parameter family of strongly hyperbolic evolution systems is
identified for which the characteristic speeds coincide with the
physical speeds of the system. 

Recalling the discussion of hyperbolic first-order systems in
section~\use{sec.HypIntro}, the evolution equations of
figure~\use{Fg.HypFREqns} should be rewritten in the form of
equation~\EqNum{HypPDESystem}{} to allow their characteristic structure
to be investigated.
Neglecting undifferentiated terms this gives
$$\eqalign{
  ( \partial_t - N^n \partial_n ) & h^{ij}
    = \ldots 
  \, , \cr
  ( \partial_t - N^n \partial_n ) & M^{ij}{}_{k}
       + N C^{n\,ij}{}_{k\,pq}
         \partial_n P^{pq}
    = \ldots 
  \, , \cr
  ( \partial_t - N^n \partial_n ) & P^{ij}
       + N D^{n\,ij}{}_{\,pq}{}^{r}
         \partial_n M^{pq}{}_{r}
    = \ldots 
  \, , \cr
} \EQN HypFRPrinc
$$
where
$$\eqalign{
  C^{n\,ij}{}_{k\,pq} = {} &
      2 \eta \, \delta_{k}{}^{(i} \delta_{p}{}^{j)} \delta_{q}{}^{n}
    - \delta_{k}{}^{n} \delta_{p}{}^{i} \delta_{q}{}^{j}
    \, , \cr         
  D^{n\,ij}{}_{\,pq}{}^{r} = {} &
      2 (1-\Theta) h^{n(i} \delta_{p}{}^{j)} \delta_{q}{}^{r}
    + 2 \Theta h^{r(i} \delta_{p}{}^{j)} \delta_{q}{}^{n}
    - h^{nr} \delta_{p}{}^{i} \delta_{q}{}^{j}
    + 2 (\gamma-1) h^{ij} \delta_{p}{}^{n} \delta_{q}{}^{r}
    \, . \cr
}$$
The symbol of the FR~system, in the form~\EqNum{HypFRPrinc}{},
is a $30\times30$ matrix~$P(\omega_n)$ defined by
equation~\EqNum{HypPDESymbol}{} where~$\omega_n$ is a three-tuple of
real values.
While the symbol~$P(\omega_n)$ could be written out in full, it is
more convenient here to allow it to be defined by the effect that it
has on an arbitrary column 
vector~$u = [ \bar{h}^{ij}, \bar{M}^{ij}{}_{k}, \bar{P}^{ij} ]^T$:
$$
  P(\omega_n)
  \left[
  \matrix{ \bar{h}^{ij} \cr
           \bar{M}^{ij}{}_{k} \cr
           \bar{P}^{ij} \cr }
  \right]
  = N
  \left[
  \matrix{ 0 \cr
           \omega_n C^{n\,ij}{}_{k\,pq} \bar{P}^{pq} \cr
           \omega_n D^{n\,ij}{}_{\,pq}{}^{r} \bar{M}^{pq}{}_{r} \cr }
  \right]
  - \omega_n N^n 
  \left[
  \matrix{ \bar{h}^{ij} \cr
           \bar{M}^{ij}{}_{k} \cr
           \bar{P}^{ij} \cr }
  \right]
  \, , \EQN HypFRSymbol
$$
where the ordering of the thirty components of the vector~$u$ corresponds
to the ordering of the evolution equations in the
system~\EqNum{HypFRPrinc}{}. 

To determine the hyperbolicity and the characteristic speeds of the
FR~formulation the eigenvalues and eigenvectors of the
symbol~$P(\omega_n)$ must be evaluated.
While this may seem to be a task requiring the use of a computer
algebra package (and, indeed, the following results were verified
using the REDUCE package), the calculation can in fact be performed by
hand following the approach of Friedrich~1996.
Using equation~\EqNum{HypFRSymbol}{}, solutions~$(\lambda,u)$ are
sought to the eigenvalue equation
$$
  P(\omega_n) u = \lambda u
  \quad\Leftrightarrow\quad
  N\left[
   \matrix{ 0 \cr
            \omega_n C^{n\,ij}{}_{k\,pq} \bar{P}^{pq} \cr
            \omega_n D^{n\,ij}{}_{\,pq}{}^{r} \bar{M}^{pq}{}_{r} \cr }
   \right]
  = ( \lambda + \omega_n N^n )
   \left[
   \matrix{ \bar{h}^{ij} \cr
            \bar{M}^{ij}{}_{k} \cr
            \bar{P}^{ij} \cr }
   \right]
  \, , \EQN HypFREigen
$$
where it is important to note that the values~$\bar{h}^{ij}$ used to
denote six of the components of the eigenvector~$u$ are different from
the values~$h^{ij}$ of the components of the intrinsic metric which
appear in the quantities~$C^{n\,ij}{}_{k\,pq}$
and~$D^{n\,ij}{}_{\,pq}{}^{r}$. 
For simplicity the value of the direction vector~$\omega_n$ is fixed
at $\omega_n = (1,0,0)$ corresponding to the $x^1$-direction of the
system. 
Although the FR~equations are not spatially covariant, their principal
part maintains its form under rotations of the spatial axes, and
results for the $x^1$-direction generalize to other spatial
directions.

To solve the eigenvalue equation~\EqNum{HypFREigen}{} a number of
different cases must be considered.
The simplest case is for the eigenvalue $\lambda = -\omega_n N^n = -N^1$, 
and it can be seen that~$\bar{h}^{ij}$ may then take any value
while~$\bar{P}^{ij}$ must be zero, and six equations must be satisfied
by the eighteen components of~$\bar{M}^{ij}{}_{k}$. 
There are thus eighteen linearly independent eigenvectors~$u$
associated with this eigenvalue. 
For~$\lambda \neq -N^1$ it can be seen that the~$\bar{h}^{ij}$
components of the eigenvector~$u$ must be zero, and that
equation~\EqNum{HypFREigen}{} can be written as a set of six equations
for the components~$\bar{P}^{ij}$, with the values
of~$\bar{M}^{ij}{}_{k}$ being determined from these.
Analysis of this set of equations for~$\bar{P}^{ij}$ is
straightforward but made tedious by the number of special cases that
must be considered. 
The seven possible eigenvalues~$\lambda$ of the symbol~$P(\omega_n)$
are found to be
$$\eqalign{
  \lambda_0 = {} & - N^1
  \, , \cr
  \lambda_{\pm A} = {} & - N^1
    \pm N \sqrt{h^{11}}
  \, , \cr
  \lambda_{\pm B} = {} & - N^1
    \pm N \sqrt{3 \eta (1-\Theta) h^{11}}
  \, , \cr
  \lambda_{\pm C} = {} & - N^1
    \pm N \sqrt{( 1 + 2 \gamma [4\eta-1] - 2\eta [1+2\Theta] )h^{11}}
  \, , \cr
} \EQN HypFRSpeeds
$$
for~$\omega_n$ the unit vector in the $x^1$-direction, with analogous
expressions holding for other spatial directions.
(The lapse~$N$ is assumed to be positive, and the spatial
metric~$h^{ij}$ is assumed to be Riemannian so that~$h^{11}$ is also
positive.) 
If it is the case that the eigenvalues of
equation~\EqNum{HypFRSpeeds}{} are all real and distinct then in
addition to the eighteen linearly independent eigenvectors associated
with~$\lambda_0$, 
the eigenvalues~$\lambda_{\pm A}$ are each found to be associated with
a set of three independent eigenvectors (for which the
components~$\bar{P}^{22}$, $\bar{P}^{23}$ and~$\bar{P}^{33}$ can be
chosen arbitrarily and determine the values of~$\bar{M}^{ij}{}_{k}$;
the components~$\bar{P}^{i1}$ and~$\bar{h}^{ij}$ are zero),
the eigenvalues~$\lambda_{\pm B}$ are each associated with
a set of two eigenvectors (for which the
components~$\bar{P}^{12}$ and~$\bar{P}^{13}$ can be chosen arbitrarily
and determine the values of~$\bar{P}^{22}$, $\bar{P}^{23}$,
$\bar{P}^{33}$ and~$\bar{M}^{ij}{}_{k}$; the
components~$\bar{P}^{11}$ and~$\bar{h}^{ij}$ are zero), and
the eigenvalues~$\lambda_{\pm C}$ are each associated with one
eigenvector (for which the component~$\bar{P}^{11}$ can be chosen
arbitrarily and determines the values of~$\bar{M}^{ij}{}_{k}$ and the
remaining~$\bar{P}^{ij}$; the components~$\bar{h}^{ij}$ are zero),
and in this case the eigenvectors form a complete set.
However, if the parameter values~$\eta$, $\gamma$ and~$\Theta$ are
such that some of the eigenvalues in equation~\EqNum{HypFRSpeeds}{}
coincide, then it is possible that the set of eigenvectors may
degenerate such that it no longer has thirty linearly independent
members; this is found to occur if any of the values~$\lambda_{\pm B}$
or~$\lambda_{\pm C}$ coincides with~$\lambda_0$, or if 
$\lambda_{\pm A} = \lambda_{\pm C} \neq \lambda_{\pm B}$.
One particularly interesting case in which the set of eigenvectors is
complete occurs when 
$\lambda_{\pm A} = \lambda_{\pm B} = \lambda_{\pm C}$: the set then
comprises eighteen eigenvectors associated with~$\lambda_0$ (for which
the components~$\bar{P}^{ij}$ are zero), and two sets of six
eigenvectors associated with each of the~$\lambda_{\pm A}$ (for which
the components~$\bar{P}^{ij}$ can be chosen arbitrarily).

As the parameters~$\eta$, $\gamma$ and~$\Theta$ are varied, it can be
seen from the above analysis of the eigenvalues and eigenvectors of
the symbol~$P(\omega_n)$ that the properties
of the evolution system of figure~\use{Fg.HypFREqns} change
substantially. 
If the eigenvalues in equation~\EqNum{HypFRSpeeds}{} are all real
then, recalling section~\use{sec.HypIntro}, the evolution system is at
least weakly hyperbolic, with it being strongly hyperbolic if in
addition the set of eigenvectors is complete.
The values in equation~\EqNum{HypFRSpeeds}{} are the characteristic
speeds of the system in the $x^1$-direction, and three of these speeds
can be considered as being physically significant: the characteristic
speed~$\lambda_0$ describes motion that is normal to the spacelike
hypersurfaces of the ADM foliation, while~$\lambda_{+A}$
and~$\lambda_{-A}$ are the speeds (relative to the spacetime
coordinates) of photons travelling in the $x^1$-direction.
For an evolution system to have only physically relevant
characteristic speeds, the values of~$\lambda_{\pm B}$
and~$\lambda_{\pm C}$ must coincide with the values of~$\lambda_{\pm A}$
or~$\lambda_0$.

Within the generalized FR~formulation certain evolution systems can be
singled out as being of particular interest.
The {\it original\/} FR~system described by Frittelli and Reula~(1996)
is reproduced by the parameter choice $\eta=\gamma=\Theta=1$.
From equation~\EqNum{HypFRSpeeds}{} it can be seen that the
characteristic speeds of this system are all physically relevant:
$\lambda_{\pm B} = \lambda_0$ and $\lambda_{\pm C} = \lambda_{\pm A}$.
However as a consequence of this the eigenvectors of the principal
part degenerate and, in contrast to what Frittelli and Reula claim in
their paper, the system is only weakly hyperbolic.
If the original FR~system is altered so that Frittelli and Reula's
proof of symmetric hyperbolicity can be correctly applied, then what
results is the {\it modified\/} FR~system (as used in Stewart~1998) for
which the parameters are $\eta=\gamma=1$ and $\Theta=0$.
This system has seven distinct characteristic speeds, four of which
are faster than the local light speed: from
equation~\EqNum{HypFRSpeeds}{}, 
$\lambda_{\pm B} = -N^1 \pm N \sqrt{3 h^{11}}$
and $\lambda_{\pm C} = -N^1 \pm N \sqrt{5 h^{11}}$.
From the earlier discussion of eigenvectors it is clear that this
system is at least strongly hyperbolic.

If the parameters are chosen to be $\eta=\gamma=0$ with~$\Theta$
arbitrary, then the FR~formulation reduces to a first-order form of
the standard ADM formulation: no additional constraint quantities are
introduced when the ADM evolution equations~\EqNum{HypADMEvolve}{} are
written in terms of the FR~variables~\EqNum{HypFRVars}{}.
The characteristic structure of these first-order ADM systems is
essentially the same as for the original FR~system: the characteristic
speeds are physically relevant but the evolution equations are only
weakly hyperbolic. 

An obvious question that can be asked about the generalized
FR~formulation is whether there exist parameter choices which lead to
evolution systems that both are strongly hyperbolic and have
physically relevant characteristic speeds. 
It turns out that there is a one-parameter family of such evolution
systems, described here as the {\it causal\/} FR~systems.
The parameters~$\eta$, $\gamma$ and~$\Theta$ must be chosen such that
the eigenvalues of equation~\EqNum{HypFRSpeeds}{} satisfy 
$\lambda_{\pm A} = \lambda_{\pm B} = \lambda_{\pm C}$, and the
solution to this condition can be expressed as
$$
  \Theta = 1 - { 1 \over {3\eta} }
  \, , \qquad
  \gamma = {{9 \eta - 2} \over {12 \eta - 3}}
  \, , \qquad
  \hbox{for \ }\eta \in \RealSet \setminus \{ 0, {\textstyle{1\over4}} \}
  \, . \EQN HypFRCausal
$$

The causal FR~systems described by the equations of
figure~\use{Fg.HypFREqns} with the parameter
settings~\EqNum{HypFRCausal}{} are well suited to numerical
applications.
Their strong hyperbolicity allows them to be used in conjunction with
sophisticated numerical methods (for instance those described in
section~\use{sec.PdeVolume}), and the causal nature of their
characteristic speeds 
should facilitate the imposition of well-behaved boundary conditions.
A key feature of the FR~systems which makes them particularly
appealing from the point of view of numerical work is that they use
only thirty unknown variables; by way of comparison the Bona-Mass\'o
hyperbolic formulation (reviewed at the end of the previous section)
which in many respects is similar to the causal FR~system and which has
been used in many numerical simulations, requires three additional
dependent variables (seven if the lapse and its derivatives are treated
as unknowns).

(Another way by which the individual evolution systems of the
generalized FR~formulation could be classified is by the hyperbolicity
of their constraint propagation equations.
The constraint quantities~$\displaystyle {\cal C}^\mu$ of
equations~\EqNum{HypFRConstr}{}, together with similar quantities
based on equation~\EqNum{HypFRCommute}{}, can be considered as
dependent variables in a first-order evolution system which is linear
with coefficients based on the values of~$h^{ij}$, $M^{ij}{}_{k}$
and~$P^{ij}$, which are assumed to be known.
Although for exact solutions of the Einstein equations the constraint
quantities will be identically zero, in numerical work constraint
violations are inevitable, and the question then arises as to whether
the system describing the propagation of the constraint violations is
well posed or not.
In Stewart~1998, as part of a discussion of the mathematical
properties of the Einstein equations, it is shown that for the
modified FR~system the constraints evolve according to a strongly
hyperbolic system of equations.
It may be speculated that the hyperbolicity of the constraint
evolution system could have some bearing on the stability of numerical
simulations, but to date there is no empirical evidence to support
this, and evolution systems for the constraints are not examined in the
present work.)

In the next section a numerical code is described which implements the
generalized FR~formulation.
Results produced using this code are presented in
section~\use{sec.MonFR}, and in particular a comparison is made
between the numerical performances of the original, modified and
causal FR~systems and the first-order ADM system.

\Section{Numerical Implementation of the Frittelli-Reula Formulation
         \label{sec.HypNumerical}}
This section discusses the implementation of a computer code for the
numerical evolution of the generalized Frittelli-Reula formulation of
Einstein's equations, described in section~\use{sec.HypFR}. 
The code is two-dimensional in the sense that it assumes the spacetime
metric to depend on only two of the three spatial coordinates, and is
restricted to the investigation of spacetimes in which no matter
fields are present.
It has been used to evolve the planar and $U(1)$-symmetric
cosmological models described in chapters~6 and~7, and
section~\use{sec.MonFR} discusses its performance on these problems in
comparison to codes which use specialized sets of evolution equations.
Section~\use{sec.MonFR} also examines how the numerical performance of
the generalized Frittelli-Reula evolution system is affected when its
parameters~$\eta$, $\gamma$ and~$\Theta$ are varied.

In addition to a general discussion of the implementation of the
Frittelli-Reula code, this section considers in detail two technical
aspects of it.
Firstly, while the evolution equations of the Frittelli-Reula
formulation can be simplified based on the symmetry assumed for the
spacetimes, care must be taken to ensure that this is not done in a
way that affects the hyperbolicity of the system;
although it is not immediately apparent that it should be the case,
some seemingly trivial simplifications of the evolution equations can
lead to significant changes in their characteristic speeds.
Secondly, to enable the code to use the high-resolution numerical
integration scheme of chapter~2, a method for solving the Riemann
problem for the principal part of the evolution system must be
devised, and in the present work both exact and approximate Riemann
solvers for the Frittelli-Reula equations are implemented. 
While numerical methods based on Riemann solvers have been used
extensively in work on relativistic hydrodynamics (with the spacetime
metric either being given by an exact solution or evolved separately
using a standard finite difference method; see Font et al.~1998 for a
recent review), this is, to the author's knowledge, the first time
such methods have been used to evolve gravitational fields.

\SubSection{Overview of the Frittelli-Reula Code}
The code presented here for the numerical evolution of the
Frittelli-Reula~(FR) equations is built upon the numerical methods
described in chapters~2 and~3, and can be used to simulate any
spatially two-dimensional vacuum spacetime.
The restriction of the code to two spatial dimensions (discussed at
length in the next subsection) significantly reduces the complexity of
the numerical integration routines and the demands they make on
computational resources, but does not simplify the evolution equations
in any fundamental way: the generalization of the code to three
spatial dimensions is conceptually straightforward.
Since the motivation for the construction of the code is to
investigate the behaviour of the FR~formulation and its usefulness in
numerical relativity, the additional complexity of including matter
fields in the code is forgone in the present work; however it 
should be noted that the numerical approach used here would be
particularly appropriate for simulations in which the FR~equations are
coupled to hydrodynamical sources.

The set of equations presented in figure~\use{Fg.HypFREqns} represents
an evolution system for thirty unknown variables (although, as pointed
out in the next subsection, a set of twenty-seven variables is
sufficient given the symmetry of the spacetimes considered) which has
the flux-conservative form of equation~\EqNum{PdeConsLaws}{}.
Since only vacuum spacetimes are considered here, the
quantities~$\rho$, $J^i$ and~$S^{ij}$ are taken to be identically
zero.
The values of the lapse~$N$ and the shift vector~$N^i$ are defined by
equation~\EqNum{HypFRGauge}{}, and thus the code uses generalized
harmonic time slicing with a given shift vector.
Arbitrary values for~$Q(t,x,y)$ and~$N^i(t,x,y)$ and their first and
second spatial derivatives can be specified in the code through
user-defined functions. 

The Strang splitting approach, described in
section~\use{sec.PdeDifference}, is used to separate the transport and
the source parts of the FR~evolution system.
The transport part of the system is evolved using either a
high-resolution wave-propagation method (discussed in
section~\use{sec.PdeVolume} and later in this section)
or the standard two-step Lax-Wendroff method of
section~\use{sec.PdeDifference}. 
The Lax-Wendroff method is simple to apply since it only requires the
values of the flux vectors to be programmed into the code, and is
useful in the present work because, unlike high-resolution methods, it
can be used to evolve systems of equations which are only weakly
hyperbolic. 
Time steps of variable size can be used by the code, with the
Courant-Friedrichs-Lewy condition (equation~\EqNum{PdeCFL}{}) limiting
the size of the time steps to
$$
  \Delta t \le 
  \min \left\{
    {{ \Delta x }\over{ \vert N^1 \vert + N \sqrt{h^{11}} }} 
  \, , \,
    {{ \Delta y }\over{ \vert N^2 \vert + N \sqrt{h^{22}} }}
  \right\} 
  \, , \EQN HypNumerCFL
$$
for an evolution system which has only physically relevant
characteristic speeds, or to a value smaller than this for evolution
systems (such as the modified FR~system) in which gauge information
can propagate faster than the local light speed.

The source part of the FR~evolution system can be evolved using
standard numerical methods for solving ordinary differential
equations, and in the present work a second-order Runge-Kutta method,
as discussed in section~\use{sec.PdeDifference}, is employed.
As can be seen from the right-hand sides of the equations in
figure~\use{Fg.HypFREqns}, the FR~system includes a large number of
source terms, and programming the code to evaluate these efficiently
is not a trivial matter.
A convenient approach is to use the natural factorization present in the
tensor notation: from a knowledge of the values of the
variables~$h^{ij}$, $M^{ij}{}_{k}$ and~$P^{ij}$, a range of
intermediate quantities such as~$h_{ij}$, $M^{i}{}_{j}{}^{k}$
and~$P_{i}{}^{j}$ are constructed by the code, and these quantities
are then added together in the combinations required to construct the
source terms. 
As an illustration of this approach, if a quantity 
$A^{ij} = M^{n}{}_{k}{}^{i} M^{k}{}_{n}{}^{j}$ is to be evaluated from
the values~$M^{ij}{}_{k}$, then it is much more
efficient (in terms of the total number of multiplication operations
performed) to construct the quantity in several stages, for example
$$
  M_{i}{}^{j}{}_{k} := \sum_{n} h_{in} M^{nj}{}_{k} 
  \quad \longrightarrow \quad
  M_{i}{}^{jk} := \sum_{n} h^{kn} M_{i}{}^{j}{}_{n} 
  \quad \longrightarrow \quad
  A^{ij} := \sum_{n,m} M_{n}{}^{mi} M_{m}{}^{nj}
  \, ,
$$
than it is to evaluate the quantity in one step,
$$
  A^{ij} := \sum_{k,l,m,n,p,q}
            h^{ik} h^{jl} h_{mp} h_{nq} M^{mn}{}_{k} M^{pq}{}_{l}
  \, .
$$
Since the source terms in the evolution equations are evaluated by the
code independently at each spatial point, the temporary construction
of a range of intermediate quantities involves no significant memory
overhead.

As a precaution against errors being introduced when programming in
the source and flux terms of the FR~equations, prototypes of the
routines for evaluating these terms were first implemented in the
REDUCE computer algebra system before being straightforwardly
converted into the \Cplusplus\ language.
This allowed the evolution equations implemented in the FR~code to be
verified against equations which were derived in a manner much less
prone to programming errors.

The FR~code has been thoroughly tested to ensure that it can reproduce
a range of exact solutions to the vacuum Einstein equations.
In addition to the standard Minkowski, Schwarzschild, Kasner,
stationary cylindrical, and polarized Gowdy metrics, the code has
also been tested on versions of these metrics transformed to unusual
coordinate systems.
In particular, a useful test of the code has been provided by the
Minkowski metric written in coordinates for which its ADM
representation has a shift vector~$N^i$ which depends in a non-trivial
way on the spacetime coordinates.
The test of the FR~code is then whether, given exact initial data and
exact values for the gauge functions~$Q(t,x^k)$ and~$N^i(t,x^k)$, it
is able to evolve an approximate solution in which the errors decrease
in magnitude at second order as the numerical grid is refined.
(In practice the order of convergence may be less than the
ideal value near to the edges of the grid where the exact solution is
used to provide boundary data.)
In all cases the FR~code has demonstrated second-order convergence,
both of the error values and of the violations of the constraint
equations. 

In chapters~6 and~7, inhomogeneous cosmological models are examined
for which exact solutions are not known in general.
The FR~code has been used to evolve numerical solutions to these
models, and second-order convergence of the constraint violations in
these solutions provides evidence that the code is performing
correctly.
Furthermore, alternative sets of evolution equations have been used to
evolve the same spacetimes and the results have been found to converge
at second order to the results of the FR~code.
Section~\use{sec.MonFR} discusses the performance of the FR~code in
more detail.

\SubSection{Two-Dimensional Reduction of the Frittelli-Reula System}
The Frittelli-Reula~(FR) code described in the previous subsection is
implemented based on the assumption that any spacetime under
investigation has a single spacelike Killing vector and is described
using a coordinate system in which the spacetime metric~$g_{\mu\nu}$
has no dependence on the spatial $x^3$-coordinate:
$\partial_{3} g_{\mu\nu} = 0$.
The code thus deals with the time evolution of fields defined on a
two-dimensional spatial domain.
This subsection discusses some technical points that arise in the
reduction of the FR~formulation to two spatial dimensions,
and in particular shows that care must be taken to perform the
reduction in such a way that the hyperbolicity of the evolution system
is not compromised.

Some of the simplifications that can be applied to the FR~equations of
figure~\use{Fg.HypFREqns} in two dimensions are obvious.
Since all of
the quantities that describe the spacetime are independent of the
$x^3$-coordinate, the flux vector corresponding to that direction can
be removed and the derivatives with respect to that coordinate of the
gauge functions~$Q$ and~$N^i$ in the source terms can be set to zero;
by considering how a fully three-dimensional code would evolve
two-dimensional solutions, it is clear that these simplifications will
have no effect on the behaviour of the evolution equations.
However, this turns out not to be the case for simplifications of the
formulation based on the removal of some seemingly redundant variables.
From their definition~\EqNum{HypFRVars}{} it is clear that the six
variables~$M^{ij}{}_{3}$ must be identically zero for exact
two-dimensional solutions, and naively it would seem that the
evolution system could be simplified by removing these quantities from
the set of dependent variables; but doing this in fact changes the
hyperbolicity and the characteristic speeds of the system.
Examining the evolution equations for the variables~$M^{ij}{}_{3}$ in
figure~\use{Fg.HypFREqns} it can be seen that the constraint
quantities which have been added to the equations contribute terms to
the principal part of the system which do not vanish when the
$x^3$-coordinate symmetry is imposed; the transport part of the system
(through which hyperbolicity is determined) drives the
variables~$M^{ij}{}_{3}$ away from zero but the source part
counteracts the effect, and while for exact solutions there is precise
cancellation, in numerical simulations the variables will not in
general remain zero.

The failure of the FR~formulation to maintain its hyperbolic character
under imposition of symmetry conditions is quite surprising, and to
investigate this behaviour further a simple one-dimensional reduction
of the evolution system is considered.
The analysis of section~\use{sec.HypFR} of the eigenvalues and
eigenvectors of the symbol~$P(\omega_n)$ of the equations of
figure~\use{Fg.HypFREqns} is repeated under the assumption that the
spacetime metric~$g_{\mu\nu}$ is independent of both the $x^2$-~and
the $x^3$-coordinates: the twelve variables~$M^{ij}{}_{2}$
and~$M^{ij}{}_{3}$ are taken to be identically zero, and only the flux
in the $x^1$-direction is considered.
In this case, in contrast to the equations~\EqNum{HypFRSpeeds}{}, the
eigenvalues of the symbol are found to be
$$\eqalign{
  \lambda_0 = {} & - N^1
  \, , \cr
  \lambda_{\pm A} = {} & - N^1
    \pm N \sqrt{h^{11}}
  \, , \cr
  \lambda_{\pm D} = {} & - N^1
    \pm N \sqrt{(2 \gamma - 1)(2 \eta - 1) h^{11}}
  \, . \cr
} 
$$
There is thus one less pair of eigenvalues in the one-dimensional case
as compared to the three-dimensional case, and the parameter~$\Theta$
no longer has any influence on the system.
Strong hyperbolicity here requires (in addition to the eigenvalues
being real) the parameter choice~$\eta=1$, and in contrast to the
three-dimensional case two fields related to the extrinsic curvature
($\bar{P}^{12}$~and~$\bar{P}^{13}$) propagate in the system normal to
the ADM foliation.
An analogue of the causal FR~systems described in
section~\use{sec.HypFR} (for which strong hyperbolicity is combined
with physically relevant characteristic speeds) is only possible in
the one-dimensional case for the parameter choice $\eta=\gamma=1$.
It is clear from this analysis that the removal of the~$M^{ij}{}_{2}$
and~$M^{ij}{}_{3}$ variables leads to a one-dimensional reduction of
the FR~system which has substantially different properties to the
original. 

Returning to the two-dimensional reduction, it can be seen that
changes in hyperbolicity are not a phenomenon restricted only to the
one-dimensional case: with the~$M^{ij}{}_{3}$ variables removed, the
eigenvalues of the symbol in the $x^1$-direction are
(cf.~equation~\EqNum{HypFRSpeeds}{}) 
$$\eqalign{
  \lambda_0 = {} & - N^1
  \, , \cr
  \lambda_{\pm A} = {} & - N^1
    \pm N \sqrt{h^{11}}
  \, , \cr
  \lambda_{\pm E} = {} & - N^1
    \pm N \sqrt{\eta (1-\Theta) h^{11}}
  \, , \cr
  \lambda_{\pm F} = {} & - N^1
    \pm N \sqrt{2 \eta (1-\Theta) h^{11}}
  \, , \cr
  \lambda_{\pm G} = {} & - N^1
    \pm N \sqrt{( 1 + 2 \gamma [3\eta-1] - 2 \eta [1+\Theta] )h^{11}}
  \, . \cr
} 
$$
From the point of view of constructing a two-dimensional FR~code there
are thus two possibilities: either the variables~$M^{ij}{}_{3}$ are
retained in the evolution system, or they are removed and the
hyperbolicity of the resulting set of equations is re-analysed to
determine which parameter choices lead to well-posed systems.
In the present work, since the main aim is to study the suitability of
the full FR~formulation for performing numerical simulations, the
former approach is taken.

While without fundamentally changing the nature of the FR~formulation
it is not possible to remove the variables~$M^{ij}{}_{3}$ completely from
the evolution system, some simplifications based on the two-dimensional
symmetry of the spacetimes can still be made.
First of all, note that $M^{ij}{}_{3} = 0$ can be thought of as an
additional constraint equation for the system: it is imposed on the
initial data and conserved in exact solutions, and like the other
constraints multiples of it may be added to the evolution equations
without changing their physical solutions.
Since this new constraint contains no (explicit) derivatives of the
dependent variables it can be used to modify the evolution system
without affecting its principal part, and in particular it can be used
to cancel out all of the~$M^{ij}{}_{3}$ terms that appear on the
right-hand sides of the evolution equations.
This means that when the source terms of the FR~formulation are
evaluated by the two-dimensional code, the values of the
variables~$M^{ij}{}_{3}$ can be assumed to be zero without altering
the hyperbolicity or the physical solutions of the system.

A second simplification that can be applied to the two-dimensional
FR~formulation follows from the observation that the evolution
equations for the particular three variables~$M^{11}{}_{3}$,
$M^{12}{}_{3}$($=M^{21}{}_{3}$) and~$M^{22}{}_{3}$ do not in fact
involve any constraint quantities; these variables can be omitted
altogether from the evolution system without affecting its
hyperbolicity. 
Relabelling the remaining variables
$\{ M^{13}{}_{3} , M^{23}{}_{3} , M^{33}{}_{3} \}$ as
$\{ m_1 , m_2 , m_3 \}$, the reduced evolution equations for these
quantities can be seen to take the form
$$
  \partial_t m_i - \partial_n ( N^n m_i )
  = \hbox{(combinations of ${\cal C}^k$ quantities with 
           all $M^{ij}{}_{3}$ terms removed)}
  \, ,
$$
with the only other appearance of the variables~$m_i$ in the reduced
formulation being in the flux terms of the evolutions equations for
the variables~$P^{ij}$.
Analytically the variables~$m_i$ will remain zero if they are zero
initially, but in numerical simulations truncation errors and
constraint violations will prevent this from being the case, and the
overall accuracy of the numerical solution may suffer because of
errors in the variables~$m_i$.
One possible way in which the growth of errors in these variables
could be limited would be to add damping terms to their evolution
equations:
$$
  \partial_t m_i - \partial_n ( N^n m_i )
  = ( \, \ldots \, ) - \varepsilon m_i
  \, ,
$$
where~$\varepsilon$ is a suitably chosen constant.
However, it turns out that a simpler approach can be used in practice:
empirical evidence shows that the values of the variables~$m_i$
can be set to zero intermittently by the code (possibly at the end of
each time step) without affecting its convergence, and this has been
observed to lead to an improvement in simulation accuracy.

To summarize, the two-dimensional FR~code uses a slightly simplified
version of the FR~evolution system which nonetheless has the same
characteristic structure as the original.
(This turns out to be a much less trivial point than at first it may
seem.) 
The six variables~$M^{ij}{}_{3}$ (which encode the derivatives of the
intrinsic metric with respect to the $x^3$-coordinate) are removed from
the formulation and replaced by three variables~$m_i$ which are
identically zero in exact solutions but may deviate from this in
numerical solutions.

\SubSection{The Frittelli-Reula Riemann Problem 
            I: The Case of Zero Shift}
In chapter~2 a multi-dimensional high-resolution wave-propagation
method for the 
numerical integration of strongly hyperbolic flux-conservative
evolution systems is described.
The method, which was developed for use in computational fluid
dynamics, exploits information about the characteristic structure of
the evolution equations to give second-order accurate integration of
smooth solutions while also accurately resolving any discontinuities
that may develop.
In the present work a standard implementation of the wave-propagation
method (the \Clawpack\ package; LeVeque~1998) is incorporated into the
numerical code described above for evolving the Frittelli-Reula~(FR)
formulation of Einstein's equations. 

As described earlier in this section, a splitting approach is used in
the FR~code to separate the evolution of the source terms of the
equations from the evolution of the transport part, and the
wave-propagation algorithm applies to the latter stage only; for the
remainder of this section the right-hand sides of the FR equations of
figure~\use{Fg.HypFREqns} will in general be assumed to be zero.
As discussed in section~\use{sec.PdeVolume}, the wave-propagation
method operates by considering a series of Riemann problems for the
system, and the FR~code is required to provide solutions to these.
In the present work both exact and approximate solutions to the
general Riemann problem for the FR~equations are derived; this
subsection discusses the Riemann problem for the special case when the
shift vector~$N^i$ is identically zero while the next subsection
extends the discussion to the fully general case.
The implementation of the Riemann solvers described here has enabled
the FR~code to use the wave-propagation method to evolve a range of
known solutions to the Einstein equations with second-order accuracy.
Further evidence for the correctness of the Riemann solvers is seen
from the agreement between numerical results produced using the
wave-propagation methods and results produced using the standard
Lax-Wendroff method.

For the purposes of the discussion here only FR~evolution systems
which have physically relevant characteristic speeds are considered.
(Given that an evolution system must be strongly hyperbolic for 
high-resolution numerical methods to be applied, this narrows the choice
down to the family of causal FR~systems defined by
equation~\EqNum{HypFRCausal}{}.) 
This restriction to systems with only three characteristic speeds
significantly simplifies the construction of a Riemann solver since
each distinct speed corresponds to a wave in the solution to the
Riemann problem. 
For historical reasons, the Riemann solver that has actually been
implemented in the present work only applies to the modified
FR~system and thus has to deal with the complicated situation in
which seven waves are produced in each Riemann problem.

\figure{HypRiemann}
 \DrawFig{
  \epsfxsize=0.65\hsize
  \epsfbox{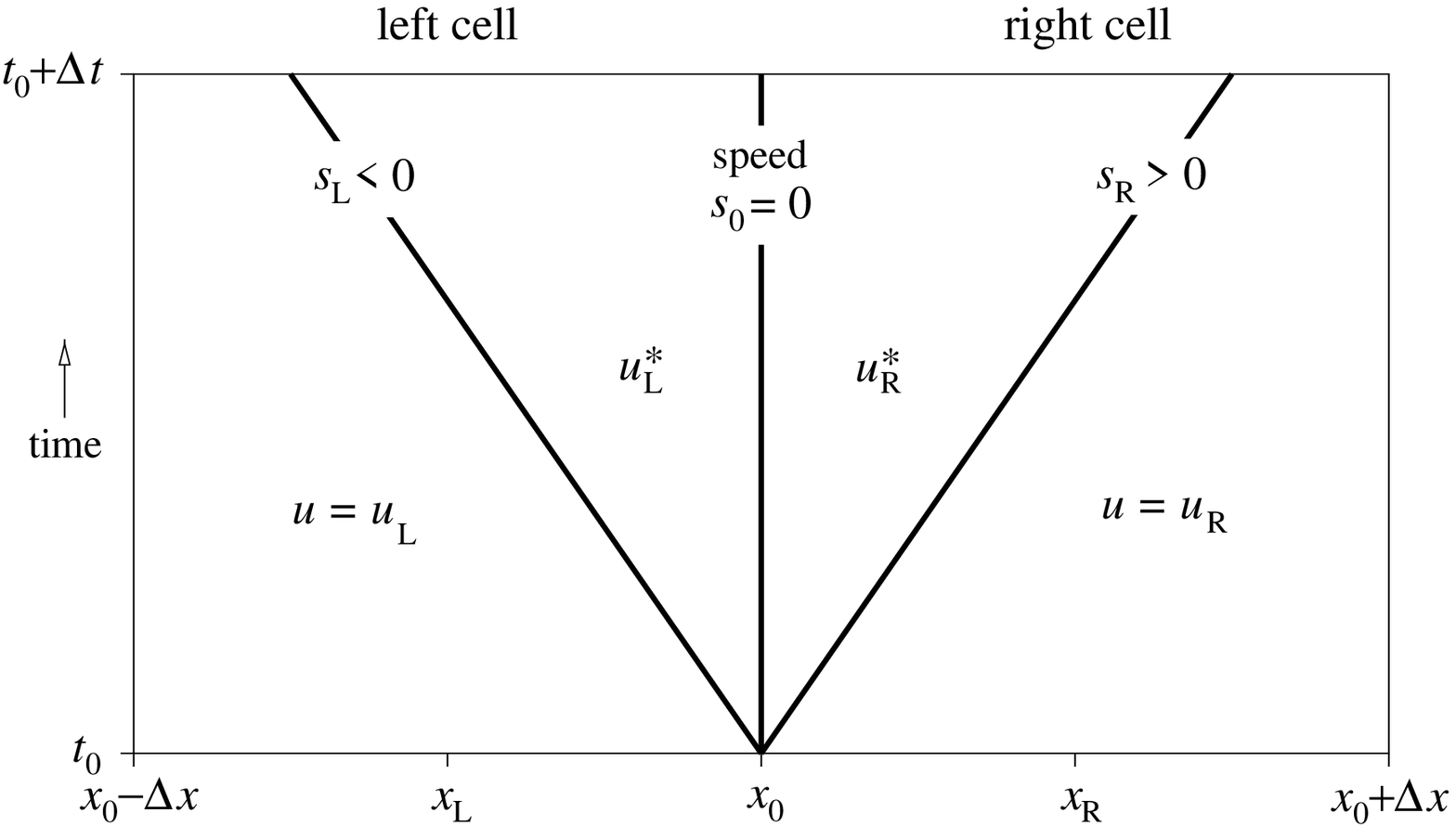}
 }
\Caption
Solution of the Riemann problem for the Frittelli-Reula system
with zero shift.
Two computational cells are shown with constant states~$u_L$ and~$u_R$
at initial time~$t=t_0$.
The discontinuity at the interface between the cells splits into three
waves (assuming the evolution system to have only physically relevant
characteristic speeds): one left-going, one right-going, and one
stationary. 
Two constant states~$u_L^{\star}$ and~$u_R^{\star}$ separate the three
waves.
\bigskip
\endCaption
\endfigure

Figure~\use{Fg.HypRiemann} illustrates the solution to the Riemann
problem for the FR~system under the assumptions that the shift vector
is zero, only physically relevant characteristic speeds are present,
and the evolution equations include no source terms.
(An additional assumption, that the slicing density~$Q$ is a constant,
is discussed below.)
The plane~$x=x_0$ separates two constant states~$u=u_L$ and~$u=u_R$ at
an initial time~$t=t_0$.
At subsequent times the initial discontinuity breaks up into three
waves, each travelling at a constant speed.
Two of the waves propagate (at what is effectively the speed of light)
in opposite directions into the initially constant regions,
leaving in their wakes constant states~$u^{\star}_L$
and~$u^{\star}_R$, while the third wave is a stationary discontinuity
separating these two new states.
(Strictly speaking the three waves are `contact discontinuities', and
this fact makes it feasible to try to derive an exact solution to the
Riemann problem.)
A Riemann solver for the FR~system must determine from a knowledge of
the initial states~$u_L$ and~$u_R$ the values of the speeds of the
three waves, $s_L<0$, $s_0=0$ and $s_R>0$, and the jumps they produce
between the constant states,
$$
  {\cal W}_L = u^{\star}_L - u_L
  \, , \qquad
  {\cal W}_0 = u^{\star}_R - u^{\star}_L
  \qquad \hbox{and} \qquad
  {\cal W}_R = u_R - u^{\star}_R
  \, . \EQN HypRiemJumps
$$
In the discussion here only the Riemann problem in the $x$-direction
is considered; the solutions for the other spatial directions can be
derived following analogous arguments.

The following notation is used here.
The state vector~$u$ for the FR~system consists of thirty dependent
variables~$h^{ij}$, $M^{ij}{}_{k}$ and~$P^{ij}$ which are written with
their indices suppressed: $u=[h,M,P]^T$, such that~$h$ and~$P$ are column
vectors with six components, and~$M$ is a column vector with eighteen
components. 
(The full FR~system is considered here although the discussion applies
equally to the two-dimensional reduction of the system described in
the previous subsection for which the state vector has three less
components.) 
It will also be convenient to define here a reduced state
vector~$v=[M,P]^T$ from which the six variables~$h^{ij}$ are omitted.
The FR~evolution equations of figure~\use{Fg.HypFREqns} (in the
absence of source terms and shift) can be written as
$$
  \partial_t u + \partial_x f(u) 
  = \partial_t \! \left[\matrix{ h \cr M \cr P \cr}\right]
    + \partial_x \! \left[\matrix{ 0 \cr f_M(h,P) \cr 
                                   f_P(h,M) \cr }\right]
  = 0
  \, , \EQN HypRiemEvolve
$$
where~$f(u)$ is the flux vector in the $x$-direction, and the other
spatial directions play no part in the one-dimensional Riemann
problem.
A reduced Jacobian matrix for the system is also defined as
$$
  A = \left[\matrix{ 0 & C \cr  
                     D & 0 \cr }\right]
    = \left[\matrix{ 0 & \partial f_M / \partial P \cr
                     \partial f_P / \partial M & 0 \cr }\right]
  \, , \EQN HypRiemJacob
$$
where the $18\times6$ matrix~$C$ and the $6\times18$ matrix~$D$ both
depend only on the variables~$h^{ij}$ (and the slicing density~$Q$).
The analysis of section~\use{sec.HypFR} shows that the matrix~$A$ has
three eigenvalues: the two eigenvalues 
$\lambda^{\pm} = \pm N \sqrt{h^{11}}$ each corresponds to six
independent eigenvectors, and the eigenvalue $\lambda^{0} = 0$
corresponds to twelve independent eigenvectors.

The exact solution to the Riemann problem illustrated in
figure~\use{Fg.HypRiemann} can now be derived.
(An approximate solution, which has proved to be useful in the present
work, is also derived later in this subsection.)
It is convenient in the following argument to treat the slicing
density~$Q(t,x^i)$ as a constant; variations in the value of~$Q$ are
taken into account in the implementation of the Riemann solver in the
FR~code, and are considered below.
The key point to note in deriving the exact Riemann solution is that,
from equation~\EqNum{HypRiemEvolve}{}, the variables~$h^{ij}$ remain
constant in time, and so, using the subscript and superscript notation
of figure~\use{Fg.HypRiemann}, 
$$
  h^{\star}_{L} = h_{L}
  \qquad \hbox{and} \qquad
  h^{\star}_{R} = h_{R}
  \, . 
$$
Since the flux vector~$f$ is linear in the variables~$M^{ij}{}_{k}$
and~$P^{ij}$ (with coefficients depending on~$h^{ij}$) it follows that
the left-going and right-going waves of the Riemann solution both
propagate as if they were produced by two different linear evolution
systems, and the theory of shock waves in such systems (see
LeVeque~1992, and the discussion in section~\use{sec.PdeRiemann}) is
straightforward to apply. 
The left- and right-going waves have speeds
$$
  s_L = \lambda^-_L = - N_L \sqrt{ h^{11}_L } < 0
  \qquad \hbox{and} \qquad
  s_R = \lambda^+_R = + N_R \sqrt{ h^{11}_R } > 0
  \, , \EQN HypRiemSpeeds
$$
and (omitting the variables~$h^{ij}$ from the state vectors) the
intermediate states~$v^{\star}_L$ and~$v^{\star}_R$ are related to the
initial states~$v_L$ are~$v_R$ by
$$
  v^{\star}_{L} = v_L + \sum_{i=1}^{6} \alpha_i r^i_L
  \qquad \hbox{and} \qquad
  v^{\star}_{R} = v_R - \sum_{i=1}^{6} \beta_i r^i_R
  \, , 
$$
where $r^1_L$,~\dots,~$r^6_L$ are the six eigenvectors of~$A_L$
associated with the eigenvalue~$\lambda^-_L < 0$, 
and $r^1_R$,~\dots,~$r^6_R$ are the six eigenvectors of~$A_R$
associated with the eigenvalue~$\lambda^+_R > 0$, and these
eigenvectors are assumed to be known.
The twelve real values $\alpha_1$,~\dots,~$\alpha_6$ 
and $\beta_1$,~\dots,~$\beta_6$ must be determined in order to solve
the Riemann problem.
(The subscripts~`L' and~`R' are used here to distinguish between
quantities which are evaluated based on the values~$h^{ij}_L$ and
those which are based on the values~$h^{ij}_R$.)

From consideration of the Rankine-Hugoniot jump conditions across
the individual waves (equation~\EqNum{PdeRHJump}{}) it can be seen
that the intermediate states must satisfy 
$$
  f(u_R) - f(u_L) = s_R ( u_R - u^{\star}_R )
                  - s_L ( u_L - u^{\star}_L )
  \, , \EQN HypRiemJumpU
$$
and from equations~\EqNum{HypRiemEvolve}{} and~\EqNum{HypRiemJacob}{}
this condition can be written as 
$$
  A_R v_R - A_L v_L = 
    s_L \sum_{i=1}^{6} \alpha_i r^i_L
  + s_R \sum_{i=1}^{6} \beta_i r^i_R
  \, . \EQN HypRiemJumpV
$$
All of the quantities in this equation are known (given the initial
states~$u_L$ and~$u_R$) except for the twelve values~$\alpha_i$
and~$\beta_i$, and it turns out that the equation, viewed as a linear
system of twenty-four equations, can be inverted in a relatively
straightforward manner to determine the exact solution of the Riemann
problem.
If the components of the eigenvectors~$r^i_{L/R}$ are separated as 
$r = [\mu,\pi]^T$ in an analogous manner to the variables in the state
vector~$v$, then equation~\EqNum{HypRiemJumpV}{} can be re-expressed
in block matrix form as
$$
  { {\left[\matrix{ s_L \Boldmu_L & s_R \Boldmu_R \cr
                    s_L \Boldpi_L & s_R \Boldpi_R \cr }\right]}
    \atop {\mathstrut\hbox{\ninepoint ($24\times12$)}} }
  \!\!
  { {\left[\matrix{ \alpha \cr \beta \cr }\right]}
    \atop {\mathstrut\hbox{\ninepoint ($12\times1$)}} }
  \!\! { = \atop {\vrule width0pt height16pt} } \,
  { {\left[\matrix{ C_R P_R - C_L P_L \cr 
                    D_R M_R - D_L M_L \cr }\right]}
    \atop {\mathstrut\hbox{\ninepoint ($24\times1$)}} }
  \, { , \atop {\vrule width0pt height16pt} } \,
  \EQN HypRiemSystBig
$$
where~$\Boldmu$ and~$\Boldpi$ are (respectively)
$18\times6$ and $6\times6$~matrices with columns equal to the vectors
$\mu^1$,~\dots,~$\mu^6$ and
$\pi^1$,~\dots,~$\pi^6$, and the unknowns~$\alpha_i$
and~$\beta_i$ are written here combined as a column vector.
The form of the FR~evolution equations is such that a
$6\times18$~matrix~$\xi$ with constant coefficients can be found
which multiplies the matrix~$C$ to give $\bar{C} = \xi C$, a
non-singular $6\times6$ matrix; in fact~$\xi$ can be chosen such
that the new matrix~$\bar{C}$ is proportional to the identity matrix.
Multiplying through the upper part of
equation~\EqNum{HypRiemSystBig}{} by~$\xi$ and defining 
$\bar{\mu} = \xi \mu$ then gives
$$
  { {\left[\matrix{ s_L \bar{\Boldmu}_L & s_R \bar{\Boldmu}_R \cr
                    s_L \Boldpi_L & s_R \Boldpi_R \cr }\right]}
    \atop {\mathstrut\hbox{\ninepoint ($12\times12$)}} }
  \!\!
  { {\left[\matrix{ \alpha \cr \beta \cr }\right]}
    \atop {\mathstrut\hbox{\ninepoint ($12\times1$)}} }
  \!\! { = \atop {\vrule width0pt height16pt} } \,
  { {\left[\matrix{ \bar{C}_R P_R - \bar{C}_L P_L \cr 
                    D_R M_R - D_L M_L \cr }\right]}
    \atop {\mathstrut\hbox{\ninepoint ($12\times1$)}} }
  \, { . \atop {\vrule width0pt height16pt} } \,
  \EQN HypRiemSystSmall
$$
Recalling that $r=[\mu,\pi]^T$ is defined to be an eigenvector of the
matrix~$A$ of equation~\EqNum{HypRiemJacob}{}, it is easy to show that
each of the vectors~$\bar{\mu}^i$ must be proportional to its
corresponding vector~$\pi^i$.
Further noting that the eigenvectors of the matrix~$A$ can always be
chosen such that the matrix~$\Boldpi$ is the identity matrix, it
becomes clear that the $12\times12$~matrix on the left-hand side of
equation~\EqNum{HypRiemSystSmall}{} is non-singular and can be
straightforwardly inverted to give the values of~$\alpha_i$
and~$\beta_i$ in terms of the initial states~$u_L$ and~$u_R$.

The exact solution to the Riemann problem is thus found, with the
speeds~$s_L$, $s_0$ and~$s_R$ known (see
equation~\EqNum{HypRiemSpeeds}{}) and the wave jumps~${\cal W}_L$,
${\cal W}_0$ and~${\cal W}_R$ of equation~\EqNum{HypRiemJumps}{} being
simple to construct once the intermediate states~$u^{\star}_L$
and~$u^{\star}_R$ are determined from the values~$\alpha_i$
and~$\beta_i$. 
(For FR~systems in which characteristic speeds other than the three
physically relevant ones are present, the solution to the Riemann
problem can be found by following essentially the same argument as
above, but the algebra involved will be considerably more
complicated. 
In particular the matrix inversion required to solve the analogue of
equation~\EqNum{HypRiemSystSmall}{} is no longer straightforward.)

An approximate solution to the Riemann problem can also be derived by
following an approach similar to the one described in the classic~1981
paper by Roe (see also section~\use{sec.PdeRiemann}) with the nonlinear
problem~\EqNum{HypRiemEvolve}{} being in effect replaced by a linear
problem
$$
  \partial_t u + \tilde{A}_0 \partial_x u = 0
  \, , \EQN HypRiemLinear
$$
for a suitable $30\times30$ matrix~$\tilde{A}_0$ with constant
coefficients. 
The matrix~$\tilde{A}_0$ is chosen to be the Jacobian matrix 
$\tilde{A}(u) = \partial f / \partial u$ evaluated at a state $u=u_0$,
where in calculating~$\tilde{A}(u)$ the lapse~N (which of course
depends on the variables~$h^{ij}$ and the slicing density~$Q$) is
treated as having a constant value $N=N_0$.
The state~$u_0$ is chosen here to be the average of the two initial
states in the Riemann problem: $u_0 = {1\over2}(u_L + u_R)$.
If the lapse~$N$ is truly a constant then the components of the flux
vector~$f$ in the FR~system~\EqNum{HypRiemEvolve}{} are at most
quadratic in the unknowns~$u$, and it can be shown (Roe~1981) that the
following property holds:
$$
  \tilde{A}_0 (u_R - u_L) = f(u_R) - f(u_L)
  \, , 
$$
which is required in order for the approximate solution to the Riemann
problem to be conservative.
In the general case however this equation will not be satisfied,
although in the next subsection a reformulation of the FR~equations is
described which admits a conservative approximate Riemann solver of
this form.

The three waves in the approximate Riemann solution have speeds equal
to the eigenvalues of the matrix~$\tilde{A}_0$, that is 
$s_L = - N_0 \sqrt{h^{11}_0}$, $s_0 = 0$ and 
$s_R = + N_0 \sqrt{h^{11}_0}$, and jumps~${\cal W}_L$, ${\cal W}_0$
and~${\cal W}_R$ which are eigenvectors of~$\tilde{A}_0$ corresponding
to these eigenvalues and which satisfy 
$$
  \Delta u = {\cal W}_L + {\cal W}_0 + {\cal W}_R 
  \, , \EQN HypRiemDeltaU
$$
where $\Delta u = u_R - u_L$.
That three such wave jumps can be found is a consequence of the strong
hyperbolicity of the FR~system, and their values may be determined by
noting that, by definition,
$$
  \tilde{A}_0 \Delta u = s_L {\cal W}_L + s_R {\cal W}_R
  \, ,
$$
and that this can be solved in basically the same way as
equation~\EqNum{HypRiemJumpU}{}, or that
$$
  {\cal W}_L = {{ \tilde{A}_0 ( \tilde{A}_0 - s_R I ) \Delta u }
           \over{ s_L ( s_L - s_R ) }}
  \qquad \hbox{and} \qquad
  {\cal W}_R = {{ \tilde{A}_0 ( \tilde{A}_0 - s_L I ) \Delta u }
           \over{ s_R ( s_R - s_L ) }}
  \, , 
$$
where~$I$ is the identity matrix, and the final wave jump~${\cal W}_0$
can be calculated from equation~\EqNum{HypRiemDeltaU}{}.

Either of the two Riemann solvers described above can be implemented
to allow solutions of the FR~equations to be evolved using the
wave-propagation algorithm.
The Riemann problem at the interface between two computational cells
(figure~\use{Fg.HypRiemann}) which are centred at coordinate positions
$x=x_L$ and $x=x_R$ (with fixed values for the other spatial
coordinates~$y$ and~$z$) forms the basic component of the evolution
procedure that advances a numerical solution from a time $t=t_0$ to a
time $t=t_0+\Delta t$.
The wave speeds and jumps that comprise the solution to the Riemann
problem are calculated using the cell-averaged data~$u_L$ and~$u_R$
together with values~$N_L$, $N_0$ and~$N_R$ for the lapse which are
evaluated according to 
$$
  \textstyle
  N_\gamma = Q(t_0 + {1\over2}\Delta t, x_\gamma, y, z) 
             \sqrt{ \det (h_{ij})_\gamma }
  \qquad \hbox{for } \gamma = L, 0, R
  \, ,
$$
where $x_0 = {1\over2}(x_L + x_R)$.
The wave-propagation algorithm also requires the `fluctuations'
between the cells to be specified, and these are calculated following
equation~\EqNum{PdeFluxDiff}{} as
$$
  {\cal A}^+ = \sum_{\gamma=L,0,R} \max(s_\gamma,0) \, {\cal W}_\gamma
  \qquad \hbox{and} \qquad
  {\cal A}^- = \sum_{\gamma=L,0,R} \min(s_\gamma,0) \, {\cal W}_\gamma
  \, . \EQN HypRiemFlux
$$
For simulations that depend on only one spatial dimension, this
information is sufficient for a numerical solution to the FR~equations
to be evolved using the wave-propagation method---the update step is
shown in equation~\EqNum{PdeWavePropOne}{}---and when the exact
Riemann solver is used the solution is found to converge at second
order as the numerical grid is refined. 
In contrast, the approximate Riemann solver, because of the ad~hoc way
in which it treats the lapse, does not in general produce solutions
that are second-order convergent.

The usefulness of the approximate Riemann solver is realized when
simulations in more than one spatial dimension are performed.
As described in section~\use{sec.PdeVolume}, for two-dimensional
evolutions the wave-propagation algorithm applies correction terms to
the solution of each one-dimensional Riemann problem to account for
effects of wave propagation transverse to the original direction: 
as figure~\use{Fg.PdeTransverse} illustrates, the waves produced by a
Riemann problem at the interface between two adjacent computational
cells are propagated in a multi-dimensional way so that they affect
the data in all of the cells surrounding the interface.
The transverse correction terms are calculated by splitting the
original fluctuations ${\cal A}^+$~and~${\cal A}^-$ of
equation~\EqNum{HypRiemFlux}{} for the $x$-direction according to an
approximate Riemann solution for the $y$-direction (with similar
transverse corrections being also calculated with the spatial
directions switched around).
A matrix~$\tilde{B}_0$ is evaluated at the interface between the cells
in the same way as the matrix~$\tilde{A}_0$ is evaluated in the
approximate Riemann solver described above (and using the same
values of~$u_0$ and~$N_0$) but in this case it is based on the Jacobian
matrix~$\tilde{B}(u)$ for the flux vector in the transverse direction.
Instead of calculating waves based on the vector~$\Delta u$, the
transverse approximate Riemann solver is applied in turn to the two
fluctuations~${\cal A}^\pm$ to construct, in analogy with
equation~\EqNum{HypRiemFlux}{}, four transverse 
fluctuations~${\cal B}^\pm \! {\cal A}^\pm$.
These then contribute to the second-order correction terms which are
used to update the data in the surrounding cells according to the
two-dimensional wave-propagation step of
equation~\EqNum{PdeWavePropTwo}{}. 

This use of an approximate Riemann solver to produce transverse
corrections to exact Riemann solutions has been found to lead to the
production of second-order accurate numerical solutions to the
FR~evolution equations when using the wave-propagation algorithm.
Unfortunately the method does not generalize to the case when the
shift vector~$N^i$ is not identically zero, and some alternative
approaches for dealing with that case are considered in the next
subsection.

\SubSection{The Frittelli-Reula Riemann Problem 
            II: The Case of General Shift}
The incorporation of a non-trivial shift vector into
the Riemann solvers described in the previous subsection has proven 
to be problematic.
The status of the shift vector~$N^i$, together with the slicing
density~$Q$, as given functions of the spacetime coordinates rather
than either constants or dependent variables causes complexities when
attempting to use the wave-propagation algorithm to produce
second-order accurate solutions to the FR~equations. 
In this subsection two approaches for constructing Riemann solvers
which give second-order convergence in the presence of a non-zero shift
vector are discussed. 
(As an aside, it may be noted that alternative methods for dealing
with the shift vector are available.
For example, in Scheel et al.~1997 a `causal differencing' method is
developed in which the influence of the shift vector is removed from
the time advancement step of the integration algorithm and accounted
for by shifting the numerical solution using spatial interpolation.
Such a method has the practical advantage that, as
equation~\EqNum{HypNumerCFL}{} shows, the Courant-Friedrichs-Lewy
condition allows large time steps to be taken if the shift vector is
zero during the time evolution phase.)

As can be seen from the equations of figure~\use{Fg.HypFREqns}, the
shift vector in effect introduces an advection term to the transport
part of the FR~system.
If~$N^i$ has a constant value then it is straightforward to take its
presence into account in the Riemann solvers described in the previous
subsection: for the Riemann problem in the $x$-direction the wave
speeds are simply modified as
$$
  s_\gamma^{\SmallWord{new}} = s_\gamma - N^1
  \qquad \hbox{for } \gamma = L,0,R
  \, ,
$$
with these new speeds being used to calculate the fluctuations of
equation~\EqNum{HypRiemFlux}{}. 
An extension to this approach which accounts for spatial and temporal
variations in the value of the shift vector is to include
`second-order' corrections in the modifications made to the wave speeds: 
$$
  \left.\eqalign{
    \hat{s}_\gamma & {} 
      = s_\gamma - N^1(t_0, x_0, y, z) 
      \quad \cr
    \hat{x} & {} \textstyle
      = x_0 + {1\over2} \Delta t \, \hat{s}_\gamma
      \quad \cr 
    s_\gamma^{\SmallWord{new}} & {} \textstyle
      = s_\gamma - N^1(t_0+{1\over2}\Delta t, \hat{x}, y, z)
      \quad \cr 
  }\right\}
  \quad \hbox{for } \gamma = L,0,R
  \, . \EQN HypRiemShift
$$
However, in modelling the influence of the shift vector on the
transport part of the FR~equations in this way the evolution system
is in effect rewritten so that derivatives of the shift vector no longer
appear on the left-hand side; schematically,
$$
  \partial_t u + \partial_i [ f^i(u) - N^i u ] = s(u)
  \quad \longrightarrow \quad
  \partial_t u + \partial_i f^i(u) - N^i \partial_i u 
               = s(u) + N^i{}_{\!,i} u
  \, , \EQN HypRiemRewrite
$$
and the numerical implementation of the equations must be adjusted to
take into account the appearance of additional terms in the source
part of the system. 

Inclusion of the above modifications in the Riemann solvers
described in the previous subsection is not sufficient to allow the
wave-propagation algorithm to evolve second-order 
accurate numerical solutions to the FR~equations when the shift vector
is non-trivial.
However two methods which do give second-order accuracy
have been found: in both cases the FR~evolution system is reformulated
so that the slicing density~$Q$ does not appear in the transport part
of the equations.
In the first approach the lapse~$N$, which in the evolution equations
of figure~\use{Fg.HypFREqns} is used as a shorthand for the
expression~$Q\sqrt{\det h_{ij}}$, is raised to the status of an
unknown variable which evolves according to
$$
  \textstyle
  \partial_t N - \partial_n (N^n N) = 
  N ( {1\over2} N P 
    + Q_{,t} Q^{-1}
    - Q_{,k} Q^{-1} N^{k} )
  \, , \EQN HypRiemLapse
$$
(cf.~equation~\EqNum{HypHarmEvolve}{}) and an extended FR~system of
thirty-one variables is produced. 
With the lapse treated as an unknown in this way the explicit
dependence of the FR~flux vectors on the slicing density~$Q$ is
removed, although it is still present in the source terms.
The exact expression for the lapse in terms of the slicing density and
the spatial metric may be used to intermittently update the value of
the variable~$N$ during the course of numerical simulations;
however doing this is not found to lead to any substantial improvement
in the performance of the code.

The second method considered here to simplify the transport part of
the FR~equations redefines the dependent variables so that the lapse
is removed from the system altogether.
If a new set of FR~variables is defined in terms of the standard
variables and powers of the lapse according to
$$\eqalign{
  \tilde{h}^{ij} = {} & N^2 h^{ij}
  \, , \cr
  \tilde{M}^{ij}{}_{k} = {} & N^\alpha M^{ij}{}_{k}
  \, , \cr
  \tilde{P}^{ij} = {} & N^{\alpha+1} P^{ij}
  \, , \cr
} \EQN HypRiemHidden
$$
then the transformed evolution system is found to have exactly
the same transport part as in figure~\use{Fg.HypFREqns} (with the new
variables replacing the standard variables) but with the lapse in
effect set to~$N=1$.
Furthermore, if the constant in equation~\EqNum{HypRiemHidden}{} is
chosen to be~$\alpha=2$, then the source terms of the new evolution
system are also independent of~$N$, although they do still depend on
the slicing density~$Q$.
(If non-vacuum spacetimes are considered then a similar
redefinition of the variables~$\rho$, $J^i$ and~$S^{ij}$ can be used
to remove the lapse from the matter terms in the equations.)
Although this `hidden lapse' version of the FR~formulation is slightly
more awkward to interpret than the standard version, the simplicity of
its transport part makes it well suited for use with the
wave-propagation method: there are no problems in evaluating the
slicing density~$Q$ to give second-order accuracy, and
since the flux vectors are no more than quadratic in the dependent
variables an approximate Riemann solver constructed as described in
the previous subsection will be flux-conservative.

These two alternative versions of the FR~system, one in which the
lapse is evolved as an unknown variable and one in which the lapse is
hidden by a redefinition of the variables, have both been successfully
used to give second-order convergence with the wave-propagation
algorithm in two-dimensional simulations of vacuum spacetimes for
which the shift vector has a non-trivial dependence on the spacetime
coordinates.
The exact and approximate Riemann solvers of the previous subsection
are used (these generalize straightforwardly to the new systems)
with the latter providing transverse corrections to the former, and
the shift vector is treated using the modifications shown in
equations~\EqNum{HypRiemShift}{} and~\EqNum{HypRiemRewrite}{}. 
(In fact, for the two alternative FR~systems second-order convergence
is still obtained if the approximate Riemann solver is used
exclusively in the wave-propagation method.)

In all cases for which the evolution is second-order accurate, the
wave-propagation method has been found to produce results which are
comparable in accuracy to those produced using the standard
Lax-Wendroff method. 
No significant differences in accuracy have been observed between
simulations based on the three different versions of the FR~equations,
and, when both are second-order convergent, the exact and the
approximate Riemann solvers have been found to give nearly
indistinguishable results. 
Unsurprisingly given its complexity, the wave-propagation method is
found to be much slower than the Lax-Wendroff method: for the
implementation of the FR~equations used here, evolving the transport
part of the system based on exact Riemann solutions takes about twelve
times longer than if the Lax-Wendroff method is used, and for
approximate Riemann solutions it takes about ten times longer.
For comparison, more than twice as much time is spent by the code
evolving the source part of the system than is spent evolving the
transport part by the Lax-Wendroff method.
(It should be noted however that these performance results concern the
{\it modified\/} FR~system---using the terminology of
section~\use{sec.HypFR}---for which seven rather than three waves are 
present in the solution to each Riemann problem.
This makes the operation of a Riemann solver significantly more
complex, and undoubtedly makes a sizeable contribution to the running
times for the wave-propagation method.)

In this and the previous subsection a number of approaches have been
described that allow the equations of the FR~formulation to be
numerically evolved using the high-resolution wave-propagation method
of chapter~2.
As might be expected, for spacetimes in which the metric fields are
suitably smooth the results produced using the wave-propagation method
are no more accurate than those produced using a standard finite
difference method.
In chapters~6 and~7, cosmological models in which steep gradients
develop in the components of the metric are investigated, and while it
may be hoped that the use of high-resolution numerical methods would
lead to improvements in performance for these spacetimes, in practice
this is not found to be the case. 
Besides their possible usefulness in implementing well-posed boundary
conditions, the main motivation for exploring the use of 
high-resolution methods to evolve Einstein's equations comes from the 
gains in performance that may be expected when simulating spacetimes in
which the gravitational fields are coupled to hydrodynamical sources:
both metric and matter variables can be evolved then in a unified
manner with shock waves in the fluid (and any knock-on effects in the
gravitational fields) receiving a sophisticated numerical treatment.
The extension of the FR~code to non-vacuum spacetimes is, however,
beyond the scope of the present work.


\Chapter{Homogeneous Cosmologies}
The preceeding three chapters discuss the development and
implementation of a computer code capable of generating approximate
solutions to the field equations of general relativity.
The code has been used to study inhomogeneous cosmologies, and the two
chapters that follow present results from numerical simulations of
planar and $U(1)$-symmetric cosmological models. 
In the present chapter consideration is given to some problems that
are often encountered in work on numerical relativity resulting from
freedom in the choice of spacetime coordinates, and the
discussion is illustrated using numerical simulations of simple
homogeneous cosmological models.

Homogeneous and inhomogeneous cosmologies are discussed at length in
the books by Ryan and Shepley~(1975) and Krasi\'nski~(1997); the
latter includes a discussion of the relevance of inhomogeneous models
given the widespread acceptance of the homogeneous standard
cosmological model.
Analytic work on inhomogeneous cosmologies is typically limited to
spacetimes which have a high degree of symmetry or which are within
the `linear regime' of a known solution.
In the general case
progress in understanding the behaviour of inhomogeneous cosmologies
seems only to be possible through the use of numerical simulations
(see Anninos~1998b for a review).
In the present work (chapters~6 and~7) inhomogeneous cosmological
models are studied numerically with the aim being to develop and
test numerical approaches to solving the Einstein equations.
The symmetry, topology, and lack of matter content of the spacetimes
considered here rules them out as realistic models of the physical
universe; however a number of aspects of the behaviour of
inhomogeneous cosmologies are investigated in the following chapters
including the propagation of linear and nonlinear gravitational waves,
the formation of fine-scale structure during gravitational collapse,
and the nature of cosmological singularities.

Even for cases in which exact solutions are known, homogeneous
models can still be of interest in numerical relativity.
When written in a coordinate system for which its homogeneity is not
explicit, a homogeneous cosmology is indistinguishable from an
inhomogeneous one for the purposes of performing numerical
simulations, and in this way provides a non-trivial test case for
numerical codes; furthermore, homogeneous cosmologies can be used to
assess the performance of different gauge conditions that could be
employed in simulations of inhomogeneous models. 
The first of these points motivates the discussion in
section~\use{sec.HomCompare}.
There the question is considered as to how a numerical solution for a
homogeneous cosmology which is constructed using an inhomogeneous time
slicing condition can be compared to a standard solution for the
model, and, conversely, how a genuinely inhomogeneous spacetime can be
distinguished from a homogeneous one evaluated in a non-standard
gauge. 
In addition, an algorithm is presented for comparing different time slicings
of the same spacetime: the coordinate system of a numerically evolved
solution is reconstructed relative to a known metric using the values
taken by the foliation's lapse function.

Section~\use{sec.HomHarmonic} considers time slicing conditions for
numerical simulations, with particular attention being given to
the harmonic slicing condition introduced in chapter~4.
Analytic and numerical approaches are used to investigate the
behaviour of foliations of a homogeneous cosmological model which are
generated using a harmonic time coordinate, and it is demonstrated that
such foliations may, under quite general circumstances, fail to cover
the whole of the spacetime due to the formation of coordinate
singularities.  
The behaviour of the metric components during the formation of
coordinate singularities is examined and compared to the behaviour
seen in numerical simulations of inhomogeneous cosmologies in
chapters~6 and~7, and the implications that these results have for
numerical work in general are discussed.

\Section{Comparing Numerical Spacetimes with Different Gauge Conditions
         \label{sec.HomCompare}}
Chapter~4 describes the standard `three-plus-one' approach to
producing numerical solutions to the Einstein equations, and two
quantities---the lapse function~$N$ and the shift vector~$N^i$---are
introduced there which encode the freedom in the choice of spacetime
coordinates for a numerical simulation. 
Although a number of standard ways of specifying the lapse and the
shift exist (properties of some of these are considered in
section~\use{sec.HomHarmonic}), in general two independent numerical
simulations of the same spacetime will produce results which are with
respect to completely different coordinate systems.
In the present section consideration is given to the question of how
such results can be compared in practice, and two points in particular
are discussed: how coordinate independent quantities can be used to
describe spacetimes, and how different coordinate systems for a
spacetime can be compared.
In the next section these ideas are applied to simulations of a
homogeneous cosmological model.

\SubSection{Scalar Quantities and Curvature Invariants}
In presenting results from numerical simulations in chapters~6 and~7
an attempt has been made to use (where possible) scalar quantities
to describe the evolved spacetimes.
This subsection considers some of the types of scalar quantities that
can be used in this way, and examines the possible advantages their
use has in practice.

Typically the unknown variables that are used in numerical simulations
have a non-trivial dependence on the gauge conditions employed: they
are not in general four-dimensional scalars, and often they obey
complicated transformation laws under changes of spacetime
coordinates. 
For the purposes of describing the evolved spacetimes it is more
practical to use the variables to construct scalar quantities which
are independent of the spacetime coordinate system than it is to use the
variables themselves, and there are several approaches to doing this.
For spacetimes which have symmetries, the properties of Killing vectors
can be used to define scalar quantities which have simple physical
interpretations.
This is done for the inhomogeneous cosmologies
investigated in chapters~6 and~7, with, for example, the proper area
of the symmetry plane providing scalar information about the planar
models. 
(The `areal radius' defines an analogous quantity for spherically
symmetric spacetimes.)
Furthermore, for non-vacuum spacetimes the matter fields may naturally
define scalar quantities, with examples being, depending on the type
of matter present, wave functions and fluid rest densities.

The types of scalar quantities described above are clearly not
appropriate for presenting results for vacuum spacetimes which have no
simplifying symmetries, and in such cases the invariants of the
Riemann curvature tensor may provide useful alternatives.
It is well known (see for example Witten~1959, York~1989) that
fourteen independent (pseudo-)scalars can be constructed from the
Riemann tensor~$R_{\mu\nu\rho\sigma}$, and that for a vacuum
spacetime all but four of these curvature invariants vanish
identically. 
In the present work, the four curvature invariants which are
non-trivial for vacuum spacetimes are expressed as
$$
  \eqalign{
    \InvA = {} & 
                 \frac{1}{8} C_{\mu\nu\rho\sigma}
                             C^{\mu\nu\rho\sigma}
    \, , \cr
    \InvC = {} & 
               - \frac{1}{16} C_{\mu\nu}{}^{\rho\sigma}
                               C_{\rho\sigma}{}^{\eta\omega}
                               C_{\eta\omega}{}^{\mu\nu}
    \, , \cr
  } \qquad \eqalign{
    \InvB = {} & 
                 \frac{1}{8} C^{\textstyle\ast}_{\mu\nu\rho\sigma}
                             C^{\mu\nu\rho\sigma}
    \, , \cr
    \InvD = {} & 
               - \frac{1}{16} C^{\textstyle\ast}_{\mu\nu}{}^{\rho\sigma}
                              C_{\rho\sigma}{}^{\eta\omega}
                              C_{\eta\omega}{}^{\mu\nu}
    \, , \cr
  } \EQN HomInvars
$$
where $C_{\mu\nu\rho\sigma}$ is the Weyl tensor (which for vacuum
spacetimes is equal to the Riemann tensor),
and $C^{\textstyle\ast}_{\mu\nu\rho\sigma} = 
{1\over2} \varepsilon_{\mu\nu\eta\omega} C^{\eta\omega}{}_{\rho\sigma}$
where $\varepsilon_{\mu\nu\rho\sigma}$ is the Levi-Civita tensor.
(These definitions of the curvature invariants are the
same as those used in the GRTensorII package for MapleV; see Musgrave,
Pollney and Lake~1996.)
Curvature invariants are used in the presentation of numerical results
for the inhomogeneous cosmological models studied in chapters~6 and~7.
These quantities can be particularly useful in interpreting the
behaviour of evolved spacetimes since their divergence to infinity
indicates the presence of a curvature singularity; however it should
be noted that non-trivial (`pp~wave') spacetimes exist for which the
curvature invariants never vary from zero.
As an example of the behaviour of the curvature invariants, for the
axisymmetric Kasner metric
$$
  ds^2 = t^{-1/2} ( - dt^2 + dx^2 ) + t \, ( dy^2 + dz^2 )
  \, , \EQN HomKasner
$$
which is considered further in the next section,
the invariants take the values
$$
  \InvA = {3 \over 32 \, t^3} \, , \qquad
  \InvB = 0 \, , \qquad
  \InvC = - {3 \over 256 \, t^{9/2}} \, , \qquad
  \InvD = 0 \, . \qquad
  \EQN HomKasInvars
$$

The use of scalar quantities to describe numerically generated
spacetimes goes some way to allowing genuine physical behaviour to be
distinguished from effects caused by unfortunate choices
of spacetime coordinates or metric variables.
For the inhomogeneous cosmologies studied in chapters~6 and~7,
consideration of the values taken by the symmetry scalars and the
curvature invariants makes it immediately clear that the spacetimes
being evolved are genuinely inhomogeneous and are not simply
homogeneous models constructed using inhomogeneous gauge conditions.
(Further comparisons are made between features of numerically evolved
homogeneous and inhomogeneous spacetimes in
section~\use{sec.HomHarmonic}.) 

Presenting results in terms of scalar quantities simplifies the
comparison of numerical simulations of the same spacetime for
which different gauge conditions are used.
This idea can be
extended to the construction of test problems for verifying the
correctness of a numerical code.
While exact solutions are often used as test bed calculations for
codes, it may not always be practical to write a solution in a form
appropriate to the gauge conditions used by a code.
(For examples of how complex such transformations of exact solutions
can be see Petrich, Shapiro and Teukolsky~1985,~1986.)
However, if the values of two or more scalar quantities are known for
an exact solution, then the correlation between them can be used as a
measure of the accuracy of a numerical simulation of that spacetime,
even if the gauge conditions used are different from those of the
exact solution; for example, for the axisymmetric Kasner
model~\EqNum{HomKasner}{} mentioned above, it follows from the
equations~\EqNum{HomKasInvars}{} for the curvature invariants that the
relationship $(\InvA)^3 = 6(\InvC)^2$ should hold at every spacetime
point regardless of how a coordinate system for the spacetime is
constructed. 

Besides judging the accuracy of a computer code, comparisons between
two numerical solutions or between an exact and a numerical solution can
be useful for investigating properties of different gauge conditions
that may be used in numerical simulations, such as how much of a
spacetime they cover or whether they develop coordinate singularities.
The direct approach to computing the effects of different gauge
conditions by explicit construction of a transformation between two
sets of spacetime coordinates is, as was mentioned above, not
practical in general; however an example of such a transformation is
given for a simple case in section~\use{sec.HomHarmonic} where
properties of different gauge conditions are considered in more
detail.
The evaluation of scalar quantities provides an alternative way of
comparing coordinate systems.
As an example, the curvature invariant~$\InvA$, evaluated at a point
in a numerical solution for the axisymmetric Kasner spacetime,
determines via equation~\EqNum{HomKasInvars}{} the label~$t$ of the
time slice of the metric~\EqNum{HomKasner}{} which includes that
point; if more independent scalar quantities are known for a
spacetime, then more information can be determined about the locations
of points in the numerical solution.
A third way of comparing different coordinate systems for a spacetime
is discussed in the next subsection where an algorithm is presented
for numerically evolving a coordinate chart with respect to a
known background chart.

\SubSection{Numerical Integration of Observer World Lines}
In the previous subsection some approaches are discussed for comparing
the results and the coordinate systems of different numerical
simulations of the same spacetime. 
The present subsection expands on the idea of comparing coordinate
systems and describes an algorithm for numerically integrating the
world lines of one coordinate system with respect to another.
Some of the ideas discussed in both this and the previous subsection
are put into practice in section~\use{sec.HomHarmonic}.

The problem can be specified as follows.
Consider a spacetime for which the metric~$g_{\mu\nu}$ is known
(either as an exact solution or as a numerical solution of sufficient
accuracy) with respect to a coordinate system~$(t,x^i)=(t,x,y,z)$.
Suppose that there is also, as described in chapter~4, an ADM
foliation $\{ \Sigma_T : T \in \RealSet \}$ of the spacetime, the
spatial hypersurfaces of which are unrelated to the 
$t=\hbox{\it constant\/}$ surfaces of the first coordinate system.
If one of the spatial slices, $\Sigma_{T_0}$~say, of the foliation is
known to have the description~$t = f_0(x^i)$ with respect to the
original coordinates, then the question is: given a knowledge of the
ADM quantities defined on the foliation, what are the coordinate
locations of the other hypersurfaces in the foliation?

In practice the foliation~$\Sigma_T$ will have been generated as part
of a numerical simulation.
Two simplifying assumptions are made for the purposes of the
discussion here: that the shift vector~$N^i$ of the foliation is
identically zero, and that the spatial hypersurfaces of the foliation
do not have boundaries.
Both of these assumptions are valid for the example problems
considered in this chapter; the latter point holds
because the spacetime models investigated here are all spatially
periodic.

Suppose that each hypersurface~$\Sigma_T$ of the spacetime foliation
is described by a spatial coordinate system $X^i = (X,Y,Z)$, and that
the timelike world line of an observer who remains stationary with
respect to these spatial coordinates is given by~$p^\mu(T;X^i)$ where
the parametrization is such that the spacetime point~$p^\mu(T_{\ast})$
falls within the spatial slice~$\Sigma_{T_{\ast}}$.
Since the shift vector~$N^i$ of the foliation is assumed to be zero,
the world line of this observer must run normal to the slices of the
foliation: if~$n^\mu$ is the vector field normal to the foliation,
normalized such that $n^\mu n_\mu = -1$, then the velocity four-vector
of the observer~$p^\mu(T)$ is
$$
  v^\mu = {{ dp^\mu }\over{ dT }} \propto n^\mu
  \, .
$$
With respect to the spacetime coordinate system~$(T,X^i)$ defined by
the foliation, it is clear that $v^\mu=(1,0,0,0)$, and it is also
known (see section~\use{sec.HypIntro}) that $n^\mu=(1/N,0,0,0)$
where~$N \equiv N(T,X^i)$ is the lapse function of the foliation.
Thus,
$$
  {{ dp^\mu }\over{ dT }} = N n^\mu
  \EQN HomObserver
$$
describes the motion of the observer in terms of the properties of the
foliation.

Now consider a single slice~$\Sigma_{T_0}$ of the foliation with
observers~$p^\mu(T_0;X^i)$ at each point.
With respect to the original coordinate system~$(t,x^i)$, assuming
that the spacetime metric~$g_{\mu\nu}$ is known, it is possible to
determine the vector field~$n^\mu$ normal to~$\Sigma_{T_0}$ from a
knowledge of the positions of the observers.
If in addition the lapse function~$N(T_0,X^i)$ is known on the slice,
then equation~\EqNum{HomObserver}{} is sufficient to determine the
location of a second spatial slice $\Sigma_{T_0 + \delta T}$
infinitesimally removed from the first one.
In this way a description with respect to the coordinates~$(t,x^i)$ of
the foliation~$\Sigma_T$ can be built up from a knowledge of the lapse
function and the location of an initial slice~$\Sigma_{T_0}$.

\figure{HomObserve}
 \DrawFig{
  \epsfxsize=0.65\hsize
  \epsfbox{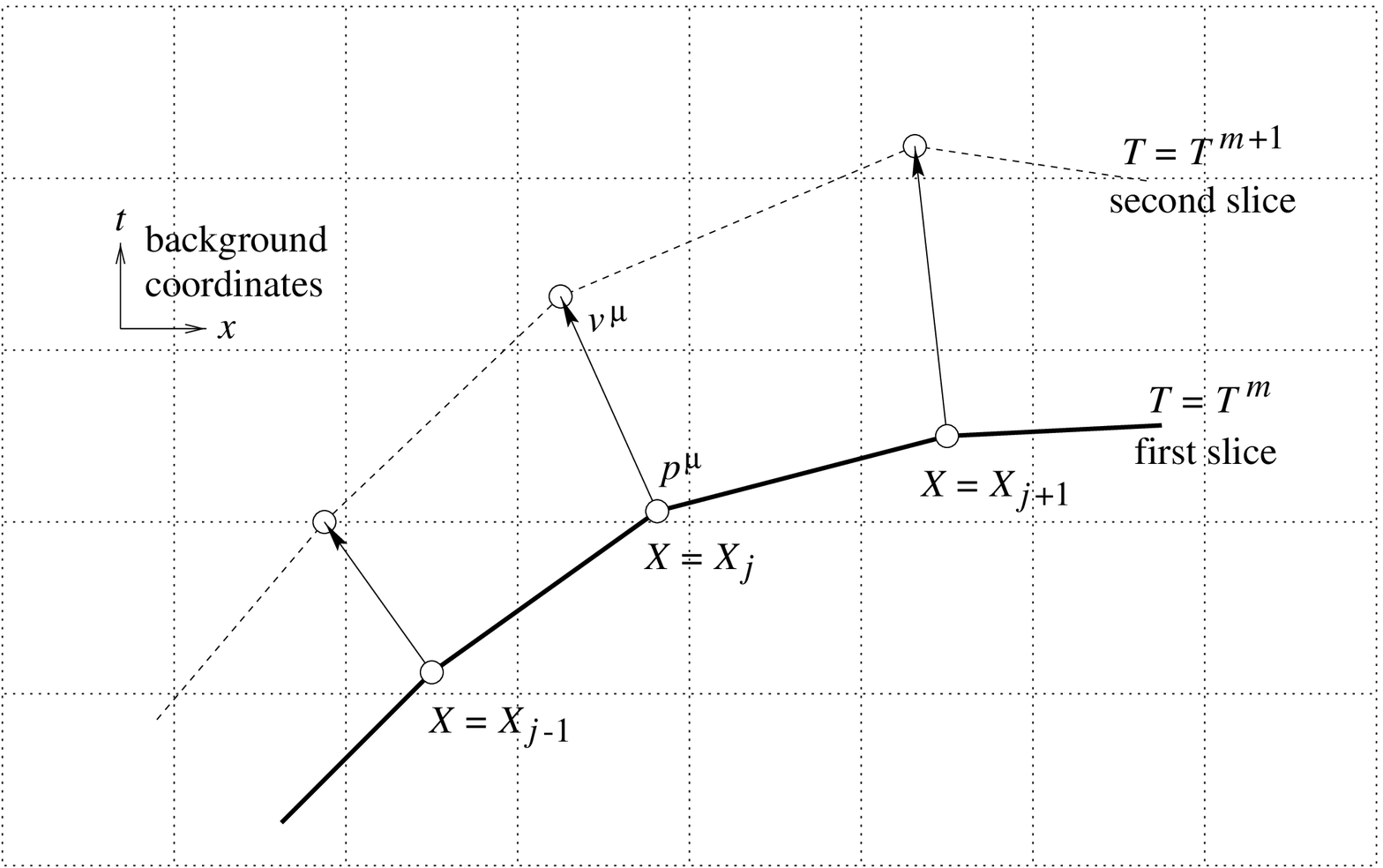}
 }
\Caption
Numerical evolution of spatial slices with respect to a background
coordinate system. 
A mesh of points $\{X_j,Y_k,Z_l\}$ describing a spacelike hypersurface
is evolved as part of a numerical simulation into a spacetime
foliation with slices labelled by time values~$\{T^m\}$.
The positions~$p^\mu$ of the mesh points with respect to a background
coordinate system $(t,x,y,z)$ can be tracked using a time-stepping
approach based on the velocity vector~$v^\mu$ which points in the
direction normal to the slices.
Only one spatial coordinate is shown here.
\bigskip
\endCaption
\endfigure

If the foliation~$\Sigma_T$ is generated as the result of a numerical
simulation, then equation~\EqNum{HomObserver}{} can be implemented as
part of a computer code to locate slices of the foliation with respect
to a reference coordinate system.
The numerical simulation produces results at discrete time intervals
corresponding to a set of slices
$$
  \{ \Sigma_T : T = T^0, T^1, \ldots, T^m, \ldots \}
  \, ,
$$
where the sizes of the time steps are not necessarily uniform, and in
addition each spatial slice is discretized by a mesh of points
$$
  \{ \, (X_j, Y_k, Z_l) = ( X_0 + j \, \Delta X, \, 
                            Y_0 + k \, \Delta Y, \,
                            Z_0 + l \, \Delta Z )
  \, : \, j,k,l \in \IntegSet
  \, \} \, .
$$
The value of a quantity~$f$ which is known at a point
$(T^m,X_j,Y_k,Z_l)$ of the foliation is denoted here
by~$f \vert^m_{jkl}$.
The lapse function~$N$ of the foliation is assumed to be known at each
point: in practice it is output as part of the results of the
numerical simulation. 

Figure~\use{Fg.HomObserve} illustrates the evolution of the
discretized foliation with respect to a background coordinate
system~$(t,x,y,z)$.
The metric~$g_{\mu\nu}$ is assumed to be known in terms of these
background coordinates, either as an exact solution or an accurate
numerical solution. 
The position of a mesh point $(T^m,X_j,Y_k,Z_l)$ of the foliation with
respect to the background coordinates is denoted by 
$p^\mu \vert^m_{jkl} = (t,x,y,z) \vert^m_{jkl}$,
and it is assumed that the positions~$p^\mu \vert^0_{jkl}$ of all of
the mesh points in the initial slice~$\Sigma_{T^0}$ are known.
A numerical algorithm for finding the approximate positions (with
respect to the background coordinate system) of the other mesh points
in the foliation then proceeds by iteration of the following procedure.

Given that the positions~$p^\mu \vert^m_{jkl}$ of observers in a
spatial slice~$\Sigma_{T^m}$ are known, approximations to the
positions~$p^\mu \vert^{m+1}_{jkl}$ of the same observers in a later
slice can be made using a discretized version of
equation~\EqNum{HomObserver}{}. 
The first step is to construct a value for the velocity (with respect
to the foliation time~$T$) of each observer: $v^\mu = N n^\mu$.
As mentioned above, the value of the lapse function~$N$ is assumed
to be known at each point of the mesh, and so the task here is to
determine a value for the vector~$n^\mu$ normal to the slice.
In the continuous case the location of the slice~$\Sigma_{T^m}$
is given by the expression $t = f(x,y,z)$ where $f = p^0(T^m;X,Y,Z)$,
and so the normal vector has the covariant form
$$
  n_\mu = \kappa \nabla_{\!\mu} (f-t)
        = \kappa ( -1, \partial f / \partial x^i )
  \, , \EQN HomNormal
$$
where~$\kappa$ is a constant (different for each point on the slice),
the value of which is determined below.
The partial derivatives in equation~\EqNum{HomNormal}{} can be
expressed in terms of the functions~$\partial p^\mu / \partial X^i$
using the chain rule for differentiation.
For example, in the
one-dimensional case for which only the spatial coordinates~$X$
and~$x$ are significant, the normal vector is given by
$$
  n_\mu = \kappa ( -1, {{\partial p^0 / \partial X} \over
                        {\partial p^1 / \partial X}}, 0, 0 )
  \, .
$$
Returning to the case in which only the positions of discrete points in
the spatial slice are known, finite differences can be used to give
approximate values for the derivatives in equation~\EqNum{HomNormal}{}
in the obvious way; for example, to second order,
$$
  ( \partial p^\mu / \partial X ) \vert^m_{jkl}
  = { p^\mu \vert^m_{j+1\,kl} - p^\mu \vert^m_{j-1\,kl}
      \over 2 \, \Delta X }
  \, .
$$
The normal vector~$n^\mu$ can now be determined by using the
metric~$g_{\mu\nu}$ evaluated at the spacetime point~$p^\mu$ to raise
the index in equation~\EqNum{HomNormal}{}, and the value of the
constant~$\kappa$ can be fixed by requiring that~$n^\mu n_\mu = -1$.

At this stage an approximate velocity vector 
$v^\mu \vert^m_{jkl} = N \vert^m_{jkl} \, n^\mu \vert^m_{jkl}$
is known at each mesh point~$p^\mu \vert^m_{jkl}$ of the spatial
slice~$\Sigma_{T^m}$.
The next step is to use a discretized version of
equation~\EqNum{HomObserver}{} to advance the mesh points to the next
slice~$\Sigma_{T^{m+1}}$.
A simple first-order approximation to the time derivative in the
equation yields a discrete evolution equation for the observers:
$$
  p^\mu \vert^{m+1}_{jkl} =
  p^\mu \vert^m_{jkl} + \Delta T \, v^\mu \vert^m_{jkl}
  \, , \EQN HomPLinear
$$
where $\Delta T = T^{m+1} - T^m$.
Alternatively, if the values of the velocity vectors on the previous
spatial slice~$\Sigma_{T^{m-1}}$ are known, then second-order
corrections based on quadratic extrapolation can be added to the
evolution equation: 
$$
  p^\mu \vert^{m+1}_{jkl} =
  p^\mu \vert^m_{jkl} + \Delta T \, v^\mu \vert^m_{jkl}
  + { (\Delta T)^2 \over 2 \, \Delta T_{-} }
    ( v^\mu \vert^m_{jkl} - v^\mu \vert^{m-1}_{jkl} )
  \, , \EQN HomPQuadratic
$$
where $\Delta T_{-} = T^m - T^{m-1}$.
However, this latter equation may overshoot in its extrapolation of
the world line of the observer; in particular, if the value of~$p^0$
decreases during the evolution step then the new position for that
observer should be recalculated using equation~\EqNum{HomPLinear}{}.

Using a sequence of evolution steps based on a discretized form of
equation~\EqNum{HomObserver}{} with the normal vector~$n^\mu$
determined through equation~\EqNum{HomNormal}{}, approximate values
for the positions~$p^\mu$ of observers can be found on a range of
spatial slices~$\Sigma_{T^0}$ through to~$\Sigma_{T^M}$ for some~$M$.
Truncation errors in the difference equations, together with errors in
the values for the lapse~$N$ produced by the original numerical
simulation, build up during the course of the evolution and may cause
it to terminate before the observers reach the final slice of the
simulation; in particular the normal vector~$n^\mu$ must be timelike
for the evolution to proceed.
Aside from this gradual build-up of errors, the algorithm has been
found to perform well in practice and has not shown signs of
instability. 

The algorithm has been implemented in a one-dimensional form suitable
for studying different slicings of planar cosmological models.
In tests using simple spacetimes for which exact solutions can be written
with respect to two different coordinate systems (so that the values
used for the lapse~$N$ are exact rather than the product of a
numerical simulation) the code has been found to converge at the
appropriate order (first or second, depending on which of the
equations~\EqNum{HomPLinear}{} or~\EqNum{HomPQuadratic}{} is used).
Results produced with the algorithm for non-standard slicings of a
homogeneous cosmological model are presented in
the next section; see in particular figure~\use{Fg.HomSlices} there.

\Section{Harmonic Time Slicing and Coordinate Singularities
         \label{sec.HomHarmonic}}
This section examines properties of the harmonic time slicing
condition introduced in chapter~4 and considers its usefulness from
the point of view of numerical relativity.
A simple homogeneous cosmological model provides a focal point for
this investigation, and analysis of the behaviour of harmonic slices
of that spacetime makes use of some of the ideas presented in the
previous section together with an analytic approach similar to that of
Geyer and Herold~(1995, 1997).
(The latter makes extensive use of Bessel functions, and these are
briefly reviewed as part of the discussion.)
The results of this investigation suggest that coordinate
singularities, much like those originally found by Alcubierre~(1997), can
develop under reasonably general conditions in numerical simulations
which use harmonic time slicing, and some of the implications of this
for numerical work are discussed.

\SubSection{On the Choice of Time Slicing Condition}
As discussed in chapter~4, most of the hyperbolic formulations of
Einstein's equations which have been developed in recent years
require a spacetime to be foliated according to a harmonic time slicing
condition if the characteristic speeds of
the equations are to coincide with physical speeds of the system.
In this subsection some basic features of harmonic time slicing are
considered. 

Traditionally, work in numerical relativity has been based on maximal
or constant mean curvature slicing conditions, initially developed for
this purpose by Smarr and York~(1978), York~(1979), and 
Eardley and Smarr~(1979), among others.
A spatial slice has constant mean curvature if~$K$, the trace of the
extrinsic curvature tensor, takes a constant value on that slice, and
if values for~$K$ are specified across a range of slices as a function
of the time coordinate~$t$, then the ADM evolution
equation~\EqNum{HypADMEvolve}{b} for the extrinsic curvature yields
(when combined with the Hamiltonian
constraint~\EqNum{HypADMConstr}{a}) an elliptic equation determining
the value of the lapse function~$N$ on each slice: 
$$
  \textstyle
  \Delta N - N [ K_{ij} K^{ij} + {1\over2} \kappa (\rho + S) ]
  = - K'(t)
  \, , \EQN HomMaximal
$$
where $\Delta = h^{ij} \nabla_{\! i} \nabla_{\! j}$.
If the value of~$K$ is zero on a slice then that slice is described as
maximal, and if the function~$K(t)$ is identically zero then the
spacetime is maximally sliced.
Maximal slicing has been found to work well in numerical simulations
of asymptotically flat spacetimes, while for closed cosmologies (in
which at most one maximal slice can exist) the more general constant
mean curvature slicing condition is appropriate.
Based on their behaviour in simple examples and their success in
numerical simulations, it is believed that the maximal and constant
mean curvature slicing conditions will in general produce foliations
which cover most, if not all, of a spacetime being investigated, and
which at the same time avoid getting too close to any singularities
that may form in that spacetime.

The generalized harmonic time slicing condition contrasts in several
ways with the maximal and constant mean curvature conditions.
The lapse function~$N$ in a
harmonically sliced spacetime is determined either through an
algebraic condition or a simple evolution equation, and not
through an elliptic condition as in equation~\EqNum{HomMaximal}{}.
This is advantageous from the point of view of implementing a
numerical code to evolve solutions to Einstein's equations since
elliptic partial differential equations are computationally very
expensive to solve.
The harmonic time slicing condition also differs from the maximal and
constant mean curvature time slicing conditions in that it specifies a
relationship between two adjacent slices in a foliation rather than
determining properties of each individual slice: while an isolated
spacetime slice can be characterized as maximal or of constant mean
curvature, there is no such thing as an individual harmonic slice.
One consequence of this is that the choice of lapse function on the
initial slice of a foliation is arbitrary if the harmonic time slicing
condition is used.

To simplify the discussion in this section, all spacetime
foliations considered here are assumed to have shift vectors~$N^i$
which are identically zero; since the shift vector controls only the
positioning of the spatial coordinates and not the slicing of a
spacetime, this assumption does not limit the generality of the
following discussion in any essential way.
Recalling equations~\EqNum{HypHarmFixed}{} and~\EqNum{HypHarmEvolve}{}
of the previous chapter, the generalized harmonic time slicing
condition can be expressed in either algebraic form, 
$$
  N = Q(t,x^k) \sqrt{\det h_{ij}}
  \, , \EQN HomHarmFixed
$$
or as an evolution equation,
$$
  { \partial N \over \partial t } = N (\ln Q)_{,t} - N^2 K
  \, , \EQN HomHarmEvolve
$$
where the slicing density~$Q(t,x^i)$ is an arbitrary function of the
spacetime coordinates.
If the slicing density is independent of the time coordinate~$t$,
then the harmonic slicing is described as simple, and in this case a
choice of the value of the lapse~$N$ on an initial slice completely
determines the value of the slicing density~$Q$.
If the slicing density has only a simple dependence on the time
coordinate, $Q(t,x^i) = f(t) \bar{Q}(x^i)$, then the foliation
that develops has the same slices as the foliation produced by
using~$\bar{Q}(x^i)$ as the slicing density, but with a different time
coordinate labelling the slices.
In general it is not obvious how a useful slicing density~$Q$ which
has a non-trivial dependence on the time coordinate may be chosen, and
so the simple form of harmonic slicing is most commonly used in
practice; this point is returned to later in the section.

The main question addressed in this section is that of how
suitable the harmonic time slicing condition is for numerical work.
Several authors have already considered this question from various
different viewpoints.
Bona and Mass\'o~(1988) have shown that the (simple) harmonic time
slicing condition can be used to foliate several standard spacetimes,
and also that `focusing singularities' are avoided by the slicing
condition in much the same way as they are in maximally sliced
spacetimes with the slices of the foliation not reaching the
singularity in a finite coordinate time.
(It should be noted that for a harmonically sliced spacetime a focusing
singularity is essentially a point at which the lapse~$N$ becomes
zero---this is in contrast to the behaviour found at the `gauge
pathologies' described below.)
Cook and Scheel~(1997) have investigated the construction of
well-behaved harmonic foliations for Kerr-Newmann black hole
spacetimes, and present numerical evidence that (in the uncharged,
spherically symmetric case) a particular time-independent harmonic
foliation acts in some sense as an attractor to other harmonic
foliations. 

The work of Alcubierre~(1997) and Alcubierre and Mass\'o~(1998) is of
particular relevance to the present discussion.
They have shown that `gauge pathologies' (described as `coordinate
shocks' in the earlier paper) can occur in numerical simulations based
on hyperbolic formulations of Einstein's equations which use harmonic
time slicing. 
(In fact, a range of gauge conditions are considered by the authors,
with simple harmonic slicing---the only gauge choice of interest
here---corresponding to the special case~$f=1$ in their formulation.
None of the alternative gauge choices they use corresponds to the
generalized harmonic time slicing condition.)
These gauge pathologies manifest as a loss of continuity at points in
the evolved solution, with in particular large spikes appearing
in the lapse.
After the time at which the pathologies appear the
numerical solution no longer converges at the expected order.
Alcubierre and Mass\'o explain the appearance of gauge pathologies in
terms of nonlinear behaviour in the hyperbolic evolution equations;
however this fails to adequately answer questions about how common the
gauge pathologies are, what happens to the foliation at the points
where pathologies appear, and what approaches can be used to prevent
the pathologies from occurring.

Harmonic slicings for the Schwarzschild and Oppenheimer-Snyder
spacetimes have been investigated by Geyer and Herold~(1995, 1997).
The approach they use is based on an alternative representation of
equations~\EqNum{HomHarmFixed}{} and~\EqNum{HomHarmEvolve}{} in the
simple harmonic case: if~$T$ is a scalar function on a spacetime, the
level surfaces of 
which represent the spatial slices of a foliation~$\{\Sigma_T\}$, then
the spacetime will be (simple) harmonically sliced if
$$
  \square T 
  \equiv g^{\mu\nu} \nabla_{\! \mu} \nabla_{\! \nu} T
  = 0
  \, . \EQN HomHarmTime
$$
(This equation explains, of course, how the harmonic slicing condition
acquires its name.)
No simple analogue of equation~\EqNum{HomHarmTime}{} exists for the
generalized harmonic time slicing condition.
By numerically integrating equation~\EqNum{HomHarmTime}{} with respect
to known background metrics, Geyer and Herold construct simple harmonic
slicings which they compare to maximal slicings of the spacetimes.
Some properties of a slicing can be determined straightforwardly from
its time function~$T$, and in particular the lapse~$N$ can be evaluated
through the equation 
$$
  g^{\mu\nu} ( \nabla_{\! \mu} T ) ( \nabla_{\! \nu} T )
  = - 1 / N^2
  \, , \EQN HomHarmLapse
$$
where it is clear that the vector field normal to the foliation must
remain timelike if the lapse is to have a positive real value.
In fact, Geyer and Herold~(1997) find that for simple harmonic
slicings of the Oppenheimer-Snyder spacetime, foliations which are
initially timelike can at later times become null or spacelike, with
the development of the foliation thus terminating at a coordinate
singularity.
The lapse becomes infinite as these singular points are reached, and
this is consistent with the behaviour found at the gauge pathologies
of Alcubierre and Mass\'o. 

In the present work equations~\EqNum{HomHarmTime}{}
and~\EqNum{HomHarmLapse}{} are used to investigate the formation of
coordinate singularities in simple harmonic slicings of cosmological
models.
The following approach is employed.
The metric~$g_{\mu\nu}$ of the spacetime being investigated is assumed
to be known with respect to a coordinate system~$(t,x,y,z)$.
An alternative foliation of the spacetime is constructed by taking as
an initial slice one of the constant time hypersurfaces of the
background coordinate system: the foliation time
coordinate~$T(t,x,y,z)$ is given an initial value
$$
  T(t_0,x,y,z) = t_0
  \, , \EQN HomHarmInitial
$$
for some value~$t_0$ of the coordinate~$t$.
(It is assumed here that the hypersurface $t=t_0$ is spacelike.)
The lapse function of the foliation can be specified arbitrarily on
the initial slice (subject to the requirement that it be positive),
and this determines via equation~\EqNum{HomHarmLapse}{} the value of
the first derivative of~$T$ away from that slice:
$$
  T_{,t} (t_0,x,y,z) = 
    { 1 \over N_0(x,y,z) \sqrt{ - g^{00}(t_0,x,y,z) } }
  \, . \EQN HomHarmInitDeriv
$$
Equations~\EqNum{HomHarmInitial}{} and~\EqNum{HomHarmInitDeriv}{}
provide initial data for the wave equation~\EqNum{HomHarmTime}{}, and
the solution~$T$ describes a new simple harmonic slicing of
the spacetime.

This method for locating the slices of a foliation with respect to a
background coordinate system is very similar to the
algorithm presented in the previous section for integrating the world
lines of coordinate observers, and both methods can be useful in
investigating the formation of coordinate singularities in harmonic
slicings.
The advantage of using equation~\EqNum{HomHarmTime}{} is its
simplicity: analytic solutions can be found in some cases, and when
this is not possible numerical methods are straightforward to apply.
However the method has the drawback that it provides no information
about the spatial coordinates describing the slices of the foliation;
these can only be found by the integration of world lines.
Furthermore, equation~\EqNum{HomHarmTime}{} is only applicable in the
case of simple harmonic time slicing whereas the approach of
section~\use{sec.HomCompare} can be used for arbitrary slicing
conditions.
Results for both of the methods are compared later in this section.

As an example of how equation~\EqNum{HomHarmTime}{} can be used to
find coordinate singularities, consider harmonic foliations of the
Minkowski spacetime, written in standard coordinates as
$$
  g_{\mu\nu} dx^{\mu} dx^{\nu} =
    - dt^2 + dx^2 + dy^2 + dz^2
  \, .
$$
Suppose that a `planar' slicing of the spacetime is constructed such
that the time function~$T$ depends only on the Minkowski
coordinates~$t$ and~$x$.
Then equation~\EqNum{HomHarmTime}{} takes the form of the
one-dimensional wave equation
$$
  T_{,tt} = T_{,xx}
  \, ,
$$
and, as is well known, this has the general solution
$$
  T(t,x) = f(t+x) + h(t-x)
  \, ,
$$
for arbitrary functions~$f(u)$ and~$h(u)$.
The lapse associated with this time function can be found from
equation~\EqNum{HomHarmLapse}{}:
$$\eqalign{
  -1/N^2 & {} = - T_{,t}{}^2 + T_{,x}{}^2 \cr
         & {} = - 4 f'(t+x) h'(t-x) \, , \cr
}$$
and the lapse will be well behaved as long as the function
$f'(t+x)h'(t-x)$ is positive.
For most problems the next step in the analysis would be to use
equations~\EqNum{HomHarmInitial}{} and~\EqNum{HomHarmInitDeriv}{} to
specify a value for the lapse on an initial slice of the foliation;
however the present case is sufficiently simple that the appearance of
coordinate singularities can be studied without needing to specify an
initial slice.
If the lapse becomes infinite at a point in the foliation then the
function $f'(t+x)h'(t-x)$ must be zero there, and it is clear that the
function must then be zero at all points along a line
$t+x=\hbox{\it constant\/}$ or $t-x=\hbox{\it constant\/}$.
Consequently, any spacelike slice of that foliation must include a
point in it at which the lapse is infinite.
It follows that for simple harmonic time slicings of the `planar'
Minkowski spacetime no coordinate singularities will be present
provided that the initial slice of the foliation is everywhere
spacelike.
(This result is consistent with the numerical simulations of Minkowski
spacetime reported on in Alcubierre~1997.
No gauge pathologies are discovered there for `planar' harmonic
slicings of flat spacetime, although they are found in the spherically
symmetric case.)

The remaining part of this section considers the appearance of
coordinate singularities in harmonic slicings of the homogeneous
Kasner cosmological model.
Analysis of the foliation equation~\EqNum{HomHarmTime}{} in this case
makes use of Bessel functions, and these are briefly reviewed in the
next subsection.

\SubSection{Some Properties of Bessel Functions}
At various points in this and the following chapters Bessel functions
are used to solve equations describing aspects of homogeneous and
inhomogeneous cosmologies, and for convenience the present subsection
summarizes some of their basic properties.
A complete exposition of Bessel functions can be found in chapter~9
of Abramowitz and Stegun~1964.

The Bessel functions of the first kind,~$J_\nu(t)$, and the second
kind,~$Y_\nu(t)$, are solutions to the differential equation
$$
  t^2 { d^2 A \over d t^2 } + t { d A \over d t }
    + (t^2 - \nu^2) A = 0
  \EQN BessEqn
$$
for~$A(t)$.
While the argument~$t$ and the order~$\nu$ may be complex in general,
for the present work it suffices to consider only real values with~$t$
being positive and~$\nu$ being non-negative.

The derivatives of the Bessel functions are
$$\EQNalign{
  J_\nu'(t) & {} = - J_{\nu+1}(t) + (\nu / t) J_\nu(t) 
  \, , \EQN BessDerivs;a \cr
  Y_\nu'(t) & {} = - Y_{\nu+1}(t) + (\nu / t) Y_\nu(t) 
  \, , \EQN BessDerivs;b \cr
}$$
and the following identities prove to be useful,
$$\EQNalign{
  J_{\nu-1}(t) + J_{\nu+1}(t) & {} = (2\nu / t) J_\nu(t)
  \, , \EQN BessIden;a \cr
  Y_{\nu-1}(t) + Y_{\nu+1}(t) & {} = (2\nu / t) Y_\nu(t)
  \, , \EQN BessIden;b \cr
}$$
together with
$$
  J_{\nu+1}(t) Y_{\nu}(t) - J_{\nu}(t) Y_{\nu+1}(t) = {2 \over \pi t}
  \, . \EQN BessWronk
$$

The asymptotic expansions of the Bessel functions for large values of
the argument~$t$ are
$$\EQNalign{
  J_\nu(t) & {} = \sqrt{2 \over \pi t} 
    \Big( P(t) \cos \Phi(t) - Q(t) \sin \Phi(t) \Big)
  \, , \EQN BessAsymp;a \cr
  Y_\nu(t) & {} = \sqrt{2 \over \pi t} 
    \Big( P(t) \sin \Phi(t) + Q(t) \cos \Phi(t) \Big)
  \, , \EQN BessAsymp;b \cr
}$$
where
$$\eqalign{
  \Phi(t) & \textstyle {} = t - ({1\over2} \nu + {1\over4}) \pi
  \, , \cr
  P(t) & {} \sim 1 - { (4 \nu^2 - 1)(4 \nu^2 - 9) \over 128 \, t^2 }
            + O(t^{-4})
  \, , \cr
  Q(t) & {} \sim { (4 \nu^2 - 1) \over 8 \, t }
            + O(t^{-3})
  \, , \cr
}$$
and to leading order these yield very useful approximations:
$$\EQNalign{
  J_\nu(t) & {} \sim
    \sqrt{2 \over \pi t}
    \cos \textstyle ( t - [{1\over2} \nu + {1\over4}] \pi )
    + O(t^{-3/2})
  \, , \EQN BessApprox;a \cr
  Y_\nu(t) & {} \sim
    \sqrt{2 \over \pi t}
    \sin \textstyle ( t - [{1\over2} \nu + {1\over4}] \pi )
    + O(t^{-3/2})
  \, . \EQN BessApprox;b \cr
}$$

As the argument~$t$ approaches zero the Bessel functions of the first
kind behave as
$$
  J_{\nu}(t) \sim { (t/2)^{\nu} \over \Gamma(\nu+1) }
                  + O(t^{\nu+2})
  \, , \EQN BessZero;a
$$
so that $J_0(0)=1$ and $J_{\mu}(0)=0$ for $\mu>0$.
In addition, $\vert J_{\nu}(t) \vert \le 1$ for all~$\nu$ and~$t$.
The Bessel functions of the second kind diverge as~$t$ tends to zero,
and for orders zero and one the asymptotic forms are
$$\EQNalign{
  Y_0(t) \sim {} & (2/\pi) [ \ln (t/2) + \gamma ]
                   + O(t^2 \ln t)
  \, , \EQN BessZero;b \cr
  Y_1(t) \sim {} & {-2 / (\pi t)}
                   + O(t \ln t)
  \, , \EQN BessZero;c \cr
}$$
for the constant~$\gamma\simeq0.577$.

Values of the Bessel functions can be calculated numerically using
standard procedures in mathematical software packages and programming
libraries; see also chapter~6 of Press et al.~1992.

\SubSection{Harmonic Slicings of the Kasner Spacetime 
            I: An Analytic Approach}
In this subsection equation~\EqNum{HomHarmTime}{} is used to construct
simple harmonic foliations of the Kasner spacetime, with the main
point of interest being the question of whether or not coordinate
singularities (gauge pathologies in the terminology of Alcubierre and
Mass\'o~1998) form in the slicings.
The next subsection compares these results to numerical simulations of
the spacetime.

The Kasner cosmological models are described in most standard
references (see, for example, chapter~30 of Misner, Thorne and
Wheeler~1973) and the present work uses the axisymmetric model
described by the metric
$$
  ds^2 = t^{-1/2} ( - dt^2 + dx^2 ) + t \, ( dy^2 + dz^2 )
  \, . \EQN HomKasnerTwo
$$
This spacetime is vacuum and explicitly homogeneous, and features a
cosmological singularity at time~$t=0$.
The form of the metric used here is in fact the expanding
version~\EqNum{GowMetricT}{} of the Gowdy~$T^3$ metric studied in
chapter~6 with variables~$P$, $Q$ and~$\lambda$ all taken to be
identically zero.
The connection between the homogeneous model~\EqNum{HomKasnerTwo}{}
and the inhomogeneous cosmologies of chapters~6 and~7 is further
strengthened by imposing a three-torus spatial topology on the Kasner
spacetime: periodicity is assumed in each of the coordinates~$x$, $y$
and~$z$ over the range~$[0,2\pi)$.

As in the simple example for the Minkowski spacetime given earlier in
this section, here `planar' slicings of the Kasner spacetime are
considered for which the time function~$T$ depends only on the
coordinates~$t$ and~$x$.
A straightforward calculation of connection coefficients then shows
that for the metric~\EqNum{HomKasnerTwo}{},
equation~\EqNum{HomHarmTime}{} for the time function of a harmonically
sliced foliation is equivalent to
$$
  T_{,tt} - T_{,xx} + t^{-1} T_{,t} = 0
  \, . \EQN HomKasTimeEqn
$$
This equation can be solved by decomposing~$T$ spatially as a sum of
Fourier modes (recalling that, since the model is spatially closed,
the function~$T$ must be periodic in the coordinate~$x$) and the
general solution is found to be
$$\eqalign{
  T & {} = a_0 \ln t + b_0
           + \sum_{n=1}^{\infty} \Big( Z_{+n}\!(t) \cos(nx)
                                     + Z_{-n}\!(t) \sin(nx) \Big)
  \, , \cr
  & \hbox{with \ \ }
  Z_{\pm n}\!(t) = a_{\pm n} J_0( \vert n \vert t )
                 + b_{\pm n} Y_0( \vert n \vert t )
  \hbox{ \ \ for } n \neq 0
  \, , \cr
} \EQN HomKasTime
$$
where~$a_{\pm n}$ and~$b_{\pm n}$ are constants, and the Bessel
functions~$J_0$ and~$Y_0$ are discussed in the previous subsection.

Equation~\EqNum{HomHarmLapse}{} for the lapse function~$N$ of the
foliation can be written in the present case as
$$
  \sqrt{t} \, ( T_{,t}{}^2 - T_{,x}{}^2 ) = 1/N^2
  \, , \EQN HomKasLapse
$$
and the requirement that the foliation be well behaved can then be
expressed as
$$
  T_{,t} > \vert T_{,x} \vert
  \, , \EQN HomKasSing
$$
where it is assumed that the orientation of the time function~$T$ of
the foliation is the same as that of the Kasner time~$t$, and thus
that~$T_{,t}$ is positive.
If the time function~$T$ of equation~\EqNum{HomKasTime}{} fails to
satisfy the condition~\EqNum{HomKasSing}{} at a point, then the
harmonic slicing must have a coordinate singularity there.

The constants~$a_{\pm n}$ and~$b_{\pm n}$ in
equation~\EqNum{HomKasTime}{} can be determined by specifying a
value~$N_0(x)$ for the lapse on an initial hypersurface~$t=t_0$ as in
equations~\EqNum{HomHarmInitial}{} and~\EqNum{HomHarmInitDeriv}{}.
In fact, rather than choosing the initial value for the lapse
directly, it proves to be more convenient to choose instead an
initial value~$I(x)$ for the time derivative of~$T$, with this being
specified as a Fourier sum,
$$
  T_{,t}(t_0,x) = I(x) 
  \equiv k_0 + \sum_{n=1}^{\infty} \Big( k_{+n} \cos (nx) 
                                       + k_{-n} \sin (nx) \Big)
  \, , \EQN HomKasInitT
$$
where the constants~$k_{\pm n}$ must be such that~$I(x)$ is everywhere
positive.
The value of the lapse on the initial slice can then be determined
through equation~\EqNum{HomHarmInitDeriv}{}:
$$
  N_0(x) = { 1 \over t_0{}^{1/4} I(x) }
  \, . \EQN HomKasInitN
$$
The particular time function~$T$ of equation~\EqNum{HomKasTime}{}
which fits the initial data of equations~\EqNum{HomHarmInitial}{}
and~\EqNum{HomKasInitT}{} is then found to be
$$\eqalign{
  T = {} & t_0 + k_0 t_0 \ln (t/t_0) \cr
  & - { \pi t_0 \over 2 } \sum_{n=1}^{\infty}
      \Big( J_0(nt) Y_0(nt_0) - Y_0(nt) J_0(nt_0) \Big)
      \Big( k_{+n} \cos(nx) + k_{-n} \sin(nx) \Big)
  \, . \cr
} \EQN HomKasT
$$

The simplest harmonic foliations of the Kasner
metric~\EqNum{HomKasnerTwo}{} are the homogeneous ones for which the
time function~$T$ is independent of the coordinate~$x$.
Then, to within a linear rescaling, the function~$T$ equals the
logarithm of~$t$, and so the cosmological singularity in the spacetime
occurs in the harmonic foliation only in the limit as~$T$ tends to
negative infinity.
Clearly the foliation covers the entire Kasner spacetime and (as can
be seen from equation~\EqNum{HomKasSing}{}) contains no coordinate
singularities.

\figure{HomFoliation}
 \DrawFig{
  \vskip-5mm
  \epsfxsize=0.9\hsize
  \epsfbox{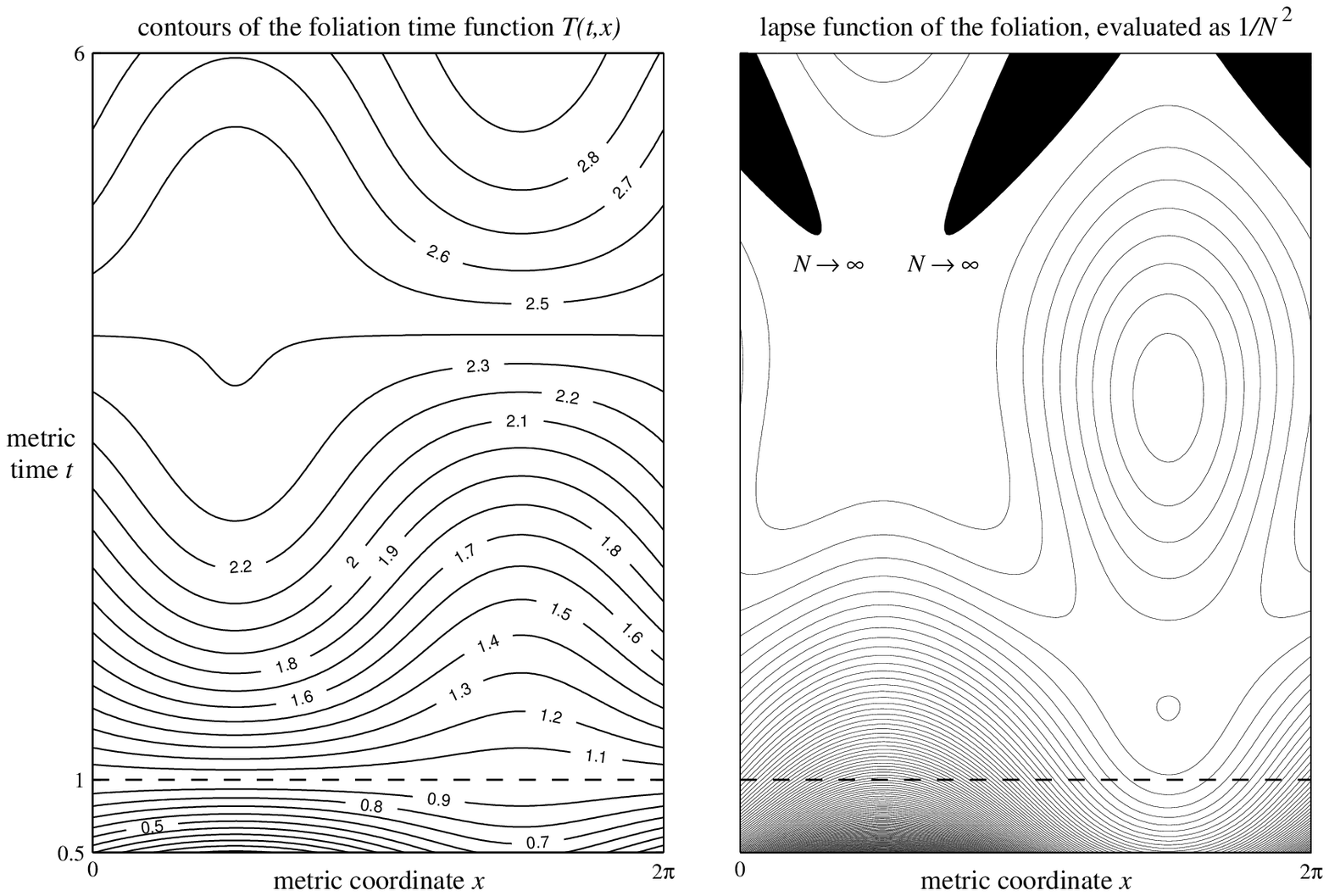}
 }
\Caption
A simple harmonic foliation and its lapse function as seen in the
background Kasner metric.
The initial slice of the foliation coincides with the $t=1$
hypersurface of the metric~\EqNum{HomKasnerTwo}{} and is given
an inhomogeneous initial lapse by equation~\EqNum{HomKasExamp}{}.
The time function~$T$ for the foliation, and its lapse~$N$, are
evaluated numerically from equations~\EqNum{HomKasT}{}
and~\EqNum{HomKasLapse}{}, and are presented as contour plots with
respect to the coordinates~$(t,x)$ of the original metric.
The left-hand plot shows level surfaces of the time function~$T$ which
represent slices of the foliation.
The dashed line marks the position of the initial slice.
The right-hand plot shows evenly spaced contours of the
function~$1/N^2$ for the foliation as determined through
equation~\EqNum{HomKasLapse}{}. 
The filled-in regions at the top of the plot show where~$1/N^2$ ceases
to be a positive function, and thus where coordinate singularities
must appear in the slices of the foliation.
\bigskip
\endCaption
\endfigure

The behaviour of homogeneous harmonic slicings is, however, not
representative of the general inhomogeneous case, as is demonstrated
below. 
As an example, figure~\use{Fg.HomFoliation} shows the foliation which
results from setting the lapse on the initial slice to be
$$
  N_0(x) = { 1 \over {\textstyle 1 + {1\over2} \sin x} }
  \qquad \hbox{for } t_0 = 1
  \, , \EQN HomKasExamp
$$
which corresponds to the choice of coefficients~$k_0=1$
and~$k_{-1}=1/2$ (the others being zero) in
equation~\EqNum{HomKasInitT}{}.
When equation~\EqNum{HomKasLapse}{} is evaluated to determine the
lapse on this foliation, it is found that in some spacetime regions to
the future of the initial slice the quantity~$1/N^2$ is non-positive;
the right-hand plot of figure~\use{Fg.HomFoliation} is filled in where
this happens.
Coordinate singularities must appear in any slices of the foliation
which intersect these regions, and the development of the foliation
cannot proceed beyond a time $T=T_{\SmallWord{sing}}\simeq2.45$ with
the lapse~$N$ 
tending to infinity at points as the limiting slice is approached.
In contrast, if the foliation is extended backwards in time from the
initial slice towards the cosmological singularity, then the lapse
appears to be well behaved on all of the slices. 

It can in fact be shown that, in general, harmonic foliations of the
Kasner spacetime always develop coordinate singularities at
sufficiently late (future) times.
From equation~\EqNum{HomKasT}{} it follows that
$$\EQNalign{
  T_{,t} = {} & 
      { \pi t_0 \over 2 } \sum_{n=1}^{\infty} n
        \Big( J_1(nt) Y_0(nt_0) - Y_1(nt) J_0(nt_0) \Big)
        \Big( k_{+n} \cos(nx) + k_{-n} \sin(nx) \Big)
      + { k_0 t_0 \over t }
  \, , \qquad \EQN HomKasDerivs;a \cr
  T_{,x} = {} & 
      { \pi t_0 \over 2 } \sum_{n=1}^{\infty} n
        \Big( J_0(nt) Y_0(nt_0) - Y_0(nt) J_0(nt_0) \Big)
        \Big( k_{+n} \sin(nx) - k_{-n} \cos(nx) \Big)
  \, , \EQN HomKasDerivs;b \cr
}$$
and the first of these expressions can be written for a fixed spatial
coordinate~$x=x_0$ as
$$
  T_{,t}(t,x_0) = 
    { \pi t_0 \over 2 } \sum_{n=1}^{\infty} n c_n
      \Big( J_1(nt) Y_0(nt_0) - Y_1(nt) J_0(nt_0) \Big)
    + { k_0 t_0 \over t }
  \, ,
$$
where it is assumed that the value~$x_0$ has been chosen such that at
least one of the constants~$c_n$ is non-zero.
For sufficiently large values of the coordinate~$t$, the
approximations~\EqNum{BessApprox}{} can be applied to the Bessel
functions in this equation for~$T_{,t}$ with the result
(which disregards the possibility of convergence problems occurring
when the summation is over an infinite number of modes)
$$\eqalign{
  T_{,t}(t,x_0) & {} \simeq { k_0 t_0 \over t }
    + \sqrt{ \pi \over 2 \, t } t_0 \sum_{n=1}^{\infty} \sqrt{n} \, c_n
        \Big( Y_0(nt_0) \cos (nt-{\textstyle{3\over4}}\pi) 
            - J_0(nt_0) \sin (nt-{\textstyle{3\over4}}\pi) \Big)
  \cr
  & {} = { k_0 t_0 \over t } + { \Theta(t) \over \sqrt{t} }
  \, , \cr
}$$
where~$\Theta(t)$ is a periodic function which (by consideration of its
Fourier decomposition) can be seen to take both positive and negative
values. 
It then follows that, for some sufficiently large value of the
coordinate~$t$, the value of~$T_{,t}$ must be negative at a point.
Since this violates the condition~\EqNum{HomKasSing}{}, it must
be the case that a coordinate singularity is present in the foliation.

The above argument shows that coordinate singularities must in general
appear in harmonic slicings of an expanding Kasner cosmology but gives
no indication of how much of the spacetime a slicing will cover before
it becomes pathological.
To investigate this, consider the simple case of a time function~$T$
from equation~\EqNum{HomKasT}{} which, like the foliation produced by
the initial data~\EqNum{HomKasExamp}{}, contains only one non-trivial
mode: 
$$
  T = t_0 + k_0 t_0 \ln (t/t_0)
    - k { \pi t_0 \over 2 }
      \Big( J_0(nt) Y_0(nt_0) - Y_0(nt) J_0(nt_0) \Big) \sin(nx)
  \, , \EQN HomKasSingle
$$
where $0 < k < k_0$.
For a fixed value of the Kasner time~$t$ the foliation will be well
behaved (in that condition~\EqNum{HomKasSing}{} is satisfied) if and
only if 
$$
  {k_0 t_0 \over t} >
    { k n \pi t_0 \over 2}
    \max_{x}
    \Big\{ \vert A(t) \cos (nx) \vert - B(t) \sin (nx) \Big\}
  \, ,
$$
where
$$\eqalign{
  A(t) & {} = J_0(nt) Y_0(nt_0) - Y_0(nt) J_0(nt_0) \, , \cr
  B(t) & {} = J_1(nt) Y_0(nt_0) - Y_1(nt) J_0(nt_0) \, , \cr
}$$
and this condition is equivalent to
$$
  F(t) \equiv t^2 [ A(t)^2 + B(t)^2 ]
  < \left( { 2 k_0 \over k n \pi } \right)^2
  \, . \EQN HomKasSingF
$$
It is straightforward to show (using the formulae from the previous
subsection) that~$F'(t) \ge 0$.
Furthermore, if the time~$t$ is assumed to be sufficiently large that
the approximations~\EqNum{BessApprox}{} can be applied to~$A(t)$
and~$B(t)$, then $F(t) \simeq t \, \Pi(t)$ where~$\Pi(t)$ is a positive,
periodic function.
An estimate of the time~$t=t_{\SmallWord{sing}}$ at which a coordinate
singularity develops can then be seen to obey
$$
  t_{\SmallWord{sing}} \propto ( k_0 / k )^2
  \, , \EQN HomKasEstimate
$$
and so, for a harmonic slicing which is perturbed from homogeneity by a
single mode of amplitude~$k$, the amount of the spacetime covered by
the slicing becomes infinite as~$k$ tends to zero.

Considering now the behaviour of the slicing in the region of spacetime
between the initial slice and the cosmological singularity, it follows
from equation~\EqNum{HomKasSingF}{} that, for the single mode case,
coordinate singularities never appear.
To see this, note that the function~$F(t)$ must satisfy the
inequality at time~$t=t_0$ since the lapse is
required to be well behaved on the initial slice, and since it is
an increasing function of time it must then also satisfy the
inequality at all earlier times.
Moreover, it can be seen from
equation~\EqNum{HomKasSingle}{} together with the
equations~\EqNum{BessZero}{} that in the limit as~$t$ tends to zero
while~$x$ remains constant, the time function~$T$ must have the form
$$
  T \sim t_0 [ k_0 + k J_0(nt_0) \cos(nx) ] \ln t 
   + \hbox{(terms that tend to a constant)} 
  \, , \EQN HomKasZero
$$
and since the factor multiplying the logarithm is always positive, the
value of~$T$ must tend to negative infinity as the cosmological
singularity is approached.
Thus a single mode harmonic slicing will always cover the whole of the
Kasner spacetime to the past of the initial slice.
 
Returning to the general case in which an arbitrary number of modes
are present in the time function~$T$, the above argument can be used
to derive a simple condition on the initial data~\EqNum{HomKasInitT}{}
which ensures that a slicing is well behaved in the region of
spacetime between the initial slice and the cosmological singularity.
Suppose that equation~\EqNum{HomKasT}{} is split up as
$$
  T = t_0 + \sum_{n=1}^{\infty} \left( T^{(+n)} + T^{(-n)} \right)
  \, , \EQN HomKasSplit
$$
where
$$\eqalign{
  T^{(+n)} = {} &
    k_0^{(+n)} t_0 \ln (t/t_0)
    - k_{+n} { \pi t_0 \over 2 } 
      \Big( J_0(nt) Y_0(nt_0) - Y_0(nt) J_0(nt_0) \Big) \cos(nx)
    \, , \cr
  T^{(-n)} = {} &
    k_0^{(-n)} t_0 \ln (t/t_0)
    - k_{-n} { \pi t_0 \over 2 } 
      \Big( J_0(nt) Y_0(nt_0) - Y_0(nt) J_0(nt_0) \Big) \sin(nx)
    \, , \cr
}$$
and the values~$k^{(\pm n)}_0$ are arbitrary subject to the condition
$$
  \sum_{n=1}^{\infty} \left( k_0^{(+n)} + k_0^{(-n)} \right) = k_0
  \, .
$$
The condition~\EqNum{HomKasSing}{} will then certainly be satisfied if
$$
  T^{(+n)}{}_{\!,t} > \vert T^{(+n)}{}_{\!,x} \vert
  \qquad \hbox{and} \qquad
  T^{(-n)}{}_{\!,t} > \vert T^{(-n)}{}_{\!,x} \vert
  \qquad \hbox{for all }n 
  \, ,
$$
and as the analysis of the single mode case shows, this will be true
for $0<t<t_0$ if it is true for~$t=t_0$.
The time derivative of each function~$T^{(\pm n)}$ must therefore be
positive on the initial hypersurface $t=t_0$, and for this to be the
case the values~$k_0^{(\pm n)}$ must be chosen such that
$$
  k^{(+n)}_0 > \vert k_{+n} \vert
  \qquad \hbox{and} \qquad
  k^{(-n)}_0 > \vert k_{-n} \vert
  \qquad \hbox{for all }n 
  \, .
$$
This can always be done if
$$
  k_0 > \sum_{n=1}^{\infty} \Big( \vert k_{+n} \vert 
                                + \vert k_{-n} \vert \Big)
  \, , \EQN HomKasSufficient
$$
and so any initial value for the lapse which satisfies this condition
must produce a foliation which is free of coordinate singularities to
the past of the initial slice.
In addition to this, if the behaviour of the time function~$T$ is
considered as~$t$ tends to zero, then it is straightforward to show
(in analogy with equation~\EqNum{HomKasZero}{} for the single mode case)
that condition~\EqNum{HomKasSufficient}{} ensures that~$T$ tends
to negative infinity for all values of~$x$ as the cosmological
singularity is approached, and hence that the foliation covers all of
the spacetime up to the singularity.
The condition~\EqNum{HomKasSufficient}{} on the initial lapse is
sufficient but not necessary for the harmonic foliation to be well
behaved to the past of the initial slice; in fact, experimentation
with different choices of initial data~\EqNum{HomKasInitT}{} suggests
that, even when condition~\EqNum{HomKasSufficient}{} is not satisfied,
harmonic foliations may never develop coordinate singularities as the
cosmological singularity is approached, although no proof (or
disproof) of this has yet been constructed.

To summarize the above results, foliations of the Kasner
spacetime~\EqNum{HomKasnerTwo}{} based on the simple harmonic slicing
condition~\EqNum{HomHarmTime}{} have been investigated under the
assumptions that one slice of the foliation coincides with a
$t=\hbox{\it constant\/}$ hypersurface of the original metric, and that
the foliation is independent of two of the standard spatial
coordinates.
It is found that when the Kasner cosmology is expanding, the slices of
the foliation must eventually develop coordinate singularities, with
this happening at late times for foliations which are initially close
to homogeneous.
In contrast, when the Kasner cosmology is collapsing, the results
suggest that coordinate singularities never develop in the foliation
and that the slices extend all the way to the cosmological
singularity.
(This is certainly true if the lapse on the initial slice is chosen
such that the coefficients of equation~\EqNum{HomKasInitT}{} satisfy
the condition~\EqNum{HomKasSufficient}{}, and it may in fact be the case
for all reasonable choices of initial lapse.)

The behavioural differences found in harmonic foliations according to
whether the Kasner spacetime is expanding or collapsing can be
explained by considering the relationship between the lapse
function~$N$ and the mean curvature~$K$ of the slicing.
In any foliation, the mean curvature provides a measure of the local
convergence of world lines running normal to the spatial slices:
$$
  K = - \nabla_{\!\mu} n^{\mu}
    = - \Lie_{\!n} ({ \textstyle \ln \sqrt{\det h_{ij}} })
  \, ,
$$
where~$n^{\mu}$ is the vector normal to the slices, $\Lie_{\!n}$~is
the Lie derivative along that vector, and~$\sqrt{\det h_{ij}}$ is the
spatial volume element.
The value of~$K$ is positive when the foliation world lines are
locally converging, and negative when they are expanding.
The harmonic time slicing condition can be written in terms of the
mean curvature of the foliation using
equation~\EqNum{HomHarmEvolve}{}:
$$
  { \partial (1/N) \over \partial t } = K
  \, ,
$$
where it is assumed here that the shift vector~$N^i$ of the foliation
is zero, and that the harmonic slicing is of the simple
type~($Q_{,t}=0$).
It then follows that when the mean curvature~$K$ is positive (as it
may be expected to be for slices of a collapsing cosmology) the value
of~$1/N$ will increase, and so the lapse will approach (but usually
not reach) zero.
(The ability of the harmonic slicing condition to avoid `focusing'
singularities at which the lapse vanishes is discussed in 
Bona and Mass\'o~1988.) 
Conversely, when the mean curvature is negative (as it typically will
be in an expanding spacetime) the value of~$1/N$ will decrease, and if
it reaches zero then a coordinate singularity of the type investigated
in this subsection will develop in the foliation.

\SubSection{Harmonic Slicings of the Kasner Spacetime 
            II: Numerical Implications}
The results of the previous subsection show that coordinate
singularities are a generic feature of harmonic foliations of the
expanding axisymmetric Kasner cosmology.
It follows then that numerical simulations of that spacetime which use
the harmonic slicing condition will be forced to terminate after a
finite number of time steps, regardless of their accuracy, and the
practical details of this are examined in the present subsection.
Some of the methods for comparing spacetime slicings described in
section~\use{sec.HomCompare} are put into practice here, and the
implications of these results for numerical simulations of other
spacetimes are discussed.

In chapter~4 a computer code for numerically evolving solutions to the
vacuum Einstein equations is described.
Here the code is used to evolve initial data for the Kasner spacetime
given by the metric~\EqNum{HomKasnerTwo}{} using the harmonic time
slicing condition and the initial value~\EqNum{HomKasExamp}{} for the
lapse.
The foliation of the Kasner spacetime which results is the one
pictured in figure~\use{Fg.HomFoliation}, and the numerical simulation
thus allows the prediction made in the previous subsection---that
coordinate singularities prevent the foliation from developing beyond a
time $T=T_{\SmallWord{sing}}\simeq2.45$---to be verified.
The metric variables~\EqNum{HypFRVars}{} evolved by the code are given
homogeneous initial values appropriate to a slice $t=t_0$ of the
metric~\EqNum{HomKasnerTwo}{}:
$$
  \eqalign{ h^{11} & {} = t_0{}^{1/2} \, , \cr 
            P^{11} & {} = t_0{}^{-1/4} \, , \cr }
  \qquad
  \eqalign{ h^{22} = h^{33} & {} = t_0{}^{-1} \, , \cr 
            P^{22} = P^{33} & {} 
                   = {\textstyle{1\over4}}t_0{}^{-7/4} \, , \cr }
  \qquad \hbox{for } t_0 = 1
  \, , \EQN HomNumInit
$$
with the other variables all being initially zero.
The initial value of the lapse is given the inhomogeneous form of
equation~\EqNum{HomKasExamp}{} by setting the slicing density of the
foliation to be
$$
  Q(T,X,Y,Z) = { t_0{}^{-3/4} \over 1 + {1\over2} \sin X }
  \, , \EQN HomNumDensity
$$
where $(T,X,Y,Z)$ is the coordinate system of the numerical
simulation, and is distinct from the coordinate system $(t,x,y,z)$ of
the original Kasner metric, except on the initial slice which is
labelled~$T=t_0$ and on which $(X,Y,Z)=(x,y,z)$.
The shift vector~$N^i$ is taken to be zero throughout the simulation,
and the lack of dependence of the slicing density~$Q$ on the foliation
time~$T$ then means that the harmonic slicing is of the simple type.
Since all of the quantities involved in the numerical simulation are
independent of the spatial coordinates~$Y$ and~$Z$, the evolved
spacetime in effect has a planar symmetry, and a one-dimensional grid
with a cell size $\Delta X = 2 \pi / 200$ is used to discretize the
spatial domain.
The adaptive mesh refinement capabilities of the code are described in
chapter~3, and are used here to allow the local resolution of the
simulation to increase as needed to a maximum 
of $\Delta X = 2 \pi / 3200$.

\figure{HomSlices}
 \DrawFig{
  \vskip-5mm
  \epsfxsize=0.8\hsize
  \DrawHiLoRes{\epsfbox{slices.eps}}
              {\epsfbox{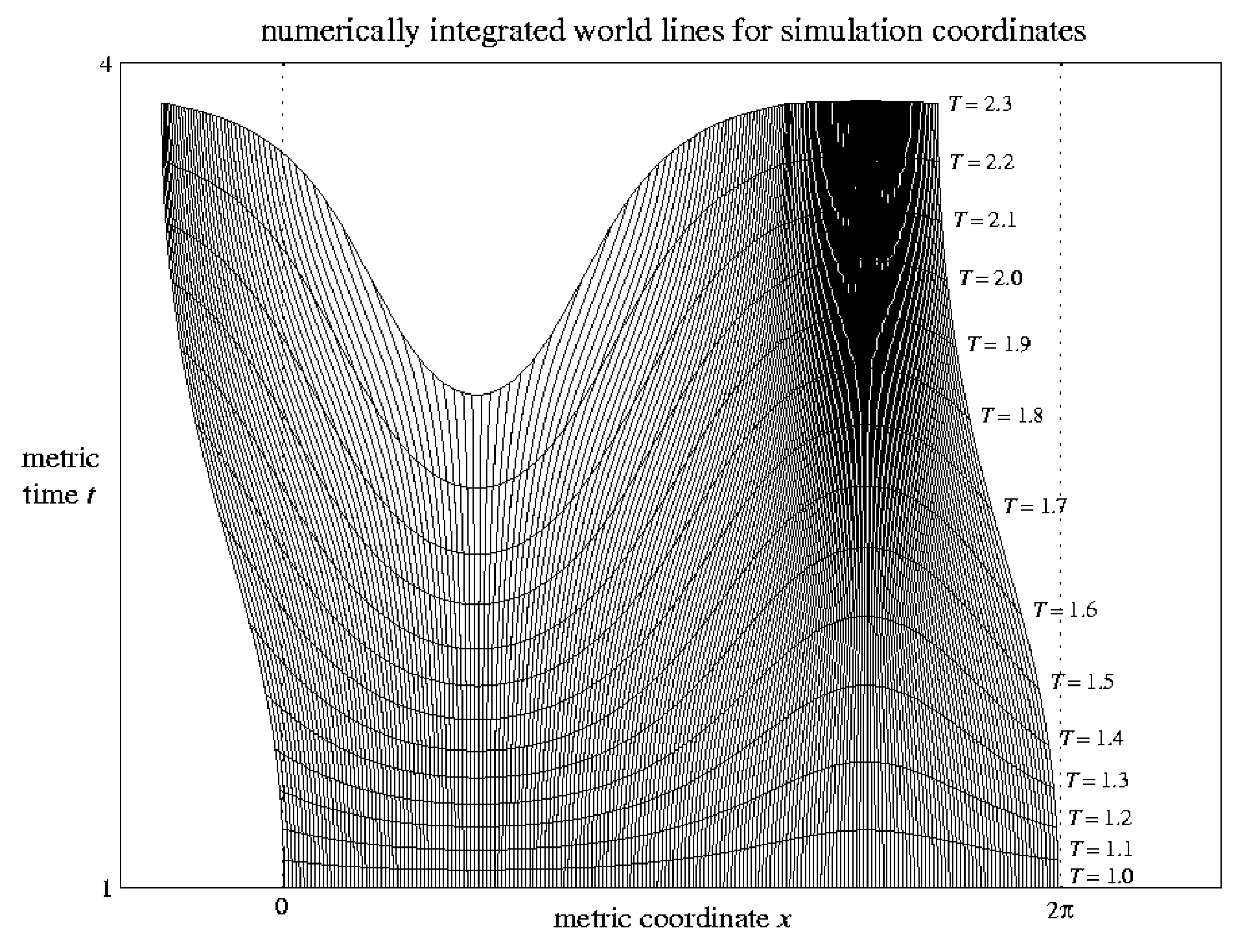}}
 }
\Caption
Coordinate world lines for a numerical simulation as seen in the
background Kasner metric.
The homogeneous initial data set~\EqNum{HomNumInit}{} is numerically
evolved using harmonic slicing based on the inhomogeneous gauge
function~\EqNum{HomNumDensity}{}, and the world lines of the
simulation's grid points are tracked relative to the coordinate
system~$(t,x)$ of the background metric~\EqNum{HomKasnerTwo}{} using
the algorithm described in section~\use{sec.HomCompare}.
Two hundred grid points are evenly spaced on the $t=1$
hypersurface of the background metric at the initial simulation
time $T=1$, and the paths they subsequently follow are plotted up
until the end of the simulation at time~$T=2.3$.
Periodicity of the spatial domain means that observers who `go off'
at $x=0$ will `come on' at $x=2\pi$.
This figure is to be compared with the left-hand plot of
figure~\use{Fg.HomFoliation}.
\bigskip
\endCaption
\endfigure

Figure~\use{Fg.HomSlices} plots the~$(T,X)$ coordinate system of the
numerical simulation as it appears in the~$(t,x)$ coordinates of the
homogeneous background metric.
This comparison between coordinate systems is made using the algorithm
described in section~\use{sec.HomCompare} for tracking the
positions~$p^\mu$ of the observers at rest in the slices of a
foliation, and is based on the values~$N(T,X)$ taken by the lapse
during the course of the simulation.
The observers are positioned at the centres of the cells in the
simulation grid, and their initial positions with respect to
the~$(t,x)$ coordinates are
$$
  p^\mu \vert^0_j = (t,x)\vert^0_j = (t_0,X_0+j\,\Delta X)
  \, .
$$
The observer world lines are plotted up until a simulation
time~$T=2.3$; although the inaccuracies in the numerical solution
which are present at that time are not sufficient to stop the
simulation, they do make it increasingly difficult to accurately track
the positions of the observers.
The $T=\hbox{\it constant\/}$ surfaces reconstructed in
figure~\use{Fg.HomSlices} can be seen to be in good agreement with the
exact results shown in the left-hand plot of
figure~\use{Fg.HomFoliation}. 

\figure{HomLapseArea}
 \DrawFig{
  \vskip-3mm
  \epsfxsize=\hsize
  \epsfbox{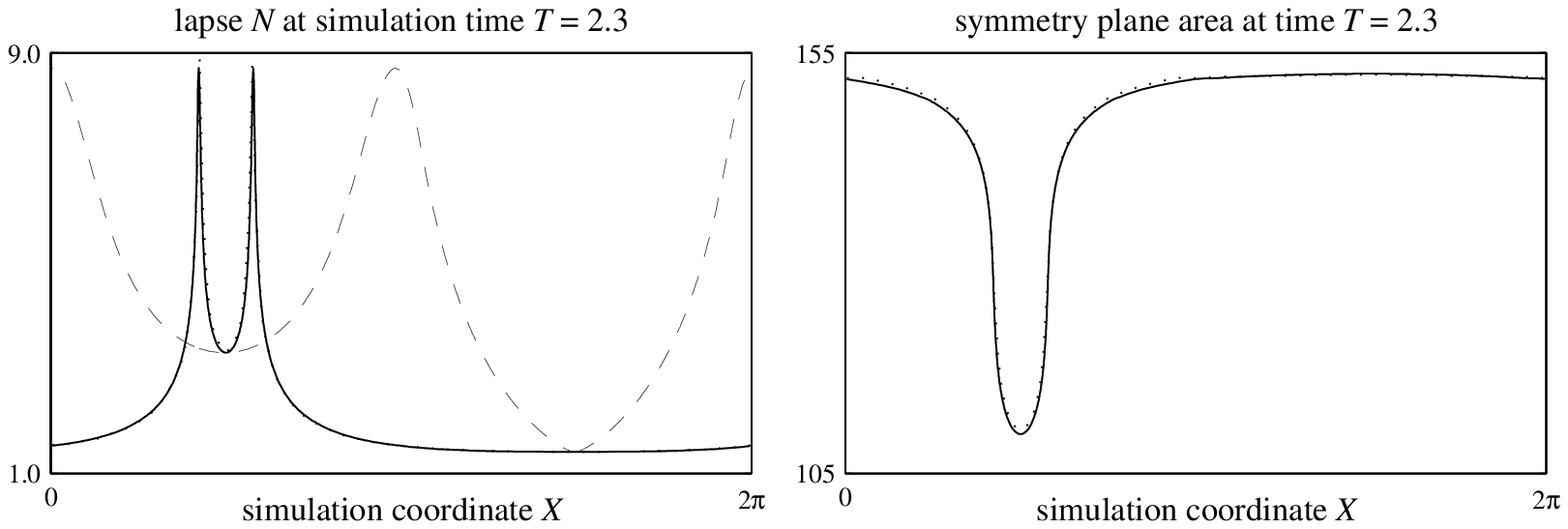}
 }
\Caption
The lapse and the symmetry plane area for a numerical simulation of an
inhomogeneously sliced Kasner spacetime.
The quantities plotted here as solid lines are evaluated based on the
evolved variables at the final time of the simulation shown in
figure~\use{Fg.HomSlices}. 
Dotted lines in the plots show the same quantities evaluated
indirectly using the final positions $p^\mu(X) = (t,x)$ estimated for
the simulation grid points with respect to the known
metric~\EqNum{HomKasnerTwo}{}. 
The simple relation~\EqNum{HomKasArea}{} between the coordinate~$t$
and the symmetry plane area allows an alternative approximation for
the latter to be calculated, while an alternative value for the lapse
can be estimated from the positions~$p^\mu$ together with the data
plotted in the second half of figure~\use{Fg.HomFoliation}.
The dashed line in the left-hand plot shows the lapse as a function of
 proper distance along the $X$-axis rather than coordinate distance.  
(The measure of proper distance is rescaled to be consistent with the
horizontal axis of the plot.)
\bigskip
\endCaption
\endfigure

The analysis of the previous subsection (as presented in
figure~\use{Fg.HomFoliation}) shows that at a time
$T=T_{\SmallWord{sing}}\simeq2.45$ the lapse~$N$ must become infinite
at points of the evolved foliation. 
The left-hand plot of figure~\use{Fg.HomLapseArea} shows the actual
value taken by the lapse at the earlier time~$T=2.3$ in the
simulation, and it can be seen that two sharp spikes are already
present in the solution.
As the evolution progresses these spikes grow in size, and if the
evolved solution were exact they would reach infinite heights within a
short time.
As they grow, the two spikes also narrow and become closer together.
The reason for this can be seen from the paths of the coordinate
observers plotted in figure~\use{Fg.HomSlices}: in the
region~$0<x<\pi$ where the coordinate singularities eventually form, 
the world lines normal to the foliation are diverging (with respect to
the background coordinates), and so the region is resolved by a
diminishing number of grid points.
It should be noted that this divergence of observers is not itself the
cause of the coordinate singularities; it could in principle be
counteracted by an appropriate choice of shift vector for the
foliation, but this would not prevent the lapse from becoming
infinite.
The narrowing and effective coalescence of the spikes is problematic
for the simulation since features which are smaller than the grid
spacing~$\Delta X$ cannot be accurately resolved, and while the
adaptive mesh refinement capabilities of the numerical code prove to
be very effective in extending the useful running time of the
simulation, eventually the numerical solution fails to accurately
model the behaviour of the exact solution.
In fact, with the spikes being inadequately resolved, the evolved
value of the lapse fails to become infinite (or even the computer
representation of this) as the coordinate singularities are reached, and
it is possible for the simulation to continue beyond the time at which
the slices of the foliation cease to be spacelike in the exact solution.
However, as pointed out by Alcubierre and Mass\'o~(1998), the
numerical solution has no physical meaning past the points at which
coordinate singularities form, and in particular it will no longer
converge as the grid is refined. 

A check on the accuracy of both the numerical solution produced by the
simulation and the world line reconstruction shown in
figure~\use{Fg.HomSlices} can be made by utilizing some of the ideas
presented in section~\use{sec.HomCompare}.
For a spatially closed spacetime with planar symmetry (which for the
purposes of the present discussion the inhomogeneously sliced Kasner
cosmology is taken to be), the proper area of the symmetry plane
defines a scalar field on the spacetime.
For the metric~\EqNum{HomKasnerTwo}{} this can be seen to take the value
$$
  \hbox{proper area of $y$-$z$ symmetry plane}
  = 4 \pi^2 t
  \, . \EQN HomKasArea
$$
A value for the area can also be readily calculated from the metric
components at any point~$(T,X)$ of a numerically evolved solution, in
which case equation~\EqNum{HomKasArea}{} can be used to determine how
much of the Kasner spacetime the simulation has covered.
The right-hand plot of figure~\use{Fg.HomLapseArea} shows two values
calculated for the symmetry plane area at a fixed time in the
simulation: one of the values is calculated directly from the metric
variables evolved by the simulation, while the other is calculated by
substituting the observer positions plotted in
figure~\use{Fg.HomSlices} into the equation~\EqNum{HomKasArea}{}.
The two values can be seen to be in good agreement.
(Figure~\use{Fg.HomLapseArea} also shows an alternative value for the
lapse calculated in a similar though more convoluted manner by using
the observer positions of figure~\use{Fg.HomSlices} to evaluate the
exact expression~\EqNum{HomKasLapse}{}.
Again, good agreement with the evolved value is seen.)

The spikes that develop in the lapse function as the coordinate
singularities are approached are at face value very similar to some
features that are seen in numerical solutions for collapsing
inhomogeneous cosmologies, as studied in chapters~6 and~7 (see for
example figure~\use{Fg.GowVarP}).
The question then arises as to whether the features seen in the
inhomogeneous cosmologies have any physical relevance or whether they
are, like the features in the Kasner simulation, entirely a coordinate
effect.
While a full investigation of this point is postponed until the
following chapter, a comment is worth making here about the nature of
the spikes seen in the lapse in figure~\use{Fg.HomLapseArea}.
If the metric component~$h_{11}$, which measures the proper distance in
the $X$-direction of the simulation, is examined for the data plotted
in figure~\use{Fg.HomLapseArea}, it is found that sharp spikes
(towards positive infinity) are coincident there with the spikes in
the lapse, and hence that the apparently narrow spikes are in fact
spread over large physical distances. 
As the dashed line in figure~\use{Fg.HomLapseArea} demonstrates, 
if instead of being plotted against the coordinate~$X$, the lapse is
shown as a function of the proper distance along the $X$-axis, then
the spikes no longer appear as notable features of the solution.

One of the main conclusions of the present section is that coordinate
singularities can appear in numerical simulations which use the simple
harmonic time slicing condition under reasonably general circumstances
(seemingly whenever the foliation locally undergoes a protracted
period of expansion).
As was shown in the previous subsection, the coordinate singularities
are a feature of the exact solution, and cannot be avoided by
employing alternative numerical methods.
This is clearly a drawback to the use of harmonic slicing in
numerical simulations, and by extension to the use of hyperbolic
formulations of Einstein's equation which, as discussed in chapter~4,
typically rely on this type of slicing.
It should be realized though that no known slicing condition is
guaranteed to completely cover an arbitrary spacetime, and potential
problems with coordinate singularities do not necessarily outweigh the
advantages in using hyperbolic formulations for numerical work.
Of course, the formation of coordinate singularities has only been
discussed here for simple harmonic slicing, and it is
possible that the freedom in choosing the time dependence of the
slicing density in generalized harmonic slicing could be put to use in
controlling the development of the spacetime foliation such that
coordinate singularities are avoided. 
Although when used in the context of a hyperbolic formulation the slicing
density is formally required to be independent of the evolved
variables, in practice there seems to be no problem in allowing
occasional `corrections' to be made to its value in response to the
behaviour of the foliation; in effect this amounts to intermittently
halting the simulation and choosing a new value for the lapse on the
current slice.
(An important point here is that changes to the slicing density should
be made only at fixed time intervals, rather than after a fixed number
of time steps, since otherwise the exact solution being sought will
depend on the resolution of the simulation.)
This `piecewise harmonic' form of time slicing has been used in some
preliminary tests employing simple heuristics for making alterations to
the value of the slicing density, with the resulting foliations being
examined using the world line integration algorithm of
section~\use{sec.HomCompare} (which was in fact originally developed 
for this purpose); however no definite conclusions regarding the
effectiveness of the approach have yet been reached.

The present chapter has considered some of the difficulties that may
be encountered because of gauge effects in numerical simulations of
even very simple spacetimes.
In the two chapters that follow numerical methods are used to study
inhomogeneous cosmological models (many aspects of the behaviour of
which are largely unknown), and the ideas developed in the present
chapter help to distinguish coordinate effects in the numerical
results from genuine physical behaviour.
For the majority of the work done on the inhomogeneous models,
specialized reductions of the Einstein equations are used rather than
the hyperbolic formulation described in chapter~4; however, these
reductions are all still based on harmonic time slicings, with fixed
forms being assumed for the slicing densities.
While little is known about the behaviour of harmonic slicings of the
$U(1)$-symmetric cosmologies discussed in chapter~7, for the Gowdy
cosmologies of chapter~6 it is known that the 
$t=\hbox{\it constant\/}$ slices of the metric~\EqNum{GowMetricT}{}
cover the entire spacetime for $0<t<\infty$ (see
section~\use{sec.GowGowdy} for further details) and hence that
coordinate singularities are not an issue for simulations based
on this form of the metric.
Following on from the work of the present section, it is natural to
enquire as to the extent to which more general harmonic slicings of these
inhomogeneous cosmologies are well behaved.
A partial answer can be readily obtained in the case of the Gowdy~$T^3$
model.
If the analysis of the previous subsection for harmonic slicings of
the axisymmetric Kasner spacetime~\EqNum{HomKasnerTwo}{} is applied to
the Gowdy metric~\EqNum{GowMetricT}{} then it is found that the time
function~$T$ must satisfy
$$
  T_{,tt} - T_{,\theta\theta} + t^{-1} T_{,t} = 0
  \, ,
$$
while the lapse~$N$ of the associated foliation is defined through
$$
  e^{-\lambda/2} \sqrt{t} \, ( T_{,t}{}^2 - T_{,\theta}{}^2 ) 
  = 1/N^2
  \, , 
$$
where~$\lambda(t,\theta)$ is one of the variables describing the Gowdy
spacetime.
Comparing these expressions to equations~\EqNum{HomKasTimeEqn}{}
and~\EqNum{HomKasLapse}{}, it is clear that the Gowdy spacetimes admit
the same harmonic foliations as the Kasner spacetime, and furthermore
that the condition that foliations must satisfy to be free of
coordinate singularities is the same.
It thus turns out that all the results of the present section on the
behaviour of harmonic slicings of a Kasner cosmology carry over
directly to a fairly general class of planar cosmological models.


\Chapter{Planar Cosmologies}
Even under assumptions of high degrees of symmetry, the general
behaviour of inhomogeneous cosmological spacetimes cannot readily be
described using analytic methods.
The simplest inhomogeneous cosmological models assume the existence of
two spacelike Killing vectors and have either planar, spherical or
cylindrical symmetry.
In the present chapter a class of spatially closed vacuum planar
cosmologies are investigated using the numerical methods described
earlier in this work.

Planar cosmologies have been an area of interest in numerical
relativity since its early days.
Since the late seventies
Centrella, Matzner, Wilson, Anninos and co-workers have
developed four generations of planar codes which have been
used to investigate a variety of cosmological phenomena.
The first code (Centrella~1979, 1980a, 1980b; Centrella and
Matzner~1979, 1982) evolved gravitational shock waves in vacuum
spacetimes. 
It used geodesic slicing and unconstrained evolution, and restricted
the metric to diagonal form so that only one polarization of
gravitational radiation was present in the simulations.
Centrella and Wilson~(1983, 1984) developed a second code capable of
evolving planar cosmologies containing a single polarization of
gravitational radiation together with matter in the form of a perfect
fluid. 
This code used constant mean curvature slicing and fully constrained
evolution. 
It was subsequently used to study nucleosynthesis (Centrella et
al.~1986) and inflation (Kurki-Suonio et al.~1987) in inhomogeneous
model universes.
The third generation planar code (Anninos, Centrella and 
Matzner~1991a, 1991b) updated the algorithms used by Centrella and
Wilson while discarding the hydrodynamical sources and was used
to investigate nonlinear wave propagation in the vacuum Einstein
equations. 
Recently, Anninos~(1998a) has developed a new planar code which evolves
a perfect fluid together with both polarizations of gravitational
radiation.
The code is capable of using a variety of slicing conditions and its
evolution is unconstrained.

Other numerical work investigating planar cosmologies has been carried
out by Shinkai and Maeda~(1993, 1994) who considered inflationary 
scenarios involving a positive cosmological constant and an inhomogeneous
inflaton field.
They allowed for both polarizations of gravitational radiation, and
used geodesic slicing and unconstrained evolution.
Ove~(1990a, 1990b) used a two-dimensional vacuum cosmology code
restricted to planar symmetry to investigate nonlinear effects in
gravitational wave propagation.
His code implemented Moncrief's~(1986) reduction of the Einstein
equations, discussed in depth in chapter~7.

In contrast to the investigations cited above which are concerned
with expanding spacetimes, Berger and Moncrief~(1993) have performed
simulations of collapsing planar cosmologies, their
aim being to study the behaviour of spacetimes close to `big bang' 
singularities. 
Berger and Moncrief's work is based on the Gowdy~$T^3$ family of
vacuum spacetimes, and these models have been adopted for the work
presented in this chapter; they are described in
section~\use{sec.GowGowdy}. 

In this chapter vacuum planar cosmological models are evolved
numerically both forwards and backwards in cosmic time.
The work of Berger and Moncrief~(1993) provides a reference for the
investigation of the collapsing case, and the fine-scale structure
that develops in the evolved solutions proves to be an excellent test
of the capabilities of the adaptive mesh refinement code described in
chapter~3. 
Results from numerical simulations are presented in
section~\use{sec.GowCollapse}, and in section~\use{sec.GowSing} these
are used to assess well-known conjectures regarding the nature of
inhomogeneous cosmological singularities.
(Much of the work in these sections was previously published in Hern
and Stewart~1998.)

Expanding planar cosmologies are considered in
section~\use{sec.GowExpand}; there the results from numerical
simulations of Gowdy spacetimes are compared to the studies of other
planar cosmological models mentioned above.
Three types of behaviour are investigated in particular: the formation
of spatial structure in expanding spacetimes, the late-time fate
of initially inhomogeneous cosmologies, and the differences between
linear and nonlinear gravitational waves.

Some further work making use of the Gowdy cosmological models is
presented in chapter~7. 
Section~\use{sec.MonCollapse} addresses the question of how
the fine-scale spatial structure found in collapsing Gowdy spacetimes
generalizes to spacetimes with less symmetry.
In section~\use{sec.MonFR} Gowdy spacetimes are used as test cases for
the Frittelli-Reula evolution system described in chapter~4.

\Section{The Gowdy Cosmological Models
         \label{sec.GowGowdy}}
In Gowdy~1974 a family of vacuum metrics having closed spacelike
hypersurfaces and two-parameter isometry groups is constructed.
These metrics are interpreted as describing closed
inhomogeneous cosmologies containing two polarizations of
gravitational waves.
The Einstein equations for the Gowdy spacetimes reduce to a simple
set of partial differential equations which can be studied by analytic
methods, but for which exact solutions can in general only be
constructed numerically.
The simplicity of the equations, together with the body of existing
analytic work describing their global properties, makes the Gowdy
metrics well-suited to numerical investigations of the behaviour of
inhomogeneous cosmologies.

This section presents an overview of the Gowdy models and considers in
detail the subclass which are spatially of three-torus~($T^3$) topology.
The Gowdy~$T^3$ metrics describe closed vacuum planar cosmologies,
and their behaviour is investigated numerically in the remainder of
this chapter.

\SubSection{Topologies of the Gowdy Models}
Gowdy~1974 considers spacetimes which have two-parameter isometry
groups with spacelike Killing vectors (which are assumed, without loss
of generality, to commute) together with compact spacelike
hypersurfaces of finite volume.
The natural coordinates of the isometry group provide two spatial
coordinates, angles~$\sigma$ and~$\delta$, for the spacetime; the
components of the metric are independent of these coordinates.
Two other spacetime coordinates,~$\theta$ and~$t$, label the
orbits of the isometry group, and Gowdy assumes a `two-surface
orthogonality' condition to ensure that the coordinate system is a
global one.
(Details of $T^2$-isometric spacetimes without such restrictions are
given in Berger et al.~1997.)
It follows from the original compactness assumption that the spacelike
hypersurfaces of the spacetime must be homeomorphic to the
three-torus~($T^3$), the three-handle~($S^1 \times S^2$), the
three-sphere~($S^3$), or to a manifold covered by one of these.
(Alternative topologies are possible if assumptions about the global
nature of the Killing vectors are weakened; see Tanimoto~1998.)
By applying the vacuum Einstein equations, and making use of the
available coordinate freedom, the Gowdy metrics are written in a form
that depends only on three functions of~$\theta$ and~$t$
satisfying simple systems of evolution and constraint equations.

If the Killing vector fields which generate the isometry group are
everywhere mutually orthogonal then the Gowdy model is said to be
polarized and in such cases exact solutions can be obtained.
Gowdy~1971 discusses polarized spacetimes having~$S^1 \times S^2$
and~$S^3$ spatial topologies.
Polarized Gowdy~$T^3$ spacetimes are considered later in this section.

The Gowdy~$S^1 \times S^2$ and Gowdy~$S^3$ spacetimes have similar
global structures.
In both models the spacelike hypersurfaces connect two timelike
regions where the trajectories of the isometry group degenerate; this
manifests in the metric as two `cylindrical' coordinate axes at the
limits of the~$\theta$ coordinate range.
In addition, these models have both initial and final
spacelike singularities with each spacetime first expanding, then
collapsing during the course of its evolution.
In contrast, the Gowdy~$T^3$ spacetimes have no degenerate
trajectories on their spacelike hypersurfaces, and their metric
components are periodic in the angle~$\theta$.
These spacetimes begin at initial singularities and expand forever.

The Gowdy cosmological models have been studied in a variety of
contexts. 
They have been used as background spacetimes in which to investigate
quantum effects and approaches to quantization (Misner~1973, 
Berger~1974 and~1984, McGuigan~1991, Ashtekar and Husain~1998).
They have provided example cases for tests of the strong cosmic 
censorship conjecture (Moncrief~1981, Chrusciel, Isenberg and
Moncrief~1990, Grubi{\u s}i\'c and Moncrief~1993) and investigations of 
the nature of cosmological singularities (Isenberg and Moncrief~1990,
Berger and Moncrief~1993, Kichenassamy and Rendall~1998,
Garfinkle~1999; also see section~\use{sec.GowSing} of the present work).
They have been used as backgrounds on which to visualize gravitational
wave motion (Berger, Garfinkle and Swamy~1995) and,
most recently, as known metrics against which to
test the performance of numerical methods and formulations
(van Putten~1997, New et al.~1998).

The Gowdy spacetimes with~$T^3$ spatial topology are the simplest
kind, in terms of both equations and boundary conditions, and for this
reason they have received most attention in the literature.
For the remainder of this chapter the three-torus models are considered
exclusively. 
The Gowdy~$T^3$ models provide a simple, practical  framework within
which the behaviour of planar cosmologies can be investigated, and the
results presented here can be compared with the numerical studies of
planar cosmologies cited in the introduction to this chapter.
(Since those studies invariably use periodic boundary
conditions the topologies of the different models are effectively
the same.)
The advantage of studying the Gowdy~$T^3$ models is that their Einstein
equations reduce to a very simple form; the disadvantage is that 
extending the models to include fluid matter sources is not
straightforward. 
(Generalized Gowdy spacetimes containing scalar and electromagnetic
fields are discussed in Carmeli, Charach and Malin~1981; for models
including matter in the form of collisionless particles see
Andr\'easson~1998.)

\SubSection{The Gowdy $T^3$ Cosmology}
The Gowdy~$T^3$ line element is written here following
Berger and Moncrief~1993 as
$$
  ds^2 =
  e^{(\lambda + \tau)/2}
    ( -e^{-2\tau} d\tau^2 + d\theta^2 )
  + e^{-\tau}( e^P \, d\sigma^2
             + 2 e^P Q \, d\sigma \, d\delta
             + ( e^P Q^2 + e^{-P} ) \, d\delta^2 )
  \, .
  \EQN GowMetricTau
$$
The spatial coordinates~$\theta$, $\sigma$ and~$\delta$ range from~$0$
to~$2\pi$ and describe a three-torus.
The time coordinate~$\tau$ can take any real value;~$\tau = +\infty$
marks the cosmological singularity and~$\tau = -\infty$ corresponds to
an infinitely expanded universe.
The functions~$\lambda$, $P$ and~$Q$ depend on the coordinates~$\theta$
and~$\tau$ only and are required to be periodic in~$\theta$.

The vacuum Einstein equations for the metric reduce to two
wave equations for~$P$ and~$Q$,
$$\EQNalign{
  P_{,\tau\tau} & {}=
    e^{-2\tau} P_{,\theta\theta}
  + e^{2P} ( Q_{,\tau}{}^2 - e^{-2\tau} Q_{,\theta}{}^2 )
  \, ,
  \EQN GowEinst;a \cr
  Q_{,\tau\tau} & {}=
    e^{-2\tau} Q_{,\theta\theta}
  - 2 ( P_{,\tau} Q_{,\tau} - e^{-2\tau} P_{,\theta} Q_{,\theta} )
  \, ,
  \EQN GowEinst;b \cr
}$$
and two equations for~$\lambda$ deriving from the Hamiltonian and
momentum constraints,
$$\EQNalign{
  - \lambda_{,\tau} & {}=
    P_{,\tau}{}^2
  + e^{-2\tau} P_{,\theta}{}^2
  + e^{2P} ( Q_{,\tau}{}^2 + e^{-2\tau} Q_{,\theta}{}^2 )
  \, ,
  \EQN GowEinst;c \cr
  - \lambda_{,\theta} & {}=
    2 ( P_{,\tau} P_{,\theta} + e^{2P} Q_{,\tau} Q_{,\theta} )
  \, .
  \EQN GowEinst;d \cr
}$$
It is important to note that equations~\EqNum{GowEinst}{a}
and~\EqNum{GowEinst}{b} do not
involve~$\lambda$; they form an independent sub-system within
which~$P$ and~$Q$ are completely determined from appropriate initial
data. 
Subsequently, if~$P$ and~$Q$ are known, then 
equation~\EqNum{GowEinst}{c} can be integrated to find~$\lambda$.
Equation~\EqNum{GowEinst}{d} then remains as a constraint which 
determines (up to addition of a constant) the initial value 
of~$\lambda$ from the
initial values of~$P$ and~$Q$ (and their first time derivatives).
This constraint is conserved in the sense that, provided the evolution
equations~\EqNum{GowEinst}{a,b,c} hold, equation~\EqNum{GowEinst}{d}
is satisfied for all time if it is satisfied initially.
Arbitrary (periodic) values of~$P$, $Q$, $P_{,\tau}$ and~$Q_{,\tau}$
thus provide sufficient initial data for the system, subject to one
condition: periodicity in~$\lambda$ is required, and
from equation~\EqNum{GowEinst}{d},
$$
  \lambda\vert_{\theta=0} = \lambda\vert_{\theta=2\pi}
  \quad\iff\quad
  \int_{0}^{2\pi}
    ( P_{,\tau} P_{,\theta} + e^{2P} Q_{,\tau} Q_{,\theta} ) 
  \, d\theta
  = 0
  \, .
  \EQN GowMoment
$$
Gowdy~(1974) notes that this is equivalent to demanding that the
total $\theta$-momentum of the system be zero.

The choice of time coordinate in the metric~\EqNum{GowMetricTau}{} 
is well suited to numerical investigations of a Gowdy~$T^3$ spacetime
as it collapses:
the cosmological singularity is approached arbitrarily closely but
not reached in finite coordinate time.
However, for studying expanding cosmologies Gowdy's original time
coordinate~$t$ is more appropriate.
It is defined by
$$
  t = e^{-\tau}
  \, ,
  \EQN GowTimes
$$
and transforms the metric~\EqNum{GowMetricTau}{} to
$$
  ds^2 =
  e^{\lambda/2} t^{-1/2}
    ( - dt^2 + d\theta^2 )
  + t ( e^P \, d\sigma^2
      + 2 e^P Q \, d\sigma \, d\delta
      + ( e^P Q^2 + e^{-P} ) \, d\delta^2 )
  \, ,
  \EQN GowMetricT
$$
with evolution and constraint equations
$$\EQNalign{
  P_{,tt} & {}=
    P_{,\theta\theta}
  + e^{2P} ( Q_{,t}{}^2 - Q_{,\theta}{}^2 )
  - (1/t) P_{,t}
  \, ,
  \EQN GowNewEinst;a \cr
  Q_{,tt} & {}=
    Q_{,\theta\theta}
  - 2 ( P_{,t} Q_{,t} - P_{,\theta} Q_{,\theta} )
  - (1/t) Q_{,t}
  \, ,
  \EQN GowNewEinst;b \cr
  \lambda_{,t} & {}=
    t ( P_{,t}{}^2 + P_{,\theta}{}^2 )
  + t e^{2P} ( Q_{,t}{}^2 + Q_{,\theta}{}^2 )
  \, ,
  \EQN GowNewEinst;c \cr
  \lambda_{,\theta} & {}=
    2 t ( P_{,t} P_{,\theta} + e^{2P} Q_{,t} Q_{,\theta} )
  \, .
  \EQN GowNewEinst;d \cr
}$$
The cosmological singularity occurs at time~$t=0$ and the universe
expands forever through positive values of~$t$.

Some important results can be read straight off the 
metrics~\EqNum{GowMetricTau}{} and~\EqNum{GowMetricT}{}.
The areas of the trajectories of the isometry group define a scalar
field on the spacetime equal to
$$
  \hbox{area of $\sigma$-$\delta$ symmetry plane}
  = 4 \pi^2 e^{-\tau}
  = 4 \pi^2 t
  \, .
  \EQN GowArea
$$
(This field is independent of the~$\theta$ coordinate by construction;
see Gowdy~1974.)
The constant time hypersurfaces for both 
metrics~\EqNum{GowMetricTau}{} and~\EqNum{GowMetricT}{} slice the
spacetime harmonically (see section~\use{sec.HypIntro}; the slicing
density depends on the time coordinate only) and the slices are not in
general of constant mean curvature.
The coordinate speeds of photons moving orthogonal to the trajectories
of the isometry group are
$$
  \hbox{coordinate light speed in $\theta$-direction}
  = \left\{\eqalign{ 
    \pm e^{-\tau}  &\hbox{ for metric~\EqNum{GowMetricTau}{}}\,,\cr
    \pm 1 \>\>\> &\hbox{ for metric~\EqNum{GowMetricT}{}}\,,\cr
  }\right.
  \EQN GowSpeeds
$$
and these are the (non-trivial) characteristic speeds of the evolution
equations of the two metrics.
For the expanding form~\EqNum{GowMetricT}{} of the metric, light rays 
complete an orbit of the universe in a time period~$\Delta t = 2 \pi$.
In the collapsing case~\EqNum{GowMetricTau}{} the coordinate light
speed tends to zero as the cosmological singularity is approached, and
the total distance that can be travelled by a light ray released at
time~$\tau = \tau_0$ before it hits the singularity at~$\tau=+\infty$
is~$\Delta \theta = \exp(-\tau_0)$; in the numerical work performed
here with the collapsing model, a starting time of~$\tau_0 = 0$ is
typical and light rays never travel more than a fraction of the
distance around the universe.

Moncrief~(1981) has shown that no singularities appear in the
Gowdy~$T^3$ spacetimes for $0<t<+\infty$, and that this coordinate
range describes the maximal Cauchy development of the models.
(In particular, no coordinate singularities of the type investigated
in section~\use{sec.HomHarmonic} can be present in the standard Gowdy
slicing.) 
The singularity at~$t=0$ is `crushing' (the trace of the extrinsic
curvature of the $t=\hbox{\it constant\/}$ slices blows up uniformly 
as~$t \rightarrow 0+$; see Eardley and Smarr~1979) and is conjectured
to be a curvature singularity.
Isenberg and Moncrief~(1982) have further shown that time slices of
constant mean curvature can be used to foliate the maximally extended
Gowdy~$T^3$ spacetimes.

\SubSection{Exact and Approximate Solutions}
If the time slices of the Gowdy~$T^3$ spacetimes are assumed to be
homogeneous then the reduced Einstein equations~\EqNum{GowEinst}{}
can be solved exactly to give (from Berger and Moncrief~1993)
$$
  P = \left\{ \eqalign{
    & \ln( \alpha e^{-\beta\tau} ( 1 + \xi^2 e^{2\beta\tau} ) ) \, , \cr
    & \ln( \alpha t^\beta ( 1 + \xi^2 t^{-2\beta} ) ) \, , \cr
  }\right. \qquad
  Q = \left\{ \eqalign{
    & \eta - { \xi e^{2\beta\tau}
             \over \alpha ( 1 + \xi^2 e^{2\beta\tau} ) } \, , \cr
    & \eta - { \xi t^{-2\beta}
             \over \alpha ( 1 + \xi^2 t^{-2\beta} ) } \, ,\cr
  }\right. \qquad
  \lambda = \left\{ \eqalign{
    & \chi - \beta^2 \tau \, , \cr
    & \chi + \beta^2 \ln t \, , \cr
  }\right.
  \EQN GowHomog
$$
where~$\alpha$, $\beta$, $\eta$, $\chi$ and~$\xi$ are constants,
with~$\alpha$ strictly positive,
and~$\beta$ assumed non-negative without loss of generality.
The resulting Gowdy metric~\EqNum{GowMetricTau}{}
or~\EqNum{GowMetricT}{} then describes a Kasner spacetime (with
imposed three-torus spatial topology) which can be written in the
standard form
$$
  ds^2 = - dT^2 + T^{2p_1} dX^2 + T^{2p_2} dY^2 +  T^{2p_3} dZ^2
  \, ,
  \EQN GowKasner
$$
where~$T$ is positive and~$p_1$, $p_2$ and~$p_3$ are constants
(see, for example, section~7.2 of Wald~1984).
If the $X$-axis of the metric~\EqNum{GowKasner}{} is assumed to
correspond to the $\theta$-axis of the Gowdy metric, then the
constant~$\beta$ determines the Kasner expansion parameters:
$$
  p_1 = { ( \beta^2 - 1 ) \over ( \beta^2 + 3 ) }
  \, , \qquad
  p_2 = { 2 ( 1 + \beta ) \over ( \beta^2 + 3 ) }
  \, , \qquad
  p_3 = { 2 ( 1 - \beta ) \over ( \beta^2 + 3 ) }
  \, ,
  \EQN GowKasnerP
$$
with the values for~$p_2$ and~$p_3$ being interchangeable.
The other constants~$\alpha$, $\eta$, $\chi$ and~$\xi$ in the
solution~\EqNum{GowHomog}{} choose the scaling of the Gowdy
coordinates relative to the Kasner coordinates and the rotation of
the~$\sigma$- and~$\delta$-axes relative to the~$Y$- and~$Z$-axes.
In the case that the Gowdy variables~$P$, $Q$ and~$\lambda$ are 
constants, the parameter~$\beta$ is zero and the spacetime is
axisymmetric about the $\theta$-axis.
For $\beta = \pm 1$ the spacetime is flat (although this is not
immediately obvious from the form of the metric) and the initial
singularity at $t = 0$ is a coordinate singularity.

The homogeneous solutions~\EqNum{GowHomog}{} provide a useful
reference against which to compare the behaviour of more general
Gowdy~$T^3$ spacetimes.
In section~\use{sec.GowExpand} first-order perturbations of the
homogeneous metrics are studied and the resulting approximate
solutions are compared with numerically evolved spacetimes.
As discussed by Berger and Moncrief~(1993), linearized theory provides
an interpretation of the Gowdy metric functions~$P$ and~$Q$ as
amplitudes of orthogonal polarizations of
gravitational waves propagating in a
background spacetime described by the function~$\lambda$.
If~$P$ and~$Q$ are both close to zero then their evolution
equations~\EqNum{GowEinst}{a} and~\EqNum{GowEinst}{b} can be
approximated to first order as
$P_{,\tau\tau} = e^{-2\tau} P_{,\theta\theta}$ and
$Q_{,\tau\tau} = e^{-2\tau} Q_{,\theta\theta}$,
two decoupled wave equations.
The metric~\EqNum{GowMetricTau}{} can then be expanded as
$$
  g_{\mu\nu}
  = g_{\mu\nu}^{\SmallWord{background}}
  + e^{-\tau} P \, \varepsilon_{\mu\nu}^{+}
  + e^{-\tau} Q \, \varepsilon_{\mu\nu}^{\times}
  + \hbox{(second- and higher-order terms)}
  \, ,
$$
where the background part of the metric depends only on the
function~$\lambda$, and~$\varepsilon^{+}$
and~$\varepsilon^{\times}$ are the two polarization tensors for
gravitational waves propagating along the $\theta$-axis
(see Misner, Thorne and Wheeler~1973, section~35.6).

The Gowdy~$T^3$ metrics are said to be polarized if the
function~$Q$ is identically zero.
(If~$Q$ is zero on an initial time slice then the evolution
equation~\EqNum{GowEinst}{b} ensures that it remains zero
throughout the spacetime; the corresponding statement does not
hold in general for the polarization of radiation represented by~$P$.)
The polarized Gowdy~$T^3$ spacetimes have been extensively
studied (see Berger~1974 and~1984, and Misner~1973) and are of 
particular interest because their evolution equations can be solved
exactly.
If~$Q$ is zero then equation~\EqNum{GowEinst}{a} is linear in~$P$
and has the general solution
$$
  P = a_0 \tau + b_0
    + \sum_{n=1}^{\infty} \Big( Z_{+n}\!(\tau) \cos(n\theta)
                          + Z_{-n}\!(\tau) \sin(n\theta) \Big)
  \, ,
  \EQN GowPolarP
$$ $$
  \hbox{with \ \ }
  Z_{\pm n}\!(\tau) = a_{\pm n} J_0( \vert n \vert e^{-\tau} )
                    + b_{\pm n} Y_0( \vert n \vert e^{-\tau} )
  \hbox{ \ \ for } n \neq 0
  \, ,
$$
where~$J_i(x)$ and~$Y_i(x)$ are Bessel functions (reviewed in
section~\use{sec.HomHarmonic}) and the~$a_{\pm n}$ and~$b_{\pm n}$ are
constants. 
The general solution for~$\lambda$ (obtained from 
equations~\EqNum{GowEinst}{c} and~\EqNum{GowEinst}{d}) is given in
Berger~1974, along with the restrictions imposed on the
constants~$a_{\pm n}$ and~$b_{\pm n}$ by the momentum
constraint~\EqNum{GowMoment}{}. 
For the current work a particular polarized solution has been chosen:
$$
\eqalign{
  P = {}&
    J_0(e^{-\tau}) \cos(\theta)
    + Y_0(2e^{-\tau}) \sin(2\theta)
  \, , \cr 
  \lambda = {}&
  - e^{-\tau} \Big(
      J_1(e^{-\tau}) J_0(e^{-\tau}) \cos^2(\theta)
    + J_1(e^{-\tau}) Y_0(2e^{-\tau}) 
        [ {\textstyle {2 \over 3}} \sin(3\theta) + 2 \sin(\theta) ]
  \cr & \qquad\quad {}
    + J_0(e^{-\tau}) Y_1(2e^{-\tau}) 
        [ {\textstyle {2 \over 3}} \sin(3\theta) - 2 \sin(\theta) ]
    - Y_0(2e^{-\tau}) Y_1(2e^{-\tau}) [ \cos(4\theta) - 1 ]
    \Big)   
  \cr &{}
  + e^{-2\tau} \Big(
      {\textstyle {1 \over 2}} [ J_0(e^{-\tau})^2 + J_1(e^{-\tau})^2 ]
    + 2 [ Y_0(2e^{-\tau})^2 + Y_1(2e^{-\tau})^2 ]
    \Big)
  \, . \cr
} \EQN GowPolar  
$$
Berger and Moncrief~(1993) note that a class of coordinate
transformations which leave the form of the
metric~\EqNum{GowMetricTau}{} unchanged can be used to generate new
exact solutions from known ones; in particular an exact polarized
solution can be transformed into a `pseudo-unpolarized' solution for
which the function~$Q$ is not identically zero (although the metrics
are physically equivalent).
A simple example of such a transformation converts a polarized
solution, $P_{\SmallWord{pol}}$ and 
$\lambda_{\SmallWord{pol}}$, into a pseudo-unpolarized solution,
$$
  P_{\SmallWord{pseudo}} 
  = \ln( \cosh P_{\SmallWord{pol}} )
  \, , \qquad
  Q_{\SmallWord{pseudo}} 
  = \tanh P_{\SmallWord{pol}}
  \, , \qquad
  \lambda_{\SmallWord{pseudo}} 
  = \lambda_{\SmallWord{pol}}
  \, .
  \EQN GowPseudo
$$
Polarized and pseudo-unpolarized solutions play an important part in
the work described in this chapter: they provide non-trivial exact 
solutions against which numerical codes developed to evolve genuinely
unpolarized Gowdy~$T^3$ metrics can be tested.
(Recent work in Griffiths and Alekseev~1998 shows that polarized
solutions can also be transformed to produce a class of genuinely
unpolarized solutions, and in principle these too could be used as
code tests.)

Another class of Gowdy solutions is discussed by Grubi{\u s}i\'c and 
Moncrief~(1993) who use different variables to describe the
metric~\EqNum{GowMetricTau}{}.
Their `circular loop' solutions have simple spatial profiles which
evolve according to ordinary differential equations.

\Section{Collapsing Planar Cosmologies
         \label{sec.GowCollapse}}
In this section the results of numerical simulations of collapsing
unpolarized Gowdy~$T^3$ spacetimes are presented.
The starting point for this work is the~1993 paper of Berger and
Moncrief. 
They used a symplectic integration scheme (described in their paper)
to evolve the Gowdy equations~\EqNum{GowEinst}{a,b}, and
found that the metric variables developed structure on very fine
scales which their code could not reliably resolve.
The present work uses the adaptive mesh refinement code 
described in chapter~3 to repeat and extend Berger and Moncrief's
calculations. 

Work following on from Berger and Moncrief~1993 has been published by
Berger~(1997) and, recently, Berger and Garfinkle~(1998).
Berger et al.~1998 summarizes the findings of a long-term investigation
into the behaviour of collapsing cosmological models which includes
results from the Gowdy~$T^3$ studies. 

The metric variables in the collapsing Gowdy~$T^3$ simulations show
complicated behaviour which, at a first glance, appears to be evidence
of numerical instabilities in the code.
However careful observation of
the variables at high resolutions (made possible through the use of 
adaptive mesh refinement) reveals the behaviour to be a genuine
nonlinear effect.
Numerical results are presented in this section in sufficient detail
to allow the unpolarized Gowdy~$T^3$ spacetimes to be used as a
non-trivial test bed calculation in the development of other codes for
numerical relativity.
The following discussion concentrates on each metric variable in turn,
and attempts to distinguish covariant behaviour from coordinate
effects by examining the curvature invariants of the simulated
spacetimes. 
In section~\use{sec.GowSing} the results of these simulations are
re-examined to see what evidence they provide as to the nature of
the initial singularity.
In section~\use{sec.MonCollapse} a similar investigation of the
behaviour of collapsing $U(1)$-symmetric cosmologies is undertaken.

\SubSection{Numerical Simulations of Collapsing Gowdy~$T^3$ Spacetimes}
The variables~$P(\tau,\theta)$, $Q(\tau,\theta)$ and
$\lambda(\tau,\theta)$ of the collapsing Gowdy
metric~\EqNum{GowMetricTau}{} are evolved here using the numerical
methods of chapters~2 and~3.
New variables are defined as
$$
  A \equiv P_{,\tau}
  \, , \quad
  B \equiv Q_{,\tau}
  \, , \quad
  C \equiv P_{,\theta}
  \, , \quad
  D \equiv Q_{,\theta}
  \, , \EQN GowNewVars
$$
and the evolution equations~\EqNum{GowEinst}{a,b,c} become a
one-dimensional first-order system suitable for
numerical integration (recall equation~\EqNum{PdeConsLaws}{}):
$$\EQNalign{
  A_{,\tau} - e^{-2\tau}C_{,\theta} &{} = e^{2P}(B^2 - e^{-2\tau}D^2)
  \, , \EQN GowEvolve;a \cr
  B_{,\tau} - e^{-2\tau}D_{,\theta} &{} = - 2(AB - e^{-2\tau}CD)
  \, , \EQN GowEvolve;b \cr
  C_{,\tau} - A_{,\theta} &{} = 0
  \, , \EQN GowEvolve;c \cr
  D_{,\tau} - B_{,\theta} &{} = 0
  \, , \EQN GowEvolve;d \cr
  P_{,\tau} &{} = A
  \, , \EQN GowEvolve;e \cr
  Q_{,\tau} &{} = B
  \, , \EQN GowEvolve;f \cr
  \lambda_{,\tau} &{} = - A^2 - e^{-2\tau}C^2
                  - e^{2P}( B^2 + e^{-2\tau}D^2 )
  \, . \EQN GowEvolve;g \cr
}$$
For the present work the variable~$\lambda$ is evolved according
to equation~\EqNum{GowEvolve}{g} and consequently
$$
  \lambda_{,\theta} = -2(AC + e^{2P}BD)
  \EQN GowConstrain
$$
from equation~\EqNum{GowEinst}{d} acts as a constraint on the
evolution.
An alternative approach would be to reconstruct~$\lambda$ at each
time level using equation~\EqNum{GowConstrain}{} so that the
numerical evolution is constrained.
This idea is returned to later in the section.

The initial data set for the simulations is 
$$\eqalign{
  &P = 0 \, , \cr
  &P_{,\tau} = v_0 \cos\theta \, , \cr
} \qquad \eqalign{
  &Q = \cos\theta \, , \cr
  &Q_{,\tau} = 0 \, , \cr
} \qquad \eqalign{
  &\lambda = 0 \, , \cr
  &\hbox{at time }\tau = 0 \, , \cr
} \EQN GowInitial
$$
which (with variables~$A$, $B$, $C$ and~$D$ defined according to
equations~\EqNum{GowNewVars}{}) clearly satisfies the
constraint~\EqNum{GowConstrain}{}.
This choice of initial data was used by Berger and
Moncrief~(1993) and can be interpreted as two (initially) standing
gravitational waves in a homogeneous background.
The magnitude of the parameter~$v_0$ affects the amount of fine-scale 
structure that develops in the spacetime.
Here, as in Berger and Moncrief~1993, the choice of $v_0=10$ is made,
and the behaviour of the evolved spacetime is complicated.
The results presented in Berger~1997 take~$v_0$ to be~$5$ and include
significantly less fine-scale structure.
The choice~\EqNum{GowInitial}{} of initial data has been found to be
reasonably generic: other initial data sets produce qualitatively
similar behaviour, but with varying amounts of fine-scale structure.

The numerical code used to evolve the system~\EqNum{GowEvolve}{} has
been thoroughly tested.
Section~\use{sec.GowGowdy} presents three types of exact
Gowdy~$T^3$ solution---homogeneous, equation~\EqNum{GowHomog}{},
polarized, equation~\EqNum{GowPolar}{}, and pseudo-unpolarized,
equation~\EqNum{GowPseudo}{}---which the code successfully
reproduces.
As with all of the numerical results presented here, the evolved
quantities have been checked to ensure that they converge
at second order as the resolution increases (see
section~\use{sec.PdeDifference}); in particular, the error in the
constraint equation~\EqNum{GowConstrain}{} converges to zero at the
expected rate. 
During the course of this work, codes for evolving a variety of
formulations of Einstein's equations have been developed, and wherever
possible the different codes have been used to evolve data for Gowdy
spacetimes. 
The consistency between the results of these codes is further evidence
that the system of equations~\EqNum{GowEvolve}{} is being evolved
correctly. 

Two different numerical integration schemes have been used in this
work: a high-resolution wave-propagation method and the standard two-step
Lax-Wendroff finite difference method, both of which are discussed 
in chapter~2.
(In fact, two different implementations of the Lax-Wendroff method
were used; see sections~\use{sec.PdeDifference}
and~\use{sec.AmrStructure}.) 
The schemes have been found to perform comparably well in evolving
numerical solutions to the Gowdy equations.
The wave-propagation algorithm uses the solution to the Riemann
problem for the transport part (left-hand side) of the
system~\EqNum{GowEvolve}{}, which is simply a combination of linear
wave equations, discussed in chapter~2 (equation~\EqNum{PdeWaveEq}{}).
Using operator splitting (section~\use{sec.PdeDifference}) the source
part (right-hand side) of the system~\EqNum{GowEvolve}{} can be
treated using standard 
numerical techniques for solving ordinary differential equations, and
in the present case second-order Runge-Kutta steps are used (see
equation~\EqNum{PdeRungeKutta}{}).
Periodic boundary conditions are used for the Gowdy simulations.

The Gowdy equations have been evolved using the adaptive mesh
refinement code described in chapter~3.
A base grid with a mesh spacing of $\Delta \theta = 2 \pi / 2000$
is refined by the code up to a maximum resolution of
$\Delta \theta = 2 \pi / 512\,000$ (the result of four stages of
refinement each by a factor of four).
Such small mesh spacings are needed to resolve the fine-scale features
that appear during the simulations, but lower resolution simulations
can still adequately determine the overall profile of the solution:
a test of the adaptive mesh refinement code using a base grid of
mesh size $\Delta \theta = 2 \pi / 8000$ produces similar
results to those plotted in figures~\use{Fg.GowVarP}, \use{Fg.GowVarQ}, 
and~\use{Fg.GowVarL} even when the mesh refinement is switched off.
Figure~\use{Fg.AmrDepths} shows an example grid hierarchy used
by the code during a Gowdy simulation.

The time steps taken by the code are allowed to vary in size according
to the characteristic speeds of the equations~\EqNum{GowEvolve}{}
(see the discussion in section~\use{sec.AmrDetails}).
From equation~\EqNum{GowSpeeds}{} it is clear that the time steps can
increase in size exponentially without breaking the
Courant-Friedrichs-Lewy condition, but in practice an upper limit (of
about five times the mesh spacing) is imposed.

In section~\use{sec.AmrDetails} several different interpolation
methods used for mesh refinement are discussed.
To obtain the results presented here, linear and quadratic piecewise
interpolation methods were used.
Because of the spiky features that develop during the
simulations at scales too small to fully resolve, smooth spline
interpolation methods cannot be used effectively here:
they produce spurious effects in the vicinity of the spikes (recall
figure~\use{Fg.AmrInterpDiscon}). 

The use of adaptive mesh refinement in the simulations
serves two main purposes: it increases the overall accuracy of the
results on the coarse base grid without substantially increasing the
times taken to produce them, and it allows fine-scale features of the
solution to be resolved without requiring a high density of grid
points everywhere in the spatial domain.
The code uses estimates of local truncation errors to decide where
refined regions should be positioned.
While this approach works
well in increasing the overall accuracy, it is found to be inefficient
in tracking regions of fine-scale structure.
A simple heuristic algorithm for identifying spiky features in the
solution has been used to improve the code's performance, 
but the most effective way to make the code refine a particular region
of emerging fine-scale structure is to program it to do so explicitly.

By using (parallelized) adaptive mesh refinement, together with
variable time steps, the Gowdy simulations 
take minutes rather than hours to run.
It is interesting to note that a significant portion (around~25\%) of
the code's total running time is spent doing nothing other than
calculating values of exponentials.

\SubSection{Behaviour of the First Gowdy Wave Variable}
In this and the following two subsections the behaviour of the
variables~$P$, $Q$ and~$\lambda$ during the evolution of
equations~\EqNum{GowEvolve}{} from initial data set~\EqNum{GowInitial}{}
is discussed.
Figures~\use{Fg.GowVarP}, \use{Fg.GowVarQ} and~\use{Fg.GowVarL} show
the variables at various times, with data from three levels of
refinement being plotted.
(The first two of these figures can be compared with figures~4 and~5
of Berger and Moncrief~1993.)

\figure{GowVarP}
\DrawFig{
  \epsfxsize=\hsize
  \epsfbox{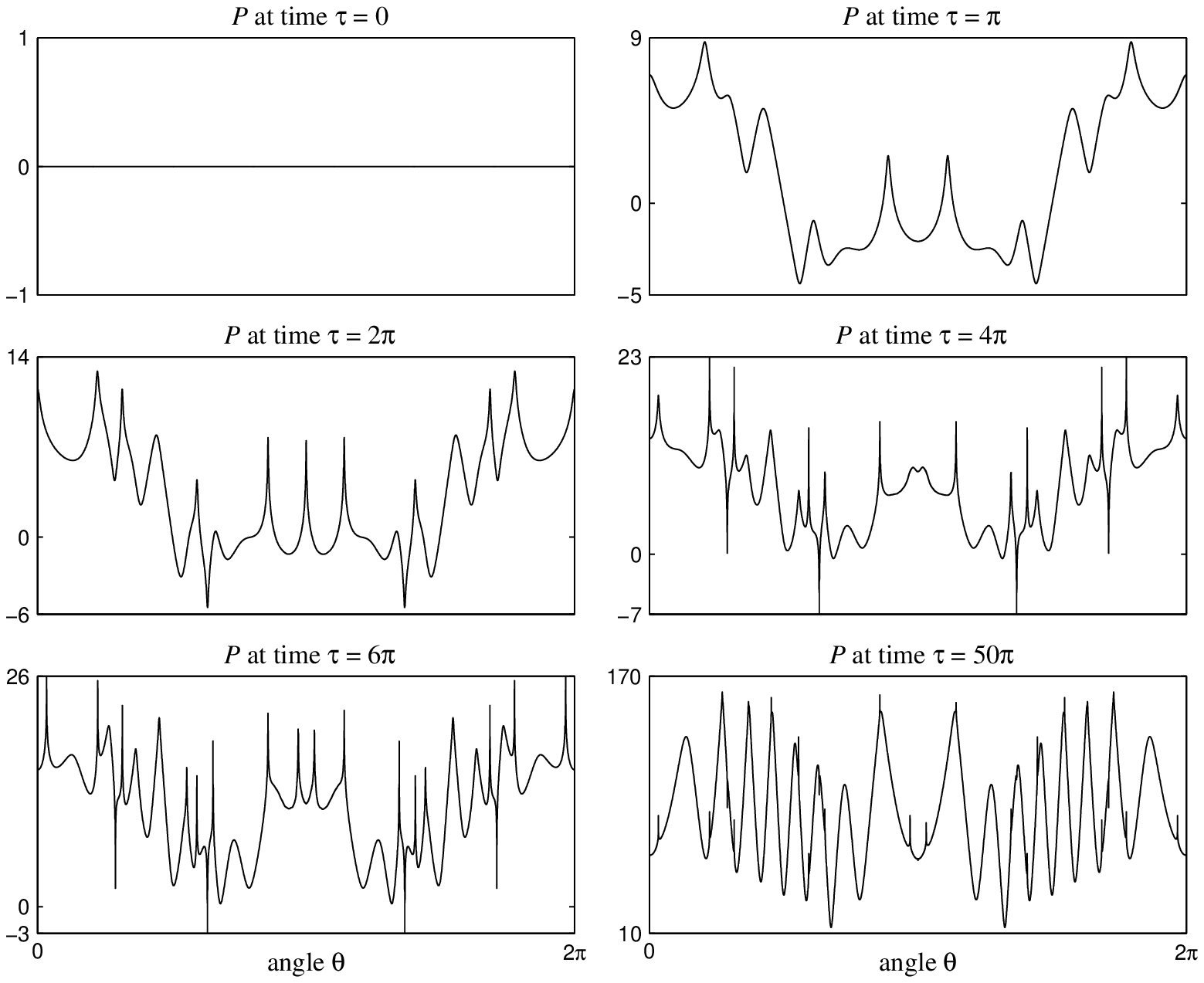}
}
\Caption
Wave variable~$P(\tau,\theta)$ of metric~\EqNum{GowMetricTau}{}
at six different times.
Initial data set~\EqNum{GowInitial}{} is evolved using
equations~\EqNum{GowEvolve}{}. 
The vertical scales chosen for the plots truncate some of the larger
spikes. 
\bigskip
\endCaption
\endfigure

\figure{GowVarQ}
\DrawFig{
  \epsfxsize=\hsize
  \epsfbox{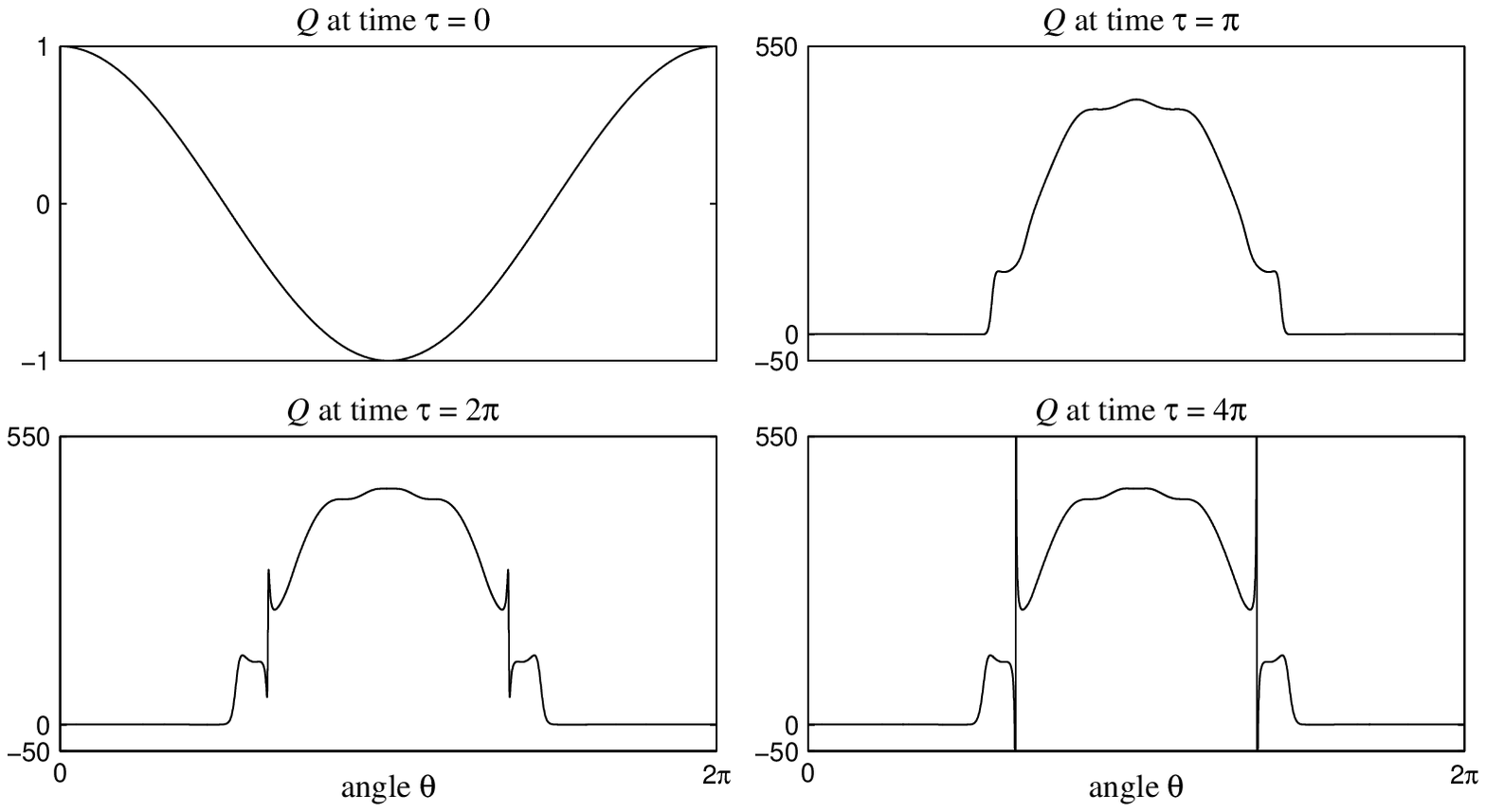}
}
\Caption
Wave variable~$Q(\tau,\theta)$ of metric~\EqNum{GowMetricTau}{}
at four different times.
Initial data set~\EqNum{GowInitial}{} is evolved using
equations~\EqNum{GowEvolve}{}. 
The vertical scale of the final plot truncates the large spikes.
The profile of the variable does not change noticeably after the time
of the final plot (except for the heights of the spikes).
\bigskip
\endCaption
\endfigure

\figure{GowVarL}
\DrawFig{
  \epsfxsize=\hsize
  \epsfbox{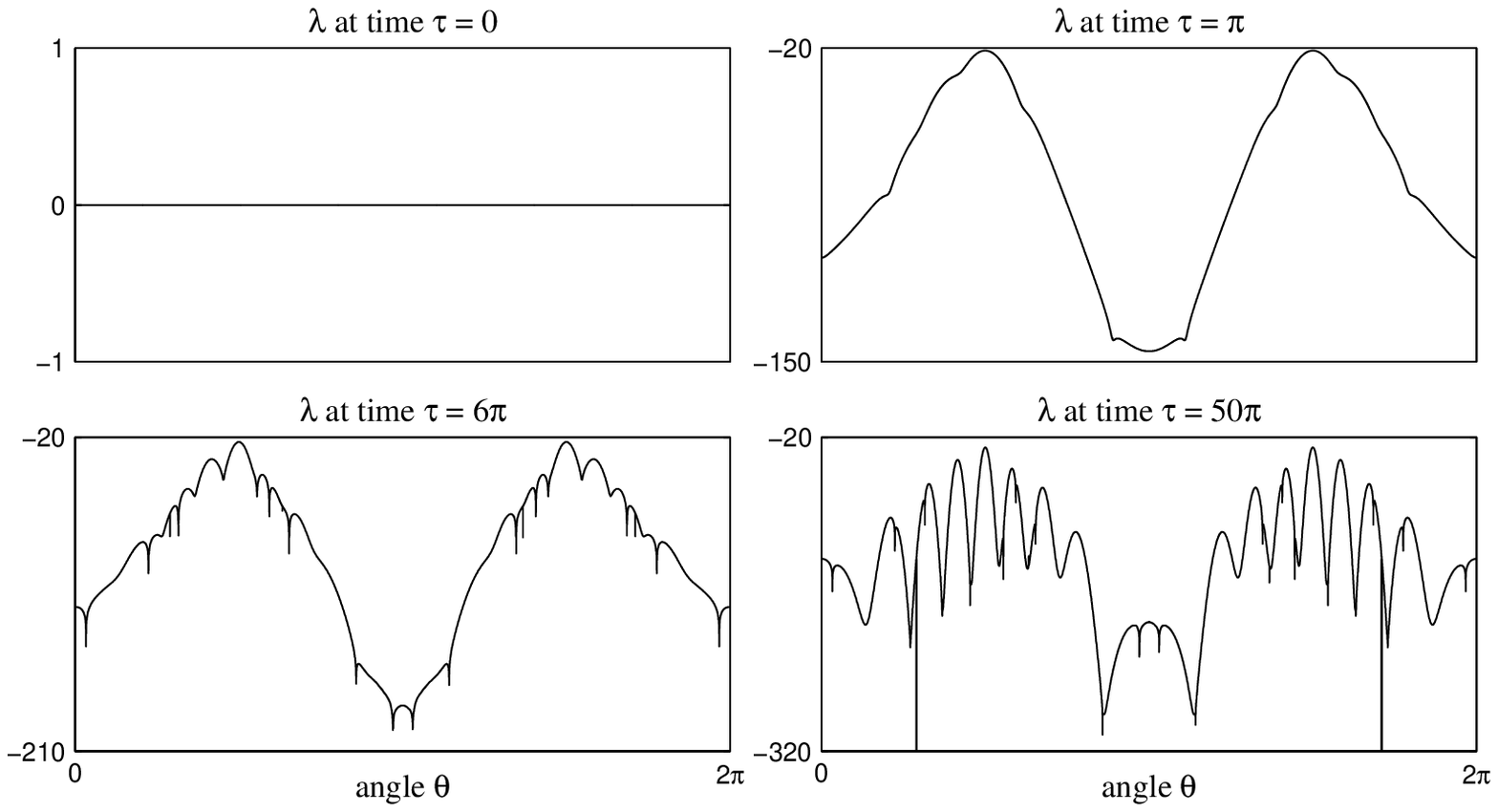}
}
\Caption
Background variable~$\lambda(\tau,\theta)$ of 
metric~\EqNum{GowMetricTau}{} at four different times.
Initial data set~\EqNum{GowInitial}{} is evolved using
equations~\EqNum{GowEvolve}{}. 
The vertical scale of the final plot truncates some of the 
larger spikes.
\bigskip
\endCaption
\endfigure

The structure of the variable~$P$ becomes complicated soon after
the start of the evolution
and by time $\tau = 4 \pi$ (as figure~\use{Fg.GowVarP} shows)
it includes a number of fine spikes.
At first sight these spikes might seem to be errors in the
evolution, and indeed when Berger and Moncrief first found spikes
in their data they considered them unreliable and used spatial
averaging to remove them from figures~4 and~5 of their paper~(1993).
However, the adaptive mesh refinement code of chapter~3 automatically
refines the regions where spikes start to form, increasing the grid
resolution so that the spikes appear smooth and the evolution
can continue without loss of accuracy. 
This is demonstrated in figure~\use{Fg.GowSpike} where an emerging
spike in~$P$ is shown for the coarse base grid together with two
levels of refined sub-grids.

\figure{GowSpike}
\DrawFig{
  \epsfxsize=12cm
  \epsfbox{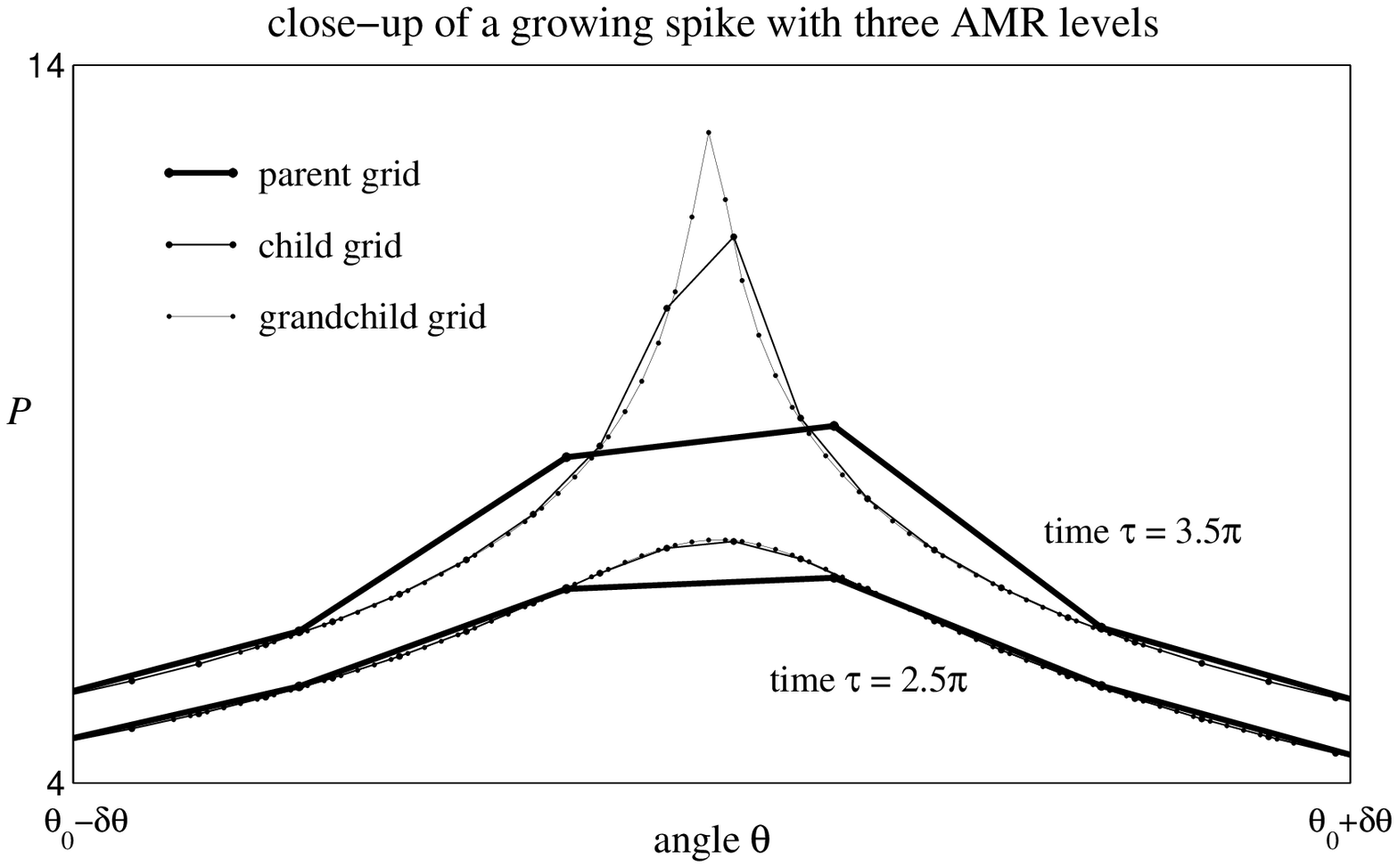}
}
\Caption
Data from three generations of grids at an emerging spike in~$P$.
Each sub-grid has a resolution four times greater than that of its
parent grid.
The spike is located at coordinate $\theta_0 = 1.8631$, and the plot
shows a region of size $\delta\theta = 7.5 \times 10^{-3}$ to either
side.
Two time levels are shown, with the spike well resolved by the two
finer grids at the earlier time, but only adequately resolved by the
finest grid at the later time.
\bigskip
\endCaption
\endfigure

Spikes form in~$P$ in both the positive and negative directions.
Examining the development of the positive spikes (the
negative spikes are discussed in the following subsection) it is
found that their peaks
grow linearly at rates $P_{,\tau} = \hbox{{\it constant}} > 1$,
while at the same time they decrease in width.
As a spike narrows (and there is no indication that the spikes ever
stop growing or narrowing) the mesh refinement code is forced to
use finer and finer grids to keep it well resolved.
Eventually, it becomes narrower than the smallest grid spacing
that the code has been instructed to use,
and from that point on it cannot be assumed that the code is 
accurately evolving the spike.
It is therefore inevitable that, as the evolution progresses, the
calculated heights of the spikes in~$P$ must be significantly
in error, but this does not imply
that there must be large errors elsewhere in the data: since the
characteristic speed of the model is $\exp(-\tau)$ (from
equation~\EqNum{GowSpeeds}{}),
deviations from the exact solution that develop at some time
not too soon after the start of the
evolution can only propagate very small
distances from their points of origin. 

At late times $P_{,\tau\tau} \equiv A_{,\tau}$ is seen to become
very small,
and so~$A$ freezes and~$P$ grows linearly in time with the spatial 
profile shown in the final plot of figure~\use{Fg.GowVarP}.
Except at the points where spikes formed, the spatial profile of~$P$
at late times has quite a simple form, much less intricate
than early on in its evolution.
For the reasons discussed above, the numerical simulation is
unable to accurately determine the late-time behaviour of the spikes.
While small spikes and lumps appear in the plot of~$P$ at time
$\tau = 50 \pi$ in figure~\use{Fg.GowVarP}, these are just the
residue left by
spikes that have narrowed beyond the resolution of the simulation.
The spikes themselves, assuming they continued to grow at their
initial rates, would all go off the scale of the graph.

\fullfigure{GowDominAt}
\vss
\DrawFig{
  \epsfxsize=0.9\hsize
  \epsfbox{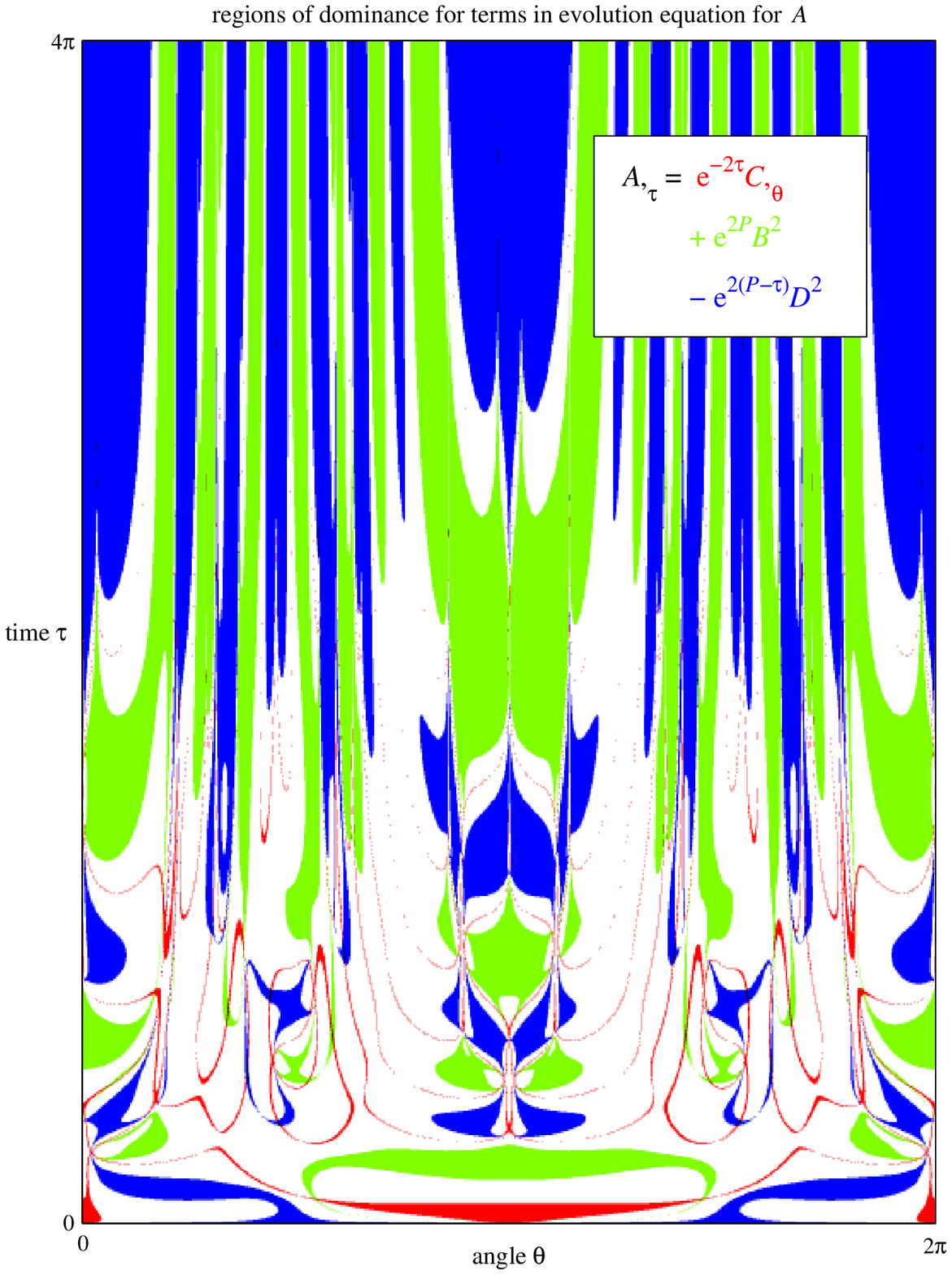}
}
\Caption
Dominant terms in the evolution equation~\EqNum{GowEvolve}{a} 
for~$A$.
Regions where one of the three terms in the equation
dominates the other terms (in the sense described in the text)
are coloured red, green or blue as shown in the key.
Uncoloured (white) regions are where no terms dominate.
\endCaption
\endfigure

\fullfigure{GowDominBt}
\vss
\DrawFig{
  \epsfxsize=0.9\hsize
  \epsfbox{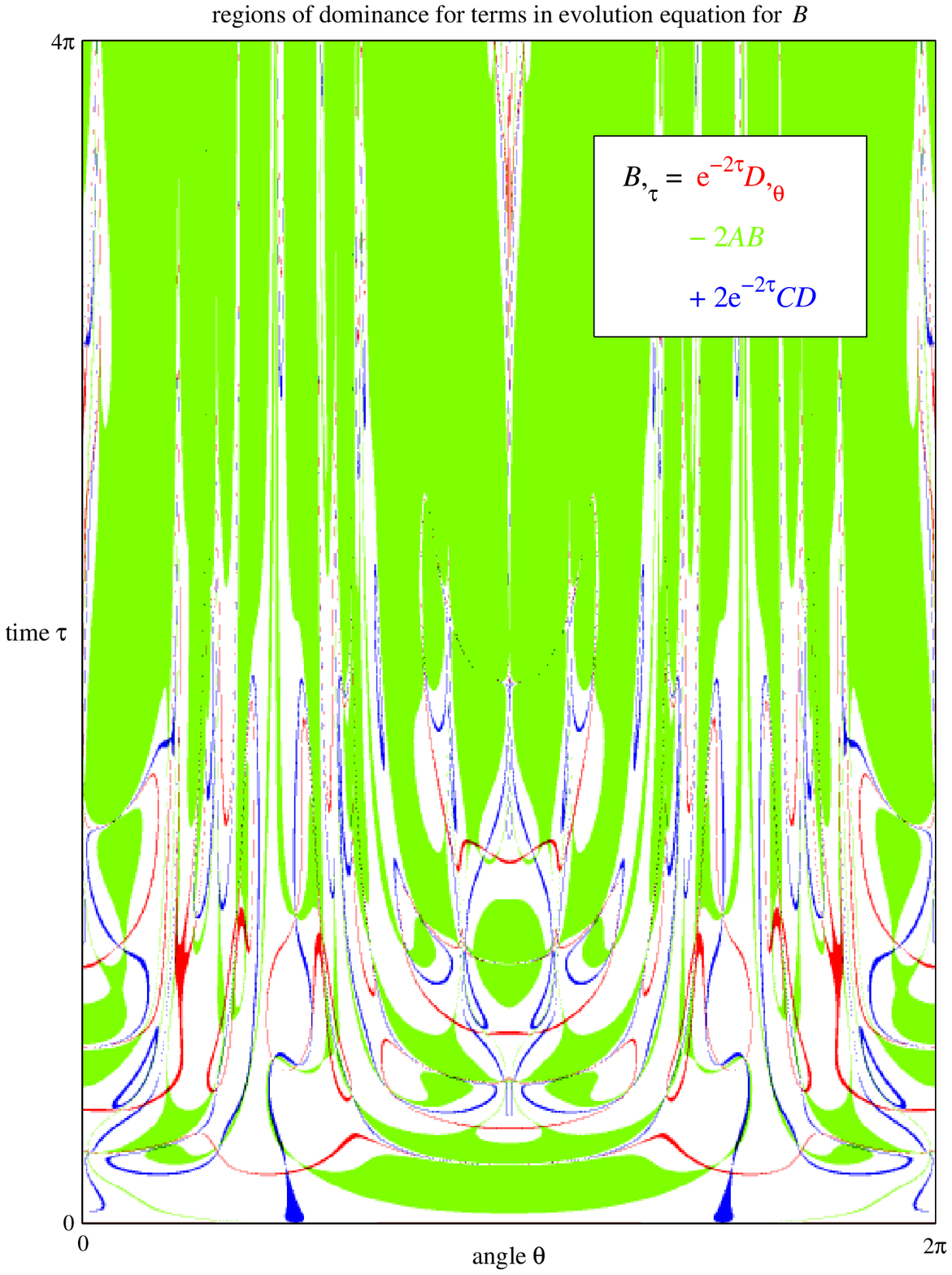}
}
\Caption
Dominant terms in the evolution equation~\EqNum{GowEvolve}{b} 
for~$B$.
Regions where one of the three terms in the equation
dominates the other terms (in the sense described in the text)
are coloured red, green or blue as shown in the key.
Uncoloured (white) regions are where no terms dominate.
\endCaption
\endfigure

In suggesting reasons for the observed behaviour it is
useful to know which are the dominant terms in the governing
equations.
Here a term on the right-hand side of an equation is defined as being
{\it dominant\/} if its absolute value exceeds ten times the absolute
value of the sum of the other terms on the right-hand side of the equation.
A dominant term effectively determines the behaviour of the equation
it is in.
Figures~\use{Fg.GowDominAt} and~\use{Fg.GowDominBt} illustrate
this for the wave equations~\EqNum{GowEvolve}{a} for
$A_{,\tau} \equiv P_{,\tau\tau}$ and~\EqNum{GowEvolve}{b} for
$B_{,\tau} \equiv Q_{,\tau\tau}$. 
The red-coloured regions in $(\tau, \theta)$-space are where
terms involving~$C_{,\theta}$ or~$D_{,\theta}$ dominate the
right-hand sides.
The green regions are where terms involving~$A$ or~$B$ dominate, and
the blue regions are where terms involving~$C$ or~$D$ dominate.
In the white regions no terms dominate.
(A slight peculiarity of this definition of dominance is that, since
terms can cancel, it is possible for more than one term to dominate an
equation at the same time.
In figures~\use{Fg.GowDominAt} and~\use{Fg.GowDominBt} colours merge
to black in regions where this happens; it is obvious from the
figures that regions of multiple dominance do not play a significant
part in the present work.)

It is clear from figures~\use{Fg.GowDominAt}
and~\use{Fg.GowDominBt} that there is very complicated interplay
between terms during the early stages of the evolution, and it seems
unlikely that the overall behaviour can be traced to simple causes.
Berger~(1997) notes that spikes form at points where
$D \equiv Q_{,\theta} = 0$ (though not all such zeros produce spikes)
and discusses their formation in terms of bounces off of potentials
in the evolution equations.
In terms of dominance behaviour, spike formation occurs in regions
which are blue in figure~\use{Fg.GowDominAt} and green in
figure~\use{Fg.GowDominBt}.
Berger and Garfinkle~(1998) show that approximate versions of the
evolution equations (which in effect ignore all but the dominant terms
just mentioned) can produce spiky behaviour very much like that
observed in the full Gowdy equations.

\SubSection{Behaviour of the Second Gowdy Wave Variable}
Sharp features also form in the variable~$Q$ as it evolves.
Figure~\use{Fg.GowVarQ} shows that between times $\tau = 2\pi$
and $\tau = 4\pi$ there develops in~$Q$ a pair of features both 
comprising a negative spike and a positive spike very close together 
with a steep gradient between their two peaks.
A second, similar pair of double-spike features is also present, but
not visible in figure~\use{Fg.GowVarQ} because of the scale.
The positions of the four double spikes in~$Q$ coincide with the
positions of the four negative spikes in the variable~$P$
at time $\tau = 4\pi$ (see figure~\use{Fg.GowVarP}).
A close examination of the data reveals that both the positive
and negative peaks of each double spike in~$Q$ grow at
exponential rates, while the overall width of the feature
becomes smaller,
and that the negative spikes in~$P$ grow linearly
in much the same way that their positive counterparts do.
The discussion in the previous subsection regarding the numerical
problems caused by the narrowing of spikes applies equally here.

An explanation of how double-spike features form follows from a close
examination of the dominance behaviour of the evolution equations
for~$A$ and~$B$.
From figures~\use{Fg.GowDominAt} and~\use{Fg.GowDominBt},
together with dominance data produced using finer sub-grids,
it can be seen that the region around a double spike
is green for both evolution equations, and consequently
in such a region the terms proportional
to $\exp(-2\tau)$ are irrelevant and
equations~\EqNum{GowEvolve}{a} and~\EqNum{GowEvolve}{b} may be
approximated as
$$
  A_{,\tau} \simeq e^{2P}B^2, \qquad B_{,\tau} \simeq -2AB.
$$
A double spike then forms under the conditions of~$B$ crossing
zero at the same time that~$A$ is negative
and~$P$ is negative or small and positive.
It follows that $A_{,\tau} \geq 0$ and is small, so that~$A$
remains negative and~$P$ forms a negative spike.
On both sides of the zero~$B$ then grows
exponentially, but in opposite directions, and the result
is a double spike in~$Q$.
Since $A_{,\tau} \geq 0$ the region in which $A < 0$ becomes
smaller, and the double spike narrows.

Away from the double spikes, $Q_{,\tau} \equiv B$ quickly approaches
zero and the variable~$Q$ freezes not long after the start of the
evolution:
except at the double spikes, $Q$~at late times does not differ from
the final plot of figure~\use{Fg.GowVarQ}.
Because of the narrowing effect the second pair of double spikes
never becomes a dominant feature in the numerically evolved
profile of~$Q$, although, extrapolating the spikes' initial growth,
they quickly become larger than the vertical scale of the
figure~\use{Fg.GowVarQ}.

Considering the steepness of the gradients that develop in the
variables~$P$ and~$Q$ it might be expected that the high-resolution
wave-propagation scheme for evolving the system~\EqNum{GowEvolve}{}
would perform better than the standard two-step
Lax-Wendroff method (see chapter~2 and the discussion at the start
of this section).
However it turns out that the two integration methods produce
nearly identical results.
In fact, if wave limiters are used
with the high-resolution scheme (see section~\use{sec.PdeVolume}) its
performance suffers;
the loss of second-order accuracy in isolated regions is not
compensated for by any improvement in shock resolution.
The Gowdy spikes are a different kind of feature to the
fluid shock waves that the high-resolution method is designed to
track---they are produced by the source rather than the transport part
of the evolution equations, as can be deduced from
figures~\use{Fg.GowDominAt} and~\use{Fg.GowDominBt}.
It is the adaptive mesh refinement code that proves to
be the more useful computational tool for this work.
The use of mesh refinement here also helps to counteract a
shortcoming of the Lax-Wendroff time-stepping scheme: whereas it
might be expected that the fine-scale structure in the evolved
Gowdy data would be smeared out by the artificial viscosity of the
Lax-Wendroff method, the magnitude of the viscosity decreases as the
spatial resolution is increased, and regions of fine-scale structure
are exactly where the code positions the highest resolution meshes.

\SubSection{Behaviour of the Gowdy Background Variable
            and Conservation of the Constraint}
The evolution of the variable~$\lambda$, figure~\use{Fg.GowVarL}, 
shows similar features to the evolution of~$P$, figure~\use{Fg.GowVarP}.
Negative spikes develop in~$\lambda$ at the same points at which
positive spikes develop in~$P$ and, apart from pointing
in a different direction, they show the same behaviour:
they grow at a constant rate and they narrow until they no
longer can be resolved by the simulation.
In contrast, no unusual behaviour seems to develop in~$\lambda$
at the points where double spikes form in~$Q$ (although the
errors in the data at the double spikes are sufficient to
produce erratic behaviour in~$\lambda$ there
at late times, as can be seen in figure~\use{Fg.GowVarL}).

Equation~\EqNum{GowEvolve}{g} shows that~$\lambda$ is a
decreasing function of time, and it is clear from 
figure~\use{Fg.GowVarL}
that~$\lambda$ becomes large and negative as it evolves.
At late times, $\lambda_{,\tau} \simeq - A^2$ 
(cf.~equation~\EqNum{GowEvolve}{g}) and~$A$ is approximately constant,
so~$\lambda$ grows linearly with a fixed spatial profile.

The constraint equation~\EqNum{GowConstrain}{} relates the 
spatial derivative of~$\lambda$ to the values of~$P$ and~$Q$ and 
their first derivatives.
For an exact solution of the evolution equations~\EqNum{GowEvolve}{}
the constraint equation is satisfied at all times if it is 
satisfied initially, but for a numerical solution this is not
necessarily the case.
The degree to which the constraint equation fails to
hold can be judged by calculating the total percentage error,
defined as
$$
  \hbox{error(\%)} = 100 
    { \Vert \lambda_{,\theta} + 2(AC + e^{2P}BD) \Vert
      \over
      \Vert \lambda_{,\theta} \Vert } 
  \, , \qquad
  \hbox{for }
  \Vert f \Vert =
    \int_{0}^{2\pi} \vert f(\theta) \vert \, d\theta
  \, ,
$$
where the integrals are evaluated using the trapezium rule
and~$\lambda_{,\theta}$ is estimated from~$\lambda$ using
second-order finite differencing.
Since the errors are known to be disproportionately large
at the spikes these points are not included when 
measuring the constraint.
(A region of width~0.02 is excised at each spike, with
about ten percent of the total domain being removed.)
At time $\tau = 6 \pi$ the error in the constraint 
is~0.27\%, rising to~0.84\% by time $\tau = 50 \pi$.

Instead of evolving~$\lambda$ with equation~\EqNum{GowEvolve}{g}
it could be evaluated at every time step by integrating
equation~\EqNum{GowConstrain}{}.
This would obviously ensure that the constraint equation is
satisfied, and such `fully constrained' approaches are often
applied to problems in numerical relativity.
However, for this particular problem enforcing the constraint
equation in this way would be a very bad idea:
the large errors at the spikes would affect the whole
of~$\lambda$ rather than being confined to isolated regions.

\SubSection{Coordinate Independence of Fine-Scale Structure}
As demonstrated in chapter~5, fine-scale spatial structure can
be found even in numerical simulations of homogeneous spacetimes
if inappropriate gauge choices are made.
The question then arises: is the fine-scale structure found in
the unpolarized Gowdy models a coordinate effect?
This subsection examines gauge-independent properties of the spacetime
and presents evidence for the physical nature of the fine-scale
structure (though this evidence is found to be stronger
for the positive spikes in~$P$ than it is for the double spikes
in~$Q$). 

The first point to note is that the spiky behaviour in the
variables~$P$ and~$Q$ is reflected in the proper lengths of the 
orbits of the Killing vectors~$\partial / \partial\sigma$
and~$\partial / \partial\delta$ which generate the 
spacetime's~$T^2$ isometry group, and so the spikes are not simply a
consequence of the choice of variables used to describe the metric.
From equation~\EqNum{GowMetricTau}{},
$$\EQNalign{
  \hbox{proper length of $\sigma$-orbits} 
  & {}= 2 \pi e^{(P-\tau)/2}
  \, , 
  \EQN GowOrbit;a \cr
  \hbox{proper length of $\delta$-orbits} 
  & {}= 2 \pi e^{-\tau/2} \sqrt{ e^P Q^2 + e^{-P} }
  \, . 
  \EQN GowOrbit;b \cr
}$$
At positive spikes in~$P$ there are spikes in the lengths of both
the~$\sigma$- and the $\delta$-orbits~\EqNum{GowOrbit}{a,b}.
At negative spikes in~$P$, corresponding to double-spike features
in~$Q$, there are spikes in the lengths of the
$\delta$-orbits~\EqNum{GowOrbit}{b}.
That the spikes are not a consequence of some kind of local divergence
of the spatial coordinates can be seen by examining the 
proper distance along the $\theta$-axis:
from the metric~\EqNum{GowMetricTau}{},
$$
  \hbox{proper distance element in $\theta$-direction}
  = e^{(\lambda + \tau)/4} \, d\theta
  \, ,
$$
and since~$\lambda$ either becomes more negative (at positive spikes
in~$P$) or shows no unusual behaviour (at double spikes in~$Q$)
in the regions of fine-scale structure, the structure exists on small
scales in terms of proper $\theta$-distance as well as
coordinate $\theta$-distance.
(This is in contrast to the fine-scale behaviour discovered in
section~\use{sec.HomHarmonic}.)

As discussed in chapter~5 (equations~\EqNum{HomInvars}{} in
particular), the curvature invariants~$\InvA$, $\InvB$,
$\InvC$ and~$\InvD$ provide
coordinate-independent information about a spacetime.
For the Gowdy metric~\EqNum{GowMetricTau}{} the values of the
invariants are calculated using expressions generated by the
GRTensorII package (Musgrave, Pollney and Lake~1996) for MapleV.
Figure~\use{Fg.GowInvars} shows two of the curvature
invariants,~$\InvA$ and~$\InvB$, evaluated at time $\tau = 6 \pi$.
The following discussion concentrates on these two quantities
since~$\InvC$ and~$\InvD$ behave qualitatively the same as,
respectively,~$\InvA$ and~$\InvB$.

\figure{GowInvars}
\DrawFig{
  \epsfxsize=0.95\hsize
  \epsfbox{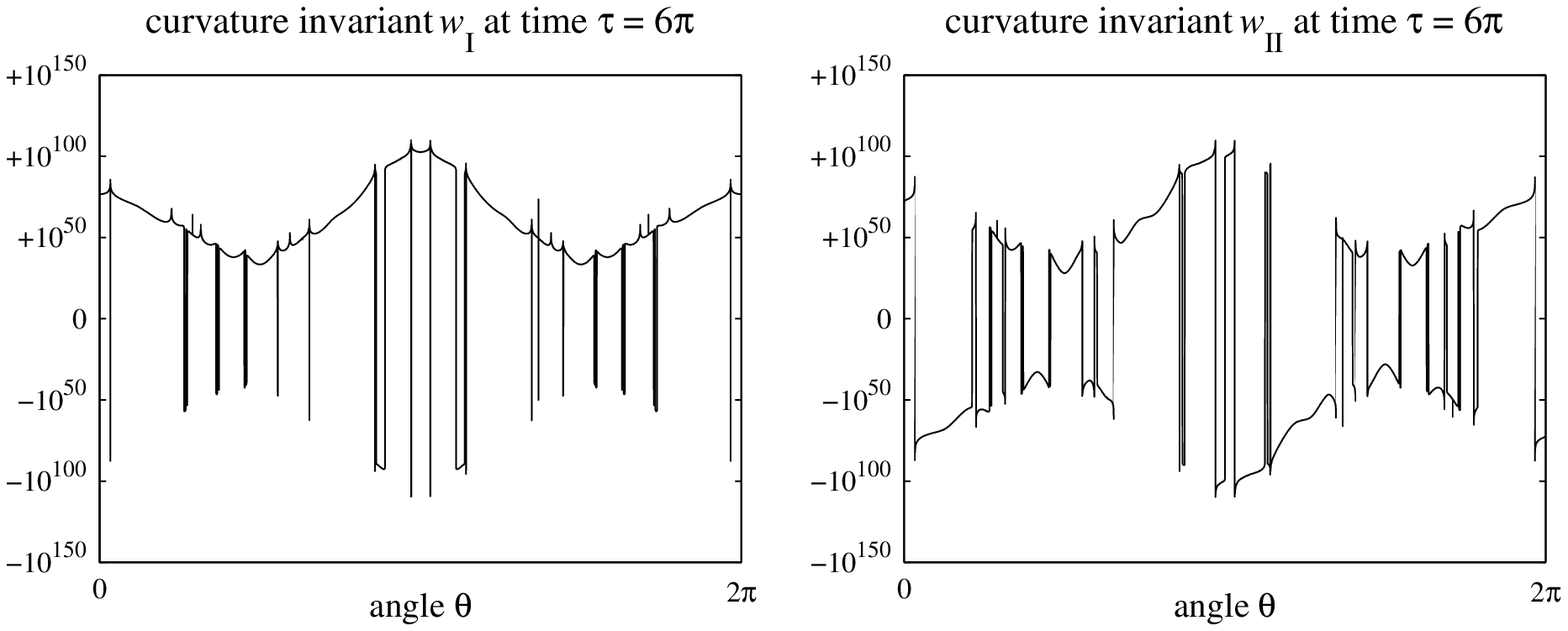}
}
\Caption
Curvature invariants~$\InvA$ and~$\InvB$ at time $\tau = 6\pi$
for initial data set~\EqNum{GowInitial}{} evolved using
equations~\EqNum{GowEvolve}{}. 
The curvature invariants are defined by equations~\EqNum{HomInvars}{}.
(A logarithmic vertical scale has been produced by use of the
transformation $w \rightarrow (1 / \ln 10) \sinh^{-1} (w/2)$.)
\bigskip
\endCaption
\endfigure

The most striking feature of the curvature invariants is
their magnitude:~$\InvA$ and~$\InvB$ at $\tau = 6 \pi$
reach both positive and negative values of size greater than
ten to the hundredth power.
At the start of the evolution~$\InvA$ takes values from $-50$
to $+30\,000$, and~$\InvB$ from $-2500$ to $+2500$.
Their subsequent growth is rapid, and shows no signs of reversing.
The main influence on the size of the curvature invariants
is the factor $\exp(-\lambda)$ in the
expressions for~$\InvA$ and~$\InvB$.
The influence of~$\lambda$ can be seen by comparing the profiles of~$\InvA$
in figure~\use{Fg.GowInvars} and~$\lambda$ at $\tau = 6\pi$ in
figure~\use{Fg.GowVarL}.
The linear growth (towards minus infinity) of~$\lambda$ 
at late times corresponds to an
exponential growth of the curvature invariants.

As figure~\use{Fg.GowInvars} shows, the curvature
invariants change sign suddenly at various values of the spatial 
coordinate~$\theta$.
These sign changes tend to form into clusters which contract
in width as the evolution progresses.
Figure~\use{Fg.GowInvCloseup} shows a sample of the behaviour
of~$\InvA$ close up.
In~$\InvA$ the clusters consist of even numbers of sign changes
and the function is predominantly positive, while in~$\InvB$ they are
odd numbered and the function changes sign between clusters.
The sign change clusters become narrow in just the same way that
the positive spikes in~$P$ do, and indeed the two types of feature 
coincide spatially.
The narrowing, together with the inaccuracies in the data at
the spikes, makes determining the precise behaviour of sign 
changes within a cluster difficult.
However, careful examination (figure~\use{Fg.GowInvCloseup})
suggests that while the width of a
cluster tends to zero, there is no cancellation between the sign
changes.

\figure{GowInvCloseup}
\DrawFig{
  \epsfxsize=0.95\hsize
  \epsfbox{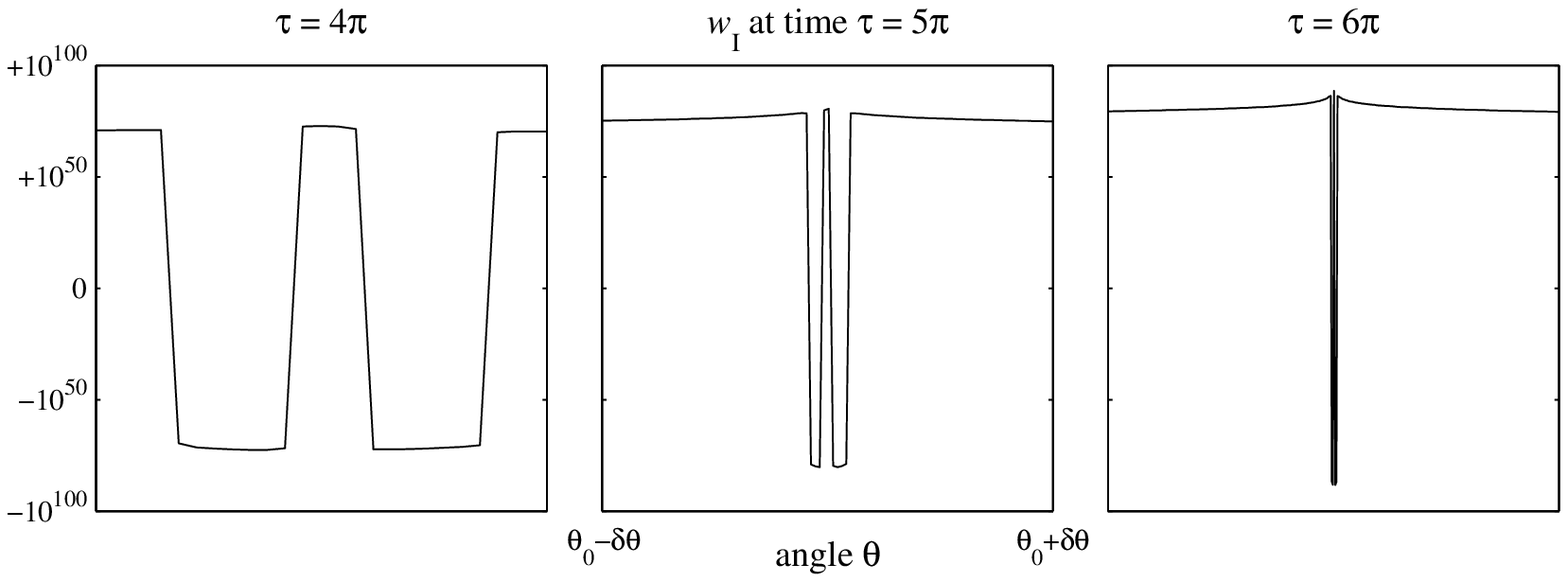}
}
\Caption
Close-ups of the curvature invariant~$\InvA$ at a positive spike
in~$P$ at three different times.
The spike is located at coordinate $\theta_0 = 0.1049$, and the plots
show regions of width $\delta\theta = 1.0 \times 10^{-2}$ to either
side of that point. 
The axes are the same for all three plots.
\bigskip
\endCaption
\endfigure

Given the large numbers of terms in the expressions for the
curvature invariants, relating their behaviour to features
of~$P$, $Q$ and~$\lambda$ is not a simple matter.
However, by comparing the magnitudes of individual terms, it is found
that at late times the curvature invariants can be approximated by
the simplified expressions
$$\EQNalign{
    \InvA
    &{} \simeq
    (1/32)
    e^{3\tau-\lambda} (A^2+3) (A^2-1)^2
    \, ,
    \EQN GowApproxInv;a \cr
    \InvB
    &{} \simeq
    - (1/16)
    e^{2\tau+P-\lambda} D A (A+3) (A-1) (A+1)^2
    \, .
    \EQN GowApproxInv;b \cr
}$$
These approximations fail when the values they predict are 
small: other terms in the curvature invariants clearly cannot be 
neglected then.
The sign changes in~$\InvA$ are all clustered around the positive
spikes in~$P$, and this makes sense in terms of the approximation
since $-1 < A < +1$ is found everywhere except at
these spikes.
(The terms neglected in equation~\EqNum{GowApproxInv}{a}
turn out to be negative at the spikes.)
The behaviour of~$\InvB$ is similar, except for the influence
of~$D$ in equation~\EqNum{GowApproxInv}{b}.
At each positive spike in~$P$ there is a zero of~$D$ together with
an odd number of sign changes in~$\InvB$.
Zeros of~$D$ at points where no spikes develop produce
isolated sign changes in~$\InvB$.
However, although a pair of zeros appears in~$D$ where a double spike
forms in~$Q$, no unusual behaviour occurs in~$\InvB$ there.
It seems that zeros in~$A$ counteract the effect of the zeros in~$D$
sufficiently so that terms neglected in
equation~\EqNum{GowApproxInv}{b} become significant and
prevent~$\InvB$ from changing sign.

(It should be noted that the sign change clusters
in~$\InvA$ and~$\InvB$
evolve at different rates, and this leads to some discrepancies
between figure~\use{Fg.GowInvars} and the behaviour described
in the paragraph above.
By time $\tau = 6\pi$ some of the clusters are already too narrow
to be resolved and inaccuracies at the double spikes are large
enough to cause the invariants to behave erratically there.)

The observed behaviour of the curvature invariants makes it clear that
the positive spikes that develop in the variable~$P$ represent
physical features of the spacetime and are not simply artifacts of the
coordinate system used.
However the curvature invariants provide no evidence suggesting any
physical significance to the double-spike features that appear in the
variable~$Q$. 
(Chapter~7 presents results from numerical simulations of
cosmological models which have less symmetry than the Gowdy spacetimes,
and evidence there suggests that the positive spikes found in
the Gowdy variable~$P$ are a generic feature of gravitational collapse.)

Extrapolating the observed behaviour of the curvature
invariants forward in time to the singularity at $\tau = +\infty$
it is clear that they become infinite---except possibly at
isolated values of the spatial coordinate~$\theta$ corresponding to
the positive spikes in~$P$.
This is in agreement with the curvature behaviour at the
singularity conjectured by Moncrief~(1981).
(The same reference shows that no curvature singularities
can develop before $\tau = +\infty$,
and the results presented here do not suggest otherwise.)

\Section{Nature of the Cosmological Singularity
         \label{sec.GowSing}}
In this section standard pictures of spacetime behaviour close to
cosmological singularities are compared to numerically evolved
solutions of the Gowdy~$T^3$ model.
Evidence suggests that the Gowdy~$T^3$ singularity is asymptotically
velocity-term dominated in the sense of Isenberg and Moncrief~(1990),
but that the evolution of the metric does not spatially decouple as
the singularity is approached.

The singularity theorems of Penrose, Hawking and others (see chapter~10
of Hawking and Ellis~1973) show that in general cosmological
spacetimes must possess singularities, but they say nothing about the
nature of those singularities.
Belinskii, Khalatnikov, Lifschitz and co-workers conducted a long-term
investigation into spacetime behaviour close to cosmological
singularities (work that in fact began some time before the
singularities were recognized as being more than a coordinate effect)
and their results are collected in the papers Belinskii, Khalatnikov
and Lifschitz~1970 and~1982.
Aspects of their treatment of singularities in spatially
inhomogeneous cosmologies have been extensively criticized (Barrow and
Tipler~1979) and much of their reasoning is difficult to
follow, but nevertheless the main properties of the
Belinskii-Khalatnikov-Lifschitz model of a generic inhomogeneous
cosmological singularity have been widely quoted:
the presence of matter does not affect the nature of the singularity,
and close to it spatial points evolve as independent homogeneous
cosmologies of Bianchi type~VIII or~IX (the `mixmaster' models).

Other early work on the nature of inhomogeneous cosmological
singularities was performed by Eardley, Liang and Sachs~(1972).
They studied families of exact solutions for irrotational dust
cosmologies which were found to have `velocity-dominated' behaviour at
the singularity: as they approach the singularity the metrics can be
approximated by solutions to modified Einstein equations with
spatial curvature terms removed.
More recently Isenberg and Moncrief~(1990) have reformulated this idea
in their definition of asymptotically velocity-term dominated (AVTD)
behaviour near a spacelike singularity.
A cosmology has AVTD behaviour in the vicinity of the singularity if
it admits a foliation by spacelike hypersurfaces on which the metric
asymptotically approaches a VTD (velocity-term dominated) solution,
the latter being defined as a solution to a modified set of Einstein
equations in which spatial derivative terms have been neglected.
Although this type of behaviour cannot occur at all cosmological
singularities (the mixmaster spacetimes mentioned above are a
counter example) the class of spacetimes with AVTD singularities is
conjectured to be non-trivial.

\figure{GowDecoupled}
\DrawFig{
  \epsfxsize=0.7\hsize
  \epsfbox{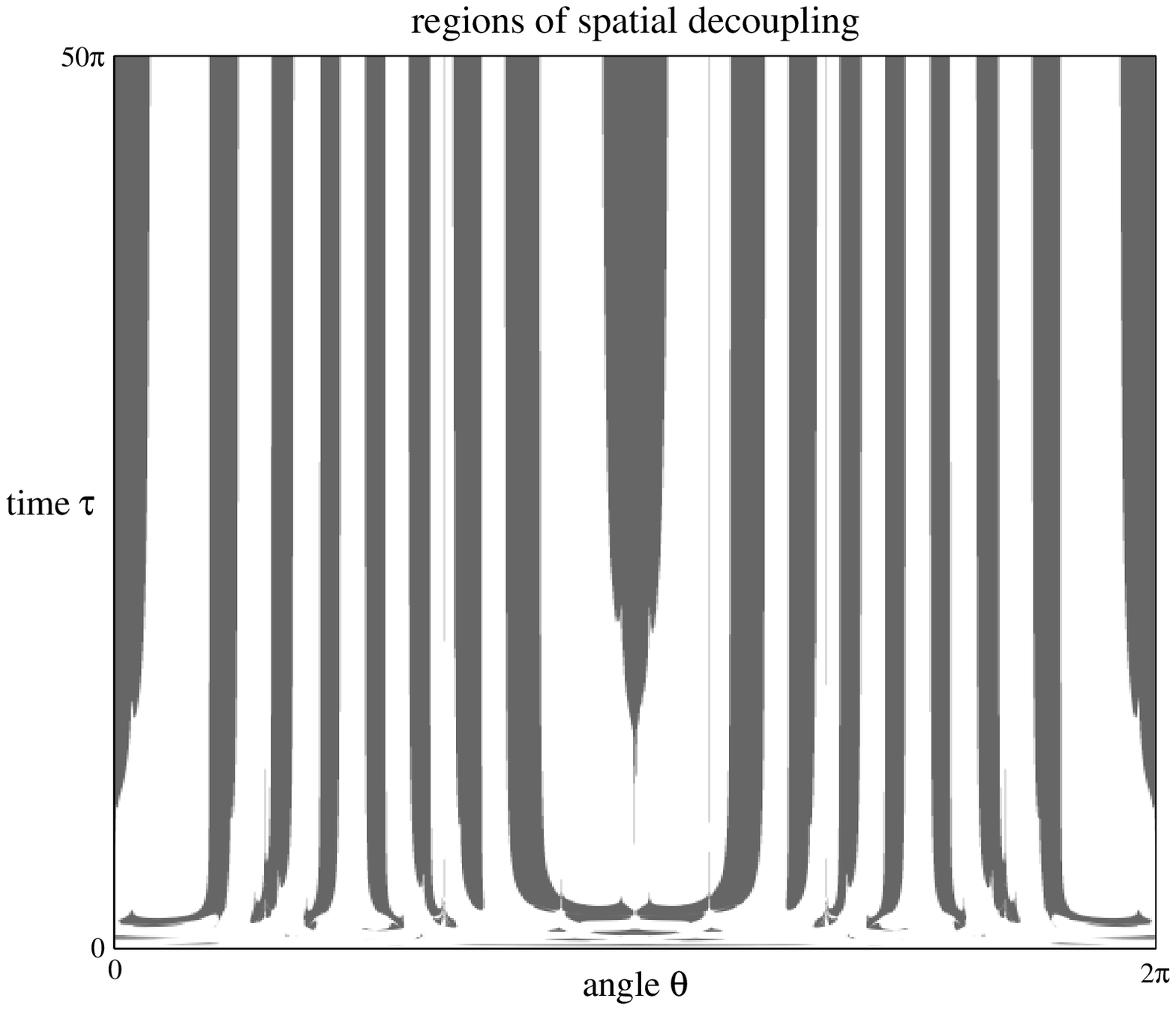}
}
\Caption
Shaded regions show where spatial derivative terms in the evolution
equations~\EqNum{GowEvolve}{a,b} for the Gowdy~$T^3$ metric 
are negligible.
These regions correspond to the intersection of the green-coloured
regions of the two figures~\use{Fg.GowDominAt} and~\use{Fg.GowDominBt}.
The evolution does not become completely spatially decoupled within 
the time range of the simulation.
\bigskip
\endCaption
\endfigure

Isenberg and Moncrief~(1990) have rigorously shown that their
definition of AVTD behaviour applies to the polarized Gowdy models
with spatial slices defined by the time coordinate~$\tau$ introduced
in section~\use{sec.GowGowdy}.
Perturbative studies of the unpolarized Gowdy~$T^3$ cosmologies led 
Grubi{\u s}i\'c and Moncrief~(1993) to conjecture that these models
also have AVTD behaviour at the singularity, and the numerical
simulations of Berger and Moncrief~(1993) provide supporting evidence
for this conjecture.
(See Berger~1998 and Berger et al.~1998 for summaries of numerical
investigations of singularity behaviour in other spacetimes. 
In particular the Gowdy~$S^1 \times S^2$ model is studied in
Garfinkle~1999, and its behaviour is found to be basically the
same as that of the Gowdy~$T^3$ model.)

The general VTD solution for the Gowdy~$T^3$ metric~\EqNum{GowMetricTau}{}
is given by the equations~\EqNum{GowHomog}{} with~$\alpha$, $\beta$,
$\eta$, $\chi$ and~$\xi$ now being functions of the spatial
coordinate~$\theta$ instead of constants.
The behaviour of the VTD solution as the singularity is approached
($\tau \rightarrow \infty$) is then described by
$$
  P_{\SmallWord{AVTD}} = p_0(\theta) + a_0(\theta) \tau
  \, , \qquad
  Q_{\SmallWord{AVTD}} = q_0(\theta)
  \, , \qquad
  \lambda_{\SmallWord{AVTD}} 
    = l_0(\theta) - a_0(\theta)^2 \tau
  \, , 
$$
and this is exactly the form of the numerically evolved solution
described in section~\use{sec.GowCollapse} at late times.
The present work thus confirms the conclusion of Berger and
Moncrief~(1993) that the behaviour of numerically evolved Gowdy~$T^3$
solutions is consistent with the presence of an AVTD singularity.
Recently Kichenassamy and Rendall~(1998) have constructed asymptotic
expansions at the singularity for a range of unpolarized Gowdy~$T^3$
solutions and found them all to show AVTD behaviour, further
strengthening the case for this being the generic behaviour of the
model.

The definitions of velocity-dominated singularities given by Eardley,
Liang and Sachs~(1972) and Isenberg and Moncrief~(1990) are
unambiguous.
However a commonly used physical picture (possibly deriving from the 
Belinskii-Khalatnikov-Lifschitz singularity studies)
of the spacetime behaviour at inhomogeneous
cosmological singularities exchanges
mathematical precision for descriptive simplicity: in the approach to
a velocity-dominated singularity spatial points are pictured as
decoupling, with the spatial derivative terms in the Einstein equations
having a negligible effect on the evolution.
Although this `spatial decoupling' interpretation of
velocity-dominated behaviour is not equivalent to Isenberg and
Moncrief's definition~(1990) of AVTD behaviour, it might be hoped that
in general the two concepts would apply to the same singularities.
Indeed, this idea seems plausible from consideration of the evolution
equations~\EqNum{GowEinst}{} for the Gowdy~$T^3$
metric~\EqNum{GowMetricTau}{}: all spatial derivative terms involve
factors of $\exp(-2\tau)$ and it is not unreasonable to expect that
they become negligible as the singularity is approached 
($\tau \rightarrow \infty$).
However this behaviour is not what is seen in numerical simulations.
Figures~\use{Fg.GowDominAt} and~\use{Fg.GowDominBt} show which terms
in the Gowdy evolution equations have most significance at which
times, and it is apparent that the terms which are independent of
spatial derivatives (the `velocity' terms) do not come to completely
dominate the evolution within the time range shown.
(Regions where spatial derivatives are negligible are taken here to be
those which are coloured green in both figures.)
A larger range of time is covered by figure~\use{Fg.GowDecoupled}, but
the conclusion is the same:
the evidence suggests that it is not the case that spatial
derivatives become negligible as the singularity is approached.
Thus, although the AVTD hypothesis of Isenberg
and Moncrief~(1990) holds for the Gowdy~$T^3$ models with the standard
choice of time slicing, the spacetime does not spatially decouple as the
singularity is approached, in contrast to the picture often presented.

\Section{Expanding Planar Cosmologies
         \label{sec.GowExpand}}
In this section the unpolarized Gowdy~$T^3$ models of
section~\use{sec.GowGowdy} form the basis of a study of expanding
vacuum planar cosmologies.
Numerical simulations of the models have been performed to investigate
three different types of cosmological behaviour.

In section~\use{sec.GowCollapse}, smooth initial fields describing
Gowdy cosmologies were evolved backwards in cosmic time towards the `big
bang' singularity and were found to develop complicated spatial
structure. 
Reversing the direction of time in these calculations would appear to
show that spatial structure is smoothed out by the process of cosmological
expansion, a result at odds with the theory of large-scale structure
formation.
However numerical simulations show that this behaviour is not generic:
smooth initial data sets for expanding Gowdy cosmologies can develop
complicated spatial structure as they evolve.

The formation of spatial structure in expanding Gowdy
cosmologies occurs at times not long after the initial singularity.
Numerical simulations have been performed to investigate the behaviour
of the Gowdy models at much later times, and it is found that
initially strong gravitational waves decrease in amplitude over time
and behave qualitatively like standing waves.
At sufficiently late times the numerically evolved Gowdy solutions are
found to have behaviour similar to that which would be expected for
perturbed Kasner spacetimes.

Anninos, Centrella and Matzner~(1991b), Shinkai and Maeda~(1994), and
Ove~(1990a) have investigated the properties of nonlinear
gravitational waves propagating in expanding cosmological models.
Their findings (in particular those concerning the appearance of wave
residue and the soliton-like behaviour of gravitational waves) are
considered here and compared to 
results from numerical evolutions of wave packet initial data for
the Gowdy metric. 
The problem of distinguishing nonlinear wave behaviour from gauge
effects is discussed.

\SubSection{Numerical Simulations of Expanding Gowdy~$T^3$ Spacetimes}
The Einstein equations for the metric~\EqNum{GowMetricT}{} are given
as equations~\EqNum{GowNewEinst}{}.
Recall that the expanding time coordinate~$t$ ranges from~$0$
to~$+\infty$, with $t = 0$ labelling the cosmological singularity.
A first-order evolution system is produced by the definition of 
new variables
$$
  A \equiv P_{,t}
  \, , \quad
  B \equiv Q_{,t}
  \, , \quad
  C \equiv P_{,\theta}
  \, , \quad
  D \equiv Q_{,\theta}
  \, , 
$$
where the variables~$A$ and~$B$ used here differ from the
analogous variables of equation~\EqNum{GowNewVars}{}.
The first-order evolution equations are then
$$\EQNalign{
  A_{,t} - C_{,\theta} &{} = e^{2P}(B^2 - D^2) - A/t
  \, , \EQN GowReverse;a \cr
  B_{,t} - D_{,\theta} &{} = - 2(AB - CD) - B/t
  \, , \EQN GowReverse;b \cr
  C_{,t} - A_{,\theta} &{} = 0
  \, , \EQN GowReverse;c \cr
  D_{,t} - B_{,\theta} &{} = 0
  \, , \EQN GowReverse;d \cr
  P_{,t} &{} = A
  \, , \EQN GowReverse;e \cr
  Q_{,t} &{} = B
  \, , \EQN GowReverse;f \cr
  \lambda_{,t} &{} = t( A^2 + C^2 )
                  + t e^{2P}( B^2 + D^2 )
  \, , \EQN GowReverse;g \cr
}$$
together with the constraint equation
$$
  \lambda_{,\theta} = 2 t (AC + e^{2P}BD)
  \, . \EQN GowRevConstr
$$
The principal part (left-hand side) of the evolution
system~\EqNum{GowReverse}{} has a particularly simple form: it
consists of two flat-space wave equations for the variable
pairs~$A$ and~$C$, and~$B$ and~$D$.

The code for evolving equations~\EqNum{GowReverse}{} has been put
through the same set of tests as the code described in
section~\use{sec.GowCollapse}.
In addition, the two codes have been compared to check that evolving
data forwards with one produces the same results as evolving data
backwards with the other.
Approximate solutions to the linearized equations discussed later in
this section provide another test of the code.

The general set-up of the numerical simulations discussed here is the
same as for those in section~\use{sec.GowCollapse}, with the exception that
variable-size time steps are not needed since the largest
characteristic speed of the system~\EqNum{GowReverse}{} is a constant
(see equation~\EqNum{GowSpeeds}{}).
Three different kinds of simulation are discussed in this section, and
further details of the numerical set-ups used are given as appropriate
in the following subsections.

\SubSection{Spatial Structure Formation in Expanding Gowdy Cosmologies}
In section~\use{sec.GowCollapse} smooth initial
data sets~\EqNum{GowInitial}{} for the collapsing Gowdy~$T^3$
metric~\EqNum{GowMetricTau}{} were numerically evolved and found to
develop complicated spatial structure on very fine scales.
Experimentation with a range of initial data sets showed 
this behaviour to be typical of the collapsing Gowdy models, with the
magnitude of the initial data determining the amount of
spatial structure that developed.
The time-reversals of these simulations provide examples of expanding
cosmologies, and present an unexpected picture of cosmological
behaviour: spacetimes which have very complicated spatial structures
a short time after the big bang lose all of that structure as they
expand. 
It is known (Moncrief~1981) that the Gowdy~$T^3$ metrics contain no
singularities apart from the cosmological singularity which is
always present.
This rules out the possibility of gravitational
waves interacting sufficiently strongly to produce black hole-like
objects in the models; however it would be surprising if strong
gravitational waves were not able to produce some kind of spatial
structure in the expanding Gowdy spacetimes.

\figure{GowRevVarP}
\DrawFig{
  \epsfxsize=\hsize
  \epsfbox{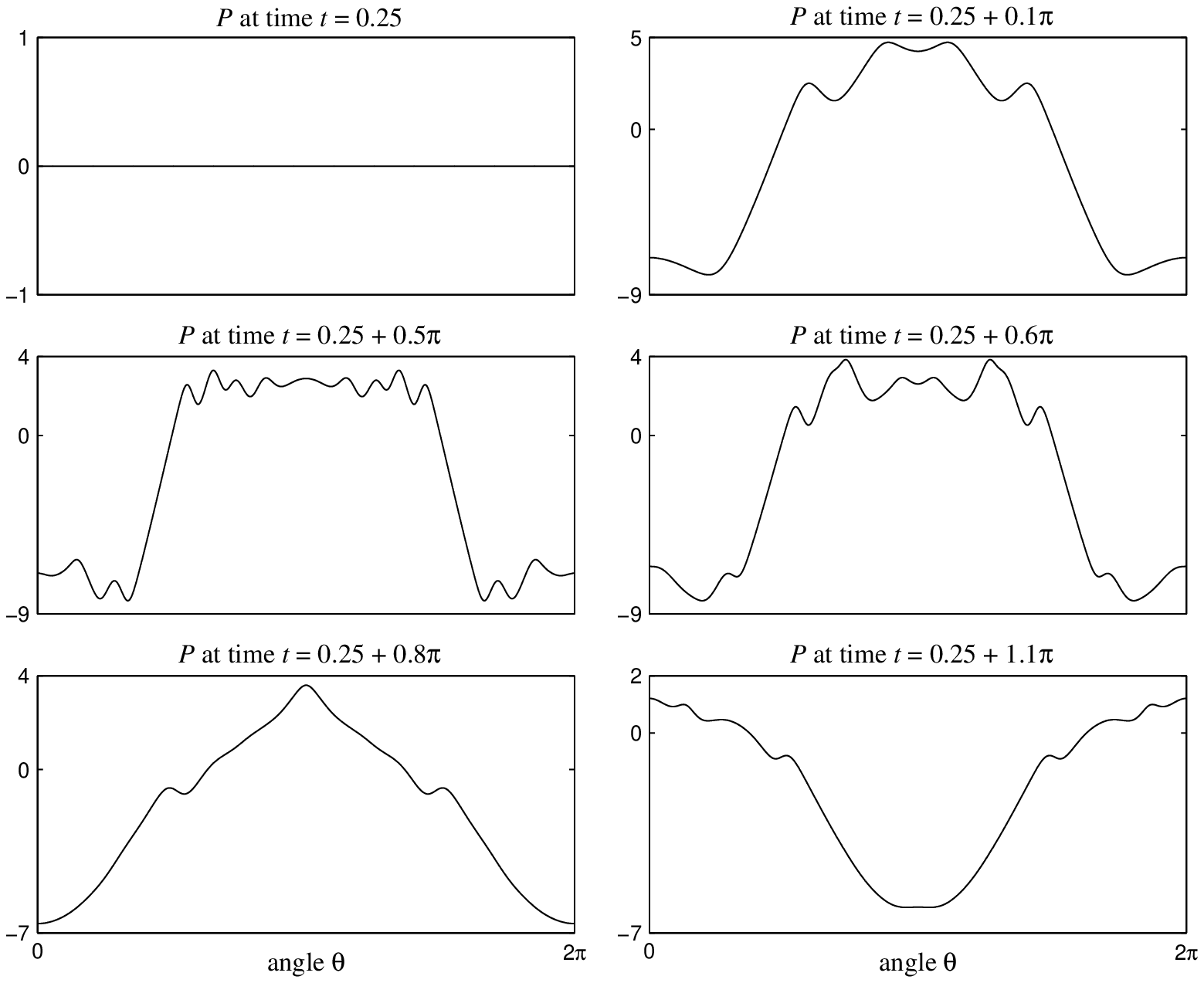}
}
\Caption
Wave variable~$P(t,\theta)$ of expanding Gowdy 
metric~\EqNum{GowMetricT}{} at six different times.
Initial data set~\EqNum{GowRevInitial}{} with $t_0 = 0.25$ and
$v_0 = 15$ is evolved using equations~\EqNum{GowReverse}{}. 
Spatial structure forms at times close to the initial
singularity.
\bigskip
\endCaption
\endfigure

\fullfigure{GowRevDomin}
\vss
\DrawFig{
  \epsfxsize=0.87\hsize
  \epsfbox{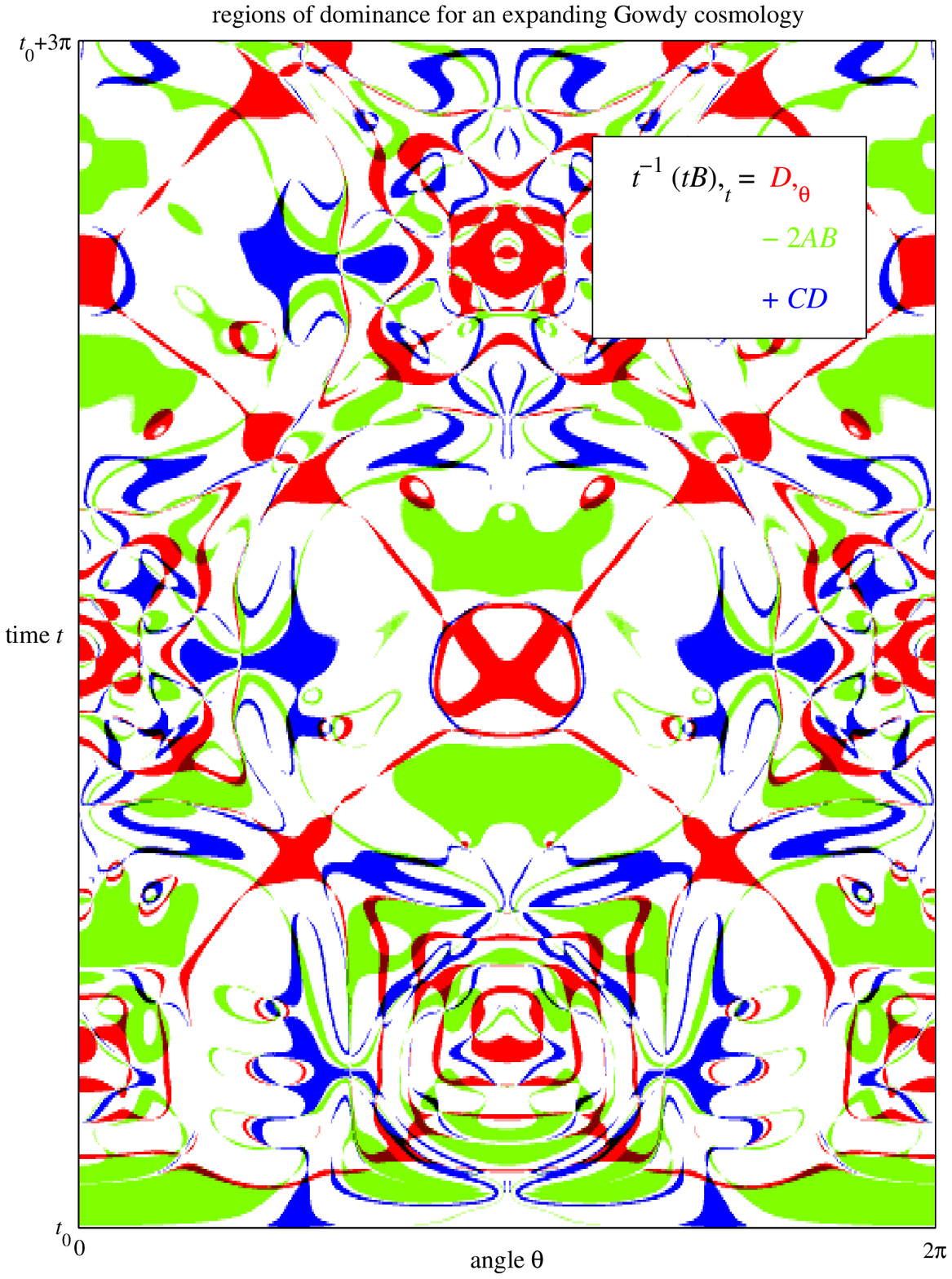}
}
\Caption
Dominant terms in the evolution equation~\EqNum{GowReverse}{b} 
for~$B$.
Initial data set~\EqNum{GowRevInitial}{} with $v_0 = 10$ and $t_0 = 0.25$
is evolved for the expanding Gowdy model.
Regions are coloured red, green or blue according to which term
dominates the evolution equation (as shown in the key), or
are white if no term dominates.
This figure shows for the expanding Gowdy model what
figure~\use{Fg.GowDominAt} shows for the collapsing model.
(The definition of dominance is discussed in
section~\use{sec.GowCollapse}.)
\endCaption
\endfigure

In this subsection results are presented from numerical simulations of
Gowdy cosmologies analogous to the simulations of
section~\use{sec.GowCollapse} but with the spacetimes expanding rather
than collapsing.
For initial data containing `strong' gravitational waves close to the
initial singularity, complicated spatial structure develops just as
in the collapsing case, and the time-reversed
results of section~\use{sec.GowCollapse} are seen not to describe 
the generic behaviour of expanding cosmologies.

The system of equations~\EqNum{GowReverse}{} is evolved numerically
using the initial data set
$$\eqalign{
  &P = 0 \, , \cr
  &P_{,t} = - v_0 \cos\theta / t_0 \, , \cr
} \qquad \eqalign{
  &Q = \cos\theta \, , \cr
  &Q_{,t} = 0 \, , \cr
} \qquad \eqalign{
  &\lambda = 0 \, , \cr
  &\hbox{at time }t = t_0 \, , \cr
} \EQN GowRevInitial
$$
which, if the initial time~$t_0$ is taken equal to~$1$, is the same
as the initial data set used in section~\use{sec.GowCollapse}
(cf.~equation~\EqNum{GowInitial}{} with the time coordinate transformed
according to equation~\EqNum{GowTimes}{}).
As in the collapsing case, the parameter~$v_0$ determines the strength
(in some sense) of the gravitational waves in the initial data.
An initial time of $t_0 = 0.25$ (corresponding to~$\tau$ equal
to~$1.386$) is used so that the influence of proximity to the
cosmological singularity is manifest in the results.
The numerical simulations are performed using a base grid of
resolution $\Delta\theta = 2\pi / 4000$ with adaptive mesh
refinement allowing the resolution to increase up 
to $\Delta\theta = 2\pi / 64 \, 000$.

Spatial structure forms (in the sense that high frequency modes are
excited) for large amplitude waves at times close
to the time of the initial singularity.
Figure~\use{Fg.GowRevVarP} shows the wave variable~$P(t,\theta)$ at
various stages during the numerical evolution of initial
data~\EqNum{GowRevInitial}{} with the choice of amplitude parameter
$v_0 = 15$, and it is clear that nonlinear effects produce complicated
behaviour in the variable.
Spatial structure is produced early on in the simulation; at later
times the behaviour of the variable~$P$ is qualitatively that of a
standing wave.
Similar behaviour is seen for the second wave variable~$Q$, while the
background variable~$\lambda$ remains relatively
smooth throughout. 
The curvature invariants decrease in magnitude very rapidly during the
course of the simulation.

Figure~\use{Fg.GowRevDomin} provides an illustration of the overall
behaviour of an expanding Gowdy~$T^3$ cosmology.
The dominant terms in the evolution equation~\EqNum{GowReverse}{b} for
$B \equiv Q_{,t}$ are shown for initial data~\EqNum{GowRevInitial}{}
with $v_0 = 10$; it is the expanding cosmology version of
figure~\use{Fg.GowDominAt} from section~\use{sec.GowCollapse}.
The two main qualitative features of the simulation are apparent in
the figure: complex behaviour occurs on small spatial scales, and the
overall pattern of the evolution has the form of a standing wave,
repeating with time period $\Delta t = 2 \pi$.

Away from the initial singularity nonlinear effects are much less
apparent in the evolution and the amplitudes of the standing waves in
the variables~$P$ and~$Q$ are seen to gradually decrease.
This motivates the study of late-time behaviour of Gowdy~$T^3$
spacetimes.

\SubSection{Late-Time Gowdy Behaviour I: A First-Order Model}
Misner~(1968) proposed a programme of cosmological research in which
spacetimes with a high degree of anisotropy close to a `big bang'
singularity would be examined to see to what extent the properties of
the presently observable universe are a typical outcome of evolution
by Einstein's equations (see the comments on `chaotic cosmology' in 
Carmeli, Charach and Malin~1981).
Although the Gowdy spacetimes do not provide realistic models of the
observable universe, they can still be used to find a partial answer
to the question (in the spirit of Misner's proposal) of what range of
initial data for Einstein's equations leads to spacetimes which are
approximately homogeneous at late times.
The next subsection examines numerically evolved cosmologies
and shows them to provide evidence that the late-time behaviour of a
general Gowdy~$T^3$ spacetime tends towards that of a perturbed 
homogeneous cosmology. 
Prior to that, in the present subsection the Gowdy evolution equations
are linearized and used to construct a model of the late-time
behaviour of an approximately homogeneous Gowdy spacetime.

Consider first the wave variables~$P$ and~$Q$ of the expanding
Gowdy~$T^3$ metric~\EqNum{GowMetricT}{}.
It is assumed that the late-time (large~$t$) behaviour of
these variables is close to the form they have in the homogeneous case
given in equations~\EqNum{GowHomog}{}.
As the time~$t$ becomes large the homogeneous
solution~\EqNum{GowHomog}{} approaches
$$
  P_{\SmallWord{homog}} = \ln(\alpha \, t^\beta)
  \, , \qquad
  Q_{\SmallWord{homog}} = \eta
  \, , \EQN GowHomogLT
$$
where~$\alpha$, $\beta$ and~$\eta$ are constants with~$\beta$, the
Kasner expansion parameter, being non-negative. 
(Note that the late-time homogeneous solution~\EqNum{GowHomogLT}{} is
an exact solution of the Gowdy equations, as can be seen by
comparing it to the general homogeneous solution~\EqNum{GowHomog}{}.)
A Gowdy solution which is approximately homogeneous at late times is
then assumed to have the form
$$
  P = P_{\SmallWord{homog}} + \delta P
  \, , \qquad
  Q = Q_{\SmallWord{homog}} + \delta Q
  \, , \EQN GowPerturb
$$
where~$\delta P$ and~$\delta Q$ are small perturbations which are
negligible at second order.
Substituting the variables~$P$ and~$Q$ of equations~\EqNum{GowPerturb}{}
into the evolution equations~\EqNum{GowNewEinst}{a,b} produces
evolution equations for the perturbations~$\delta P$ and~$\delta Q$,
$$\EQNalign{
  \delta P_{,tt} & {}=
    \delta P_{,\theta\theta}
  - (1/t) \delta P_{,t}
  \, ,
  \EQN GowApproxEinst;a \cr
  \delta Q_{,tt} & {}=
    \delta Q_{,\theta\theta}
  - ((1+2\beta)/t) \delta Q_{,t}
  \, ,
  \EQN GowApproxEinst;b \cr
}$$
where terms of second and higher order in~$\delta P$ and~$\delta Q$
have been omitted.
These equations have exact solutions
$$\EQNalign{
  \delta P & {} = 
    a_0 \ln t + b_0
    + \sum_{n=1}^{\infty} \Big( Z_{+n}\!(t) \cos(n\theta)
                              + Z_{-n}\!(t) \sin(n\theta) \Big)
  \, , 
  \EQN GowExactPert;a \cr
  & \hbox{with \ \ }
  Z_{\pm n}\!(t) = a_{\pm n} J_0( \vert n \vert t )
                 + b_{\pm n} Y_0( \vert n \vert t )
  \hbox{ \ \ for } n \neq 0
  \, , \vrule width 0pt depth 6pt \cr
  \delta Q & {} = 
    \left\{\eqalign{
           c_0 \ln t \> &\hbox{ for } \beta = 0 \cr
           c_0 t^{-2\beta} &\hbox{ for } \beta > 0 \cr
    }\right\}
    + d_0
    + t^{-\beta} 
      \sum_{n=1}^{\infty} \Big( W_{+n}\!(t) \cos(n\theta)
                              + W_{-n}\!(t) \sin(n\theta) \Big)
  \, , 
  \EQN GowExactPert;b \cr
  & \hbox{with \ \ }
  W_{\pm n}\!(t) = c_{\pm n} J_\beta( \vert n \vert t )
                 + d_{\pm n} Y_\beta( \vert n \vert t )
  \hbox{ \ \ for } n \neq 0
  \, , \cr
}$$
where the~$a_{\pm n}$, $b_{\pm n}$, $c_{\pm n}$ and~$d_{\pm n}$ are 
constants, and~$J_i(x)$ and~$Y_i(x)$ are Bessel functions.
(Note that equation~\EqNum{GowExactPert}{a} is the same as
equation~\EqNum{GowPolarP}{}, the general solution for the polarized
Gowdy model with~$Q \equiv 0$.)
The equations~\EqNum{GowExactPert}{} for~$\delta P$ and~$\delta Q$
include some zeroth-order mode terms which do not tend to zero as the
time~$t$ becomes large.
The terms~$a_0 \ln t$, $b_0$ and~$d_0$ can be absorbed into the
background homogeneous solution~\EqNum{GowHomogLT}{}, whereas the
term~$c_0 \ln t$ (which can only be present if the constant~$\beta$ is
zero) must be neglected on the grounds that it contradicts the
assumption that~$\delta Q$ is small at late times.
If in addition the Bessel functions in the expressions for~$\delta P$
and~$\delta Q$ are replaced by their approximate forms for late times
(see equations~\EqNum{BessApprox}{}) then the
expressions~\EqNum{GowPerturb}{} for~$P$ and~$Q$ can be written as
$$\EQNalign{
  P & {} \simeq
    \ln(\alpha \, t^\beta)
    + t^{-1/2} \sum \hbox{oscillations}
  \, , 
  \EQN GowApproxPert;a \cr
  Q & {} \simeq
    \eta
    + \underbrace{c_0 t^{-2\beta}}_{\hbox{\ninepoint if $\beta > 0$}}
    {} + t^{-\beta-1/2} \sum \hbox{oscillations}
  \, , 
  \EQN GowApproxPert;b \cr
}$$
where each oscillation term has the form 
$\mu \cos(n\theta) \cos(nt-\nu)$ for~$\mu$ and~$\nu$ constants,
where either or both of the cosines may be replaced by sines.

Equations~\EqNum{GowApproxPert}{} provide a first-order description of
the behaviour of the Gowdy wave variables~$P$ and~$Q$ at late times
under the assumption that the spacetime is approximately homogeneous
then.
It is straightforward to check that the
expressions~\EqNum{GowApproxPert}{} for~$P$ and~$Q$ are consistent
with the evolution equations~\EqNum{GowNewEinst}{a,b} in the sense
that the second- and higher-order terms that were omitted from
equations~\EqNum{GowApproxEinst}{} tend to zero as~$t$ tends to infinity
under substitution of these forms for~$P$ and~$Q$.

The late-time behaviour of the Gowdy background variable~$\lambda$ can
now be investigated.
As before, the variable is assumed to have the form of a perturbed
homogeneous solution, analogous to equations~\EqNum{GowHomogLT}{}
and~\EqNum{GowPerturb}{}, and so, from equation~\EqNum{GowHomog}{},
$$
  \lambda = \lambda_{\SmallWord{homog}} + \delta \lambda
  \quad \hbox{with \ }
  \lambda_{\SmallWord{homog}} = \chi + \beta^2 \ln t
  \, ,
  \EQN GowPerturbL
$$
where~$\chi$ is a constant and~$\beta$ has the same value as in
equation~\EqNum{GowHomogLT}{}.
If the evolution and constraint equations~\EqNum{GowNewEinst}{c,d}
for~$\lambda$ are expanded to first order in the 
perturbations~$\delta \lambda$, $\delta P$ and~$\delta Q$
using equations~\EqNum{GowPerturbL}{}, \EqNum{GowHomogLT}{} 
and~\EqNum{GowPerturb}{}, then it is found that 
$\delta \lambda = 2 \beta \, \delta P$ 
to within addition of a constant term which can be absorbed into the
value of~$\chi$ in equation~\EqNum{GowPerturbL}{}.
Using the expression for $\delta P$ from
equation~\EqNum{GowApproxPert}{a}, the perturbation~$\delta \lambda$
is seen to approach zero as time~$t$ becomes large.
However, the long-term behaviour of~$\delta \lambda$ is found to be
qualitatively different from this if terms at higher than first order
are considered: if the expressions~\EqNum{GowApproxPert}{} for~$P$
and~$Q$ are substituted into the evolution
equation~\EqNum{GowNewEinst}{c} for~$\lambda$, then it is found that
$$
  \delta \lambda_{,t} =
  \Big( \sum \hbox{oscillations} \Big)^2
  + \hbox{(terms that become negligible as $t \rightarrow \infty$)}
  \, ,
  \EQN GowApproxPertL
$$
and so at sufficiently late times~$\delta \lambda$ grows approximately
linearly.
As a consequence, so does the variable~$\lambda$.
The approximately homogeneous Gowdy model constructed here thus turns
out to resemble a homogeneous Gowdy solution at late times only in
terms of the wave variables~$P$ and~$Q$; not only does the background
variable~$\lambda$ have a late-time form which is different from the form
it has in a homogeneous solution (equation~\EqNum{GowPerturbL}{}), but also
the amplitudes of the oscillations in the variable do not decay to
zero as the time~$t$ tends to infinity.
(Similar behaviour has been reported for different cosmological models
by Adams et al.~1982.)

The linear growth of~$\lambda$ at late times means that perturbations
can significantly change the overall behaviour of a homogeneous
spacetime. 
Consider a homogeneous Gowdy solution~\EqNum{GowHomog}{} with 
$\beta < 1$.
As can be seen from the metric~\EqNum{GowMetricT}{} this spacetime
undergoes collapse in the direction of the $\theta$-axis as time~$t$
tends to infinity.
(Equation~\EqNum{GowKasnerP}{} shows how this behaviour relates 
to the standard Kasner cosmologies in which one spatial direction
always contracts as the other two expand.)
However, if the wave variables~$P$ and~$Q$ of the homogeneous
spacetime are perturbed slightly, then at sufficiently late times the
variable~$\lambda$ will grow linearly rather than logarithmically, and
the spacetime will expand rather than contract along the
$\theta$-axis.
In effect, an infinitesimal amount of gravitational radiation
propagating in the direction of contraction of a Kasner spacetime is
sufficient to reverse the contraction at late times and cause the
spacetime to expand in all spatial directions.
This type of behaviour is consistent with Isaacson's work~(1968a,b) on
gravitational waves, and has been previously observed in
numerically evolved planar cosmologies by Centrella~(1986); the
perturbed spacetime can be thought of as a 
homogeneous solution of Einstein's equations with an effective
energy-momentum tensor describing short wave gravitational radiation
(Isaacson~1968a,b; section~35.13 of Misner, Thorne and Wheeler~1973).
(The effective energy-momentum tensor for gravitational waves in a
polarized Gowdy~$T^3$ spacetime is constructed in Berger~1974.)

In summary, if a
Gowdy spacetime is approximately homogeneous at some time long after
the `big bang' took place, then its subsequent behaviour will be
described by equations~\EqNum{GowApproxPert}{} with the background
variable~$\lambda$ growing approximately linearly.
In the next subsection this model of late-time Gowdy behaviour is
compared to results from numerical simulations.

\SubSection{Late-Time Gowdy Behaviour II: Numerical Simulations}
Earlier in this section expanding Gowdy~$T^3$ spacetimes containing
strong gravitational waves close to the initial singularity were
investigated. 
Numerical simulations showed their behaviour to be complicated with
spatial structure forming on fine scales.
However, away from the initial singularity it was found that the
amplitudes of oscillations in the metric variables tend to decay and
their behaviour becomes 
qualitatively like that of standing waves in flat space (much as would
be expected for weak gravitational waves propagating
in an expanding spacetime).
In this subsection the behaviours of Gowdy cosmologies at times long
after the `big bang' are investigated numerically and the results are
compared to the model of a perturbed homogeneous cosmology constructed
in the previous subsection.
(Similar numerical investigations of the approach to homogeneity for
initially inhomogeneous expanding planar spacetimes are reported in
Shinkai and Maeda~1993 and~1994; there a positive cosmological
constant is assumed, and only one polarization of gravitational radiation is
considered.)

\figure{GowLateTime}
\DrawFig{
  \epsfxsize=\hsize
  \epsfbox{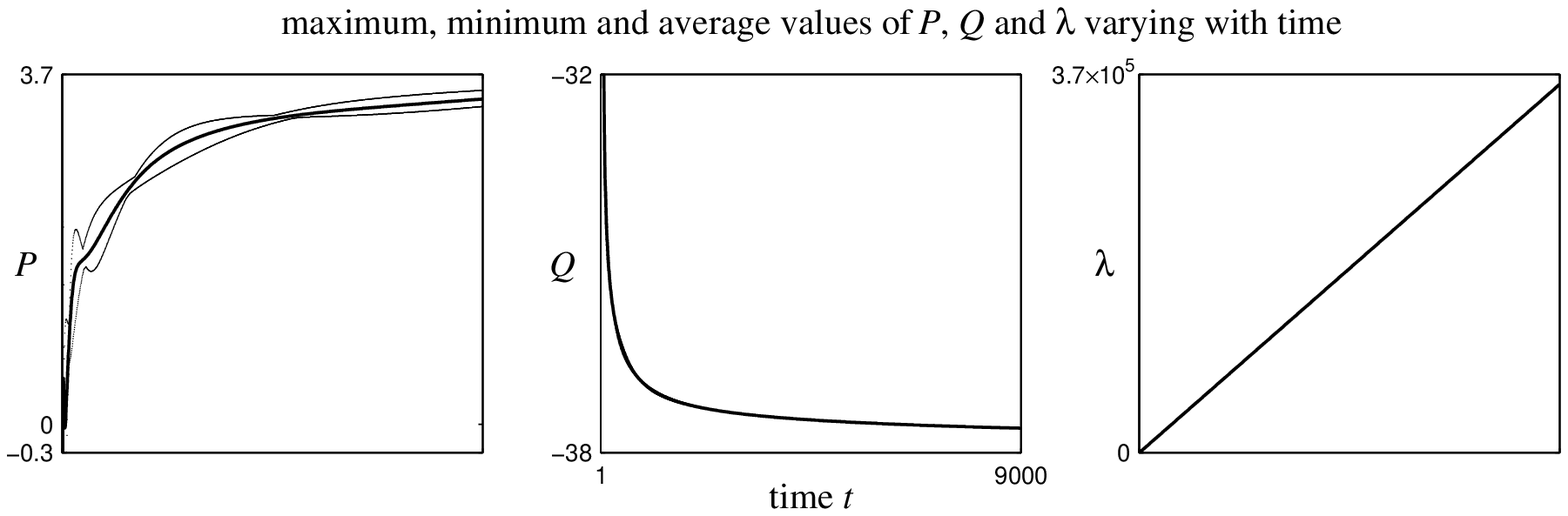}
}
\vskip -2mm
\Caption
Behaviour of the variables~$P$, $Q$ and~$\lambda$ at late times.
Initial data set~\EqNum{GowLTInitial}{} for the
metric~\EqNum{GowMetricT}{} with $v_P = v_Q = 7.5$ and 
$\beta_0 = 0.75$ is evolved numerically.
The average value of each variable is plotted as a solid line together
with its maximum and minimum values plotted as individual points.
The averages, maxima and minima are taken over all space and over time
intervals of size $\Delta t = 2\pi$.
(The time axis is the same for all three plots and is linear.)
\bigskip
\endCaption
\endfigure

Numerical simulations are performed using the initial data set
(cf.~equation~\EqNum{GowRevInitial}{})
$$\eqalign{
  &P = 0 \, , \cr
  &P_{,t} = \beta_0 - v_P \cos\theta \, , \cr
} \qquad \eqalign{
  &Q = v_Q \cos\theta \, , \cr
  &Q_{,t} = 0 \, , \cr
} \qquad \eqalign{
  &\lambda = 0 \, , \cr
  &\hbox{at time }t = 1 \, , \cr
} \EQN GowLTInitial
$$
where~$v_P$ and~$v_Q$ are viewed as measuring the amplitudes of the
initial gravitational waves and~$\beta_0$ as describing the background
spacetime in which the waves propagate (it is analogous to the
constant~$\beta$ in the homogeneous solution~\EqNum{GowHomog}{}).
The constraint on the initial data, equation~\EqNum{GowRevConstr}{},
is trivially satisfied.
When~$v_P$ and~$v_Q$ are small (much less than one) the numerically 
evolved solution can be
compared to the exact linearized solution~\EqNum{GowExactPert}{}
derived in the previous subsection for a perturbed homogeneous
spacetime.
The good agreement between the two provides evidence
for the validity both of the numerical code and of the first-order
solution. 
For larger~$v_P$ or~$v_Q$ the evolution of the numerical solution is
initially nonlinear and the late-time behaviour of the spacetime is
not easily predicted; in particular the question of whether or not
cosmological expansion makes the spacetime homogeneous at late times
(in the sense that it behaves then like a perturbed homogeneous
spacetime) is investigated here. 

Figure~\use{Fg.GowLateTime} shows results for the numerical evolution
of initial data~\EqNum{GowLTInitial}{} with $v_P = v_Q = 7.5$ and 
$\beta_0 = 0.75$ using the system of equations~\EqNum{GowReverse}{}.
(The value of~$\beta_0$ used here produces a Kasner cosmology which is
collapsing in the direction of the $\theta$-axis if both~$v_P$
and~$v_Q$ are zero.)
The simulation used a spatial resolution of~$\Delta \theta = 2\pi/3200$
and was run for approximately four million time steps.
Adaptive mesh refinement (described in chapter~3) was found to be an
impractical method for use on this problem; although the local truncation
error of the numerical evolution could be estimated in the usual way,
the variation in magnitude of the variables (over large time scales in
addition to rapid oscillations) made it difficult to define a
refinement criterion which was effective throughout the entire time
range of the simulation.

\figure{GowLateScale}
\DrawFig{
  \epsfxsize=\hsize
  \epsfbox{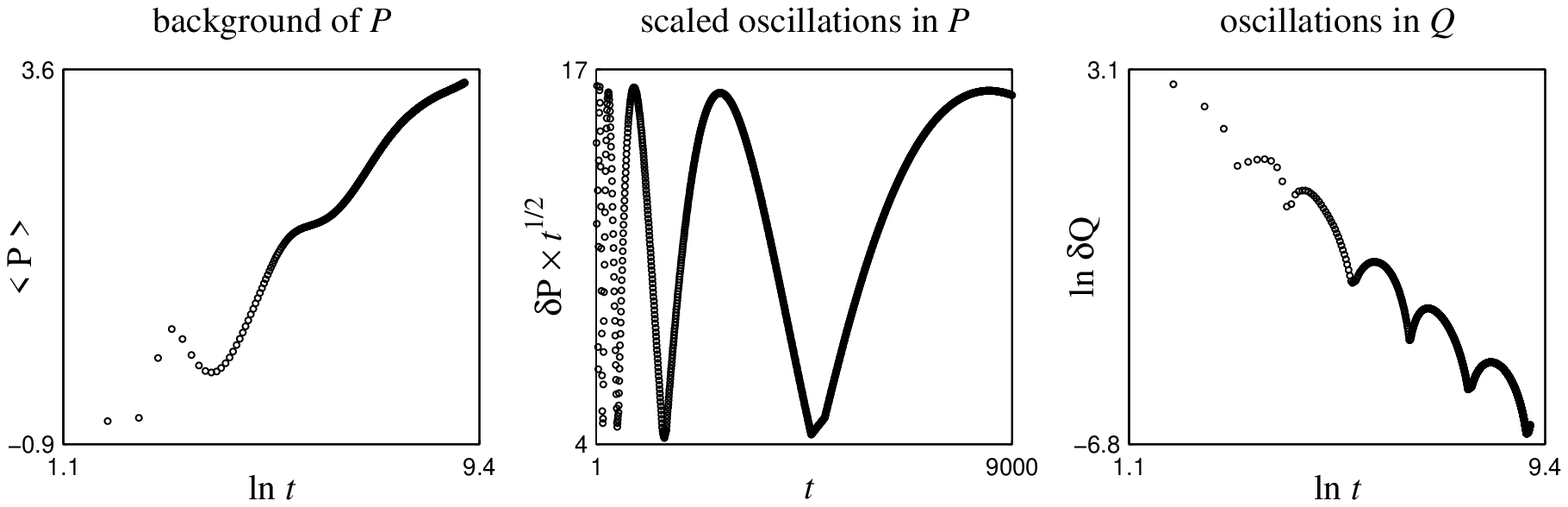}
}
\vskip -2mm
\Caption
The numerical results shown in figure~\use{Fg.GowLateTime} are 
re-plotted to allow comparison with the predicted behaviour given in
equations~\EqNum{GowApproxPert}{} for~$P$ and~$Q$ in a spacetime
which is approximately homogeneous.
The quantity~$\langle P \rangle$ is the spatial average of the
numerically evolved~$P$ (as plotted in figure~\use{Fg.GowLateTime})
while the quantity~$\delta P$ is the size of the oscillations in~$P$,
the difference between the maximum and minimum values of the
variable.
Analogous quantities are defined for the numerically evolved~$Q$.
(As in figure~\use{Fg.GowLateTime} each data point plotted here
represents a time interval of size $\Delta t = 2\pi$.)
\bigskip
\endCaption
\endfigure

It is clear from figure~\use{Fg.GowLateTime} that the late-time
behaviour of the evolved solution is broadly similar to the behaviour
of the perturbed homogeneous model considered in the previous
subsection: the variables~$P$ and~$Q$ approach the form of the
homogeneous background solution given in equation~\EqNum{GowHomogLT}{}
($P$~grows logarithmically, $Q$~tends to a constant, and the
amplitudes of the oscillations decrease in size) while the
variable~$\lambda$ grows linearly. 
In figure~\use{Fg.GowLateScale} the numerical solution is re-plotted
so as to allow a more detailed comparison with the perturbed
homogeneous solution~\EqNum{GowApproxPert}{}.
In the first plot of the figure the spatial average of~$P$ is shown
against a logarithmic time axis; for an approximately homogeneous
solution this plot is linear with gradient equal to the background
Kasner parameter~$\beta$.
The second plot shows the size of the oscillations in~$P$ scaled by
the square root of the time~$t$; for an approximately
homogeneous solution the peaks in this plot reach a constant height.
The third plot shows the size of the oscillations in~$Q$ against time
on logarithmic axes; for an approximately
homogeneous solution this plot is linear overall with gradient equal
to~$-(\beta+1/2)$.
It is apparent from figure~\use{Fg.GowLateScale} that as time passes
the behaviour of the evolved solution becomes increasingly like that
of a perturbed homogeneous cosmology.
Numerical evidence thus
supports the conjecture that strong initial inhomogeneities in a
Gowdy~$T^3$ spacetime do not prevent the spacetime from being
close to homogeneous at late times.

The first and third plots of figure~\use{Fg.GowLateScale} allow values
to be estimated for~$\beta$, the Kasner parameter for the homogeneous
background solution given in equation~\EqNum{GowHomogLT}{}.
From the first plot it is estimated that $0.55 < \beta < 0.65$ while from
the third plot it is estimated that $0.65 < \beta < 0.75$, and
the two results suggest a value for~$\beta$ of around~$0.65$.
This estimate is to be compared with the value of~$0.75$ used for the
parameter~$\beta_0$ in the initial data set~\EqNum{GowLTInitial}{}.
If~$\beta_0$ is thought of as describing an initial homogeneous
background in the same way that~$\beta$ describes the late-time
homogeneous background then it can be seen from
equations~\EqNum{GowKasner}{} and~\EqNum{GowKasnerP}{}
that the $\sigma$-$\delta$~symmetry plane of the Gowdy model is
slightly more isotropic at late times than it is initially, but that
the isotropy of the spacetime does not increase indefinitely, in
contrast to its homogeneity.

Figure~\use{Fg.GowLateScale} also reveals a feature of the late-time
behaviour of an initially strongly inhomogeneous Gowdy cosmology which
is not present in the first-order solution~\EqNum{GowApproxPert}{} (or
in numerical simulations based on initial data containing only weak
gravitational waves). 
From the second and third plots it can be seen
that the wave variables~$P$ and~$Q$ have oscillatory behaviour on time
scales which increase in size exponentially during the course of the
simulation.
By contrast, from equation~\EqNum{GowApproxPert}{}, the longest time
period that would be expected for oscillations in an approximately
homogeneous spacetime is $\Delta t = 2\pi$, and such oscillations would
not even be visible in figure~\use{Fg.GowLateScale}{}.
It would seem that these low frequency oscillations are some kind of
nonlinear effect of the Gowdy spacetime's approach to late-time
homogeneity. 
(It is interesting to speculate as to whether there might be an
underlying connection between this behaviour and the oscillations on
logarithmic time scales that occur during the spherically symmetric
collapse of massless scalar fields: there the oscillations have time
periods which decrease exponentially prior to the formation of a
central singularity, and with the direction of time reversed the third
plot of figure~\use{Fg.GowLateScale} can be seen to be similar to, for
example, figure~5 of Hamad\'e and Stewart~1996.)

\SubSection{Linear and Nonlinear Gravitational Waves}
The general behaviour of gravitational
waves in planar cosmologies is well understood only for waves which
have amplitudes that are small enough to allow the application of
linearized theory.
Various researchers have
compared perturbative models to numerical solutions at the point at
which the first-order approximation becomes inaccurate in order to 
identify new features that emerge as the behaviour becomes nonlinear.
(In addition, linearized gravitational wave models are often used as code
tests; see Centrella~1986, Anninos, Centrella and Matzner~1991a,
Anninos et al.~1997, and Anninos~1998a, among others.)
In the previous subsection the late-time behaviour of a numerically
evolved Gowdy~$T^3$ cosmology was compared to a simple model
containing linear gravitational waves;
in the present subsection linear and nonlinear gravitational waves are
compared and their differences in behaviour are highlighted.

As a first step in the investigation, a family of initial
data sets is chosen for which continuous parameters determine the strength
(in some sense) of the gravitational waves which are present.
When the parameters are zero the initial data set describes an exact
homogeneous solution, and for parameter values close to zero the
evolution can be approximated by a system of first-order
equations produced by linearization about the homogeneous solution.
As the parameters become larger the inaccuracies in the linear
approximation grow and the wave behaviour in the evolved solution
becomes nonlinear.
In the present work the axisymmetric Kasner spacetime has been chosen
as the homogeneous background solution: it is described by the
metric~\EqNum{GowMetricT}{} with variables~$P$, $Q$ and~$\lambda$
identically zero.
(As discussed in section~\use{sec.GowGowdy} wave motion in the Gowdy
spacetimes is determined by the variables~$P$ and~$Q$, and the
background variable~$\lambda$ plays no part in the following discussion.)
The linear range of initial data sets then comprises values of the
variables~$P$ and~$Q$ (and their derivatives) which are close to zero,
and for such values the evolution
equations~\EqNum{GowNewEinst}{} linearized about the homogeneous
solution have exact solutions
$$\EQNalign{
  P_{\SmallWord{linear}} & {} = 
    a_0 \ln t + b_0
    + \sum_{n=1}^{\infty} \Big( Z_{+n}\!(t) \cos(n\theta)
                              + Z_{-n}\!(t) \sin(n\theta) \Big)
  \, , 
  \EQN GowExactLinear;a \cr
  & \hbox{with \ \ }
  Z_{\pm n}\!(t) = a_{\pm n} J_0( \vert n \vert t )
                 + b_{\pm n} Y_0( \vert n \vert t )
  \hbox{ \ \ for } n \neq 0
  \, , \cr 
  Q_{\SmallWord{linear}} & {} = 
    c_0 \ln t + d_0
    + \sum_{n=1}^{\infty} \Big( W_{+n}\!(t) \cos(n\theta)
                              + W_{-n}\!(t) \sin(n\theta) \Big)
  \, , 
  \EQN GowExactLinear;b \cr
  & \hbox{with \ \ }
  W_{\pm n}\!(t) = c_{\pm n} J_0( \vert n \vert t )
                 + d_{\pm n} Y_0( \vert n \vert t )
  \hbox{ \ \ for } n \neq 0
  \, , \cr
}$$
for~$a_{\pm n}$, $b_{\pm n}$, $c_{\pm n}$ and~$d_{\pm n}$ constants,
which were derived earlier as equations~\EqNum{GowExactPert}{} (with the
parameter~$\beta$ here set to zero).

For the polarized Gowdy spacetimes in which the variable~$Q$ is
identically zero, the equation~\EqNum{GowExactLinear}{a} for~$P$
is the exact solution even when the variable is large 
(compare equation~\EqNum{GowPolarP}{}), and so gravitational waves in a
polarized Gowdy spacetime are always linear.
(It is interesting to note that many numerical studies of planar
cosmological spacetimes have, as noted in the introduction to this
chapter, only considered polarized metrics, and this may go some way
to explaining why the behaviour observed there has been
essentially linear.)
Although the linear solution~\EqNum{GowExactLinear}{b} for the
variable~$Q$ has the same form as the
solution~\EqNum{GowExactLinear}{a} for~$P$, the behaviour of~$Q$ is
genuinely nonlinear for large amplitude waves.
In addition, wave motion in~$Q$ excites wave motion in~$P$:
the variable~$P$ cannot be identically zero unless the variable~$Q$ is
trivial.

The numerical simulations described in this subsection use initial data
sets of the following form, for which the constraint
equation~\EqNum{GowRevConstr}{} is satisfied exactly:
$$\eqalign{
  &P = \ldots \, , \cr
  &P_{,t}= 0 \, , \cr
} \qquad \eqalign{
  &Q = \ldots \, , \cr
  &Q_{,t} = 0 \, , \cr
} \qquad \eqalign{
  &\lambda = 0 \, , \cr
  &\hbox{at time }t = t_0 \, , \cr
} \EQN GowPacketInitial
$$
where~$P$ and~$Q$ have initial values (specified below) based on the
Gaussian-like function
$$
  \varphi(\theta;\mu) = e^{\mu (\sin\theta-1)}
  \, , \EQN GowWavePacket
$$
in which~$\mu$ is a measure of the pulse width.
This function has a profile similar to a Gaussian distribution, but is
preferred for the present work since it is spatially periodic.
If a single wave packet having the initial form $P=\varphi(\theta)$, 
$P_{,t}=0$ is evolved with the flat space wave equation
$P_{,tt}=P_{,\theta\theta}$ then the behaviour observed is as follows:
the initial wave packet splits into two similar wave packets each of half
the original amplitude which travel in opposite directions with unit
speed; after one half cycle the two smaller wave packets collide and
recombine into their original form.
For the linearized Gowdy evolution equations the behaviour is similar
to this at late times (as can be seen from
equations~\EqNum{GowExactLinear}{} if the large parameter
approximations~\EqNum{BessApprox}{} to the Bessel functions are used)
with the difference that the amplitudes of the wave packets decrease
in time. 
The qualitative behaviour of wave packets in the linear regime can be
used as a reference against which the evolution of nonlinear wave
packets can be compared.

Ove~(1990a,b) examined the nonlinear evolution of planar Gaussian wave
packets (of both polarizations) using a numerical code for evolving
$U(1)$-symmetric cosmologies (discussed further in chapter~7).
For a particular choice of background spacetime he found that in the
linear limit of his field equations gravitational waves evolved
according to the flat space wave equation, and he subsequently
identified a qualitative difference between the evolutions of large
and small amplitude wave packets which he concluded to be a nonlinear
effect: when a large amplitude wave packet splits in two it leaves in
its wake an approximately constant value 
which differs from the value zero which would be expected for
linear behaviour.
This effect he termed `gravitational wave residue'.
Anninos, Centrella and Matzner~(1991b) also evolved nonlinear wave
packet initial data in a cosmological setting, though for a very
different choice of metric variables and gauge, and they too reported
behaviour consistent with Ove's wave residue effect.
(While similar behaviour is also described in Anninos et al.~1997 for
studies of gravitational waves in Minkowski spacetime, in that case it
is identified as being an artifact of lack of convergence in the
numerical code.)

In the present work, variable and coordinate choices differ from those
of the studies mentioned above, and while a gravitational wave residue
effect is observed in numerical simulations of wave packet initial
data, here it is seen to be more of a linear than a nonlinear effect.
Figure~\use{Fg.GowWavePacket} illustrates this point.
Two sets of initial data, both comprising large amplitude wave
packets, were evolved:
$$\eqalign{
  &\hbox{case I:} \cr
  &\hbox{case II:} \cr
} \qquad \eqalign{
  &P = 5 \, \varphi(\theta;10) \, , \cr
  &P = 0 \, , \cr
} \qquad \eqalign{
  &Q = 0 \, , \cr
  &Q = 5 \, \varphi(\theta;10) \, , \cr
} \qquad
  \hbox{for } t_0 = 1 \, ,
  \EQN GowTwoPack
$$
in conjunction with equations~\EqNum{GowPacketInitial}{}
and~\EqNum{GowWavePacket}{}.
A base grid of resolution $\Delta\theta = 2\pi/400$ was used, with
adaptive mesh refinement allowing the local resolution to increase to
$\Delta\theta = 2\pi/1600$.
As the figure shows, in both cases a residue is left behind when
the wave packet splits, and the residue is larger for the wave
variable~$P$ (case~I) than for the wave variable~$Q$ (case~II).
However, since the variable~$Q$ is identically zero in case~I, the
solution is a polarized Gowdy spacetime.
Consequently the evolution
of the variable~$P$, and the appearance of wave residue within it,
is completely determined by a linear equation with
exact solution~\EqNum{GowExactLinear}{a}.
While the formation of gravitational wave residue in the cosmological
model considered here appears to be principally a linear effect, the
residue itself can show nonlinear behaviour: in
figure~\use{Fg.GowWavePacket} the size of the residue is seen to differ
for the two cases considered, but in the linear regime (when the
amplitudes of the wave packets in equations~\EqNum{GowTwoPack}{} are
small) the behaviour, and the size of the residue, is the same for both
of the wave variables.

\figure{GowWavePacket}
\DrawFig{
  \epsfxsize=\hsize
  \epsfbox{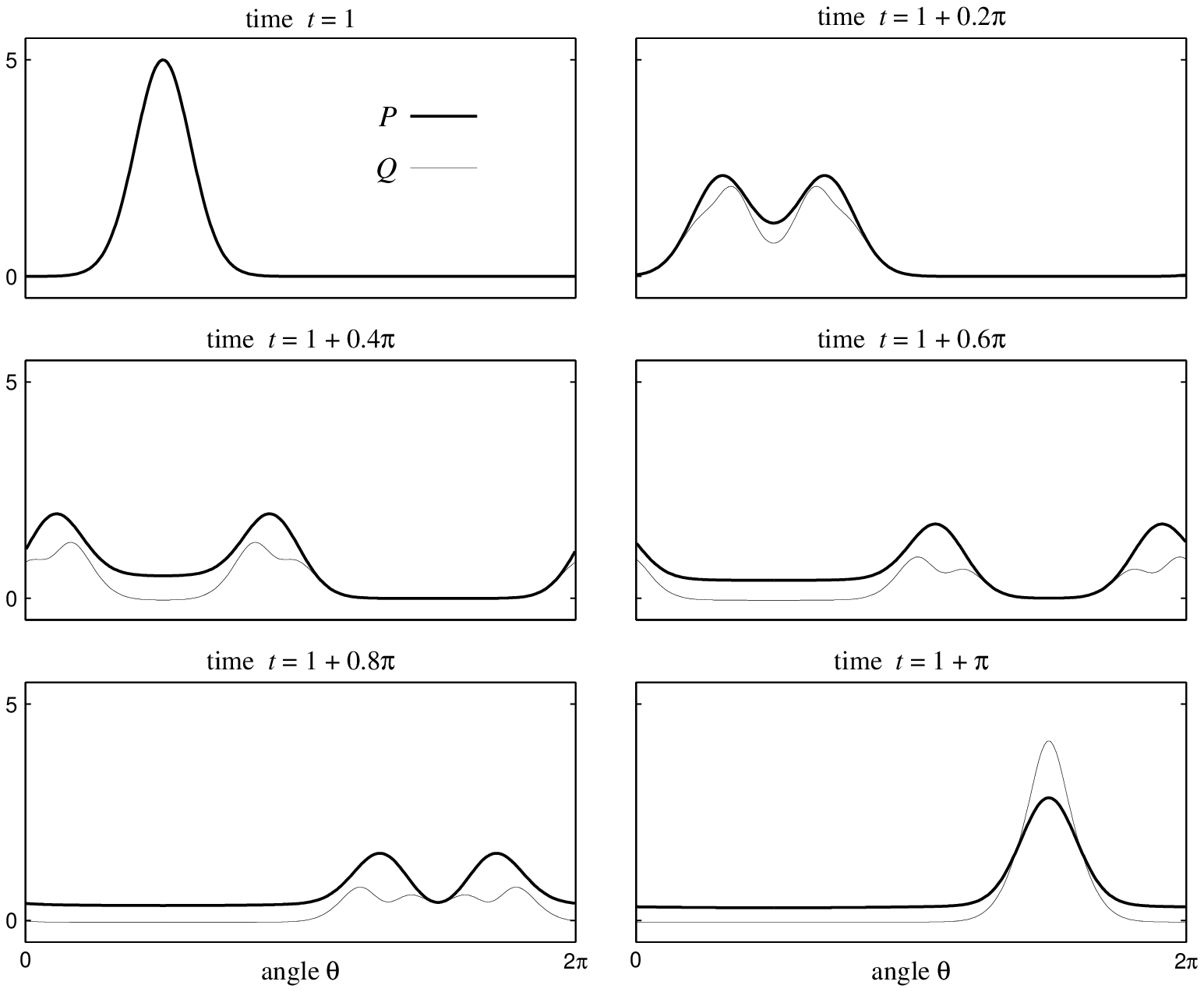}
}
\Caption
Two simulations of wave packet break-up and reconciliation.
The two initial data sets of equations~\EqNum{GowTwoPack}{} (with
equations~\EqNum{GowPacketInitial}{} and~\EqNum{GowWavePacket}{})
are evolved numerically in an expanding Gowdy~$T^3$ spacetime.
The plots show the variable~$P$ from the `case~I' simulation (wave
packet in~$P$ only) and the variable~$Q$ from the `case~II' simulation
(wave packet in~$Q$ only) on the same axes.
If the waves were linear the two variables would behave identically.
(A `residue' is left when the wave packet in~$Q$ splits apart, but its
size is small compared to the vertical scale of the plots.)
\bigskip
\endCaption
\endfigure

\figure{GowWaveShape}
\DrawFig{
  \epsfxsize=0.65\hsize
  \epsfbox{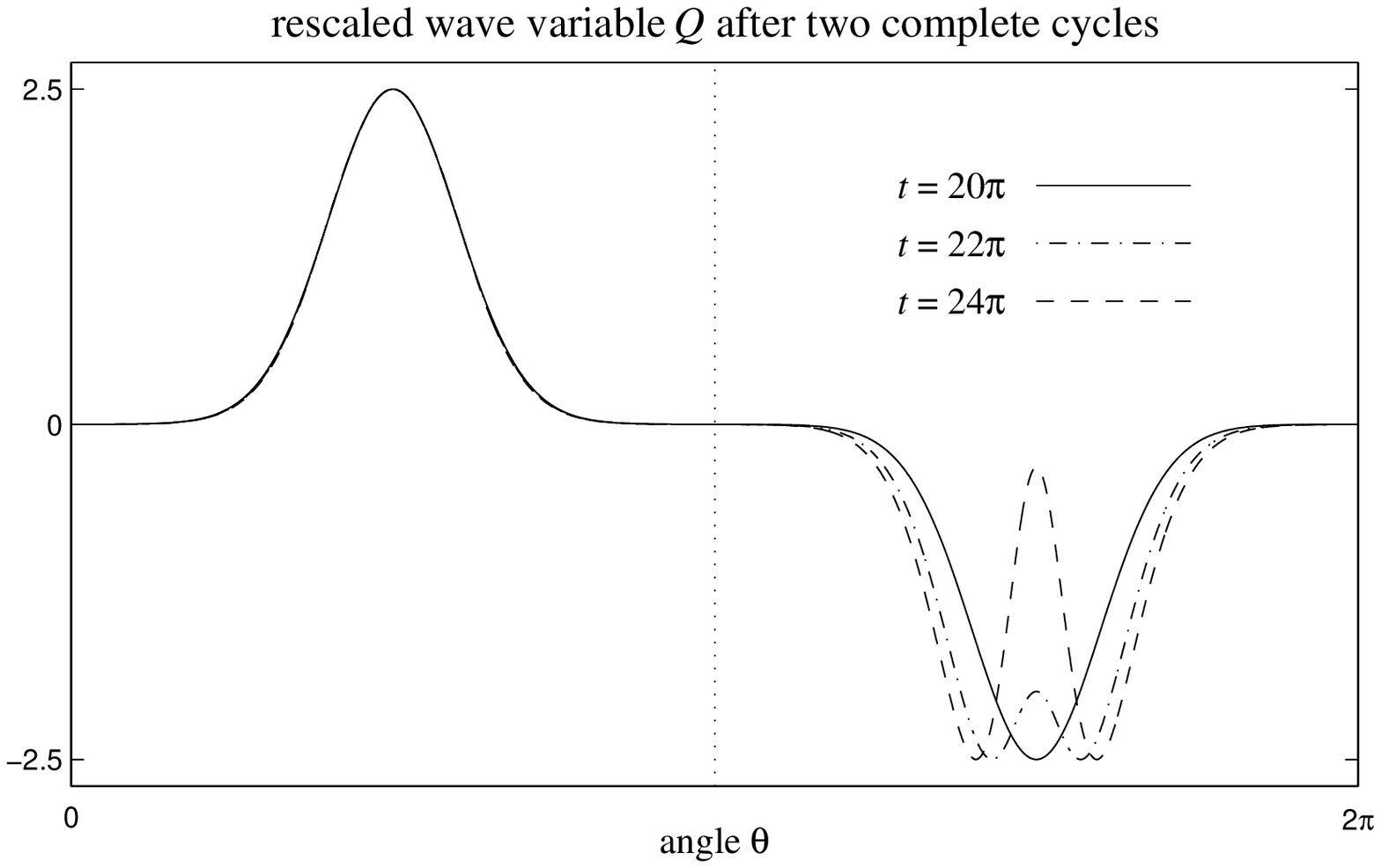}
}
\Caption
Deformation of wave packet shape during nonlinear evolution.
The initial data set~\EqNum{GowShapeInitial}{} (together with
equations~\EqNum{GowPacketInitial}{} and~\EqNum{GowWavePacket}{}) is
evolved numerically for two cycles, each of duration $\Delta t = 2\pi$.
The variable~$Q$ is plotted at the initial time and at the end of each
cycle, and to facilitate comparison it is linearly rescaled: the value
of~$Q$ at $\theta = \pi$ is made to be zero and the amplitude of each
wave packet (the two halves of the spatial domain are treated
independently) is made to equal the amplitude of the wave packets in the
initial data.
For linear wave behaviour no deformation of the wave packet shape
would be expected.
\bigskip
\endCaption
\endfigure

Another aspect of gravitational wave behaviour that can be
investigated numerically is the effect that nonlinearity
has on the shape of a propagating wave packet.
In their independent studies,
Anninos, Centrella and Matzner~(1991b) and Shinkai and Maeda~(1994)
found wave packet behaviour to be predominantly linear, and comparisons
can be made with the behaviour of soliton waves
(see Boyd, Centrella and Klasky~1991; informally, a {\it soliton} is
a nonlinearly propagating pulse wave which maintains its shape under
collision with other solitons).
Figure~\use{Fg.GowWavePacket} shows that for the cosmological model
studied here, nonlinear wave packets do not necessarily maintain their
pulse shape when they propagate; however when two sub-packets recombine
they do seem to take on a form very similar to that of the
wave packet that produced them.
The following numerical simulation tests how well the shape of a wave
packet is maintained between cycles.

It can been seen from equations~\EqNum{GowExactLinear}{} together with
the large parameter approximations for the Bessel
functions~\EqNum{BessApprox}{} that at late times the profiles of the
wave variables~$P$ and~$Q$ repeat themselves (to within linear
rescaling) with time period~$\Delta t = 2\pi$,
provided that their behaviour is linear (or that the spacetime
is polarized).
To see to what extent the same statement can be made for nonlinear
gravitational waves, the initial data set
$$
  P = Q = 2.5( \varphi(+\theta;10) - \varphi(-\theta;10) )
  \quad \hbox{for } t_0 = 20\pi
  \, , \EQN GowShapeInitial
$$
(together with equations~\EqNum{GowPacketInitial}{}
and~\EqNum{GowWavePacket}{}) is numerically evolved and the results
compared at the end of each cycle.
(The `double pulse' form of equation~\EqNum{GowShapeInitial}{} is
used to try to keep the average values of the wave variables around
zero; this avoids ambiguity if the waves are to be considered as propagating
in a particular homogeneous background spacetime.)
A basic spatial resolution of $\Delta\theta = 2\pi/8000$ is used for
the simulation with adaptive mesh refinement allowing a maximum
resolution of $\Delta\theta = 2\pi/128\,000$.
Figure~\use{Fg.GowWaveShape} shows the wave variable~$Q$ at time
intervals of $\Delta t = 2\pi$;
so that changes in shape of the wave packets can be seen clearly, the
two halves of the plot have been linearly rescaled.
The positive wave packet on the left-hand side of the figure has
maintained its shape in just the same way as would be expected for
linear wave behaviour, and this is also true of the two wave packets in
the variable~$P$ (which is not shown here).
In contrast, the influence of nonlinearity on the negative wave
packet on the right-hand side of the figure is clear: the wave packet
has been deformed from its original shape.

While some of the results of the investigations of gravitational wave
behaviour near to the linear regime presented here are in agreement
with those of earlier studies, others are not.
It is apparent that different conclusions will be drawn for studies
based on different cosmological models: in addition to the familiar
problems caused by different time slicing conditions (as discussed in
chapter~5), the choice of variables used to describe the metric also 
plays a role (in particular the identification of certain variables as
descriptors of wave behaviour), as does the path through parameter
space that is taken to relate nonlinear initial data sets to linear
ones. 
In some ways the Gowdy models are a more natural basis for studies of
gravitational wave behaviour than other models that have been used (for
example they naturally decompose the metric into wave and
background variables); however without nonlinear gravitational waves
being described in a covariant manner, results
produced using one particular model cannot be claimed to be
representative of the general case.


\Chapter{{\TbfBoldItalic U}(1)-Symmetric Cosmologies}
This chapter is a continuation of the work described in chapter~6 on
numerical simulations of inhomogeneous vacuum cosmological models.
The spacetimes considered in the previous chapter had two spacelike
Killing vectors and three-torus spatial topology, and were
interpreted as closed planar cosmologies.
In the present chapter the symmetry restrictions on the spacetimes
are relaxed slightly so that only one spacelike Killing vector
is present: the cosmologies are spatially compact with spacelike
$U(1)$~isometry groups.
In an analogous way to the use in chapter~6 of Gowdy's~(1974)
reduction of the Einstein equations to describe planar cosmologies, the
numerical simulations performed here use the equations of
Moncrief's~(1986) reduction for $U(1)$-symmetric cosmologies, and
the relevant features of the reduction are discussed in
section~\use{sec.MonIntro}.

In section~\use{sec.MonCollapse} results are presented from numerical
simulations performed to investigate the behaviour of collapsing
unpolarized $U(1)$-symmetric cosmologies.
As reported by Berger and Moncrief~(1998b), these spacetimes develop
spatial structure on very fine scales in an analogous way to the
collapsing unpolarized Gowdy cosmologies of
section~\use{sec.GowCollapse}, and here the adaptive mesh refinement
code described in chapter~3 is used to investigate the formation of
this fine-scale structure at high grid resolutions.

A series of numerical tests of the Frittelli-Reula formulation of
Einstein's equations (introduced in chapter~4) are described in
section~\use{sec.MonFR}, with planar and $U(1)$-symmetric spacetimes
providing reference solutions against which different evolution
systems are compared.
The reasons underlying the differences in numerical accuracy of a
range of formulations are investigated, and in the process an
alternative form of the Frittelli-Reula system is derived in which the
variables used are adapted to cosmological spacetimes with
$U(1)$~symmetry.

\Section{Cosmologies with Spacelike {\tbfBoldItalic U}(1) Isometry Groups
         \label{sec.MonIntro}}
In chapter~6, Gowdy's~(1974) reduction of the Einstein equations for
spacetimes with two spacelike Killing vectors and compact spatial
hypersurfaces was used as the basis for an investigation of planar
cosmological spacetimes.
The evolution equations for the Gowdy spacetimes have a particularly
simple form, attained through the removal of unnecessary metric
variables using coordinate conditions and constraint equations.
In the present chapter cosmological models with one spacelike Killing
vector are studied, and as before a special form for the metric is
assumed so that the Einstein equations can be reduced to a reasonably
simple form.
The reduction is due to Moncrief~(1986), with further specialization
by Berger and Moncrief~(1998a).
The cosmological models that result are vacuum with spacelike~$U(1)$
isometry groups and spatial slices having three-torus~($T^3$)
topology, and this section summarizes their derivation, governing
equations and basic properties.

\SubSection{Moncrief's Reduction of the Einstein Equations}
In Moncrief~1986, Gowdy's (1974) approach to the reduction of the
Einstein equations for cosmological spacetimes with two spacelike
Killing vectors is generalized to cosmologies with a single Killing
vector. 
Two variables representing the dynamical degrees of freedom of the
spacetime are identified, and the remaining variables are eliminated
by solving constraints and applying coordinate conditions.
In this subsection the main features of the reduction are summarized,
and in the next subsection the reduced Einstein equations for a
particular $U(1)$-symmetric cosmological model are presented.

Moncrief's reduction applies to vacuum spacetimes with
spacelike~$U(1)$ isometry groups on $S^1$-bundles over 
$\RealSet \times M$ where~$M$ is an arbitrary compact two-manifold.
In Moncrief~1986 the particular case of~$M$ being the two-sphere is
considered, since the analysis is then slightly simpler.
For the present work the manifold~$M$ is taken to be the two-torus,
and the $S^1$-bundle is constructed as a simple product so that
spatially the cosmologies have three-torus topology.
(For simplicity, only product $S^1$-bundles are considered in the brief
discussion below.)
A time coordinate~$\tau$ is introduced together with coordinates
$\{x^1,x^2,x^3\}$ on the spatial slices, 
with~$\partial / \partial x^3$ being the Killing vector for the
isometry group.
The metric is then written in the form
$$
  ds^2 = e^{-2\phi} d\sigma^2 + e^{2\phi} (dx^3 + \BoldBeta)^2
  \, , 
  \EQN MonGenMetric;a
$$
where
$$
  d\sigma^2 = - \tilde{N}^2 d\tau^2 
              + \tilde{g}_{ab} ( dx^a + \tilde{N}^a d\tau )
                               ( dx^b + \tilde{N}^b d\tau )
  \, ,
  \EQN MonGenMetric;b
$$
$$
  \BoldBeta = \beta_0 d\tau + \beta_a dx^a
  \, ,
  \EQN MonGenMetric;c
$$
and summation for the indices~$a$, $b$,~\dots\ is over the
range~$\{1,2\}$.
Equation~\EqNum{MonGenMetric}{b} shows~$d\sigma^2$ to have the ADM form
for a Lorentzian metric on the three-manifold~$\RealSet \times M$,
with the two-dimensional lapse~$\tilde{N}$, shift~$\tilde{N}^a$, and
induced metric~$\tilde{g}_{ab}$ all independent of the
coordinate~$x^3$.
The scalar field~$\phi$ and the one-form~$\BoldBeta$ from
equations~\EqNum{MonGenMetric}{a,c} are defined on
the same manifold.

The above decomposition of the metric bears a close resemblance to
Kaluza-Klein-Jordan theory, but involves one less `spacetime'
dimension.
The three-metric~$d\sigma^2$ satisfies the three-dimensional analogue
of Einstein's equations with source terms provided by the fields~$\phi$
and~$\BoldBeta$, and as a consequence of its dimensionality~$d\sigma^2$
carries no dynamical degrees of freedom.
In Moncrief's reduction the dynamical variables are the fields~$\phi$
and~$\BoldBeta$, and each represents one degree of freedom per
spacetime point.
The single dynamical degree of freedom present in the
one-form~$\BoldBeta$ can be made explicit by introducing a new
field~$\omega$ which represents the general solution to a constraint
imposed on~$\BoldBeta$ by the Einstein equations; with~$\beta_0$ in
equation~\EqNum{MonGenMetric}{c} set to zero without loss of
generality, the one-form~$\BoldBeta$ can always be 
reconstructed from a knowledge of the field~$\omega$.

If the three-metric~$d\sigma^2$ is given, then the dynamical
variables~$\phi$ and~$\omega$ are found to satisfy simple wave equations.
Conversely, Moncrief has shown that coordinate conditions can be imposed
on the two-dimensional lapse~$\tilde{N}$ and shift~$\tilde{N}^a$
in equation~\EqNum{MonGenMetric}{b} such
that the three-metric~$d\sigma^2$ can be uniquely determined
from~$\phi$ and~$\omega$ (and their conjugate momenta) on each
spatial slice by solving elliptic equations which derive
from the Einstein constraints.
(This is subject to the condition that certain constraint equations
hold on the initial spatial slice.) 
Thus, in effect the dynamical variables~$\phi$ and~$\omega$ propagate
in a background three-dimensional spacetime described by the
metric~$d\sigma^2$ which in turn is determined at every instant by an
elliptic system involving~$\phi$ and~$\omega$.

Moncrief's reduction of the Einstein equations was motivated in part
by its possible use as a tool in numerical relativity, and this idea
was followed up by Ove~(1986, 1989). 
Ove constructed a code implementing Moncrief's formalism for
$U(1)$-symmetric cosmologies with three-torus~($T^3$) spatial topology
(the numerical boundary conditions in this case being trivial).
The code evolved the dynamical variables~$\phi$ and~$\omega$ using a
standard time-stepping method, and reconstructed the background metric
on each new time slice by solving a system of elliptic equations.
Ove used his code to address the question of whether singularities can
form in expanding $U(1)$-symmetric~$T^3$ cosmologies, and his results
support the tentative conclusion that they cannot (Ove~1989, 1990c).
(As mentioned in chapter~6, Moncrief~1981 proves the corresponding
result for expanding Gowdy~$T^3$ models.)
Other work performed with the code (Ove~1990a, 1990b) which investigates
the nonlinear behaviour of gravitational waves has been discussed in
section~\use{sec.GowExpand}. 

The formulation of Moncrief's reduction which is used for the
numerical simulations of $U(1)$-symmetric~$T^3$ cosmologies
described in the present chapter differs from the one used by Ove; it
is described in detail in the next subsection.
The most significant difference between the formulations is in the
choice of time slicing: in the present work harmonic slicing is
used so that results can be readily compared with simulations
performed using the Frittelli-Reula system,
while in Ove's code (as in the original work by Moncrief) a
two-dimensional form of constant mean curvature slicing was
employed. 

Some analytic results have been derived using Moncrief's reduction for
$U(1)$-symmetric spacetimes on $S^1$-bundles over the two-sphere.
Choquet-Bruhat and Moncrief~(1994) have proved a local (in time)
existence and uniqueness theorem.
Moncrief~(1987) has generated in power series form a family of
$U(1)$-symmetric spacetimes which have curvature singularities.
Grubi{\u s}i\'c and Moncrief~(1994) have used perturbative techniques
to find $U(1)$-symmetric spacetimes which are related to the mixmaster
solutions and which presumably have oscillatory behaviour at the
initial singularity (in contrast to the velocity-dominated behaviour
typically found in the Gowdy models; see
section~\use{sec.GowSing}). 

The approach of Moncrief~1986 has been applied to other
$U(1)$-symmetric spacetimes.
Moncrief~1990 considers the Einstein-Maxwell equations on arbitrary
$S^1$-bundles over the two-sphere and also the Einstein-Maxwell-Higgs
equations on the three-torus (the latter case allowing interactions
between parallel cosmic strings to be studied).
Varadarajan~1995 reduces the vacuum Einstein equations for spacetimes
with one spacelike Killing vector which are asymptotically flat in an
appropriate sense.
The connection between Moncrief's reduction and the Einstein equations
for three-dimensional spacetimes has been mentioned above, and
in Moncrief~1989 (2+1)-dimensional pure gravity is considered
explicitly.

\SubSection{A $U(1)$-Symmetric Cosmological Model on the Three-Torus}
Berger and Moncrief~(1998a, 1998b) have performed numerical
simulations of vacuum $U(1)$-symmetric~$T^3$ cosmologies using a
formulation based on Moncrief's~(1986) reduction of the Einstein
equations.
Their approach differs from that of Moncrief's original work (and the
numerical work of Ove discussed in the previous subsection) in that
their evolution system does not involve any elliptic equations: the
Einstein constraint equations are not enforced during the evolution
and a local time slicing condition is used.
The formulation of Berger and Moncrief, described in their~1998a
paper, is adopted for the work in the present chapter and is
summarized in this subsection.

The metric used is the same as in
equations~\EqNum{MonGenMetric}{} but with the following conditions
imposed: 
$$
  \tilde{N} = e^{\Lambda-2\tau}
  \, , \qquad
  \tilde{N}^a = 0
  \, , \qquad
  \tilde{g}_{ab} = e^{\Lambda-2\tau} e_{ab}
  \, , \qquad
  \beta_0 = 0
  \, . \EQN MonSpecMetric
$$
The coordinate~$\tau$ denotes time while the spatial slices (of~$T^3$
topology) are described by the coordinates $\{x^1,x^2,x^3\} = \{u,v,w\}$
which are periodic on the domain~$[0,2\pi)$.
For the summation convention used in the present work,
the indices~$\mu$, $\nu$,~\dots\ range over $\{0,1,2,3\}$,
the indices~$i$, $j$,~\dots\ range over $\{1,2,3\}$,
and the indices~$a$, $b$,~\dots\ range over $\{1,2\}$.
The conformal metric~$e_{ab}$ of the $u$-$v$~plane has unit
determinant and components
$$\matrix{
  \textstyle 
  e_{uu} = {1\over2} ( e^{2z} + e^{-2z} [x+1]^2 )
  \, , \qquad
  e_{vv} = {1\over2} ( e^{2z} + e^{-2z} [x-1]^2 )
  \vrule width0pt depth12pt \, , \cr
  \textstyle 
  e_{uv} = e_{vu} = {1\over2} ( e^{2z} + e^{-2z} [x^2-1] )
  \, . \cr
} \EQN MonConfMetric
$$
Its inverse has components~$e^{uu} = e_{vv}$, $e^{vv} = e_{uu}$, 
and~$e^{uv} = e^{vu} = - e_{uv}$.

The spacetime is thus described by six variables:~$\Lambda$, $z$,
and~$x$ make up the (2+1)-dimensional `background' metric
while~$\phi$, $\beta_1$ and~$\beta_2$ carry the dynamical degrees of
freedom. 
All of the variables are periodic in~$x^1=u$ and~$x^2=v$, and
independent of the symmetry coordinate~$x^3=w$.
Combining equations~\EqNum{MonGenMetric}{} and~\EqNum{MonSpecMetric}{},
the metric of the $U(1)$-symmetric~$T^3$ model has the form
$$
  g_{\mu\nu} = \bordermatrix{
         & \tau & u & v & w \cr
    \tau & -e^{2\Lambda-4\tau-2\phi} & 0 & 0 & 0 \cr
    u    & 0 & \kappa_{11} & \kappa_{12} & e^{2\phi} \beta_1 \cr
    v    & 0 & \kappa_{21} & \kappa_{22} & e^{2\phi} \beta_2 \cr
    w    & 0 & e^{2\phi} \beta_1 & e^{2\phi} \beta_2 & e^{2\phi} \cr
  } \, ,
  \EQN MonMonMetric
$$
where
$$
  \kappa_{ab} = 
  e^{\Lambda-2\tau-2\phi} e_{ab} + e^{2\phi} \beta_a \beta_b
  \, .
$$

The ADM lapse and shift of the foliation can be read off
equation~\EqNum{MonMonMetric}{} for the metric:
$$
  \hbox{lapse } N = e^{\Lambda-2\tau-\phi}
  \, , \qquad
  \hbox{shift } N^i = 0
  \, . \EQN MonLapseShift
$$
(These quantities are not to be confused with their (2+1)-dimensional
analogues~$\tilde{N}$ and~$\tilde{N}^a$ from
equation~\EqNum{MonGenMetric}{b}.)
The determinant of the induced metric is equal to the square of the
lapse, and consequently the constant time hypersurfaces slice the
spacetime harmonically (with slicing density identically equal to one;
see equation~\EqNum{HypHarmFixed}{}).
Results produced using this formulation of Moncrief's reduction can
thus be readily compared with results from a two-dimensional
implementation of the Frittelli-Reula system, as
section~\use{sec.MonFR} demonstrates.

The number of metric variables used in equation~\EqNum{MonMonMetric}{}
is the same as the number used in the standard ADM description of a
spacetime (given that the lapse and the shift are known); however
Moncrief's reduction allows the two variables~$\beta_1$ and~$\beta_2$
to be replaced by a single variable~$\omega$ defined via
$$
  \omega_{,\tau} = e^{4\phi} \varepsilon^{ab} \beta_{a,b}
  \, , \qquad
  \omega_{,a} = - e^{4\phi+2\tau-\Lambda} e_{ab} 
                \varepsilon^{bc} \beta_{c,\tau}
  \, , \EQN MonOmegaDef
$$
where the alternating symbol~$\varepsilon^{ab}$ has non-zero
components~$\varepsilon^{12}=+1$ and~$\varepsilon^{21}=-1$.
(As a simplification of the formulation, two constants that should
appear in equation~\EqNum{MonOmegaDef}{} have been set to zero; in
Moncrief's original work these constants vanish identically because of
the two-sphere topology used.)
The variables~$\beta_1$ and~$\beta_2$ can be reconstructed
from~$\omega$ by integrating in time
$$
  \beta_{a,\tau} = e^{\Lambda-2\tau-4\phi}
                   e_{ab} \varepsilon^{bc} \omega_{,c}
  \, , \EQN MonBetaEvol
$$
with~$\beta_1$ and~$\beta_2$ being defined on an initial
slice~$\tau=\tau_0$ according to
$$\EQNalign{
  \beta_1 (u,v,\tau_0) & {} = c_1 + \lambda_{,u}
    + \int_0^v \Omega(u,\bar{v},\tau_0) \, d\bar{v}
    - {v \over 2\pi} \int_0^{2\pi} \Omega(u,\bar{v},\tau_0) \, d\bar{v}
  \, , 
  \EQN MonBetaInit;a \cr
  \beta_2 (u,v,\tau_0) & {} = c_2 + \lambda_{,v}
    - {1 \over 2\pi} \int_0^u \! \int_0^{2\pi} 
      \Omega(\bar{u},\bar{v},\tau_0) \, d\bar{v} \, d\bar{u}
  \, , 
  \EQN MonBetaInit;b \cr
}$$
where the function~$\lambda(u,v)$ and the constants~$c_a$ are
arbitrary (and here are typically set to zero) and~$\Omega$ is the
conjugate momentum of~$\omega$ defined 
by $\Omega = \exp(-4\phi) \, \omega_{,\tau}$.
The requirement that~$\beta_1$ and~$\beta_2$ be periodic translates to
a constraint on the initial value of~$\Omega$:
$$
  \int_0^{2\pi} \! \int_0^{2\pi} 
    \Omega(u,v,\tau_0) \, du \, dv = 0
  \, . \EQN MonOmegaConstr
$$

Evolution equations for the five variables of the Moncrief
reduction can be derived by varying the Hamiltonian given in
Berger and Moncrief~1998a:
$$\EQNalign{
  \Lambda_{,\tau\tau} & {} = \textstyle
    e^{\Lambda-2\tau} (
        e^{-\Lambda} [ e^{\Lambda} e^{ab} ]_{,ab}
      - 2 e^{-2z} \varepsilon^{ab} z_{,a} x_{,b}
      + e^{ab} \Pi_{ab}
    ) \, ,
  \EQN MonEvolve;a \cr
  z_{,\tau\tau} & {} = \textstyle
    {} - {1\over2} e^{-4z} x_{,\tau}{}^2
    - {1\over2} e^{\Lambda-2\tau-2z}
      \varepsilon^{ab} \Lambda_{,a} x_{,b}
    - {1\over4} e^{\Lambda-2\tau} e^{ab}{}_{\!,z} \Pi_{ab}
    \, ,
  \EQN MonEvolve;b \cr
  x_{,\tau\tau} & {} = \textstyle
    4 z_{,\tau} x_{,\tau}
    + e^{\Lambda - 2\tau + 4z} (
        2 e^{-2z} \varepsilon^{ab} \Lambda_{,a} z_{,b}
      - e^{ab}{}_{\!,x} \Pi_{ab}
      ) \, ,
  \EQN MonEvolve;c \cr
  \phi_{,\tau\tau} & {} = \textstyle
      ( e^{\Lambda-2\tau} e^{ab} \phi_{,a} )_{,b}
    + {1\over2} e^{-4\phi} (
          e^{\Lambda-2\tau} e^{ab} \omega_{,a} \omega_{,b}
        - \omega_{,\tau}{}^2
      )
    \, ,
  \EQN MonEvolve;d \cr
  \omega_{,\tau\tau} & {} = \textstyle
      e^{4\phi-2\tau} ( e^{\Lambda-4\phi} e^{ab} \omega_{,a} )_{,b}
    + 4 \phi_{,\tau} \omega_{,\tau}
    \, ,
  \EQN MonEvolve;e \cr
}$$
where
$$
  \textstyle
  \Pi_{ab} = \Lambda_{,ab} + 2 \phi_{,a} \phi_{,b}
           + {1\over2} e^{-4\phi} \omega_{,a} \omega_{,b}
  \, .
$$
It is a straightforward matter to put these equations into a
conservative first-order form by defining the first derivatives
of~$\Lambda$, $z$, $x$, $\phi$ and~$\omega$ to be new variables.
(In section~\use{sec.MonFR} this is done explicitly.)
Considering only the principal part of the evolution system, the wave
variables~$\phi$ and~$\omega$ can be seen to decouple from each other.
The wave equations~\EqNum{MonEvolve}{d} and~\EqNum{MonEvolve}{e}
are strongly hyperbolic and have non-trivial characteristic speeds
equal to the coordinate light speeds of the
metric~\EqNum{MonMonMetric}{}, these being
$$\eqalign{
  \hbox{coordinate light speed in $u$-direction} & {} =
    \pm e^{\Lambda/2-\tau} \sqrt{e^{uu}}
    \, , \cr
  \hbox{in $v$-direction} & {} =
    \pm e^{\Lambda/2-\tau} \sqrt{e^{vv}}
    \, . \cr
} \EQN MonLightSpeeds
$$
The sub-system for the evolution of the background
variables~$\Lambda$, $z$ and~$x$ has the same characteristic speeds
but is only weakly hyperbolic since the eigenvectors of the principal
part do not form a complete set.
(This lack of strong hyperbolicity is not altogether surprising
considering that in Moncrief's original reduction the background
variables were determined by a system of elliptic equations.)
One method by which the Moncrief equations can be made strongly
hyperbolic is explored in section~\use{sec.MonFR}.

The evolution system~\EqNum{MonEvolve}{} is constrained by the
Hamiltonian constraint equation, ${\cal H}_0=0$, and by two momentum
constraint equations, 
${\cal H}_u={\cal H}_v=0$, where
$$\EQNalign{
  {\cal H}_0 & {} = \textstyle
      2 z_{,\tau}{}^2
    + {1\over2} e^{-4z} x_{,\tau}{}^2
    + 2 \phi_{,\tau}{}^2
    + {1\over2} e^{-4\phi} \omega_{,\tau}{}^2
    - {1\over2} (2 - \Lambda_{,\tau})^2
  \cr & {} \textstyle \quad
    + e^{\Lambda-2\tau} ( 
        \Lambda_{,a} e^{ab}{}_{\!,b}
      + e^{ab}{}_{\!,ab}
      - 2 e^{-2z} \varepsilon^{ab} z_{,a} x_{,b}
      + e^{ab} \Pi_{ab}
    ) \, ,
  \EQN MonConstr;a \cr
  {\cal H}_a & {} = \textstyle
      4 z_{,\tau} z_{,a}
    + e^{-4z} x_{,\tau} x_{,a}
    + (2-\Lambda_{,\tau}) \Lambda_{,a}
    + \Lambda_{,\tau a}
    + 4 \phi_{,\tau} \phi_{,a}
  \cr & {} \textstyle \quad
    + e^{-4\phi} \omega_{,\tau} \omega_{,a}
    + \varepsilon^{bc} ( 
          {1\over2} e^{-2z} e_{ab,z} x_{,\tau}
        - 2 e^{2z} e_{ab,x} z_{,\tau}
      )_{,c}
  \, .
  \EQN MonConstr;b \cr
}$$

Moncrief's $U(1)$-symmetric cosmologies can be polarized in the same
way as the Gowdy spacetimes: as equation~\EqNum{MonEvolve}{e} shows,
if the wave variable~$\omega$ and its time derivative are zero on an
initial spatial slice then they are zero throughout the spacetime.
The corresponding statement does not hold for the other wave
variable~$\phi$ except when both~$\phi$ and~$\omega$ are identically
zero. 
Equations~\EqNum{MonBetaEvol}{} and~\EqNum{MonBetaInit}{} show that in
a polarized spacetime the vanishing of~$\omega$ implies that the
variables~$\beta_1$ and~$\beta_2$ are constants with respect to time
and satisfy~$\beta_{1,v}=\beta_{2,u}$.

If in a numerical simulation the five variables~$\Lambda$, $z$, $x$,
$\phi$ and~$\omega$ are evolved using the
equations~\EqNum{MonEvolve}{} then equations~\EqNum{MonBetaEvol}{} can
be evolved at the same time to find the values of~$\beta_1$
and~$\beta_2$. 
Alternatively,~$\beta_1$ and~$\beta_2$ could be used as variables
instead of~$\omega$ with the evolution equation~\EqNum{MonEvolve}{e}
being replaced by the two equations
$$
  \beta_{a,\tau\tau} = 
    (\Lambda_{,\tau} - 2 - 4\phi_{,\tau}) \beta_{a,\tau}
  + e_{ab,\tau} e^{bc} \beta_{c,\tau}
  + e^{\Lambda-2\tau} e_{ab} \varepsilon^{bc} \varepsilon^{de}
    ( 4 \phi_{,c} \beta_{d,e}
    + \beta_{d,ec} )
  \, , \EQN MonBetaVar
$$
and the derivatives of~$\omega$ in equations~\EqNum{MonEvolve}{a--d}
being replaced by derivatives of~$\beta_1$ and~$\beta_2$ using
equations~\EqNum{MonOmegaDef}{}. 
(Clearly the reintroduction of~$\beta_1$ and~$\beta_2$ as variables in
the system is a step backwards in terms of simplifying the Einstein
equations; however it is useful when making comparisons between
numerical simulations, as section~\use{sec.MonFR} demonstrates.)
If~$\beta_1$ and~$\beta_2$ are used as variables then the system
acquires an additional constraint equation, ${\cal H}_{\beta}=0$,
where
$$
  {\cal H}_{\beta} =
    e^{ab} (4\phi - \Lambda)_{,a} \beta_{b,\tau}
  + ( e^{ab} \beta_{a,\tau} )_{,b}
  \, . \EQN MonBetaConstr
$$
This constraint is equivalent to the requirement that the spatial
derivatives of~$\omega$ commute.

A useful coordinate-independent quantity defined at each point of the
spacetime is the norm of the Killing vector~$\partial / \partial w$
associated with the~$U(1)$ isometry group, and from
equation~\EqNum{MonMonMetric}{} it can be seen that
$$
  \hbox{proper length of $\partial / \partial w$ symmetry orbits}
    = 2 \pi e^{\phi}
  \, . \EQN MonOrbitLength
$$
The~$U(1)$ symmetry of the spacetime thus provides the variable~$\phi$
with a simple physical interpretation, and the numerical results
presented in section~\use{sec.MonCollapse} focus attention on this
variable.

\SubSection{Initial Data for Numerical Simulations}
A set of initial data for the system of evolution
equations~\EqNum{MonEvolve}{} comprises values for the
variables~$\Lambda$, $z$, $x$, $\phi$ and~$\omega$ and their first
time derivatives for which the constraint values~${\cal H}_0$,
${\cal H}_u$ and~${\cal H}_v$ from equations~\EqNum{MonConstr}{}
vanish, and which in addition satisfy equation~\EqNum{MonOmegaConstr}{}.
Clearly in the general case constructing sets of initial data
requires the numerical solution of elliptic equations; however
Berger and Moncrief~(1998a) have found a non-trivial family of
initial data sets for which the constraint equations can be solved
algebraically.
This family of initial data sets is used for the numerical simulations
performed in the present work, and its construction is summarized in
this subsection.

First of all the momentum constraints ${\cal H}_u={\cal H}_v=0$
(see equation~\EqNum{MonConstr}{b}) can be satisfied on an initial
hypersurface by requiring that
$$
  \Lambda_{,\tau} = 2 - c e^{\Lambda}
  \, , \qquad
  z_{,\tau} = x_{,\tau} = 0
  \, , \qquad
  \phi = \omega = 0
  \qquad \hbox{at time } \tau = \tau_0
  \, , \EQN MonInitMom
$$
for some constant value~$c$.
(In fact the requirement is that~$\phi$ and~$\omega$ be spatially
constant, not necessarily zero.
Here~$\phi$ is set to zero as a slight simplification; there is no loss
of generality in taking~$\omega$ to be zero since only its derivatives
appear in the evolution equations.)
A value for~$\omega_{,\tau}$ is then chosen such that
$$
  \int_0^{2\pi} \! \int_0^{2\pi} 
    \omega_{,\tau} \, du \, dv = 0
  \qquad \hbox{at time } \tau = \tau_0
  \, , \EQN MonInitOmega
$$
which derives from equation~\EqNum{MonOmegaConstr}{}.
The Hamiltonian constraint ${\cal H}_0=0$ (see
equation~\EqNum{MonConstr}{a}) can then be satisfied by allowing it to
determine the value of~$\phi_{,\tau}$ at time $\tau=\tau_0$ given values
for~$\Lambda$, $z$, $x$ and~$\omega_{,\tau}$ then:
$$
  \textstyle
    \phi_{,\tau}{}^2 =
      {1\over4} c^2 e^{2\Lambda}
    - {1\over4} \omega_{,\tau}{}^2
    - {1\over2} e^{\Lambda-2\tau_0} (
        \Lambda_{,ab} e^{ab}
      + \Lambda_{,a} e^{ab}{}_{\!,b}
      + e^{ab}{}_{\!,ab}
      - 2 e^{-2z} \varepsilon^{ab} z_{,a} x_{,b}
    ) \, .
  \EQN MonInitHam
$$
The right-hand side of this can always be made non-negative by
choosing a sufficiently large value for the constant~$c$.
It is convenient to choose the sign of~$\phi_{,\tau}$ to be the
same as the sign of~$c$.

A set of initial data can thus be constructed based on arbitrary
(spatially periodic) values of~$\Lambda$, $z$ and~$x$, and a
value of~$\omega_{,\tau}$ satisfying equation~\EqNum{MonInitOmega}{}.
The sign of the constant~$c$ determines whether the spacetime
initially expands ($c$~is negative) or collapses ($c$~is positive) as
the time~$\tau$ increases.
This can be seen from the mean curvature of the spatial slices,
$$
  \hbox{mean curvature }K = K^{i}{}_{i}
    = ( \phi_{,\tau} + 2 - \Lambda_{,\tau} )
      e^{\phi + 2\tau - \Lambda}
  \, , \EQN MonMeanCurv
$$
where~$K_{ij}$ is the extrinsic curvature tensor of the foliation.
The mean curvature~$K$ is a measure of the convergence of the world
lines that run normal to the spatial slices: when~$K$ is positive the
world lines are converging and when it is negative they are diverging.
It is straightforward to show that on the initial slice the sign
of~$K$ is the same as the sign of~$c$ (provided that the sign
of~$\phi_{,\tau}$ is chosen as described above).

\Section{Collapsing {\tbfBoldItalic U}(1)-Symmetric Cosmologies
         \label{sec.MonCollapse}}
In this section results are presented from numerical simulations of
vacuum $U(1)$-symmetric~$T^3$ cosmologies.
As mentioned in section~\use{sec.MonIntro} these spacetimes have also
been studied numerically by Ove~(1986, 1989,~1990b) and by Berger and
Moncrief~(1998a, 1998b).
Ove investigated nonlinear gravitational wave effects and singularity
formation in expanding spacetimes, and while his work is based on the 
same formulation of Einstein's equations as is used here (Moncrief~1986)
the time slicing condition employed is quite different.
This difference in gauge, together with differences in the approaches
used to generate initial data sets, makes it impractical to use Ove's
results to test the code used for the work described in this chapter.

Berger and Moncrief have investigated the behaviour of
$U(1)$-symmetric cosmologies as they collapse towards `big bang'
singularities.
Their findings extend previous results for the Gowdy cosmological
models (Berger and Moncrief~1993; see section~\use{sec.GowSing}): for
polarized $U(1)$-symmetric spacetimes the approach to the singularity
has the quality of a velocity-term dominated solution of the Einstein
equations (Berger and Moncrief~1998a), while in the unpolarized case
the behaviour close to the singularity is oscillatory, consistent with
the Belinskii-Khalatnikov-Lifschitz model of a generic inhomogeneous
singularity (Berger and Moncrief~1998b).
In the present work the same formulation of Einstein's equations is used
as in Berger and Moncrief~1998a, but a different numerical method is
used to evolve solutions.
(As in their earlier work, Berger and Moncrief use a symplectic
integration algorithm.)
The behaviour of polarized $U(1)$-symmetric spacetimes evolved using
the code described in the present work is found to be consistent with
the (mostly qualitative) results presented in Berger and
Moncrief~1998a.

In their studies of unpolarized $U(1)$-symmetric spacetimes, Berger and
Moncrief~(1998b) found that structure develops in the evolved solutions
on very small spatial scales in much the same way as it does in
collapsing unpolarized Gowdy~$T^3$ cosmologies (see
section~\use{sec.GowCollapse}).
However, the inability of their code to adequately resolve this spatial
structure led to it becoming unstable, and this they countered by
spatially averaging the solution data between time steps.
The advantage of this smoothing process is that it allows numerical
solutions to be evolved up to much later times than would otherwise be
possible; the disadvantage is that it prevents the numerical solution
from accurately modelling the exact solution.
Berger and Moncrief's justification for using spatial averaging is that
they are interested in the qualitative behaviour of the spacetime as it
approaches the singularity, not in the details of the local features
that develop in it.

In this section the adaptive mesh refinement code described in
chapter~3 is used to study the formation of fine-scale features in
collapsing unpolarized $U(1)$-symmetric spacetimes.
In contrast to the work of Berger and Moncrief~(1998b), the spacetimes
are evolved only for short time periods and the emphasis is on the
accurate resolution of the fine-scale features.

\SubSection{Numerical Simulations of Collapsing
            $U(1)$-Symmetric $T^3$ Cosmologies}
Compared to the numerical simulations of Gowdy spacetimes presented in
chapter~6, simulations of $U(1)$-symmetric spacetimes are considerably
more difficult to perform: more computer resources are needed because of
the larger number of variables and spatial dimensions,
and there is a greater tendency for the evolution to become unstable
because of the complexity of the governing equations.
For these reasons the numerical simulations presented in this section
explore a narrower range of behaviour than the simulations of chapter~6.
This subsection describes the numerical apparatus used to study
collapsing unpolarized $U(1)$-symmetric~$T^3$ cosmologies; numerical
results are presented in the following subsection.
Simulations of polarized and expanding spacetimes have also been
performed, but the results are less interesting and are
not discussed here.

The simulations are based on Moncrief's~(1986) reduction of the
Einstein equations, detailed in section~\use{sec.MonIntro}.
The spacetime is described by five principal variables,~$\Lambda$,
$z$, $x$, $\phi$ and~$\omega$, and by defining the first derivatives
of these with respect to the coordinates~$u$, $v$ and~$\tau$ to be
additional variables, a first-order flux-conservative form for the
evolution equations~\EqNum{MonEvolve}{} is obtained (recall
equation~\EqNum{PdeConsLaws}{}). 
Additional values~$\beta_1$ and~$\beta_2$ are required in order for
the metric~\EqNum{MonMonMetric}{} to be reconstructed, and these are
evaluated during the course of the simulation by including
equations~\EqNum{MonBetaEvol}{} as part of the evolution system.
Since the variable~$\omega$ does not explicitly appear in the
evolution equations (only its derivatives appear) it can be omitted
from the system, and so twenty-one variables in total are evolved.
The evolution system is unconstrained in the sense that the
Hamiltonian and momentum constraints from
equations~\EqNum{MonConstr}{} are only imposed at the initial
time~$\tau=\tau_0$. 

As pointed out in section~\use{sec.MonIntro} the evolution
equations~\EqNum{MonEvolve}{} are only weakly hyperbolic, and so the
high-resolution integration methods of chapter~2 are inapplicable
here.
The numerical results in this section were generated using the
standard two-step Lax-Wendroff method, with a splitting approach being
used to separate transport and source terms (the latter are treated
using a second-order Runge-Kutta method); details are given in
section~\use{sec.PdeDifference}.
As in the collapsing Gowdy case (section~\use{sec.GowCollapse}) the
size of the time step~$\Delta\tau$ is allowed to increase during the
course of the simulation, up to a maximum size of about twice the
spatial mesh width, provided that the Courant-Friedrichs-Lewy
condition is not violated (recall the discussion of variable time
steps in section~\use{sec.AmrDetails}).
Periodic boundary conditions are used as appropriate for the toroidal
spatial topology of the models under investigation.

Initial data sets for the simulations are constructed using the
approach (from Berger and Moncrief~1998a) described in
section~\use{sec.MonIntro}.
The particular initial data set used to produce the results presented
in this section has
$$\matrix{
  \textstyle 
  \Lambda = \alpha \sin u \sin v
  \, , \qquad
  z = x = \alpha \cos u \cos v
  \, , \qquad
  \omega_{,\tau} = \alpha \cos u \cos v
  \vrule width0pt depth12pt \, , \cr
  \textstyle 
  \beta_1 = \alpha \cos u \sin v
  \, , \qquad
  \beta_2 = 0
  \, , \qquad
  \hbox{with } \alpha = 0.1
  \hbox{ and } c = +1.2
  \hbox{ at time } \tau_0 = 0
  \, , \cr
} \EQN MonInitial
$$
together with equation~\EqNum{MonInitMom}{} and~$\phi_{,\tau}$
determined from equation~\EqNum{MonInitHam}{}.
The initial value for~$\omega_{,\tau}$ is consistent with the
condition~\EqNum{MonInitOmega}{}, and the initial values for~$\beta_1$
and~$\beta_2$ come from equations~\EqNum{MonBetaInit}{} with~$c_1$,
$c_2$ and~$\lambda$ taken to be zero.
The parameter~$\alpha$ measures, in a loose sense, the strength of the
gravitational waves present in the initial data: when it is zero the
spacetime is a homogeneous Kasner model, and for large values the
behaviour of the spacetime becomes complicated very quickly.
The initial data sets used in Berger and Moncrief~1998b have larger
amplitudes than those used here.

The values of~$\Lambda$, $z$ and~$x$ in the initial
data~\EqNum{MonInitial}{} can be varied without qualitatively changing
the behaviour of the evolved solution.
The appearance of interesting features in the spacetimes seems to be
largely determined by the value of~$\omega_{,\tau}$.
(The extreme case of~$\omega_{,\tau}$ being everywhere
zero results in a spacetime that is polarized.)
From figure~\use{Fg.MonPhi} it can be seen that the regions where
fine-scale spatial structure develops are close to points at
which~$\omega_{,\tau}$ is zero in the initial
data~\EqNum{MonInitial}{}, and this remains true when alternative
initial values for~$\omega_{,\tau}$ are used:
figure~\use{Fg.AmrLayout} shows a plot of the variable~$\phi$ for a
simulation in which the initial value of~$\omega_{,\tau}$ is zero at
points forming a circle in the spatial domain with centre~$u=v=\pi$. 

The numerical simulations discussed in this section make
use of the adaptive mesh refinement algorithm described in chapter~3.
The simulations use a coarse base grid of resolution
$\Delta u = \Delta v = 2\pi / 200$ which can be locally refined (using
two additional levels in the grid hierarchy) up to a maximum resolution
of $\Delta u = \Delta v = 2\pi / 3200$.
The main reason for using adaptive mesh refinement in this case is that
it allows emerging fine-scale spatial features to be resolved
without requiring a very fine mesh to be used over
the entire spatial domain (or for the whole duration of the simulation).
However, for the collapsing cosmological model studied here the usual
approach of selecting regions to be refined based on an estimate of
the local truncation error has been found to be inefficient: 
it leads to the refinement of a much larger portion of the spatial domain 
than is necessary.
To compensate for this the automatic refinement procedure has been set
to a relatively low level of sensitivity and the code has been explicitly
programmed to refine in regions where fine-scale features are known to
appear.
By using mesh refinement in this way simulations can be performed (up
to a final time of~$\tau \simeq 13$) in about five percent of the
time that would be required for a simulation based on a single uniform
grid as fine as the finest grid used here.
Interpolation is used as part of the adaptive mesh refinement
algorithm; in section~\use{sec.AmrDetails} the advantages and
disadvantages of several different interpolation methods are
considered.
The simulations described here used piecewise quadratic interpolation
to minimize problems caused by interpolating near to poorly resolved
fine-scale features, although in practice these problems only occur
at late times since the code ensures that the fine-scale features are
always refined to the highest level.

The Gowdy~$T^3$ cosmological models studied in chapter~6 are special
cases of $U(1)$-symmetric~$T^3$ spacetimes, and they provide a
useful test case for the $U(1)$-symmetric code.
Since the same slicing condition is used for both of the
metrics~\EqNum{GowMetricTau}{} and~\EqNum{MonMonMetric}{}, it is
straightforward to convert between the Gowdy variables~$P$, $Q$
and~$\lambda$ and the Moncrief variables~$\Lambda$, $x$, $z$, $\phi$
and~$\omega$.
The $U(1)$-symmetric code has been tested to check
that it reproduces the results of the Gowdy code and the exact
polarized solutions known for the Gowdy models.
Two further tests provide evidence that the $U(1)$-symmetric code
performs correctly when evolving more general spacetimes.
Firstly, the quantities evolved by the code always converge at second
order as the mesh width is reduced (in particular the constraint
values of equations~\EqNum{MonConstr}{} converge to zero).
Secondly, the results produced by the code converge to the same
solutions as the results of the independent Frittelli-Reula code
(discussed in chapter~4 and also section~\use{sec.MonFR}) which has
itself been tested on a different range of problems.
Because of the data smoothing used in their production (recall the
comments at the start of this section), the results presented in
Berger and Moncrief~1998b do not constitute a `known' solution against
which the code used in the present work can be tested; however some
qualitative similarities can be seen between figure~8 (at early times)
of Berger and Moncrief's paper and figure~\use{Fg.MonPhi} of the
present work despite different initial data sets having been used.

\SubSection{Fine-Scale Features in Collapsing
            $U(1)$-Symmetric Spacetimes}
Figure~\use{Fg.MonPhi} shows the metric variable~$\phi$ at several
stages during the evolution of the initial data~\EqNum{MonInitial}{}
using the numerical methods outlined in the previous subsection.
The variable~$\phi$ is the most easily interpreted of the variables
describing the evolved spacetime: as given in
equation~\EqNum{MonOrbitLength}{}, the proper distance around the
spacetime in the spacelike direction of symmetry is proportional
to~$\exp(\phi)$.
It is clear from the figure that during the collapse of the
unpolarized $U(1)$-symmetric spacetime, features develop in~$\phi$ on
small spatial scales: by time~$\tau \simeq 6$ sharp ridges have
formed which run parallel to the $u$-~and $v$-axes with large peaks
appearing where two ridges cross, and at later times (less obvious
in the figure) deep pits form in between the ridges, oriented
at forty-five degrees to the $u$-~and $v$-axes.
The nature of these fine-scale features and the difficulties they
cause for numerical simulations are the topics of the present
subsection. 

\fullfigure{MonPhi}
\vss
\DrawFig{
  \epsfxsize=\hsize
  \DrawHiLoRes{\epsfbox{phi.eps}}
              {\epsfbox{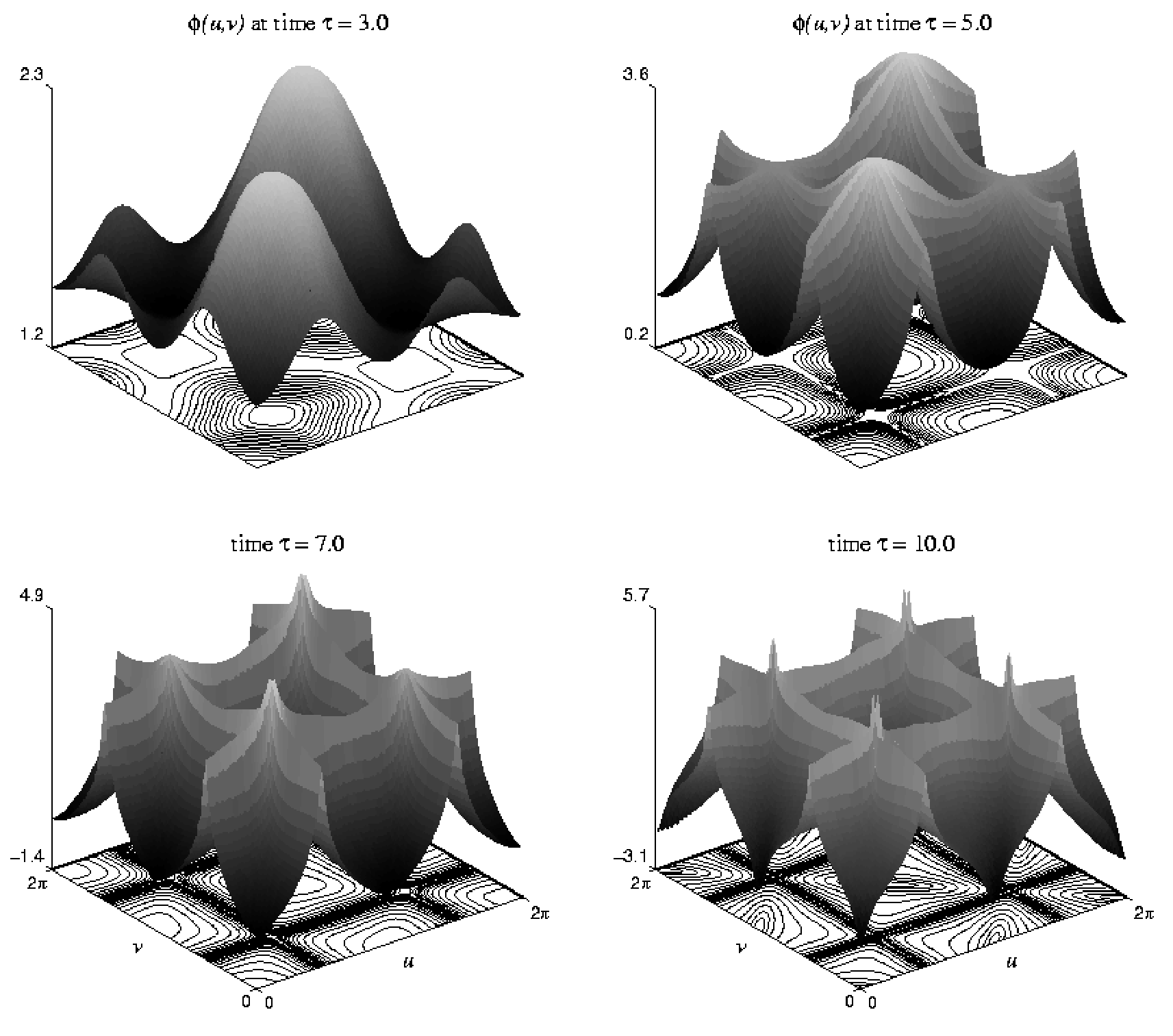}}
}
\Caption
Variable~$\phi(u,v,\tau)$ of collapsing $U(1)$-symmetric
spacetime~\EqNum{MonMonMetric}{} at four different times.
Surface and contour plots of the data are shown.
The initial data set~\EqNum{MonInitial}{} with
equations~\EqNum{MonInitMom}{} and~\EqNum{MonInitHam}{} is evolved
using equations~\EqNum{MonEvolve}{}. 
The resolution of the plots ($100\times100$ points in the
$u$-$v$~domain) is insufficient to resolve the fine-scale features
that develop at late times.
\vskip27mm
\endCaption
\endfigure

\fullfigure{MonPhiPeak}
\vss
\DrawFig{
  \epsfxsize=\hsize
  \DrawHiLoRes{\epsfbox{phipeak.eps}}
              {\epsfbox{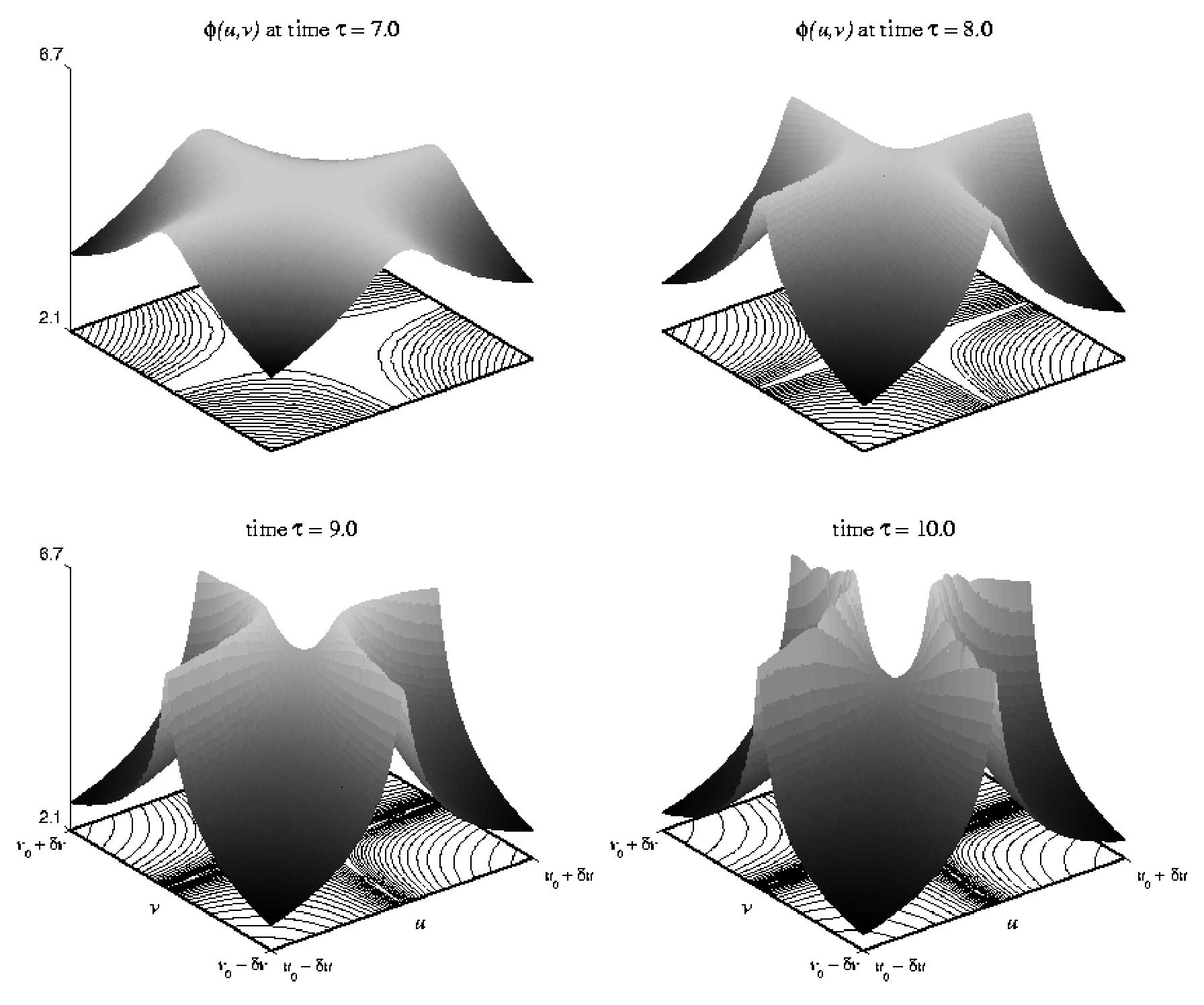}}
}
\Caption
Close-ups of a peak in the variable~$\phi(u,v,\tau)$ for the
simulation shown in figure~\use{Fg.MonPhi}.
The plots are all centred on $u_0=v_0=\pi/2$ and cover a region given
by $\delta u = \delta v = 0.12$.
The resolution of the plots ($100\times100$~points) is slightly less
than the resolution of the finest grids used in the simulation.
The same vertical axis is used for all the plots.
\vskip31mm
\endCaption
\endfigure

\fullfigure{MonOmega}
\vss
\DrawFig{
  \epsfxsize=\hsize
  \DrawHiLoRes{\epsfbox{omega.eps}}
              {\epsfbox{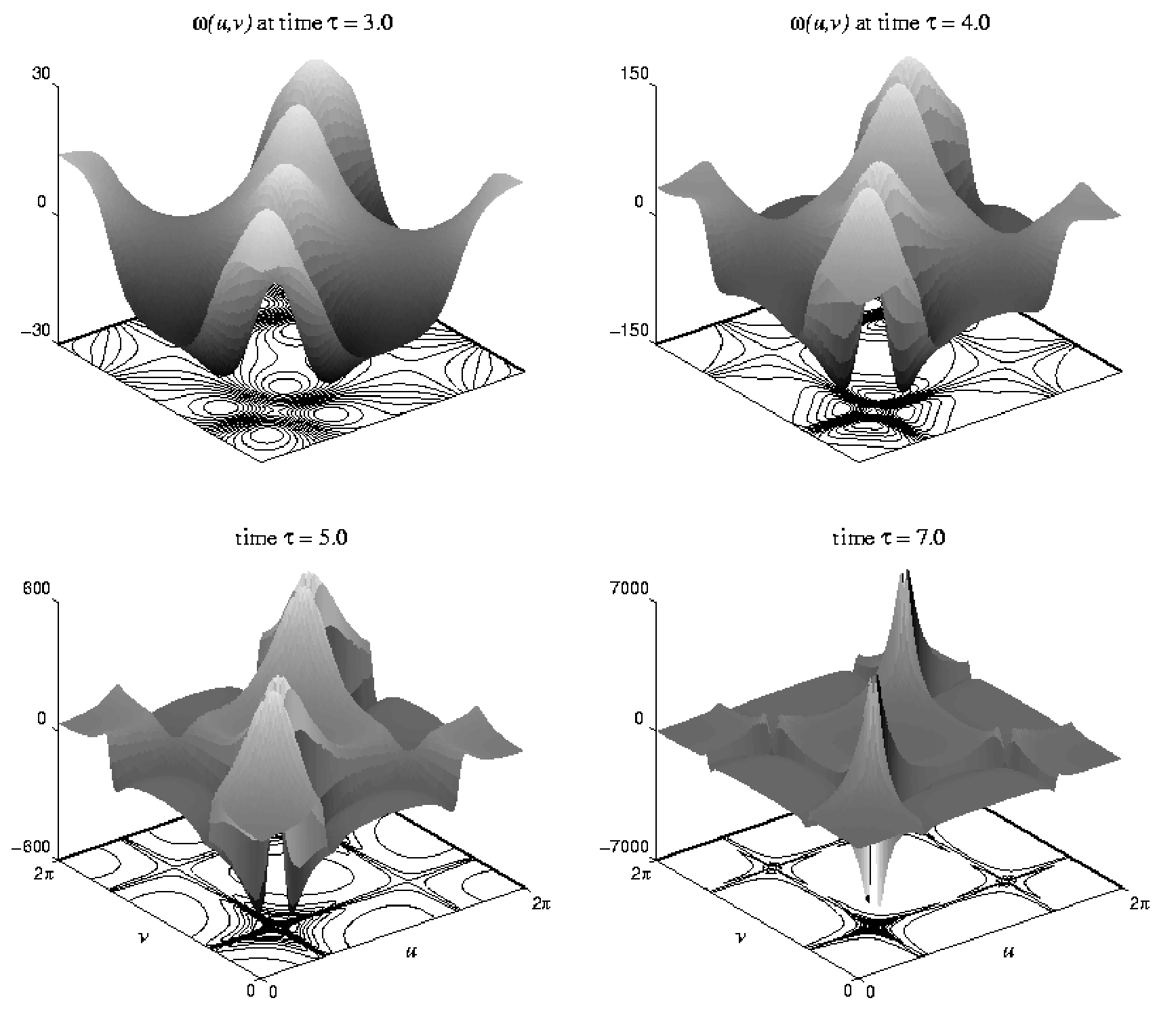}}
}
\Caption
Variable~$\omega(u,v,\tau)$ of the Moncrief
reduction~\EqNum{MonOmegaDef}{} for a collapsing $U(1)$-symmetric
spacetime~\EqNum{MonMonMetric}{} at four different times. 
Surface and contour plots of the data are shown.
The same initial data set is used as in figure~\use{Fg.MonPhi} but the
results are plotted at different times.
The resolution of the plots ($100\times100$ points in the
$u$-$v$~domain) is insufficient to resolve the fine-scale features
that develop at late times.
\vskip25mm
\endCaption
\endfigure

The sharp ridges that develop in the variable~$\phi$ bear a close
resemblance to the spikes that appear in simulations of collapsing
unpolarized Gowdy cosmologies, as described in
section~\use{sec.GowCollapse}. 
The ridges grow (towards positive infinity) at approximately linear
rates while their widths gradually decrease, and as the simulation
progresses they become increasingly difficult to resolve.
The numerical problems caused by the narrowing of the ridges are
similar to the problems caused by the spikes in the Gowdy simulations,
but they are more acute because of the additional spatial dimension.
In the Gowdy case the narrowing of the spikes causes their heights to
be inaccurately represented by the numerical solution; in the
$U(1)$-symmetric case the heights of the ridges become inaccurate in
an analogous manner, but with the error in the height varying along
the length of each ridge producing a `ripple' effect in the numerical
solution.
(Ripples along the ridges are just visible in the last plot of
figure~\use{Fg.MonPhi}.)
The use of mesh refinement techniques (as described in the previous
subsection) prolongs the time for which the simulation can keep the
ridges well resolved, but since the ridges seem never to stop
narrowing, eventually their representation in the numerical solution
must become inaccurate. 
By time~$\tau \simeq 10$ the resolution of the simulation used to
produce figure~\use{Fg.MonPhi} is insufficient to adequately resolve
the ridge features in the variable~$\phi$.

As figure~\use{Fg.MonPhi} shows, peaks appear in the
variable~$\phi$ at points where ridges cross, and like the ridges the
peaks sharpen as the simulation progresses.
However in contrast to the other fine-scale features that develop in
collapsing unpolarized $U(1)$-symmetric and Gowdy cosmologies, the
nature of the peaks changes at late times.
Figure~\use{Fg.MonPhiPeak} plots the variable~$\phi$ in close-up at
the position of one of the peaks ($u=v=\pi/2$), and it can be seen
that by time~$\tau \simeq 9$ the peak has split apart, separating the
intersecting ridges.
The value of~$\phi$ at the point where the peak used to be
subsequently decreases while the nearby ridges grow and narrow in the
usual way.
This behaviour is common to all of the peaks seen in
figure~\use{Fg.MonPhi}, although it takes place at different times for
different peaks.
Figure~\use{Fg.MonPhiPeak} is plotted using data from the finest grids
used in the simulation; at time~$\tau=10$ it is clear that the ridges
are imperfectly resolved.

A third kind of fine-scale feature appears in the variable~$\phi$ at
late times~($\tau \simeq 8$).
In between the ridges, at coordinate positions where both~$u$ and~$v$
are multiples of~$\pi$, linear growth of~$\phi$ towards negative
infinity produces steep-sided pits in the variable.
The pits are most obvious in the contour plots of
figure~\use{Fg.MonPhi}.
Like the ridge and peak features, the pits become narrow and are
difficult for the simulation to resolve.
However, unlike the ridges and peaks, the pits do not seem to represent
physical features of the spacetime: if~$\exp(\phi)$ (the
proper length of an orbit of the~$U(1)$ isometry group; see
equation~\EqNum{MonOrbitLength}{}) is plotted instead of~$\phi$, no
special features are seen at the positions where pits form.
(The pits in~$\phi$ are in some respects similar to the double-spike
features that appear in the collapsing Gowdy simulations of
section~\use{sec.GowCollapse}.) 

Figure~\use{Fg.MonOmega} shows the second wave-like variable~$\omega$
of the Moncrief reduction for the same simulation.
Like~$\phi$ the variable~$\omega$ is related to the Killing vector
field associated with the $U(1)$~symmetry of the spacetime, but
unlike~$\phi$ it does not have a direct physical interpretation.
Fine-scale features develop in the variable~$\omega$ in much the same
way as they do in the variable~$\phi$, and at the same locations.
At late times the profile of~$\omega$ is dominated by large spikes;
only earlier times are shown in the figure.

\figure{MonSpeeds}
\DrawFig{
  \epsfxsize=0.95\hsize
  \epsfbox{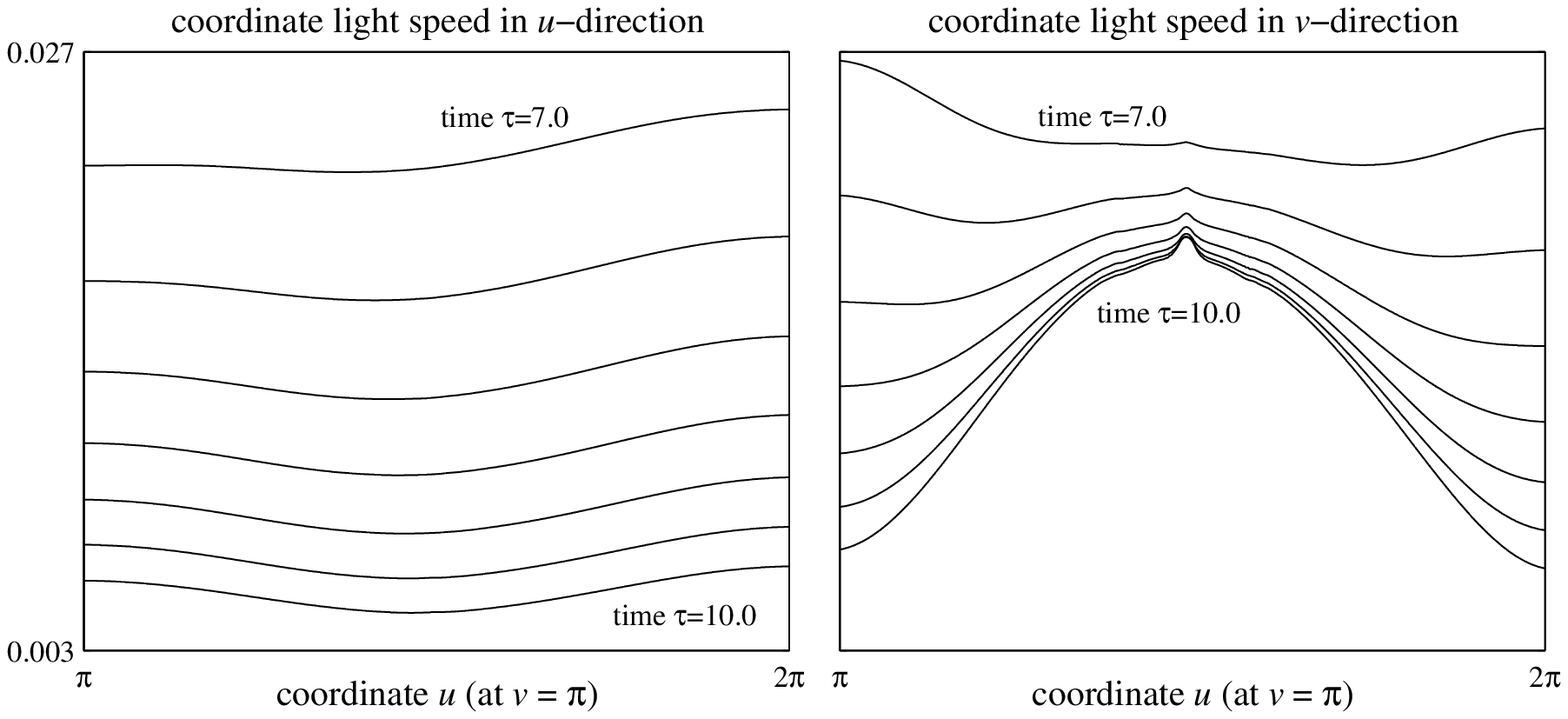}
}
\Caption
Characteristic speeds at a ridge feature in~$\phi$.
The speed parallel to the $u$-axis and the speed parallel to the
$v$-axis are plotted along a one-dimensional slice through the 
spatial domain given by $\pi \le u \le 2\pi$ and $v = \pi$.
The speeds are given by equation~\EqNum{MonLightSpeeds}{}.  
The ridge feature runs perpendicular to the slice and crosses it at
the centre of the plotted region.
The speeds are shown at equal time intervals for $7 \le \tau \le 10$.
While the characteristic speeds in the directions away from the ridge
decrease exponentially (left-hand plot), the speeds along the ridge
tend to a limiting value greater than zero (right-hand plot).
(Inaccuracies at the ridge are responsible for the lump in the middle
of the data shown in the right-hand plot.)
\bigskip
\endCaption
\endfigure

In the numerical simulations of collapsing Gowdy spacetimes
(section~\use{sec.GowCollapse}) it was found that although the spiky
features of the solution were badly resolved and greatly in error at
late times, the simulations could continue to run almost indefinitely.
This is not the case for the $U(1)$-symmetric simulations: poor
resolution of the fine-scale features at times beyond~$\tau\simeq10$
leads to the evolution becoming unstable in those regions, and the
code is forced to terminate at a time around~$\tau\simeq13$.
(The reason for this seems to be that in the $U(1)$-symmetric case lack of
resolution tends to occur over extended regions rather than at isolated
points.)
In section~\use{sec.GowCollapse} it is argued that large errors at the
Gowdy spikes do not necessarily contaminate the entire solution
because the exponential decay of the characteristic speeds of the
model, equation~\EqNum{GowSpeeds}{}, implies that the errors can 
propagate only a finite, and typically very small, distance before the
cosmological singularity is reached.
In the $U(1)$-symmetric case exact values for the characteristic
speeds, equation~\EqNum{MonLightSpeeds}{}, are not known and it is
harder to make statements about a spacetime's causal structure.
However numerical results, plotted in figure~\use{Fg.MonSpeeds}, show
that in regions where fine-scale features form the characteristic
speeds for information propagating away from the regions decrease at
approximately exponential rates, and so inaccuracies there can only
influence a very small part of the complete spacetime.
(In contrast the speeds for propagation within the regions do not
appear to tend to zero.)
This suggests that an approach like Berger and Moncrief's~(1998b) in
which fine-scale structure is removed from the solution by a
smoothing algorithm may be valid if applied selectively:
it would help stabilize the evolution beyond the time at which fine-scale
structure becomes inadequately resolved without significantly
affecting the majority of the solution.

Evidence supports the view that the ridge and peak features found in
the numerical solutions are not simply gauge effects.
As mentioned above, the proper lengths of the orbits of the~$U(1)$
isometry group (equation~\EqNum{MonOrbitLength}{}) are a physical
property of the spacetime which is affected by the presence of the
features. 
Furthermore the fine-scale nature of the features is not simply a
consequence of a poor choice of spatial coordinates: the proper
distances in the~$u$- and $v$-directions (from
equation~\EqNum{MonMonMetric}{} for the metric) are found to decrease
near to the peaks and ridges, and so the features are small in terms
of proper distance as well as coordinate distance.
The curvature invariants~$\InvA$, $\InvB$, $\InvC$ and~$\InvD$ were
presented in section~\use{sec.HomCompare} (specifically
equation~\EqNum{HomInvars}{}) as useful 
indicators of coordinate-independent behaviour in a spacetime, and
they have been evaluated for the $U(1)$-symmetric
solutions following the approach described in York~1989.
(The invariants are constructed from the four-dimensional Riemann
tensor which in turn is constructed from intermediate
quantities~$P_{ijkl}$, $Q_{ijk}$ and~$D_{ij}$ which in turn are
constructed from the induced metric, extrinsic curvature and
three-dimensional Riemann tensor of the spatial slices.)
Figure~\use{Fg.MonInvars} shows the behaviour of the curvature
invariant~$\InvA$ on a slice through the $u$-$v$~domain which crosses
an emerging ridge feature; the behaviour of the other invariants is
broadly similar. 
Two features of the behaviour of~$\InvA$ are familiar from the study
of curvature invariants in the Gowdy models
(section~\use{sec.GowCollapse}).
Firstly, the curvature invariants grow at approximately exponential
rates and their magnitudes become very large (compare
figure~\use{Fg.GowInvars}).
Secondly, the zeros of the curvature invariants are found to converge
on regions where fine-scale structure forms;
figure~\use{Fg.GowInvCloseup} shows the same behaviour occurring at a
spike in a Gowdy spacetime.

\figure{MonInvars}
\DrawFig{
  \epsfxsize=0.95\hsize
  \epsfbox{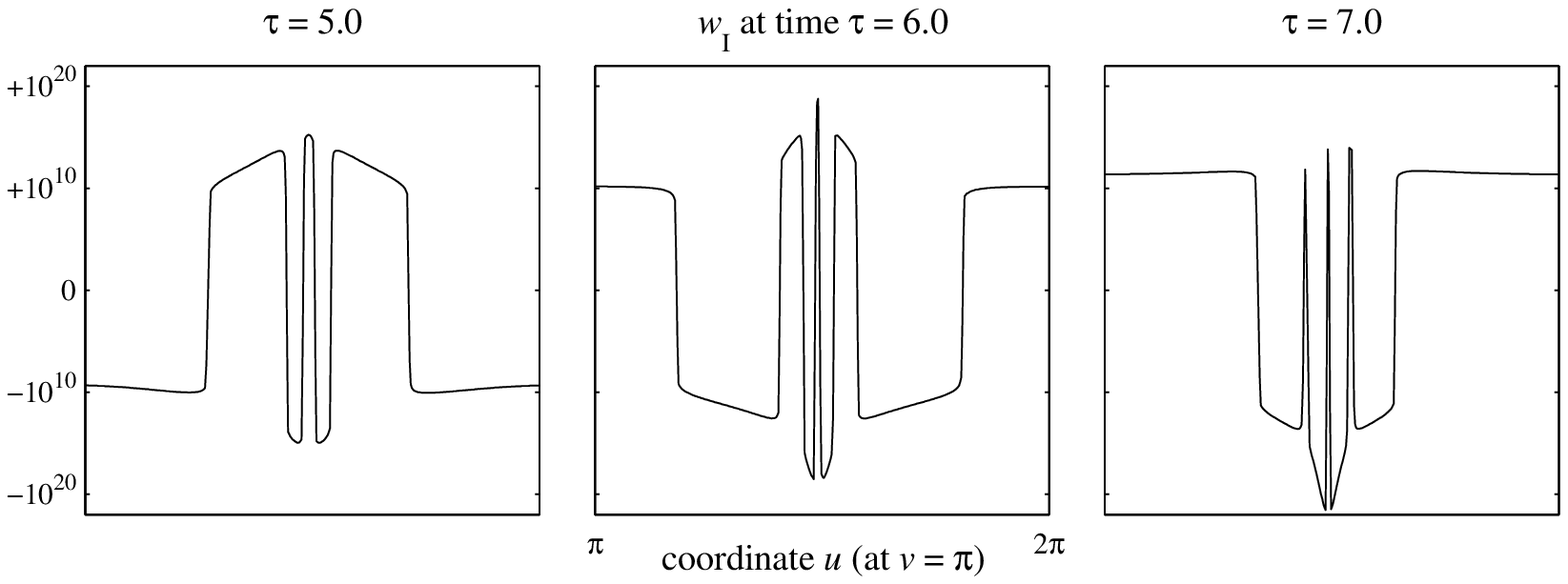}
}
\Caption
Curvature invariant~$\InvA$ at a ridge feature in~$\phi$
at three different times.
The invariant is plotted along a one-dimensional slice through the
spatial domain given by $\pi \le u \le 2\pi$ and $v = \pi$.
The ridge feature runs perpendicular to the slice and crosses it at
the centre of the plotted region.
The axes are the same for all three plots and a logarithmic scale for
the vertical axis has been produced by use of the
transformation $w \rightarrow (1 / \ln 10) \sinh^{-1} (w/2)$.
\bigskip
\endCaption
\endfigure

The ridge features in the unpolarized $U(1)$-symmetric
cosmologies appear to be the generalization of the spike features in
the unpolarized Gowdy cosmologies.
Since the Gowdy and Moncrief metrics, equations~\EqNum{GowMetricTau}{}
and~\EqNum{MonMonMetric}{}, use time coordinates defined by the same
harmonic slicing condition, it is a straightforward matter to identify
the $\theta$~coordinate of the former with the $u$~coordinate of the
latter and transform a Gowdy solution (described by variables~$P$, $Q$
and~$\lambda$) into expressions for the Moncrief variables.
(The reverse transformation is not generally possible: even if a
$U(1)$-symmetric spacetime possesses Gowdy symmetry it has more
coordinate freedom than allowed for by the Gowdy metric.)
Positive spikes in the Gowdy variable~$P$ are then seen to transform
into ridges in the Moncrief variable~$\phi$ according to
$$
  \phi(u,v,\tau) \simeq 
    ( P(u,\tau) - \tau )/2 + Q_0(u)
  \, ,
$$
for some function~$Q_0$,
where the assumptions have been made that~$P$ grows linearly and~$Q$
becomes constant at late times.
In the Gowdy spacetimes positive spikes in~$P$ are accompanied by
zeros in~$Q_{,\theta}$, and the generalization of this for
$U(1)$-symmetric spacetimes is that ridge features in~$\phi$ are
accompanied by zeros in~$\omega_{,\tau}$.
Given this connection between~$\phi$ and~$\omega_{,\tau}$ it is
possible to give a very rough sketch of how the process of ridge
formation may operate.
If~$\phi_\tau = \phi_{,\tau}$ and~$\omega_\tau = \omega_{,\tau}$ are
defined as variables and the evolution equations~\EqNum{MonEvolve}{d}
and~\EqNum{MonEvolve}{e} are simplified so that only terms involving
these variables are present then they read as
$$
  \textstyle
  \partial_\tau \phi_\tau \simeq
    - {1\over2} e^{-4\phi} \omega_\tau{}^2
  \, , \qquad
  \partial_\tau \omega_\tau \simeq
    4 \phi_\tau \omega_\tau
  \, .
$$
Making the assumptions that~$\omega_\tau$ crosses zero at the location
of a ridge and that~$\phi_\tau$ is positive there, the expected
behaviour would be for~$\phi_\tau$ to decrease except
where~$\omega_\tau$ is zero (and consequently for the ridge in~$\phi$
to sharpen) and for~$\omega_\tau$ to grow exponentially either side of
the zero; evidence of this latter behaviour can be seen in the plots
of~$\omega$ in figure~\use{Fg.MonOmega}. 
(In section~\use{sec.GowCollapse} it was noted that in regions where
spikes form in the Gowdy variable~$P$ the evolution equations are
dominated by spatial derivative terms, and this seems to contradict
the simplified model given above in which fine-scale structure is
produced by equations containing no spatial derivatives;
however equation~\EqNum{MonOmegaDef}{} shows that despite appearances the
variable~$\omega_\tau$ is a spatial rather than a temporal derivative
term.) 

From the results of this section and of section~\use{sec.GowCollapse}
it would seem that the formation of fine-scale spatial structure is a
generic feature of collapsing vacuum cosmologies, and presumably of
gravitational collapse in general since it can be inferred from the small
particle horizon sizes of the models considered here that the effect is
a local one.
The appearance of fine-scale structure is problematic for attempts
to numerically evolve collapsing spacetimes, although in
the present work mesh refinement techniques (described in chapter~3)
have been used to alleviate the problems to some extent.
For the numerical simulations of $U(1)$-symmetric cosmologies performed
here it has proven to be very difficult to reliably evolve solutions
for more than a modest interval of time; the question of what happens
to the fine-scale structure in the long-term (particularly as regards the
oscillatory behaviour of $U(1)$-symmetric spacetimes reported in Berger
and Moncrief~1998b) remains unanswered.


\Section{{\tbfBoldItalic U}(1)-Symmetric Cosmologies 
         and the Frittelli-Reula System \label{sec.MonFR}}
This section reports on a series of tests carried out to assess the
value of the Frittelli-Reula formulation (described in detail in
chapter~4) as a 
tool for the numerical evolution of Einstein's equations.
In particular the question is addressed as to whether there are gains
to be made in basing numerical simulations on evolution systems which
are strongly (or symmetric) hyperbolic or which have only physically 
relevant characteristic speeds.
The parameter range of the generalized Frittelli-Reula
formulation (presented in section~\use{sec.HypFR}) is explored here,
and an 
alternative hyperbolic evolution system is derived based on the
variables used in the Moncrief reduction of
section~\use{sec.MonIntro}. 
The tests are based on planar and $U(1)$-symmetric vacuum
cosmologies.

\SubSection{Testing the Frittelli-Reula System I: Planar Cosmologies}
Chapter~4 describes the construction of a computer code for the
evolution of
numerical solutions to the Frittelli-Reula (henceforth~FR) formulation
of the Einstein equations.
The code has been extensively tested to check that it produces correct
results: evolved solutions converge at the
expected order to solutions of the Einstein equations.
However this
says little about the overall accuracy of the code or its stability
over long time periods.
As a practical test of the suitability of the FR~formulation 
for numerical work the code has been used to evolve initial data
for planar and $U(1)$-symmetric cosmologies.
The results have been compared to numerical solutions produced by
codes based on the Gowdy 
(section~\use{sec.GowGowdy}) and Moncrief~(section~\use{sec.MonIntro})
reductions of the Einstein equations.
The analysis of the results for the $U(1)$-symmetric cosmologies is
the main topic of the present section; as a prelude to that, the
findings from the tests based on Gowdy~$T^3$ spacetimes are summarized
below.

The FR~system, figure~\use{Fg.HypFREqns}, is a fully three-dimensional
formulation of the Einstein equations---it assumes no
symmetries in the spacetimes it describes
(although as a simplification for the numerical work performed here,
simulated spacetimes are assumed to have at least one spacelike Killing
vector, as section~\use{sec.HypNumerical} explains).
For spacetimes which have a high degree of symmetry the Einstein
equations can often be reduced to very simple forms; the Gowdy~$T^3$
evolution equations~\EqNum{GowEinst}{} are a good
example of this, with coordinate conditions and constraint equations
having been applied to eliminate all but the essential variables.
It is to be expected that numerical simulations of planar cosmologies
performed using the Gowdy equations will be considerably more accurate
than equivalent simulations performed using the FR~equations 
(in addition to the former requiring significantly less
computational resources).
It is interesting to ask however exactly what the sources of the
discrepancies in accuracy are; the answers to this will guide the
comparison between the FR~system and the Moncrief reduction for
$U(1)$-symmetric cosmologies later in this section.

It is straightforward to compare results for the
FR~metric~\EqNum{HypADMMetric}{} with results for the Gowdy
metric~\EqNum{GowMetricTau}{}.
Since in both cases a harmonic time
slicing condition is used, the two coordinate systems can be made to
coincide by taking the slicing density in the FR~formulation to be
$Q(t,x^j)=1$ (see equation~\EqNum{HypFRGauge}{}) and the shift vector
to be~$N^i(t,x^j)=0$.
It is then a simple matter to convert between the Gowdy
variables~$P$, $Q$ and~$\lambda$, and the FR~variables~$h^{ij}$,
$M^{ij}{}_{k}$ and~$P^{ij}$. 
The FR~system is then tested as follows:
the exact polarized Gowdy solution~\EqNum{GowPolar}{} is numerically
evolved using both the FR~equations of figure~\use{Fg.HypFREqns} and
the Gowdy equations~\EqNum{GowEinst}{}, with the same spatial
resolution and the same time step being used in both cases.
The sizes of the errors in the numerical solutions (always measured
with respect to the FR~three-metric~$h^{ij}$) are monitored.

The test is applied first to the {\it modified\/} form of the
FR~system (figure~\use{Fg.HypFREqns} with parameter
settings~$\Theta=0$ and $\eta=\gamma=1$), and as expected the errors
in the evolved solution grow at a substantially faster rate than for
the Gowdy evolution system.
Part of the reason for this can be traced to the off-diagonal
components of the metric: while in the exact polarized solution (and
in the simulations based on the Gowdy equations) the metric always
maintains a diagonal form, truncation errors in the FR~simulations
can cause the off-diagonal metric components to take on non-zero
values.
(This effect is related to the inclusion of constraint quantities in
the FR~evolution equations; it does not occur if the
parameter~$\eta$ is zero.)
The performance of the FR~system can thus be improved by
reducing it in the following way: assume the spacetime metric to be
always diagonal, identify the variables in the system which are
identically zero given this assumption, and eliminate them from the
evolution equations.
This reduction vastly simplifies the evolution system: only nine
rather than thirty variables are needed, and as a side effect
non-physical characteristic speeds are no longer a feature of the
equations. 

The diagonal reduction of the FR~system still performs
significantly less well than the Gowdy system, and the next stage in
improving its accuracy is to change the variables used to a set that
are better adapted to the behaviour of the evolved spacetime.
Since the (non-zero) components of the metric ($h^{11}$,~$h^{22}$
and~$h^{33}$) grow or decay exponentially, the metric is better
described by the variables~$H_A$, $H_B$ and~$H_C$ where
$h^{11}=e^{H_A}$, $h^{22}=e^{H_B}$ and $h^{33}=e^{H_C}$.
The other FR~variables~$M^{ij}{}_{k}$ and~$P^{ij}$ can be changed
accordingly, and the result is a set of variables which are very similar
to the variables of the Gowdy reduction.
In addition to improving its overall performance, this
change of variables cures an explicit problem with the FR~system: 
since all real values for the variables~$H_A$, $H_B$ and~$H_C$
describe valid spacetime metrics, the possibility of errors in the
variables causing the determinant of the three-metric to become
non-positive (with the code crashing as a consequence) is excluded. 

A final reduction of the FR~system enables it to perform with an
accuracy comparable to that of the Gowdy system.
For the form of the FR~system derived above in which a diagonal metric
is represented by the variables~$H_A$, $H_B$ and~$H_C$,
the evolution equations are made
conspicuously more complex than they need to be by the presence of
constraint terms which vanish for exact solutions of the Einstein
equations; these terms can be cancelled from the system without
affecting its hyperbolicity or characteristic speeds.
The reduced form of the FR~system reached in this way is still more
general than the (polarized) Gowdy system (in particular it uses more
variables to describe the metric), but its accuracy in numerical
simulations while worse is only slightly so.

In conclusion, two differences between the FR~system and the Gowdy
system can be thought of as being responsible for the larger rate of
numerical error growth in the former. 
Firstly, the FR~evolution equations are made more complicated than
they could be by their inclusion of constraint terms.
Secondly, the variables of the FR~formulation are not as well
adapted as the Gowdy variables to the behaviour of the spacetimes
being evolved.
When assessing the performance of the FR~system in simulating
$U(1)$-symmetric spacetimes later in this section, these ideas play
a central role.
The effect that including constraint terms in the FR~evolution
equations has is investigated by varying the values of the parameters
in the generalized FR~formulation.
Also, in the next subsection, an alternative form of the
FR~system is produced in which the standard FR~variables are replaced
by the variables of the Moncrief reduction.

\SubSection{Adaption of the Frittelli-Reula Formulation to
            $U(1)$-Symmetric Spacetimes}
In this subsection a transformation is applied to the
generalized Frittelli-Reula~(FR) formulation of Einstein's equations
such that the FR~variables~$h^{ij}$, $M^{ij}{}_{k}$ and~$P^{ij}$ are
replaced by the variables~$\Lambda$, $z$, $x$, $\phi$, $\beta_1$
and~$\beta_2$ (together with their first derivatives) that describe
the Moncrief form of a $U(1)$-symmetric metric,
equation~\EqNum{MonMonMetric}{}.
The reasons for doing this are twofold.
Firstly, as suggested by the results of the previous subsection,
an evolution system having variables which are adapted to the spacetime
under investigation may lead to improved performance in numerical
simulations, and this idea is tested for $U(1)$-symmetric cosmologies in
the next subsection.
Secondly, as mentioned in section~\use{sec.MonIntro}, the evolution
equations~\EqNum{MonEvolve}{} of the Moncrief reduction for
$U(1)$-symmetric spacetimes are only weakly hyperbolic, and the
transformation of the FR~formulation described here provides a method
by which a strongly (or symmetric) hyperbolic
version of the Moncrief system can be produced.
The new form of the FR~system is given schematically by
equations~\EqNum{MonFR}{} with explicit expressions for the terms in
the system being given in figures~\use{Fg.MonFluxUCon}
to~\use{Fg.MonFluxTEvol}. 
Only an outline of the derivation is presented here; the algebra
involved in changing the variables of the FR~system is exceedingly
lengthy and was only made practical to perform in this work by the use
of the REDUCE computer algebra package (Hearn~1995, MacCallum and
Wright~1991). 

The generalized (vacuum) FR~evolution system,
figure~\use{Fg.HypFREqns}, is used here in a (2+1)-dimensional form:
$$\EQNalign{
  \partial_t \Boldh & {}
    + \partial_a \BoldF^a_{(h)\SmallWord{evol}} (\Boldh, N^i)
    + \BoldS_{(h)\SmallWord{evol}} (\Boldh, \BoldP, Q, N^i)
  = 0 \, ,
  \EQN MonFRStart;a \cr
  \partial_t \BoldM_1 & {}
    + \partial_a \Big[ \BoldF^a_{(M1)\SmallWord{evol}} 
                       (\Boldh, \BoldM_1, \BoldP, Q, N^i)
                     + \eta \times \BoldF^a_{(M1)\SmallWord{con}} 
                       (\Boldh, \BoldP, Q) \Big]
    \cr & {} 
    + \BoldS_{(M1)\SmallWord{evol}} 
             (\Boldh, \BoldM_b, \BoldP, Q, N^i)
    + \eta \times \BoldS_{(M1)\SmallWord{con}} 
             (\Boldh, \BoldM_b, \BoldP, Q)
  = 0 \, ,
  \EQN MonFRStart;b \cr
  \partial_t \BoldM_2 & {}
    + \partial_a \Big[ \BoldF^a_{(M2)\SmallWord{evol}} 
                       (\Boldh, \BoldM_2, \BoldP, Q, N^i)
                     + \eta \times \BoldF^a_{(M2)\SmallWord{con}} 
                       (\Boldh, \BoldP, Q) \Big]
    \cr & {} 
    + \BoldS_{(M2)\SmallWord{evol}} 
             (\Boldh, \BoldM_b, \BoldP, Q, N^i)
    + \eta \times \BoldS_{(M2)\SmallWord{con}} 
             (\Boldh, \BoldM_b, \BoldP, Q)
  = 0 \, ,
  \EQN MonFRStart;c \cr
  \partial_t \BoldM_3 & {}
    + \eta \times \partial_a \BoldF^a_{(M3)\SmallWord{con}} 
                             (\Boldh, \BoldP, Q)
    + \eta \times \BoldS_{(M3)\SmallWord{con}} 
                  (\Boldh, \BoldM_b, \BoldP, Q)
  = 0 \, ,
  \EQN MonFRStart;d \cr
  \partial_t \BoldP & {}
    + \partial_a \Big[ \BoldF^a_{(P)\SmallWord{evol}} 
                       (\Boldh, \BoldM_b, \BoldM_3, 
                        \BoldP, Q, N^i, \Theta)
                     + \gamma \times \BoldF^a_{(P)\SmallWord{con}} 
                       (\Boldh, \BoldM_b, \BoldM_3, Q) \Big]
    \cr & {} 
    + \BoldS_{(P)\SmallWord{evol}} 
             (\Boldh, \BoldM_b, \BoldP, Q, N^i, \Theta)
    + \gamma \times \BoldS_{(P)\SmallWord{con}} 
             (\Boldh, \BoldM_b, \BoldP, Q)
  = 0 \, ,
  \EQN MonFRStart;e \cr
}$$
where it has been assumed that the spacetime being evolved has a
symmetry such that none of the quantities in the system depend on the
third spatial coordinate~$x^3$, and the convention used is that the
indices~$a$, $b$,~\dots\ range over the values~$\{1,2\}$.
The partial derivatives~$\partial_t$, $\partial_1$ and~$\partial_2$ operate
with respect to the coordinates~$t$, $x^1$ and~$x^2$.
The vectors~$\Boldh$, $\BoldM_1$, $\BoldM_2$, $\BoldM_3$ and~$\BoldP$
each represent six variables, the independent components of~$h^{ij}$,
$M^{ij}{}_{1}$, $M^{ij}{}_{2}$, $M^{ij}{}_{3}$ and~$P^{ij}$, recalling
that all of these tensor-like quantities are symmetric in their
$i$-$j$~indices. 
The slicing density~$Q$ and the shift vector~$N^i$ are fixed functions
of the coordinates~$t$, $x^1$ and~$x^2$.
The generalized FR~system depends on three parameters: the
parameters~$\eta$ and~$\gamma$ determine the amounts of  
constraint quantities that are added to the evolution
equations~\EqNum{MonFRStart}{b--e}, while the parameter~$\Theta$
determines how commuting derivative terms are distributed in the
transport part of equation~\EqNum{MonFRStart}{e}.

The variables~$M^{ij}{}_{3}$ are related to the derivatives of the
three-metric~$h^{ij}$ with respect to the symmetry coordinate~$x^3$,
and for exact solutions of the equations~\EqNum{MonFRStart}{} they
will be identically zero.
However, as explained in section~\use{sec.HypNumerical}, the
variables~$M^{ij}{}_{3}$ 
cannot be omitted from the system~\EqNum{MonFRStart}{} without
affecting its hyperbolicity, although there is no harm in simplifying
the system by setting them to zero in the source terms (the
undifferentiated expressions~$\BoldS_{(h)\SmallWord{evol}}$, etc.).
In numerical simulations truncation errors can lead to the
variables~$M^{ij}{}_{3}$ becoming non-zero; however simply resetting
their values to zero at the end of each time step is not found to
cause adverse effects.
(Section~\use{sec.HypNumerical} discusses these points in greater depth.)

The FR~system~\EqNum{MonFRStart}{} can be assigned the same coordinate
conditions as those used in the Moncrief reduction for
$U(1)$-symmetric~$T^3$ spacetimes (section~\use{sec.MonIntro}) by
choosing the slicing density to be~$Q=1$ and the shift vector to
be~$N^i=0$. 
The FR~coordinate system $\{t,x^1,x^2,x^3\}$ then coincides with the
Moncrief coordinate system $\{\tau,u,v,w\}$.

The new variables for the FR~system are taken to be the variables used
in the first-order form of the Moncrief reduction, that is~$\BoldLam$,
$\BoldLam_u$, $\BoldLam_v$ and~$\BoldLam_\tau$ where
$$
  \BoldLam = [ \Lambda, z, x, \phi, \beta_1, \beta_2 ]^T
  \, ,
$$
and~$\BoldLam_u$, $\BoldLam_v$ and~$\BoldLam_\tau$ are the variables
defined by the first derivatives of~$\BoldLam$ with respect to the
coordinates~$u$, $v$ and~$\tau$.
Here the variables~$\beta_1$ and~$\beta_2$ are used instead of the
single variable~$\omega$ so that there is a one-to-one correspondence
between the three-metrics described by the variables~$\BoldLam$ and
those described by the FR~variables~$\Boldh$.
While the variables~$\BoldLam$ replace the variables~$\Boldh$, and the
variables~$\BoldLam_u$, $\BoldLam_v$ and~$\BoldLam_\tau$ replace
(respectively) the variables~$\BoldM_1$, $\BoldM_2$ and~$\BoldP$, the
Moncrief formulation presents no new variables to replace the
variables~$\BoldM_3$. 
As mentioned above, these variables need to be kept in the evolution
system for technical reasons, and so in adapting the FR~equations they
are simply relabelled:~$\BoldM_3 \rightarrow \Boldm$.

The change in variables from
$\{\Boldh,\BoldM_1,\BoldM_2,\BoldM_3,\BoldP\}$ to
$\{\BoldLam,\BoldLam_u,\BoldLam_v,\Boldm,\BoldLam_\tau\}$ for
equations~\EqNum{MonFRStart}{} could be performed directly using a
computer algebra package, but the results would not be in a
particularly manageable form.
However some thought about the structure of the
system~\EqNum{MonFRStart}{} reveals that it must be equal to
$$\EQNalign{  
  H_{(h)} \Big[ & \partial_\tau \BoldLam - \BoldLam_\tau \Big]
  = 0 \, , 
  \EQN MonFRMid;a \cr
  H_{(M1)} \Big[ & \partial_\tau \BoldLam_u 
                 - \partial_u \BoldLam_\tau \Big]
    + \eta \times \partial_a \BoldF^a_{(M1)\SmallWord{con}}
    + \eta \times \BoldS_{(M1)\SmallWord{con}}
  = 0 \, , 
  \EQN MonFRMid;b \cr
  H_{(M2)} \Big[ & \partial_\tau \BoldLam_v 
                 - \partial_v \BoldLam_\tau \Big]
    + \eta \times \partial_a \BoldF^a_{(M2)\SmallWord{con}}
    + \eta \times \BoldS_{(M2)\SmallWord{con}}
  = 0 \, , 
  \EQN MonFRMid;c \cr
  & \partial_\tau \Boldm 
    + \eta \times \partial_a \BoldF^a_{(M3)\SmallWord{con}}
    + \eta \times \BoldS_{(M3)\SmallWord{con}}
  = 0 \, , 
  \EQN MonFRMid;d \cr
  H_{(P)} \Big[ & \partial_\tau \BoldLam_\tau 
                + \partial_a \Boldf^a_{(\tau)\SmallWord{evol}} 
                  ( \BoldLam, \BoldLam_u, \BoldLam_v, \Theta )
                + \Bolds_{(\tau)\SmallWord{evol}}
                  ( \BoldLam, \BoldLam_u, \BoldLam_v, 
                    \BoldLam_\tau, \Theta ) \Big]
  \cr
  & \qquad\quad\; {} 
    + \partial_a \BoldF^a_{(P)m} (\Boldh, \BoldM_3, \Theta)
    + \gamma \times \partial_a \BoldF^a_{(P)\SmallWord{con}}
    + \gamma \times \BoldS_{(P)\SmallWord{con}}
  = 0 \, , 
  \EQN MonFRMid;e \cr
}$$
where~$H_{(h)}$, $H_{(M1)}$, $H_{(M2)}$ and~$H_{(P)}$ are
$6\times6$~matrices depending on the variables~$\BoldLam$.
The terms in the square brackets in these equations are simply the
first-order evolution equations for the Moncrief variables; in
particular, equation~\EqNum{MonFRMid}{e} incorporates the Moncrief
evolution equations~\EqNum{MonEvolve}{a--d} and~\EqNum{MonBetaVar}{}.
(Ignoring the constraint terms in the equation~\EqNum{MonFRStart}{e},
the FR~evolution system would be expected to be the same as the
Moncrief evolution system since they both derive from the ADM~form of
the Einstein equations; checking the final results of the change of
variables performed here using computer algebra software reveals that
this is indeed the case, and no additional constraint terms need to be
included in equation~\EqNum{MonFRMid}{e}.)

There is some flexibility in the first-order form of the Moncrief
evolution equations that appears as part of the transformed FR~system
in that terms of the form~$X_{,uv}$ can be represented 
as~$\partial_u X_v$ or as~$\partial_v X_u$ or as a combination
of the two.
A comparison between the principal parts of the FR~system and the
Moncrief system fixes the form of the latter in
equation~\EqNum{MonFRMid}{e}, and in the process introduces a dependence
of the flux~$\Boldf^a_{(\tau)\SmallWord{evol}}$ and the
source~$\Bolds_{(\tau)\SmallWord{evol}}$ on the parameter~$\Theta$.
A number of terms involving the variables~$\BoldM_3$ appear in the
FR~fluxes~$\BoldF^a_{(P)\SmallWord{evol}}$ in
equation~\EqNum{MonFRStart}{e} which are not accounted for in the
Moncrief evolution system.
This leads to the appearance of additional flux
terms~$\BoldF^a_{(P)m}$ in the equation~\EqNum{MonFRMid}{e}.

Completing the change of variables for the FR~system leads to the set
of evolution equations
$$\EQNalign{
  \partial_\tau & \BoldLam 
    - \BoldLam_\tau
  = 0 \, , 
  \EQN MonFR;a \cr
  \partial_\tau & \BoldLam_u
    + \partial_u ( \eta \times \Boldf^u_{(u)\SmallWord{con}} 
                 - \BoldLam_\tau )
    + \eta \times \partial_v \Boldf^v_{(u)\SmallWord{con}}
    + \eta \times \Bolds_{(u)\SmallWord{con}}
  = 0 \, , 
  \EQN MonFR;b \cr
  \partial_\tau & \BoldLam_v
    + \eta \times \partial_u \Boldf^u_{(v)\SmallWord{con}}
    + \partial_v ( \eta \times \Boldf^v_{(v)\SmallWord{con}} 
                 - \BoldLam_\tau )
    + \eta \times \Bolds_{(v)\SmallWord{con}}
  = 0 \, , 
  \EQN MonFR;c \cr
  \partial_\tau & \Boldm
    + \eta \times \partial_a \Boldf^a_{(m)\SmallWord{con}}
    + \eta \times \Bolds_{(m)\SmallWord{con}}
  = 0 \, , 
  \EQN MonFR;d \cr
  \partial_\tau & \BoldLam_\tau
    + \partial_a \Big[ \Boldf^a_{(\tau)\SmallWord{evol}} (\Theta)
                     + \gamma \times \Boldf^a_{(\tau)\SmallWord{con}} 
                     + \Boldf^a_{(\tau)\SmallWord{m}} 
                       (\Boldm,\Theta,\gamma) \Big]
    + \Bolds_{(\tau)\SmallWord{evol}} (\Theta)
    + \gamma \times \Bolds_{(\tau)\SmallWord{con}}
  = 0 \, , 
  \EQN MonFR;e \cr
}$$
where dependence of the flux and source terms on the
variables~$\BoldLam$, $\BoldLam_u$, $\BoldLam_v$ and~$\BoldLam_\tau$
is implicit.
These equations follow from inversion of the matrices in the
equations~\EqNum{MonFRMid}{}, which are seen to be the Jacobian
matrices of the change of variables: 
$H_{(h)} = \partial\Boldh / \partial\BoldLam$, 
$H_{(M1)} = \partial\BoldM_1 / \partial\BoldLam_u$, 
$H_{(M2)} = \partial\BoldM_2 / \partial\BoldLam_v$ and
$H_{(P)} = \partial\BoldP / \partial\BoldLam_\tau$.
(Since the variables~$\BoldM_1$, $\BoldM_2$ and~$\BoldP$ depend
on~$\BoldLam$ as well as on~$\BoldLam_u$, $\BoldLam_v$
and~$\BoldLam_\tau$, it might seem that there should be additional
Jacobian matrices in the system reflecting this; however the terms
affected are all of the form of equation~\EqNum{MonFR}{a}, and when
not in the principal parts of evolution systems these terms are
trivially zero.)
The FR~constraint terms in the equations~\EqNum{MonFRMid}{} are
converted into the new variables according to 
$$\EQNalign{
  {\cal C}^0 & {} = \textstyle
    - {1\over2} e^{2\phi + 4\tau - 2\Lambda} {\cal H}_0
    - ( \partial_u m_A + \partial_v m_B )
  \, , \cr
  {\cal C}^a & {} = \textstyle
    {1\over2} e^{3\phi + 4\tau - 2\Lambda} e^{ab} {\cal H}_b
  \, , \EQN MonFRConstr \cr
  {\cal C}^3 & {} = \textstyle
    - {1\over2} e^{3\phi + 4\tau - 2\Lambda} e^{ab} \beta_a {\cal H}_b
    - {1\over2} e^{-\phi + 2\tau - \Lambda} {\cal H}_\beta
  \, , \cr
}$$
where~${\{\cal C}^{\mu}\}$ are the FR~constraint quantities of
equations~\EqNum{HypFRConstr}{} 
and~$\{{\cal H}_0, {\cal H}_a, {\cal H}_\beta\}$ 
are the Moncrief constraint quantities of
equations~\EqNum{MonConstr}{} and~\EqNum{MonBetaConstr}{}.
Undifferentiated $\Boldm$~terms have been omitted from
equations~\EqNum{MonFRConstr}{} since they make no contribution to
the system~\EqNum{MonFR}{}; however derivatives of the~$\Boldm$
variables (specifically the variables~$m_A$ and~$m_B$
in~${\cal C}^0$,
identified below) must be incorporated into the principal part of
the system, and in equation~\EqNum{MonFR}{e} the flux
vectors~$\Boldf^a_{(\tau)m}$ represent such terms which come from the
constraint quantities~$\BoldF^a_{(P)\SmallWord{con}}$ together with
the $\BoldM_3$~terms in the fluxes~$\BoldF^a_{(P)m}$.

It turns out that only three of the variables~$\Boldm$,
labelled~$m_A$, $m_B$ and~$m_C$, have
non-trivial evolution equations as represented by
equation~\EqNum{MonFR}{d}.
The three other $\Boldm$~variables are not affected by the transport
part of the evolution system, and consequently can be omitted
altogether for the equations without affecting their hyperbolicity.
The system~\EqNum{MonFR}{} can thus be simplified by truncating
the vector of variables~$\Boldm$ to
$$
  \Boldm = [ m_A, m_B, m_C ]^T
         = [ M^{13}{}_{3}, M^{23}{}_{3}, M^{33}{}_{3} ]^T
  \, .
$$


\figure{MonFluxUCon}
\BoxTop
$$\eqalign{
  \Boldf^u_{(u)\SmallWord{con}} = {} & \ninepoint
    \left[\eqalign{
      & \textstyle 
        \Lambda_\tau + 2 x z_\tau 
        + {1\over2} e^{-2z} e^{uv}_{,z} x_\tau 
      \cr
      & \textstyle 
        {1\over2} x \Lambda_\tau 
        + z_\tau 
        - {1\over2} e^{-2z} e^{uv} x_\tau
      \cr
      & \textstyle 
        e^{2z} ( {1\over2} e^{uv}_{,z} \Lambda_\tau 
               + 2 e^{uv} z_\tau )
        + x_\tau
      \cr
      & \textstyle 
        {1\over2} \Lambda_\tau 
        + x z_\tau
        + {1\over4} e^{-2z} e^{uv}_{,z} x_\tau
        + {1\over2} \beta_1 \omega_{,v}
      \cr
      & \textstyle 
        - \beta_1 ( \Lambda_\tau + 2 x z_\tau 
                  + {1\over2} e^{-2z} e^{uv}_{,z} x_\tau )
        + e^{vv} e^{uc} \beta_{c\tau}
      \cr
      & \textstyle 
        - \beta_1 ( 2 (x-1) z_\tau 
                  - {1\over2} e^{-2z} e^{uu}_{,z} x_\tau )
        - e^{uv} e^{uc} \beta_{c\tau}
      \cr
    }\right]
\cr
  \Boldf^v_{(u)\SmallWord{con}} = {} & \ninepoint
    \left[\eqalign{
      & \textstyle 
        - 2 (x+1) z_\tau 
        + {1\over2} e^{-2z} e^{vv}_{,z} x_\tau 
      \cr
      & \textstyle 
        - {1\over2} (x+1) \Lambda_\tau
        - {1\over2} e^{-2z} e^{vv} x_\tau
      \cr
      & \textstyle 
        e^{2z} ( {1\over2} e^{vv}_{,z} \Lambda_\tau
               + 2 e^{vv} z_\tau )
      \cr
      & \textstyle 
        - (x+1) z_\tau
        + {1\over4} e^{-2z} e^{vv}_{,z} x_\tau
        - {1\over2} \beta_1 \omega_{,u}
      \cr
      & \textstyle 
        \beta_1 ( 2 (x+1) z_\tau
                - {1\over2} e^{-2z} e^{vv}_{,z} x_\tau )
        + e^{vv} e^{vc} \beta_{c\tau}
      \cr
      & \textstyle 
        \beta_1 ( 2 x z_\tau
                - \Lambda_\tau
                + {1\over2} e^{-2z} e^{uv}_{,z} x_\tau )
        - e^{uv} e^{vc} \beta_{c\tau}
      \cr
    }\right]
\cr
  \Bolds_{(u)\SmallWord{con}} = {} & \ninepoint
    \left[\eqalign{
      & \textstyle 
        Y_u
      \cr
      & \textstyle 
        - {1\over2} e^{2z} e^{cv}_{,x} \Lambda_c (2 - \Lambda_\tau)
        + {1\over2} (x_v - x_u) \Lambda_\tau
        - ( 2 e^{2z} e^{vc}_{,x} z_c 
            + x_u - x_v ) z_\tau
        + {1\over2} e^{-4z} x_u x_\tau
        \cr & \textstyle \qquad\qquad {}
        - 2 e^{2z} e^{vc}_{,x} \phi_c \phi_\tau
        + {1\over4} e^{2\tau-\Lambda} \omega_{,\tau} 
                    e^{vc}_{,z} \beta_{c\tau}
      \cr
      & \textstyle 
        {1\over2} e^{2z} e^{cv}_{,z} \Lambda_c (2-\Lambda_\tau)
        - [ 2 e^{4z} (z_v - z_u) 
            + x x_u - (x+1) x_v ] \Lambda_\tau
        - ( 4 e^{2z} e^{vc} z_c + 2 x_u ) z_\tau
        \cr & \textstyle \qquad\qquad {}
        - [ 2 (1-x) z_u + 2 (1+x) z_v + x_u - x_v ] x_\tau
        + 2 e^{2z} e^{vc}_{,z} \phi_c \phi_\tau
        + e^{4z+2\tau-\Lambda} \omega_{,\tau} e^{vc}_{,x} \beta_{c\tau}
      \cr
      & \textstyle 
        {1\over2} Y_u
        - {1\over2} e^{4\phi+2\tau-\Lambda} e^{bc} 
                    \beta_{1b} \beta_{c\tau}
      \cr
      & \textstyle 
        - \beta_1 Y_u
        + e^{bc} \beta_{b\tau} [ e^{vv} ( 4 \phi_c - \Lambda_c )
                               - e^{vv}_{,c} ]
        + \beta_{1c} ( {1\over2} e^{-2z} e^{vc}_{,z} x_\tau
                     - 2 e^{2z} e^{vc}_{,x} z_\tau )
        + \beta_{1u} \Lambda_\tau
      \cr
      & \textstyle 
        - \beta_1 Y_v
        - e^{bc} \beta_{b\tau} [ e^{uv} ( 4 \phi_c - \Lambda_c )
                               - e^{uv}_{,c} ]
        - \beta_{1c} ( {1\over2} e^{-2z} e^{uc}_{,z} x_\tau
                     - 2 e^{2z} e^{uc}_{,x} z_\tau )
        + \beta_{1v} \Lambda_\tau
      \cr
    }\right]
\cr
  \vrule width0pt height14pt
  \hbox{where} \quad & Y_a = 
    \Lambda_a (2 - \Lambda_\tau)
    + 4 z_a z_\tau
    + e^{-4z} x_a x_\tau
    + 4 \phi_a \phi_\tau
    + e^{-4\phi} \omega_{,\tau} \omega_{,a}
\cr}$$
\BoxBottom
\Caption
Flux and source terms for the evolution equation~\EqNum{MonFR}{b}
for~$\BoldLam_u$ in the transformed Frittelli-Reula system.
These terms derive from the constraint quantities added to the system
to ensure hyperbolicity.
The definitions of~$\omega_{,u}$, $\omega_{,v}$ and~$\omega_{,\tau}$ 
are given in equation~\EqNum{MonOmegaDef}{}.
\bigskip
\endCaption
\endfigure

\figure{MonFluxVCon}
\BoxTop
$$\eqalign{
  \Boldf^u_{(v)\SmallWord{con}} = {} & \ninepoint
    \left[\eqalign{
      & \textstyle 
        2 (x-1) z_\tau
        - {1\over2} e^{-2z} e^{uu}_{,z} x_\tau
      \cr
      & \textstyle 
        {1\over2} (x-1) \Lambda_\tau
        + {1\over2} e^{-2z} e^{uu} x_\tau
      \cr
      & \textstyle 
        - e^{2z} ( {1\over2} e^{uu}_{,z} \Lambda_\tau
                 + 2 e^{uu} z_\tau )
      \cr
      & \textstyle 
        (x-1) z_\tau
        - {1\over4} e^{-2z} e^{uu}_{,z} x_\tau
        + {1\over2} \beta_2 \omega_{,v}
      \cr
      & \textstyle 
        - \beta_2 ( \Lambda_\tau
                  + 2 x z_\tau
                  + {1\over2} e^{-2z} e^{uv}_{,z} x_\tau )
        - e^{uv} e^{uc} \beta_{c\tau}
      \cr
      & \textstyle 
        - \beta_2 ( 2 (x-1) z_\tau
                  - {1\over2} e^{-2z} e^{uu}_{,z} x_\tau )
        + e^{uu} e^{uc} \beta_{c\tau}
      \cr
    }\right]
\cr
  \Boldf^v_{(v)\SmallWord{con}} = {} & \ninepoint
    \left[\eqalign{
      & \textstyle 
        \Lambda_\tau
        - 2 x z_\tau
        - {1\over2} e^{-2z} e^{uv}_{,z} x_\tau
      \cr
      & \textstyle 
        - {1\over2} x \Lambda_\tau
        + z_\tau
        + {1\over2} e^{-2z} e^{uv} x_\tau
      \cr
      & \textstyle 
        - e^{2z} ( {1\over2} e^{uv}_{,z} \Lambda_\tau
                 + 2 e^{uv} z_\tau )
        + x_\tau
      \cr
      & \textstyle 
        {1\over2} \Lambda_\tau
        - x z_\tau
        - {1\over4} e^{-2z} e^{uv}_{,z} x_\tau
        - {1\over2} \beta_2 \omega_{,u}
      \cr
      & \textstyle 
        \beta_2 ( 2 (x+1) z_\tau 
                - {1\over2} e^{-2z} e^{vv}_{,z} x_\tau )
        - e^{uv} e^{vc} \beta_{c\tau}
      \cr
      & \textstyle 
        \beta_2 ( 2 x z_\tau
                - \Lambda_\tau
                + {1\over2} e^{-2z} e^{uv}_{,z} x_\tau )
        + e^{uu} e^{vc} \beta_{c\tau}
      \cr
    }\right]
\cr
  \Bolds_{(v)\SmallWord{con}} = {} & \ninepoint
    \left[\eqalign{
      & \textstyle 
        Y_v
      \cr
      & \textstyle 
        {1\over2} e^{2z} e^{uc}_{,x} \Lambda_c ( 2 - \Lambda_\tau )
        + {1\over2} (x_v - x_u) \Lambda_\tau
        + ( 2 e^{2z} e^{uc}_{,x} z_c
            - x_u + x_v ) z_\tau
        + {1\over2} e^{-4z} x_v x_\tau
        \cr & \textstyle \qquad\qquad {}
        + 2 e^{2z} e^{uc}_{,x} \phi_c \phi_\tau
        - {1\over4} e^{2\tau-\Lambda} \omega_{,\tau} 
                    e^{uc}_{,z} \beta_{c\tau}
      \cr
      & \textstyle 
        - {1\over2} e^{2z} e^{uc}_{,z} \Lambda_c ( 2 - \Lambda_\tau )
        - [ 2 e^{4z} (z_v - z_u) 
          + (x-1) x_u - x x_v ] \Lambda_\tau
        + ( 4 e^{2z} e^{uc} z_c - 2 x_v ) z_\tau
        \cr & \textstyle \qquad\qquad {}
        + [ 2 (x-1) z_u - 2 (x+1) z_v 
          + x_v - x_u ] x_\tau
        - 2 e^{2z} e^{uc}_{,z} \phi_c \phi_\tau
        - e^{4z+2\tau-\Lambda} \omega_{,\tau} e^{uc}_{,x} \beta_{c\tau}
      \cr
      & \textstyle 
        {1\over2} Y_v
        - {1\over2} e^{4\phi+2\tau-\Lambda} e^{bc} 
                    \beta_{2b} \beta_{c\tau}
      \cr
      & \textstyle 
        - \beta_2 Y_u
        - e^{bc} \beta_{b\tau} [ e^{uv} ( 4 \phi_c - \Lambda_c )
                               - e^{uv}_{,c} ]
        + \beta_{2c} ( {1\over2} e^{-2z} e^{vc}_{,z} x_\tau
                     - 2 e^{2z} e^{vc}_{,x} z_\tau )
        + \beta_{2u} \Lambda_\tau
      \cr
      & \textstyle 
        - \beta_2 Y_v
        + e^{bc} \beta_{b\tau} [ e^{uu} ( 4 \phi_c - \Lambda_c )
                               - e^{uu}_{,c} ]
        - \beta_{2c} ( {1\over2} e^{-2z} e^{uc}_{,z} x_\tau
                     - 2 e^{2z} e^{uc}_{,x} z_\tau )
        + \beta_{2v} \Lambda_\tau
      \cr
    }\right]
\cr}$$
\BoxBottom
\Caption
Flux and source terms for the evolution equation~\EqNum{MonFR}{c}
for~$\BoldLam_v$ in the transformed Frittelli-Reula system.
These terms derive from the constraint quantities added to the system
to ensure hyperbolicity.
The definitions of~$\omega_{,u}$, $\omega_{,v}$ and~$\omega_{,\tau}$ 
are given in equation~\EqNum{MonOmegaDef}{}.
See figure~\use{Fg.MonFluxUCon} for the definitions of~$Y_u$ and~$Y_v$.
\bigskip
\endCaption
\endfigure

\figure{MonFluxMCon}
\vskip5mm
\BoxTop
$$\eqalign{
  \Boldf^a_{(m)\SmallWord{con}} = {} & \textstyle
    {1\over2} e^{2\phi+2\tau-\Lambda} 
    [ k^{1a} , k^{2a}, 2k^{3a} ]^T
\cr
  \Bolds_{(m)\SmallWord{con}} = {} & \textstyle
    {1\over2} e^{2\phi+2\tau-\Lambda} \Big(
      [ k^{1} , k^{2} , 2 k^{3} ]^T
      - ( 2\phi_c - \Lambda_c ) [ k^{1c} , k^{2c}, 2k^{3c} ]^T \Big)
\cr
  \vrule width0pt height14pt
  \hbox{where} \quad 
  k^{ab} = {} & {}
    e^{ab} \Lambda_\tau
    + e^{ab}_{,\tau}
\cr
  k^{3a} = {} & {}
    - \beta_c k^{ac}
    - e^{ac} \beta_{c\tau}
\cr
  k^{a} = {} & {}
    (2 - \Lambda_\tau) e^{ac} \Lambda_c
    - \Lambda_\tau e^{ac}_{,c} 
    + 4 e^{ac} \phi_c \phi_\tau
    - e^{2\tau-\Lambda} \omega_{,\tau} \varepsilon^{ac} \beta_{c\tau}
    + 2 e^{-2z} \varepsilon^{ac} ( z_c x_\tau - x_c z_\tau )
\cr
  k^{3} = {} & {}
    - \beta_c k^{c}
    + k^{bc} \beta_{bc}
    - ( 4 \phi_b - \Lambda_b ) e^{bc} \beta_{c\tau}
\cr}$$
\BoxBottom
\Caption
Flux and source terms for the evolution equation~\EqNum{MonFR}{d}
for~$\Boldm$ in the transformed Frittelli-Reula system.
These terms derive from the constraint quantities added to the system
to ensure hyperbolicity.
The definition of~$\omega_{,\tau}$ 
is given in equation~\EqNum{MonOmegaDef}{}.
Terms of the form~$e^{ab}_{,c}$ should be expanded as
$e^{ab}_{,z} z_c + e^{ab}_{,x} x_c$.
\bigskip
\endCaption
\endfigure

\figure{MonFluxTM}
\vskip10mm
\BoxTop
$$\eqalign{
  \Boldf^u_{(\tau)m} = {} & 
  e^{2\Lambda-4\tau-2\phi} \ninepoint
    \left[\eqalign{
      & \textstyle 
        2 (2\gamma-\Theta-1) m_A
      \cr
      & \textstyle 
        (1-\Theta) ( x m_A + (x-1) m_B )
      \cr
      & \textstyle 
        (1-\Theta) e^{2z} ( e^{uv}_{,z} m_A
                          - e^{uu}_{,z} m_B )
      \cr
      & \textstyle 
        (\gamma-2\Theta) m_A 
        - {1\over2} e^{4\phi+2\tau-\Lambda} e^{uc} \beta_c 
                    ( 2 \beta_1 m_A + 2 \beta_2 m_B + m_C )
      \cr
      & \textstyle 
        - 2 e^{uc} \beta_c ( e^{vv} m_A 
                           - e^{uv} m_B )
        - 2 (1-2\Theta) \beta_1 m_A 
        - 2 (1-\Theta) ( \beta_2 m_B + m_C ) 
      \cr
      & \textstyle 
        2 e^{uc} \beta_c ( e^{uv} m_A 
                         - e^{uu} m_B )
        + 2 \Theta \beta_2 m_A
      \cr
    }\right]
\cr
  \Boldf^v_{(\tau)m} = {} &
  e^{2\Lambda-4\tau-2\phi} \ninepoint
    \left[\eqalign{
      & \textstyle 
        2 (2\gamma-\Theta-1) m_B
      \cr
      & \textstyle 
        - (1-\Theta) ( (x+1) m_A
                     + x m_B )
      \cr
      & \textstyle 
        (1-\Theta) e^{2z} ( e^{vv}_{,z} m_A
                          - e^{uv}_{,z} m_B )
      \cr
      & \textstyle 
        (\gamma-2\Theta) m_B
        - {1\over2} e^{4\phi+2\tau-\Lambda} e^{vc} \beta_c
                    ( 2 \beta_1 m_A + 2 \beta_2 m_B + m_C )
      \cr
      & \textstyle 
        - 2 e^{vc} \beta_c ( e^{vv} m_A
                           - e^{uv} m_B )
        + 2 \Theta \beta_1 m_B
      \cr
      & \textstyle 
        2 e^{vc} \beta_c ( e^{uv} m_A
                         - e^{uu} m_B )
        - 2 (1-2\Theta) \beta_2 m_B
        - 2 (1-\Theta) ( \beta_1 m_A + m_C )
      \cr
    }\right]
\cr}$$
\BoxBottom
\Caption
Fluxes for the evolution equation~\EqNum{MonFR}{e}
for~$\BoldLam_\tau$ in the transformed Frittelli-Reula system.
These terms incorporate the $\Boldm$~variables which are required for
hyperbolicity in the Frittelli-Reula formulation but which are not
originally present in the Moncrief reduction.
\bigskip
\endCaption
\endfigure

\figure{MonFluxTCon}
\vskip4mm
\BoxTop
$$\eqalign{
  \Boldf^a_{(\tau)\SmallWord{con}} = {} & \textstyle
    e^{\Lambda-2\tau} ( e^{ac} \Lambda_c
                      + e^{ac}_{,c} )
    [ 2, 0, 0, {1\over2}, 0, 0 ]^T
\cr
  \Bolds_{(\tau)\SmallWord{con}} = {} & \textstyle
    \Big( e^{\Lambda-2\tau} [ 
         e^{ab} ( 2 \phi_a \phi_b - \Lambda_a \Lambda_b
                + {1\over2} e^{-4\phi} \omega_{,a} \omega_{,b} )
         - e^{ac}_{,c} \Lambda_a
         - 2 e^{-2z} ( z_u x_v - z_v x_u ) ]
     \cr & {} \textstyle \,
     + 2 z_\tau{}^2
     + {1\over2} e^{-4z} x_\tau{}^2
     + 2 \phi_\tau{}^2
     + {1\over2} e^{-4\phi} \omega_{,\tau}{}^2
     - {1\over2} ( 2 - \Lambda_\tau )^2
     \Big) [ 2, 0, 0, {1\over2}, 0, 0 ]^T
\cr}$$
\BoxBottom
\Caption
Flux and source terms for the evolution equation~\EqNum{MonFR}{e}
for~$\BoldLam_\tau$ in the transformed Frittelli-Reula system.
These terms derive from the constraint quantities added to the system
to ensure hyperbolicity.
The definitions of~$\omega_{,u}$, $\omega_{,v}$ and~$\omega_{,\tau}$ 
are given in equation~\EqNum{MonOmegaDef}{}.
\bigskip
\endCaption
\endfigure

\figure{MonFluxTEvol}
\vskip5mm
\BoxTop
$$\eqalign{
  \Boldf^a_{(\tau)\SmallWord{evol}} = {} & e^{\Lambda-2\tau} \ninepoint
    \left[\eqalign{
      & \textstyle 
        - 2 e^{ac} \Lambda_c
        - e^{ac}_{,c}
      \cr
      & \textstyle 
        {1\over4} e^{ac}_{,z} \Lambda_c
        - {1\over2} (1-2\Theta) e^{-2z} \varepsilon^{ac} x_c
      \cr
      & \textstyle 
        e^{4z} e^{ac}_{,x} \Lambda_c
        + 2 (1-2\Theta) e^{2z} \varepsilon^{ac} z_c
      \cr
      & \textstyle 
        - e^{ac} \phi_c
      \cr
      & \textstyle 
        \varepsilon^{ab} \varepsilon^{cd} e_{uc} 
            ( \beta_{bd} - \Theta \beta_{db} )
      \cr
      & \textstyle 
        \varepsilon^{ab} \varepsilon^{cd} e_{vc} 
            ( \beta_{bd} - \Theta \beta_{db} )
      \cr
    }\right]
\cr
  \Bolds_{(\tau)\SmallWord{evol}} = {} & \ninepoint
    \left[\eqalign{
      & \textstyle 
        - e^{\Lambda-2\tau} [
            e^{ab} ( 2 \phi_a \phi_b 
                   + {1\over2} e^{-4\phi} \omega_{,a} \omega_{,b}
                   - \Lambda_a \Lambda_b )
          - e^{ac}_{,c} \Lambda_a
          - 2 e^{-2z} ( z_u x_v - z_v x_u ) ]
      \cr
      & \textstyle 
        {1\over2} e^{-4z} x_\tau{}^2
        - {1\over4} e^{\Lambda-2\tau} [
              4 e^{ab} \Lambda_a z_b
            - 2 e^{ab}_{,x} \Lambda_a x_b
            + e^{ab}_{,z} ( \Lambda_a \Lambda_b
                          - 2 \phi_a \phi_b
                          - {1\over2} e^{-4\phi} \omega_{,a} \omega_{,b} )
        \cr & \textstyle \qquad\qquad\qquad\qquad\quad
            - 4 e^{-2z} (1-\Theta) ( \Lambda_u x_v - \Lambda_v x_u )
            + 4 e^{-2z} (1-2\Theta) ( z_u x_v - z_v x_u ) ]
      \cr
      & \textstyle 
        - 4 z_\tau x_\tau
        - e^{\Lambda-2\tau+4z} [
             e^{ab}_{,x} ( \Lambda_a \Lambda_b
                         + 4 z_a \Lambda_b
                         - 2 \phi_a \phi_b
                         - {1\over2} e^{-4\phi} \omega_{,a} \omega_{,b} )
        + e^{-2z} ( \Lambda_u - \Lambda_v ) ( x_u - x_v )
        \cr & \textstyle \qquad\qquad\qquad\qquad\quad\;
        - 2 e^{ab}_{,x} \Lambda_a z_b
        + 4 (1-\Theta) e^{-2z} ( \Lambda_u z_v - \Lambda_v z_u ) ]
      \cr
      & \textstyle 
        - {1\over2} e^{-4\phi} ( 
            e^{\Lambda-2\tau} e^{ab} \omega_{,a} \omega_{,b}
          - \omega_{,\tau}{}^2 )
      \cr
      & \textstyle 
        - e_{ub,\tau} e^{bc} \beta_{c\tau}
        - (\Lambda_\tau - 2 - 4\phi_\tau) \beta_{1\tau}
        - 4 e^{\Lambda-2\tau-4\phi} \omega_{,\tau} e^{vc} \phi_c
        \cr & \textstyle \qquad\qquad {}
        + e^{\Lambda-2\tau} [
            (1-\Theta) ( E_{v}{}^{uv} \beta_{1u}
                       - E_{u}{}^{vv} \beta_{2v} )
          + E_{v}{}^{vv} ( \beta_{1v} - \Theta \beta_{2u} )
          + E_{u}{}^{uv} ( \Theta \beta_{1v} - \beta_{2u} ) ]
      \cr
      & \textstyle 
        - e_{vb,\tau} e^{bc} \beta_{c\tau}
        - (\Lambda_\tau - 2 - 4\phi_\tau) \beta_{2\tau}
        + 4 e^{\Lambda-2\tau-4\phi} \omega_{,\tau} e^{uc} \phi_c
        \cr & \textstyle \qquad\qquad {}
        + e^{\Lambda-2\tau} [ 
            (1-\Theta) ( E_{u}{}^{uv} \beta_{2v}
                       - E_{v}{}^{uu} \beta_{1u} )
          + E_{u}{}^{uu} ( \beta_{2u} - \Theta \beta_{1v} )
          + E_{v}{}^{uv} ( \Theta \beta_{2u} - \beta_{1v} ) ]
      \cr
    }\right]
\cr
  \vrule width0pt height14pt
  \hbox{where} \quad & E_{a}{}^{bc} = 
    \Lambda_a e^{bc} + e^{bc}_{,a}
\cr}$$
\BoxBottom
\Caption
Flux and source terms for the evolution equation~\EqNum{MonFR}{e}
for~$\BoldLam_\tau$ in the transformed Frittelli-Reula system.
These terms make up a first-order form of the evolution
equations~\EqNum{MonEvolve}{a--d} and~\EqNum{MonBetaVar}{} for the 
Moncrief reduction.
The definitions of~$\omega_{,u}$, $\omega_{,v}$ and~$\omega_{,\tau}$ 
are given in equation~\EqNum{MonOmegaDef}{}.
\bigskip
\endCaption
\endfigure

Complete expressions for the flux and source terms in the transformed
FR~system~\EqNum{MonFR}{} are given in figures~\use{Fg.MonFluxUCon}
to~\use{Fg.MonFluxTEvol}. 
It is apparent from equations~\EqNum{MonFR}{} that in adapting
the FR~formulation to $U(1)$-symmetric spacetimes,
an extended first-order Moncrief evolution system has in effect been
constructed, and this has the same hyperbolic properties
and characteristic speeds as the FR~system.
The original Moncrief evolution system appears as a special case in
equations~\EqNum{MonFR}{}: if the parameter choices~$\eta=\gamma=0$
and~$\Theta=1/2$ are made, then the evolution
equations~\EqNum{MonEvolve}{a--d} and~\EqNum{MonBetaVar}{} are
recovered.
(In the case that~$\eta=0$ the variables~$\Boldm$ can be omitted
altogether from the system.)
In the next subsection the two FR~systems of figure~\use{Fg.HypFREqns}
and equations~\EqNum{MonFR}{} are compared to see which performs
better in numerical simulations of $U(1)$-symmetric~$T^3$ spacetimes.

\SubSection{Testing the Frittelli-Reula System II:
            $\,U(1)$-Symmetric Cosmologies}
Earlier in this section results were summarized from a set of tests
performed on a numerical implementation of the Frittelli-Reula~(FR)
formulation of Einstein's equations (as described in chapter~4).
Those tests were based on vacuum planar spacetimes, and such
spacetimes can be described by very simple sets of equations (in
particular the Gowdy system of section~\use{sec.GowGowdy}).
On the one hand this is advantageous because it means that the
FR~formulation can be gradually `dissected' in order to identify
aspects of it which cause poor numerical performance, as was
done with respect to the Gowdy system earlier in this section.
On the other hand numerical simulations of simple spacetimes are
clearly not a fair test of an evolution system which is designed to
deal with the Einstein equations in their full generality.
In the present subsection results are presented from numerical tests
of the FR~system using vacuum $U(1)$-symmetric~$T^3$ spacetimes.
Although the symmetry of these spacetimes allows them to be described
by equations which are simpler than the full Einstein equations (see
the system~\EqNum{MonEvolve}{}), it may be hoped that they none the 
less provide a comprehensive test of the FR~formulation.

In the previous subsection an alternative form of the FR~evolution
system was derived in which the standard FR~variables~$h^{ij}$,
$M^{ij}{}_{k}$ and~$P^{ij}$ are replaced by the variables used to
describe the Moncrief form of a $U(1)$-symmetric metric,
equation~\EqNum{MonMonMetric}{}.
This new formulation consists of the equations~\EqNum{MonFR}{} with
flux and source terms 
given explicitly in figures~\use{Fg.MonFluxUCon}
to~\use{Fg.MonFluxTEvol}, and it is tested
here alongside the standard FR~system given by figure~\use{Fg.HypFREqns}.
In both versions of the FR~formulation, three parameters~$\Theta$,
$\eta$ and~$\gamma$ control the hyperbolic character of the evolution
equations and, as discussed in section~\use{sec.HypFR}, four ranges of
these parameters are of particular interest.
The {\it original\/} FR~system is given by~$\Theta=\eta=\gamma=1$;
it is the evolution system described by Frittelli and Reula in
their~1996 paper, and though it has physically relevant
characteristic speeds it is only weakly hyperbolic.
The {\it modified\/} FR~system is given by~$\Theta=0$
and~$\eta=\gamma=1$; it corrects the original system to make it
consistent with Frittelli and Reula's proof of symmetric
hyperbolicity, but in the process it introduces acausal 
(faster than light) characteristic speeds.
The {\it causal\/} FR~systems are given by~$\Theta=(3\eta-1)/3\eta$
and $\gamma=(9\eta-2)/(12\eta-3)$ where~$\eta$ can take any value
except~0 or~1/4; these systems are strongly hyperbolic and have only
physically relevant characteristic speeds.
The ADM~systems are given by~$\eta=\gamma=0$ where~$\Theta$
can take any value; these systems do not incorporate the additional
constraint terms that characterize the general FR~formulation, and
while their characteristic speeds are physically relevant they are
only weakly hyperbolic. 

The eight variations of the FR~formulation listed above (four
parameter ranges for each of two variable sets) are tested against the
Moncrief evolution system~\EqNum{MonEvolve}{} for numerical
simulations of collapsing $U(1)$-symmetric~$T^3$ cosmologies.
The initial data set used is the same as equation~\EqNum{MonInitial}{}
but with~$\alpha=0.05$ and~$c=+1.0$.
(It is straightforward to convert the initial data from Moncrief
variables to FR~variables.)
A single uniform low-resolution grid is used in all cases with a mesh
size of $\Delta u = \Delta v = 2\pi/40$, and the time steps used have
a fixed size of $\Delta \tau = 2\pi/100$.
The simulations are run from a start time of~$\tau=0$ up to
time~$\tau=5$; this is not long enough for the fine-scale structure
formation reported in section~\use{sec.MonCollapse} to be a factor in
the results.
All of the simulations use the standard two-step Lax-Wendroff
integration method described in chapter~2.
(For the modified FR~system with standard variables, results
using the Lax-Wendroff method have been compared to results using the
high-resolution method of chapter~2, and the conclusions are
essentially the same as for the Gowdy simulations of chapter~6:
despite the appearance of steep gradients at late times in the solutions, 
the high-resolution method is not found to perform significantly
better than the Lax-Wendroff method when used to evolve these vacuum
spacetimes.) 

The FR~Hamiltonian constraint quantity~${\cal C}^0$ from
equation~\EqNum{HypFRConstr}{a} is used as a measure of the error in
the simulations: 
$$
  \hbox{error} = \Delta u \, \Delta v
                 \sum \vert {\cal C}^0 \vert
  \, ,
$$
where for the Moncrief variables the value of~${\cal C}^0$ is calculated
from equations~\EqNum{MonFRConstr}{} and~\EqNum{MonConstr}{}, and
second-order estimates of derivatives are used when needed.
Each simulation is then rated according to its `crash' time, defined
as being the value of the time coordinate~$\tau$ at which the error
exceeds the value~$10^4$.
(Since the metric components tend to grow exponentially it is not
unreasonable for the error to reach values of this magnitude.)
There are of course many different ways in which errors in the
simulations could be measured, and it could be argued that constraint
quantities are not appropriate measures of error size: the
ability of an evolution system to preserve the constraint equations is
not necessarily the same as its ability to produce accurate solutions.
However it has been found that the performance results for the different
FR~formulations are qualitatively similar regardless of the error
measure used, and the Hamiltonian constraint is in many ways the most
practical choice.
Typically the `crash' time of a simulation is reached only a short while
before the errors in the evolved solution build up to the point that
the numerical code is forced to terminate.

\fullfigure{MonCompare}
\vss
\DrawFig{
  \epsfxsize=\hsize
  \epsfbox{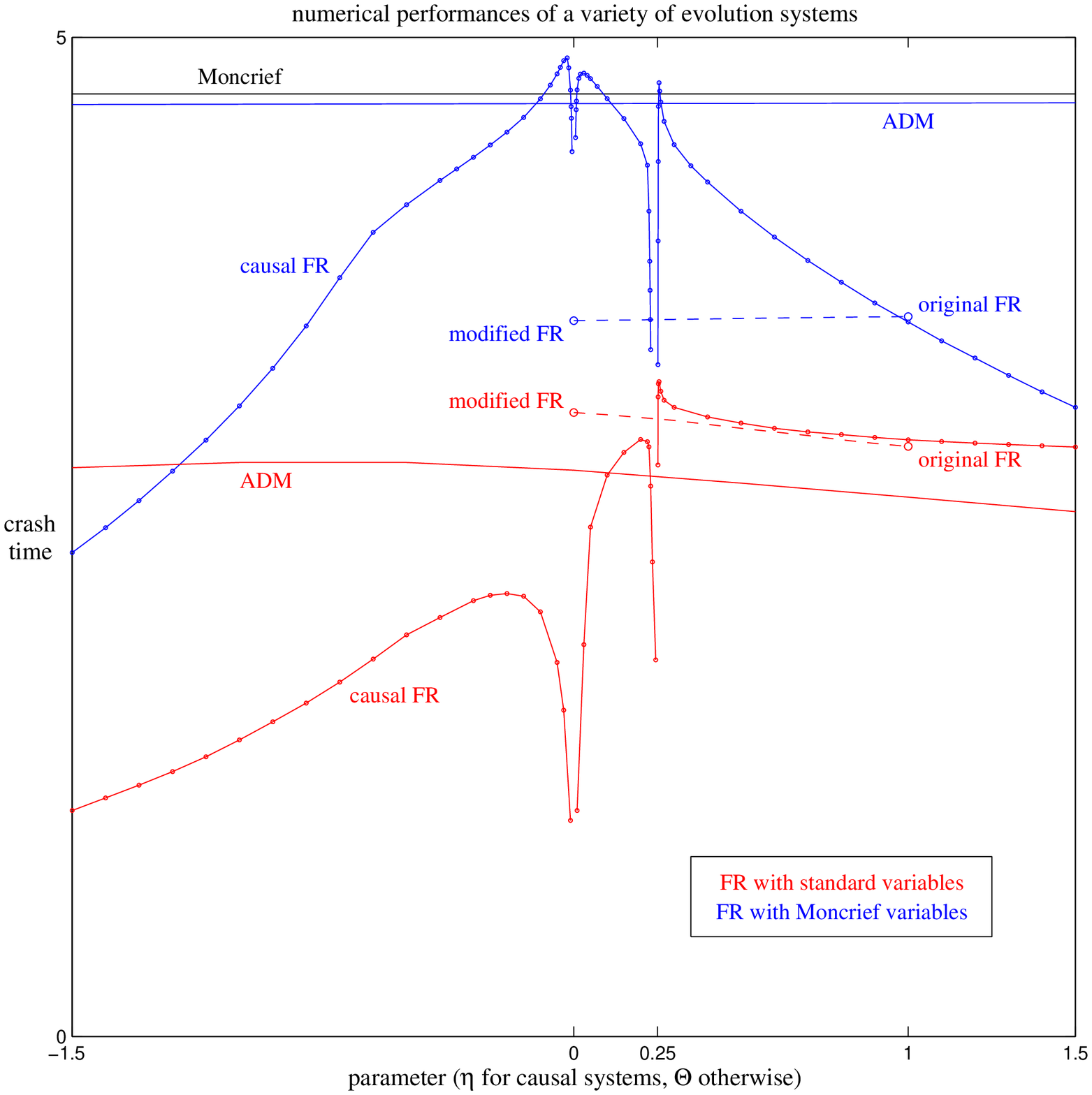}
}
\Caption
Evolution systems compared for numerical simulations of a
$U(1)$-symmetric spacetime.
For each system tested, the time at which the simulation `crashes'
(as determined by the error in the Hamiltonian constraint becoming
larger than a specified limit) is recorded.
The crash time for a simulation based on the reduced Moncrief
evolution system~\EqNum{MonEvolve}{} is shown as a black line.
Results for the standard Frittelli-Reula formulation
(figure~\use{Fg.HypFREqns}) are plotted in red while results for the
transformed formulation~\EqNum{MonFR}{} are plotted in blue.
For the ADM~family of evolution systems and the
original and modified Frittelli-Reula systems the
horizontal axis shows the parameter~$\Theta$.
For the causal Frittelli-Reula family of evolution systems
the horizontal axis shows the parameter~$\eta$.
\vskip4mm
\endCaption
\endfigure

The results of the numerical tests are presented in
figure~\use{Fg.MonCompare}.
The crash times for simulations based on the standard FR~formulation,
figure~\use{Fg.HypFREqns}, are shown in red while the crash times for
simulations based on the transformed FR~formulation,
equations~\EqNum{MonFR}{}, are shown in blue.
The crash time of a simulation using the evolution
equations~\EqNum{MonEvolve}{} of the Moncrief reduction for
$U(1)$-symmetric spacetimes is marked in the figure as a horizontal
black line, 
and the results suggest that this system of equations is in general
preferable to any of the FR~systems considered.
(While the results in figure~\use{Fg.MonCompare} show that for some
parameter settings, later crash times are observed for
FR~systems than for the Moncrief system, the parameter range for which
this is true is found to vary depending on the error measure used.)

It is clear from figure~\use{Fg.MonCompare} that of the two classes
of FR~simulations, those which are based on Moncrief's
choice of variables perform significantly better than those which use
the standard FR~variables.
This adds support to the principle suggested by the numerical studies
of planar cosmologies described earlier in this section that the use
of a set of variables which are in some sense adapted to the
spacetimes being investigated can lead to significant improvements in
the accuracy of numerical simulations.
While the FR~formulation applies to any spacetime, this generality is
not without its disadvantages.

Considering now variations in numerical performance related to changes
in the FR~parameter settings, it can be seen from
figure~\use{Fg.MonCompare} that while the variations may be large, it
is not easy to identify an underlying pattern to them.
Evidence from the~ADM and the modified and original FR~systems
suggests that changes in the parameter~$\Theta$ (with~$\eta$
and~$\gamma$ remaining constant) have relatively small effects on the
accuracies of simulations, while changes in the parameters~$\eta$
and~$\gamma$ (which are responsible for the addition of constraint
quantities to the evolution equations)
seem to have more significant effects.
Within the family of causal~FR systems sizeable variations in
numerical performance are seen as the parameter~$\eta$ is varied.
As~$\eta$ approaches the value~$0$, the associated value for the
parameter~$\Theta$ diverges, and the same
behaviour occurs in the parameter~$\gamma$ as~$\eta$ approaches~$1/4$;
in both cases the performance of the simulations deteriorates.
In general it would seem to be the case that when any of the
parameters (but particularly~$\eta$ or~$\gamma$) are large, the
numerical performance of the FR~system suffers.
However it is not generally true that small parameter values are
responsible for good simulation accuracies, as can be seen from the
results for the ADM~systems in which~$\eta=\gamma=0$.

Figure~\use{Fg.MonCompare} shows that the optimal choice of values for
the parameters~$\Theta$, $\eta$ and~$\gamma$ in the FR~formulation is
dependent on the problem being solved and the variables being used,
and experimentation may be the only method of distinguishing a `good'
evolution system from a `bad' one.
One way of picturing the growth of inaccuracies in evolved solutions
is to consider an exact solution~${\scr G}$ of Einstein's equations
which, because of numerical errors, is perturbed by an initially small
amount to give a solution~${\scr G}+\varepsilon$; while the exact
solution~${\scr G}$ is the same for all of the evolution systems, the
perturbed solution~${\scr G}+\varepsilon$ will be different in each
case.
Analysis of the linearized evolution equations for~$\varepsilon$
about~${\scr G}$ may then give an indication of how fast errors in the
numerical solutions can be expected to grow, and such an approach has
been used by Scheel et al.~(1998) to remedy a `continuum instability'
in a numerical code used to evolve the Schwarzschild solution.
However this kind of analysis may prove to be difficult to apply to
the present work: since the metric components of the cosmological
spacetimes studied here typically vary in magnitude over very large
ranges, the exponential growth of errors may be unavoidable even for
very accurate numerical simulations.

The results shown in figure~\use{Fg.MonCompare} can be used to 
answer two specific questions about the numerical performances of
evolution systems: whether there is a connection between performance and the
presence of non-physical characteristic speeds, and whether strongly
hyperbolic evolution systems perform better than weakly hyperbolic
ones.
The dashed lines in the figure show crash times for the FR~parameter
range~$0\le\Theta\le1$ with~$\eta=\gamma=1$, where the modified~FR and
the original FR~systems are at the end points of the interval.
Although there are no significant variations in numerical performance
across this parameter range, the properties of the evolution systems
vary substantially: only the original FR~system ($\Theta=1$) has
characteristic speeds which all respect causality, and the sizes of
the characteristic speeds increase up to~$\sqrt{5}$ times the speed of
light as~$\Theta$ decreases to zero; only for~$\Theta<1$ are the
evolution systems strongly hyperbolic, and the modified FR~system
($\Theta=0$) has been shown (Frittelli and Reula~1996) to be symmetric
hyperbolic.
Furthermore, figure~\use{Fg.MonCompare} compares results for the
ADM~systems, which are weakly hyperbolic, and the causal FR~systems,
which are strongly hyperbolic, with both families of evolution systems
having only physically relevant characteristic speeds, and it is
clearly not the case that the latter systems perform consistently
better that the former in numerical simulations.
In conclusion, there is no reason to believe that weakly hyperbolic
evolution systems produce less accurate numerical results than
strongly (or symmetric) hyperbolic ones in general, or that
non-physical characteristic speeds adversely affect the numerical
performance of a system.

The results of this section show that the FR~evolution system is in
general out-performed by systems which
are adapted to the symmetries of the spacetimes under investigation.
However when compared with evolution systems which use the same set of
variables, the FR~system may (for appropriate parameter settings)
produce more accurate results, although the gain in these cases is
typically small.
It is worth recalling here some of the potential advantages of hyperbolic
formulations such as the FR~system which motivate their development
for use in numerical work:
for a hyperbolic evolution system, well-posed boundary conditions can
in principle be constructed;
hyperbolicity allows analytic results concerning existence and
uniqueness of solutions to be proved;
sophisticated numerical methods can be applied to first-order systems
for which the eigenfields of the principal part are known;
and for relativistic hydrodynamics a hyperbolic formulation allows
matter fields and metric fields to be treated in a unified manner.
These advantages may be expected to become significant for numerical
simulations of spacetimes which have no simplifying symmetries or in
which realistic matter fields are present.  

(As a final remark in this section it is noted that, for the
numerical implementations of the evolution systems used here, the
simulations based on the Moncrief equations~\EqNum{MonEvolve}{} run
about twice as fast as those based on the transformed
FR~formulation~\EqNum{MonFR}{}, which in turn run about fifteen percent
faster than those based on the standard FR~formulation,
figure~\use{Fg.HypFREqns}.)


\Chapter{Summary of\/ Results}
In this work, a computer code has been developed for the generation of
numerical solutions to the vacuum Einstein equations;
different formulations of the equations have been constructed,
implemented, and tested; and
one- and two-dimensional cosmological models have been numerically
simulated and studied.

What follows is a chapter-by-chapter summary of the main results
together with suggestions for future work.

\MiniSection{Chapter 2}
The numerical simulations reported on in this work have been performed
using standard finite difference integration methods together with a
multi-dimensional high-resolution wave-propagation algorithm developed
by LeVeque~(1997).
Through the implementation of appropriate Riemann solvers, the
wave-propagation algorithm has been used to evolve one-dimensional
spacetimes described by the Gowdy equations and two-dimensional
spacetimes described by the Frittelli-Reula hyperbolic formulation of
Einstein's equations.
Second-order accuracy of the wave-propagation algorithm has been
confirmed in both cases.

While numerical methods based on Riemann solvers have previously been
used to evolve matter fields in relativistic spacetimes,  this has
been the first application of such methods to the evolution of
gravitational fields.  
The natural extension of this work is to the situation in which
gravitational fields are coupled to hydrodynamical sources;
the full potential of the high-resolution method would then be
demonstrated, with shock waves in the fluid producing
non-smooth behaviour in the spacetime metric.

\MiniSection{Chapter 3}
An adaptive mesh refinement algorithm has been implemented for
problems in one and two spatial dimensions.
The basic design of the algorithm follows the work of Berger and
Oliger~(1984) but with enhancements in grid management and
interpolation techniques, among other things. 
The implementation of the algorithm has made use of object-oriented
features of the \Cplusplus\ programming language and has been
parallelized for use on multi-processor systems.

The adaptive mesh refinement algorithm has been used to significantly
decrease the running times of simulations of Gowdy cosmologies and to
reduce the demands made on computational resources by simulations of
$U(1)$-symmetric cosmologies.
To date, this has been one of only a small number of applications of
adaptive mesh refinement to relativistic simulations in more than one
spatial dimension. 
However, the importance of adaptive mesh refinement in numerical
relativity should not be underestimated: such methods are almost
certain to be required for the accurate simulation of three-dimensional
spacetimes describing situations of astrophysical significance.

\MiniSection{Chapter 4}
Frittelli and Reula's (1996) hyperbolic formulation of the Einstein
equations has been generalized to include a one-parameter family of
evolution systems which have only physically relevant characteristic
speeds.
The generalized formulation has been implemented in a numerical code
for the evolution of vacuum spacetimes which have one spacelike Killing
vector.
This is the first time the Frittelli-Reula formulation has been used in
numerical simulations. 
Exact and approximate Riemann solvers for the principal part of the
evolution system have been developed and used to enable the code to
generate numerical solutions using the wave-propagation algorithm.

While in this work difficulties with boundary conditions have been avoided
by considering only periodic spatial domains, the use of hyperbolic
formulations in the specification of stable boundary conditions for
numerical simulations is potentially a fruitful line of research.
Analysis of the initial boundary value problem for general relativity
is simplified if the evolution system is hyperbolic; however this idea
has to date received only limited attention (for example,  
Stewart~1998 and Friedrich and Nagy~1999).

\MiniSection{Chapter 5}
Methods for comparing results from numerical simulations of the same
spacetime produced using different gauge conditions have
been investigated, and, as part of this, a numerical algorithm for
tracking observer world lines has been developed. 
An analytic approach has been used to study foliations of
Kasner and Gowdy spacetimes generated using the harmonic
time slicing condition.
Coordinate singularities of the type originally reported by
Alcubierre~(1997) have been shown to develop in these foliations
under reasonably general conditions.
They have been demonstrated to cause
fine-scale features to appear in numerical solutions; these features
are similar to but distinct from ones found in simulations of
collapsing inhomogeneous cosmologies. 
The formation of coordinate singularities is potentially a serious
drawback to the use of the harmonic time slicing condition in
numerical simulations, and more
research needs to be done to find out under what circumstances
it occurs and how in practice it can be avoided.

\MiniSection{Chapter 6}
Numerical simulations have been performed of Gowdy~$T^3$ cosmological
models.
Berger and Moncrief's~(1993) numerical studies of collapsing
cosmologies have been repeated and extended:
the fine-scale spatial features that develop in the solutions have
been tracked using adaptive mesh refinement and shown to
have a physical origin.
The evolved solutions have been found to be consistent with the
conjecture that the Gowdy models are asymptotically velocity-term
dominated at the singularity (although it is not true that the spatial
derivative terms become negligible there). 

For expanding Gowdy cosmologies, numerical evidence has been used to
show that, while complex behaviour can occur at times soon
after the initial `big bang', at late times the spacetimes behave
like perturbed homogeneous cosmologies.
Results in the literature on the properties of nonlinear gravitational
waves have been examined and found to be inconsistent with some of the
behaviour observed in the Gowdy models; the differences have been
attributed to gauge effects.

While the frontiers of research in numerical relativity lie in the
simulation of spacetimes which have no simplifying symmetries, there
is a lot still to be learnt from one-dimensional models.
Gowdy spacetimes with alternative spatial topologies and generalized
models containing matter fields have not to date been studied in as
much detail as the vacuum three-torus models.
Furthermore, it is possible that methods could be developed
for improving the resolution of spikes in collapsing metrics
without the use of adaptive mesh refinement; for example it may be
possible to introduce a non-trivial shift vector into the evolution
system which would cause grid points to migrate towards the emerging
spikes.

\MiniSection{Chapter 7}
Numerical simulations have been performed of collapsing vacuum
$U(1)$-symmetric $T^3$ cosmologies using Moncrief's~(1986) reduction
of the Einstein equations. 
Mesh refinement techniques have been used to study the
emergence in the spacetimes of fine-scale features which are related
to the ones found in collapsing Gowdy cosmologies.
However, the difficulty in evolving these spacetimes has meant that
many aspects of their behaviour are as yet unexplored; in particular,
nothing is known about the long-term behaviour of the fine-scale
features.

A comparison has been made between numerical solutions produced for
$U(1)$-symmetric spacetimes using three different evolution
systems:  Moncrief's reduction, the Frittelli-Reula formulation, and a
hybrid of these. 
The use of a set of metric variables which are adapted to the nature
of the spacetime under investigation has been shown to be a major
factor contributing to the accuracy of numerical results;
variations in the hyperbolicity and characteristic speeds of
the evolution equations (as controlled by parameters in the
generalized Frittelli-Reula system) have been shown to be a less
significant factor, although still an important one.

While many hyperbolic formulations of Einstein's equations have been
proposed in recent years, only a small number of these have to date
been used in numerical simulations, and there has been little in the
way of comparison between results produced using different
formulations. 
The development of formulations of Einstein's equations which are
well-suited to numerical work is an important line of research
in numerical relativity; it is likely that the current generation of
hyperbolic formulations represent only a first step in this direction.



\vfill\supereject
\leftline{\Tbf{References}}
\nobreak\bigskip
\addTOC{0}{References}{\folio}

\def\RefBase#1{\hangafter=1 \hangindent=\parindent \noindent
               {{#1}.}\par\smallskip}

\def\Ref#1#2#3{\RefBase{{#1} {#2}, {#3}}}
\def\RefPaper#1#2#3#4#5#6{\Ref{#1}{#2}{``#3,'' {\it #4\/}, {\bf #5\/}, {#6}}}
\def\RefBook#1#2#3#4{\RefBase{{#1} {#2}, {\it #3\/} ({#4})}}

\RefPaper{Abrahams,~A., Anderson,~A., Choquet-Bruhat,~Y., 
          and York,~J.~W.}{1995}
  {Einstein and Yang-Mills Theories in Hyperbolic Form without Gauge Fixing}
  {Phys.\ Rev.\ Lett.}{75}{3377--3381}
\RefPaper{Abrahams,~A., Anderson,~A., Choquet-Bruhat,~Y., 
          and York,~J.~W.}{1996}
  {A Non-Strictly Hyperbolic System for the Einstein Equations with 
   Arbitrary Lapse and Shift}
  {C.~R.~Acad.\ Sci.\ Paris, S{\'e}rie~II}{323}{835--841}
\RefPaper{Abrahams,~A., Anderson,~A., Choquet-Bruhat,~Y., 
          and York,~J.~W.}{1997}
  {Geometrical Hyperbolic Systems for General Relativity and Gauge Theories}
  {Class.\ Quantum Grav.}{14}{A9--A22}
\RefBook{Abramowitz,~M., and Stegun,~I.~A.}{1964}
  {Handbook of Mathematical Functions}
  {Dover Publications, Inc}
\RefPaper{Adams,~P.~J., Hellings,~R.~W., Zimmerman,~R.~L.,
   Farhoosh,~H., Levine,~D.~I., and Zeldich,~S.}{1982}
  {Inhomogeneous Cosmology: Gravitational Radiation in Bianchi Backgrounds}
  {Astrophys.~J.}{253}{1--18}
\RefPaper{Alcubierre,~M.}{1997}
  {Appearance of Coordinate Shocks in Hyperbolic Formulations of 
   General Relativity}
  {Phys.\ Rev.~D}{55}{5981--5991}
\Ref{Alcubierre,~M., Br\"ugmann,~B., Miller,~M., and Suen,~W.-M.}{1999}
  {``A Conformal Hyperbolic Formulation of the Einstein Equations,'' 
   e-print {\tt gr-qc/9903030}}
\RefPaper{Alcubierre,~M., and Mass\'o,~J.}{1998}
  {Pathologies of Hyperbolic Gauges in General Relativity and 
   Other Field Theories}
  {Phys.\ Rev.~D}{57}{4511--4515}
\RefPaper{Anderson,~A., Abrahams,~A.~M., and Lea,~C.}{1998}
  {Curvature-Based Gauge-Invariant Perturbation Theory for Gravity:
   A New Paradigm}
  {Phys.\ Rev.~D}{58}{064015 (15~pages)}
\RefPaper{Anderson,~A., Choquet-Bruhat,~Y., and York,~J.~W.}{1997}
  {Einstein-Bianchi Hyperbolic System for General Relativity}
  {Topol.\ Meth.\ Nonlinear Anal.}{10}{353--373}
\RefPaper{Anderson,~A., and York,~J.~W.}{1999}
  {Fixing Einstein's Equations}
  {Phys.\ Rev.\ Lett.}{82}{4384--4387}
\Ref{Andr\'easson,~H.}{1998}
  {``Global Foliations of Matter Spacetimes with Gowdy Symmetry,'' 
   e-print {\tt gr-qc/9812035}}
\RefPaper{Anninos,~P.}{1998a}
  {Plane-Symmetric Cosmology with Relativistic Hydrodynamics}
  {Phys.\ Rev.~D}{58}{064010 (12~pages)}
\Ref{Anninos,~P.}{1998b}
  {``Computational Cosmology: from the Early Universe to the Large
     Scale Structure,''
   Living Review,
   {\tt http://www.livingreviews.org/Articles/Volume1/1998-9anninos}}
\RefPaper{Anninos,~P., Centrella,~J., and Matzner,~R.~A.}{1991a}
  {Numerical Methods for Solving the Planar Vacuum Einstein Equations}
  {Phys.\ Rev.~D}{43}{1808--1824}
\RefPaper{Anninos,~P., Centrella,~J., and Matzner,~R.~A.}{1991b}
  {Nonlinear Wave Solutions to the Planar Vacuum Einstein Equations}
  {Phys.\ Rev.~D}{43}{1825--1838}
\RefPaper{Anninos,~P., Mass\'o,~J., Seidel,~E., Suen,~W.-M., 
          and Tobias,~M.}{1997}
  {Dynamics of Gravitational Waves in 3D: Formulations, Methods and Tests}
  {Phys.\ Rev.~D}{56}{842--858}
\RefPaper{Arbona,~A., and Bona,~C.}{1999}
  {Dealing with the Center and Boundary Problems in 1D Numerical 
   Relativity}
  {Comp.\ Phys.\ Commun.}{118}{229--235}
\RefPaper{Ashtekar,~A., and Husain,~V.}{1998}
  {Symmetry Reduced Einstein Gravity and Generalized~$\sigma$ and
   Chiral Models}
  {Int.\ J.~Mod.~Phys.~D}{7}{549--566}
\RefPaper{Barrow,~J.~D., and Tipler,~F.~J.}{1979}
  {Analysis of the Generic Singularity Studies by Belinskii,
  Khalatnikov, and Lifschitz}{Phys.\ Rep.}{56}{371--402}
\RefPaper{Belinskii,~V.~A., Khalatnikov,~I.~M., and 
          Lifschitz,~E.~M.}{1970}
  {Oscillatory Approach to a Singular Point in the Relativistic Cosmology}
  {Adv.\ Phys.}{19}{525--573}
\RefPaper{Belinskii,~V.~A., Khalatnikov,~I.~M., and 
          Lifschitz,~E.~M.}{1982}
  {A General Solution of the Einstein Equations with a Time Singularity}
  {Adv.\ Phys.}{31}{639--667}
\RefPaper{Bell,~J., Berger,~M., Saltzman,~J., and Welcome,~M.}{1994}
  {Three-Dimensional Adaptive Mesh Refinement for Hyperbolic
   Conservation Laws}{SIAM J.\ Sci.\ Comput.}{15}{127--138}
\RefPaper{Berger,~B.~K.}{1974}
  {Quantum Graviton Creation in a Model Universe}
  {Ann.\ Phys.~(N.Y.)}{83}{458--490}
\Ref{Berger,~M.~J.}{1982}
  {``Adaptive Mesh Refinement for Hyperbolic Partial Differential Equations,''
   Ph.D.\ dissertation, Department of Computer Science, Stanford University}
\RefPaper{Berger,~B.~K.}{1984}
  {Quantum Effects in the Gowdy~$T^3$ Cosmology}
  {Ann.\ Phys.~(N.Y.)}{156}{155--193}
\RefPaper{Berger,~M.~J.}{1985}
  {Stability of Interfaces with Mesh Refinement}
  {Math.\ of Comput.}{45}{301--318}
\RefPaper{Berger,~M.~J.}{1986}
  {Data-Structures for Adaptive Grid Generation}
  {SIAM J.~Sci.\ Stat.\ Comput.}{7}{904--916}
\RefPaper{Berger,~M.~J.}{1987}
  {On Conservation at Grid Interfaces}
  {SIAM J.~Num.\ Anal.}{24}{967--984}
\Ref{Berger,~B.~K.}{1997}
  {``Numerical Investigation of Singularities,'' 
   e-print {\tt gr-qc/9512003}, 
   in {\it Proceedings of the 14th International Conference on 
           General Relativity and Gravitation},
   edited by M.~Francaviglia, G.~Longhi, L.~Lusanna and E.~Sorace 
   (World Scientific)}
\Ref{Berger,~B.~K.}{1998}
  {``Numerical Approaches to Spacetime Singularities,''
   Living Review,
   {\tt http:// www.livingreviews.org/Articles/Volume1/1998-7berger/}}
\RefPaper{Berger,~B.~K., Chru\'sciel,~P.~T., Isenberg,~J., and Moncrief,~V.}
  {1997}{Global Foliations of Vacuum Spacetimes with $T^2$ Isometry}
  {Ann.\ Phys.~(N.Y.)}{260}{117--148}
\RefPaper{Berger,~M.~J., and Colella,~P.}{1989}
  {Local Adaptive Mesh Refinement for Shock Hydrodynamics}
  {J.~Comput.\ Phys.}{82}{64--84}
\RefPaper{Berger,~B.~K., and Garfinkle,~D.}{1998}
  {Phenomenology of the Gowdy Universe on $T^3 \times R$}
  {Phys.\ Rev.~D}{57}{4767--4777}
\RefPaper{Berger,~B.~K., Garfinkle,~D., Isenberg,~J., Moncrief,~V.,
          and Weaver,~M.}{1998}
  {The Singularity in Generic Gravitational Collapse is Spacelike,
   Local, and Oscillatory}
  {Mod.\ Phys.\ Lett.~A}{13}{1565--1574}
\RefPaper{Berger,~B.~K., Garfinkle,~D., and Swamy,~V.}{1995}
  {Detection of Computer Generated Gravitational Waves 
   in Numerical Cosmologies}
  {Gen.\ Rel.\ Grav.}{27}{511--527}
\Ref{Berger,~M.~J., and LeVeque,~R.~J.}{1997}
  {AMRCLAW source code, beta version 0.3, available at
   {\tt http://www.amath.washington.edu/{\Tilde}rjl/amrclaw/index.html}}
\RefPaper{Berger,~M.~J., and LeVeque,~R.~J.}{1998}
  {Adaptive Mesh Refinement using Wave-Propagation Algorithms for
   Hyperbolic Systems}{SIAM J.~Numer.\ Anal.}{35}{2298--2316}
\RefPaper{Berger,~B.~K., and Moncrief,~V.}{1993}
  {Numerical Investigation of Cosmological Singularities}
  {Phys.\ Rev.~D}{48}{4676--4687}
\RefPaper{Berger,~B.~K., and Moncrief,~V.}{1998a}
  {Numerical Evidence that the Singularity in Polarized $U(1)$
   Symmetric Cosmologies on $T^3 \times R$ is Velocity Dominated}
  {Phys.\ Rev.~D}{57}{7235--7240}
\RefPaper{Berger,~B.~K., and Moncrief,~V.}{1998b}
  {Evidence for an Oscillatory Singularity in Generic $U(1)$
   Symmetric Cosmologies on $T^3 \times R$}
  {Phys.\ Rev.~D}{58}{064023 (8~pages)}
\RefPaper{Berger,~M.~J., and Oliger,~J.}{1984}
  {Adaptive Mesh Refinement for Hyperbolic Partial Differential Equations}
  {J.~Comput.\ Phys.}{53}{484--512}
\RefPaper{Bona,~C., and Mass\'o,~J.}{1988}
  {Harmonic Synchronizations of Spacetime}
  {Phys.\ Rev.~D}{38}{2419--2422}
\RefPaper{Bona,~C., and Mass\'o,~J.}{1989}
  {Einstein's Evolution Equations as a System of Balance Laws}
  {Phys.\ Rev.~D}{40}{1022--1026}
\RefPaper{Bona,~C., and Mass\'o,~J.}{1992}
  {Hyperbolic Evolution System for Numerical Relativity}
  {Phys.\ Rev.\ Lett.}{68}{1097--1099}
\RefPaper{Bona,~C., and Mass\'o,~J.}{1993}
  {Numerical Relativity: Evolving Spacetime}
  {Int.\ J.~Mod.\ Phys.~C}{4}{883--907}
\RefPaper{Bona,~C., Mass\'o,~J., Seidel,~E., and Stela,~J.}{1995}
  {New Formalism for Numerical Relativity}
  {Phys.\ Rev.\ Lett.}{75}{600--603}
\RefPaper{Bona,~C., Mass\'o,~J., Seidel,~E., and Stela,~J.}{1997}
  {First Order Hyperbolic Formalism for Numerical Relativity}
  {Phys.\ Rev.~D}{56}{3405--3415}
\Ref{Bona,~C., Mass\'o,~J., Seidel,~E., and Walker,~P.}{1998}
  {``Three Dimensional Numerical Relativity with a Hyperbolic Formulation,''
   e-print {\tt gr-qc/9804052}}
\RefPaper{Bona,~C., Mass\'o,~J., and Stela,~J.}{1995}
  {Numerical Black Holes: A Moving Grid Approach}
  {Phys.\ Rev.~D}{51}{1639--1645}
\RefPaper{Boyd,~P.~T., Centrella,~J.~M., and Klasky,~S.~A.}{1991}
  {Properties of Gravitational `Solitons'}
  {Phys.\ Rev.~D}{43}{379--390}
\RefPaper{Brodbeck,~O., Frittelli,~S., H\"ubner,~P., and Reula,~O.~A.}{1999}
  {Einstein's Equations with Asymptotically Stable Constraint Propagation}
  {J.~Math.\ Phys.}{40}{909--923}
\RefPaper{Br\"ugmann,~B.}{1996}
  {Adaptive Mesh and Geodesically Sliced Schwarzschild Spacetime in
   $3+1$ Dimensions}{Phys.\ Rev.\ D}{54}{7361--7372}
\RefPaper{Br\"ugmann,~B.}{1999}
  {Binary Black Hole Mergers in 3D Numerical Relativity}
  {Int.\ J.~Mod.\ Phys.}{8}{85-100}
\RefPaper{Carmeli,~M., Charach,~Ch., and Malin,~S.}{1981}
  {Survey of Cosmological Models with Gravitational, Scalar and
   Electromagnetic Waves}{Phys.\ Rep.}{76}{79--156}
\Ref{Centrella,~J.}{1979}
  {``Plane-Symmetric Cosmologies: a Study of Inhomogeneities in
     the Universe using Numerical Relativity,''
   Ph.D.\ dissertation, University of Cambridge}
\RefPaper{Centrella,~J.}{1980a}
  {Gravitational Wave Perturbations and Gauge Conditions}
  {Phys.\ Rev.\ D}{21}{2776--2784}
\RefPaper{Centrella,~J.}{1980b}
  {Interacting Gravitational Shocks in Vacuum Plane-Symmetric Cosmologies}
  {Astrophys.~J.}{241}{875--885}
\Ref{Centrella,~J.~M.}{1986}
  {``Nonlinear Gravitational Waves and Inhomogeneous Cosmologies,'' 
   in {\it Dynamical Spacetimes and Numerical Relativity},
   edited by J.~M.~Centrella (Cambridge University Press)}
\RefPaper{Centrella,~J., and Matzner,~R.~A.}{1979}
  {Plane-Symmetric Cosmologies}
  {Astrophys.~J.}{230}{311--324}
\RefPaper{Centrella,~J., and Matzner,~R.~A.}{1982}
  {Colliding Gravitational Waves in Expanding Cosmologies}
  {Phys.\ Rev.\ D}{25}{930--941}
\RefPaper{Centrella,~J., Matzner,~R.~A., Rothman,~T., and Wilson,~J.~R.}
  {1986}{Cosmic Nucleosynthesis and Nonlinear Inhomogeneities}
  {Nuc.\ Phys.~B}{266}{171--227}
\RefPaper{Centrella,~J., and Wilson,~J.~R.}{1983}
  {Planar Numerical Cosmology. I.~The Differential Equations}
  {Astrophys.~J.}{273}{428--435}
\RefPaper{Centrella,~J., and Wilson,~J.~R.}{1984}
  {Planar Numerical Cosmology. II.~The Difference Equations and
   Numerical Tests}
  {Astrophys.\ J.\ Suppl.}{54}{229--249}
\Ref{Choptuik,~M.~W.}{1989}
  {``Experiences with an Adaptive Mesh Refinement Algorithm in
   Numerical Relativity,'' 
   in {\it Frontiers in Numerical Relativity},
   edited by C.~Evans, S.~Finn and D.~Hobill (Cambridge University Press)}
\Ref{Choptuik,~M.~W.}{1992}
  {``\,`Critical' Behaviour in Massless Scalar Field Collapse,'' 
   in {\it Approaches to Numerical Relativity},
   edited by R.~D'Inverno (Cambridge University Press)}
\RefPaper{Choptuik,~M.~W.}{1993}
  {Universality and Scaling in Gravitational Collapse of a Massless
   Scalar Field}{Phys.\ Rev.\ Lett.}{70}{9--12}
\RefPaper{Choptuik,~M.~W., Chmaj,~T., and Bizo\'n,~P.}{1996}
  {Critical Behaviour in Gravitational Collapse of a Yang-Mills Field}
  {Phys.\ Rev.\ Lett.}{77}{424--427}
\RefPaper{Choquet-Bruhat,~Y., and Moncrief,~V.}{1994}
  {An Existence Theorem for the Reduced Einstein Equations}
  {C.~R.~Acad.\ Sci.\ Paris, S{\'e}rie~I}{319}{153--159}
\RefPaper{Choquet-Bruhat,~Y., and York,~J.~W.}{1995}
  {Geometrical Well Posed Systems for the Einstein Equations}
  {C.~R.~Acad.\ Sci.\ Paris, S{\'e}rie~I}{321}{1089--1095}
\Ref{Choquet-Bruhat,~Y., York,~J.~W., and Anderson,~A.}{1998}
  {``Curvature-Based Hyperbolic Systems for General Relativity,'' 
   e-print {\tt gr-qc/9802027},
   in {\it Proceedings of the Eighth Marcel Grossmann Meeting on 
           General Relativity},
   edited by T.~Piran}
\RefPaper{Chrusciel,~P.~T., Isenberg,~J., and Moncrief,~V.}{1990}
  {Strong Cosmic Censorship in Polarised Gowdy Spacetimes}
  {Class.\ Quantum Grav.}{7}{1671--1680}
\RefPaper{Cook,~G.~B., and Scheel,~M.~A.}{1997}
  {Well-Behaved Harmonic Time Slices of a Charged, Rotating, Boosted
   Black Hole}{Phys.\ Rev.~D}{56}{4775--4781}
\RefPaper{Cook,~G.~B., and Teukolsky,~S.~A.}{1999}
  {Numerical Relativity: Challenges for Computational Science}
  {Acta Numerica}{8}{1--45}
\Ref{Cortesi,~D.}{1998}
  {``Origin2000 and Onyx2 Performance Tuning and Optimization Guide,''
   Silicon Graphics, Inc., Document Number 007-3430-002,
   see {\tt http://techpubs.sgi.com:/library/}}
\RefBook{Cosnard,~M., and Trystram,~D.}{1995}
  {Parallel Algorithms and Architectures}
  {International Thompson Computer Press}
\RefBook{Courant,~R., and Hilbert,~D.}{1962}
  {Methods of Mathematical Physics, Volume~II}
  {Interscience Publishers}
\Ref{Diener,~P., Jansen,~N., Khokhlov,~A., Novikov,~I.}{1999}
  {``Adaptive Mesh Refinement Approach to Construction of Initial Data
     for Black Hole Collisions,'' 
   e-print {\tt gr-qc/9905079}}
\RefPaper{Eardley,~D., Liang,~E., and Sachs,~R.}{1972}
  {Velocity-Dominated Singularities in Irrotational Dust Cosmologies}
  {J.~Math.\ Phys.}{13}{99--107}
\RefPaper{Eardley,~D.~M., and Smarr,~L.}{1979}
  {Time Functions in Numerical Relativity: Marginally Bound Dust Collapse}
  {Phys.\ Rev.~D}{19}{2239--2259}
\RefPaper{Estabrook,~F.~B., Robinson,~R.~S., and Wahlquist,~H.~D.}{1997}
  {Hyperbolic Equations for Vacuum Gravity Using Special Orthonormal Frames}
  {Class.\ Quantum Grav.}{14}{1237--1247}
\Ref{Fier,~J.}{1996}
  {``Origin2000 Performance Tuning and Optimization,''
   Silicon Graphics, Inc., Document Number 007-3430-001,
   see {\tt http://techpubs.sgi.com:/library/}}
\Ref{Font,~J.~A., Miller,~M., Suen,~W.-M., and Tobias,~M.}{1998}
  {``Three Dimensional General Relativistic Hydrodynamics.
     I.~Formulations, Methods and Code Tests,'' 
   e-print {\tt gr-qc/9811015}}
\RefPaper{Frauendiener,~J.}{1998}
  {Numerical Treatment of the Hyperboloidal Initial Value Problem for
   the Vacuum Einstein Equations.  I.~The Conformal Field Equations}
  {Phys.\ Rev.~D}{58}{064002 (10~pages)}
\RefPaper{Friedrich,~H.}{1996}
  {Hyperbolic Reductions for Einstein's Equations}
  {Class.\ Quantum Grav.}{13}{1451--1469}
\RefPaper{Friedrich,~H.}{1998}
  {Evolution Equations for Gravitating Ideal Fluid Bodies in 
   General Relativity}
  {Phys.\ Rev.~D}{57}{2317--2322}
\RefPaper{Friedrich,~H., and Nagy,~G.}{1999}
  {The Initial Boundary Value Problem for Einstein's Vacuum 
   Field Equations}
  {Commun.\ Math.\ Phys.}{201}{619--655}
\RefPaper{Frittelli,~S.}{1997}
  {Note on the Propagation of the Constraints in Standard 3+1 
   General Relativity}
  {Phys.\ Rev.~D}{55}{5992--5996}
\RefPaper{Frittelli,~S., and Reula,~O.}{1994}
  {On the Newtonian Limit of General Relativity}
  {Commun.\ Math.\ Phys.}{166}{221--235}
\RefPaper{Frittelli,~S., and Reula,~O.~A.}{1996}
  {First-Order Symmetric Hyperbolic Einstein Equations with Arbitrary
   Fixed Gauge}{Phys.\ Rev.\ Lett.}{76}{4667--4670}
\Ref{Frittelli,~S., and Reula,~O.~A.}{1999}
  {``Well-Posed Forms of the 3+1 Conformally-Decomposed Einstein 
     Equations,'' e-print {\tt gr-qc/9904048}}
\Ref{Garfinkle,~D.}{1999}
  {``Numerical Simulations of Gowdy Spacetimes on 
     $S^2 \times S^1 \times R$,'' 
   e-print {\tt gr-qc/ 9906019}}
\RefPaper{Geyer,~A., and Herold,~H.}{1995}
  {Slicing the Schwarzschild Spacetime: Harmonic versus Maximal Slicing}
  {Phys.\ Rev.~D}{52}{6182--6185}
\RefPaper{Geyer,~A., and Herold,~H.}{1997}
  {Slicing the Oppenheimer-Snyder Collapse: Harmonic versus 
   Maximal Slicing}{Gen.\ Rel.\ Grav.}{29}{1257--1268}
\RefBook{Godlewski,~E., and Raviart,~P.-A.}{1996}
  {Numerical Approximation of Hyperbolic Systems of Conservation Laws}
  {Springer-Verlag}
\RefPaper{Gowdy,~R.~H.}{1971}
  {Gravitational Waves in Closed Universes}
  {Phys.\ Rev.\ Lett.}{27}{826--829 and erratum p1102}
\RefPaper{Gowdy,~R.~H.}{1974}
  {Vacuum Spacetimes with Two-Parameter Spacelike Isometry Groups and
   Compact Invariant Hypersurfaces: Topologies and Boundary Conditions}
  {Ann.\ Phys.~(N.Y.)}{83}{203--241}
\RefPaper{Griffiths,~J.~B., and Alekseev,~G.~A.}{1998}
  {Some Unpolarized Gowdy Cosmologies and Noncolinear Colliding Plane
   Wave Spacetimes}
  {Int.\ J.~Mod.\ Phys.~D}{7}{237--247}
\RefPaper{Grubi{\u s}i\'c,~B., and Moncrief,~V.}{1993}
  {Asymptotic Behaviour of the $T^3 \times R$ Gowdy Space-Times}
  {Phys.\ Rev.~D}{47}{2371--2382}
\RefPaper{Grubi{\u s}i\'c,~B., and Moncrief,~V.}{1994}
  {Mixmaster Spacetime, Geroch's Transformation, and Constants 
   of the Motion}
  {Phys.\ Rev.~D}{49}{2792--2800}
\RefBook{Gustafsson,~B., Kreiss,~H.-O., and Oliger,~J.}{1995}
  {Time Dependent Problems and Difference Methods}
  {Wiley}
\Ref{Hamad\'e,~R.~S.}{1997}
  {``Critical Phenomena in Gravitational Collapse,''
   Ph.D.\ dissertation, University of Cambridge}
\RefPaper{Hamad\'e,~R.~S., Horne,~J.~H., and Stewart,~J.~M.}{1996}
  {Continuous Self-Similarity and $S$-Duality}
  {Class.\ Quantum Grav.}{13}{2241--2253}
\RefPaper{Hamad\'e,~R.~S., and Stewart,~J.~M.}{1996}
  {The Spherically Symmetric Collapse of a Massless Scalar Field}
  {Class.\ Quantum Grav.}{13}{497--512}
\RefBook{Hawking,~S.~W., and Ellis,~G.~F.~R.}{1973}
  {The Large Scale Structure of Space-Time}{Cambridge University Press}
\Ref{Hearn,~A.~C.}{1995}
  {``REDUCE User's Manual, Version~3.6,'' available at 
   {\tt http://www.uni-koeln .de/REDUCE/3.6/doc/reduce/}}
\RefPaper{Hern,~S.~D., and Stewart,~J.~M.}{1998}
  {The Gowdy $T^3$ Cosmologies Revisited}
  {Class.\ Quantum Grav.}{15}{1581--1593}
\RefBook{Hirsch,~C.}{1988}
  {Numerical Computation of Internal and External Flows.
   Volume~1: Fundamentals of Numerical Discretization}
  {Wiley}
\RefBook{Hirsch,~C.}{1990}
  {Numerical Computation of Internal and External Flows. 
   Volume~2: Computational Methods for Inviscid and Viscous Flows}
  {Wiley}
\Ref{Hobill,~D.~W., and Smarr,~L.~L.}{1989}
  {``Supercomputing and Numerical Relativity: 
     A Look at the Past, Present and Future'' 
   in {\it Frontiers in Numerical Relativity},
   edited by C.~Evans, S.~Finn and D.~Hobill (Cambridge University Press)}
\RefBook{Horowitz,~E., Sahni,~S., and Rajasekaran,~S.}{1998}
  {Computer Algorithms}{Computer Science Press}
\Ref{H\"ubner,~P.}{1998}
  {``How to Avoid Artificial Boundaries in the Numerical Calculation
     of Black Hole Spacetimes,''
   e-print {\tt gr-qc/9804065}}
\RefPaper{Iriondo,~M.~S., Leguizam\'on,~E.~O., and Reula,~O.~A.}{1997}
  {Einstein's Equation in Ashtekar's Variables Constitutes a Symmetric
   Hyperbolic System} 
  {Phys.\ Rev.\ Lett.}{79}{4732--4735}
\RefPaper{Iriondo,~M.~S., Leguizam\'on,~E.~O., and Reula,~O.~A.}{1998}
  {On the Dynamics of Einstein's Equations in the Ashtekar Formulation} 
  {Adv.\ Theor.\ Math.\ Phys.}{2}{1075--1103}
\RefPaper{Isaacson,~R.~A.}{1968a}
  {Gravitational Radiation in the Limit of High Frequency.
   I.~The Linear Approximation and Geometrical Optics}
  {Phys.\ Rev.}{166}{1263--1271}
\RefPaper{Isaacson,~R.~A.}{1968b}
  {Gravitational Radiation in the Limit of High Frequency.
   II.~Nonlinear Terms and the Effective Stress Tensor}
  {Phys.\ Rev.}{166}{1272--1280}
\RefPaper{Isenberg,~J., and Moncrief,~V.}{1982}
  {The Existence of Constant Mean Curvature Foliations of 
   Gowdy 3-Torus Spacetimes}
  {Commun.\ Math.\ Phys.}{86}{485--493}
\RefPaper{Isenberg,~J., and Moncrief,~V.}{1990}
  {Asymptotic Behaviour of the Gravitational Field and the Nature of
   Singularities in Gowdy Spacetimes}
  {Ann.\ Phys.~(N.Y.)}{199}{84--122}
\RefBook{John,~F.}{1982}
  {Partial Differential Equations, Fourth Edition}
  {Springer-Verlag}
\RefPaper{Kichenassamy,~S., and Rendall,~A.~D.}{1998}
  {Analytic Description of Singularities in Gowdy Spacetimes}
  {Class.\ Quantum Grav.}{15}{1339--1355}
\RefBook{Krasi\'nski,~A.}{1997}
  {Inhomogeneous Cosmological Models}
  {Cambridge University Press}
\RefBook{Kreiss,~H.-O., and Lorenz,~J.}{1989}
  {Initial-Boundary Value Problems and the Navier-Stokes Equations}
  {Academic Press}
\RefPaper{Kurki-Suonio,~H., Centrella,~J., Matzner,~R.~A., 
   and Wilson,~J.~R.}
  {1987}{Inflation from Inhomogeneous Initial Data in a
   One-Dimensional Back-Reacting Cosmology}
  {Phys.\ Rev.~D}{35}{435--448}
\Ref{Langseth,~J.~O., and LeVeque,~R.~J.}{1999}
  {``A Wave Propagation Method for Three-Dimensional Hyperbolic 
     Conservation Laws ,'' pre-print available at
  {\tt http://www.amath.washington.edu/ {\Tilde}rjl/publications.html}}
\RefBook{LeVeque,~R.~J.}{1992}{Numerical Methods for Conservation Laws}
  {Birkh\"auser-Verlag}
\RefPaper{LeVeque,~R.~J.}{1996}
  {High-Resolution Conservative Algorithms for Advection in 
   Incompressible Flow}{SIAM J.~Numer.\ Anal.}{33}{627--665}
\RefPaper{LeVeque,~R.~J.}{1997}
  {Wave Propagation Algorithms for Multi-Dimensional Hyperbolic Systems}
  {J.~Comput.\ Phys.}{131}{327--353}
\Ref{LeVeque,~R.~J.}{1998}{``CLAWPACK User Notes,'' available together
  with source code (version~3.0) at
  {\tt http://www.amath.washington.edu/{\Tilde}rjl/clawpack.html}}
\RefBook{Lippman,~S.~B., and Lajoie,~J.}{1998}{C++ Primer}
  {Addison-Wesley Publishing Company}
\RefBook{MacCallum,~M.~A.~H., and Wright,~F.~J.}{1991}
  {Algebraic Computing with REDUCE}
  {Oxford University Press}
\Ref{Mass\'o,~J., Seidel,~E., and Walker,~P.}{1995}
  {``Adaptive Mesh Refinement in Numerical Relativity,''
   e-print {\tt gr-qc/9412057}, 
   in {\it General Relativity (MG7 Proceedings)},
   edited by R.~Ruffini and M.~Keiser (World Scientific)}
\Ref{Mass\'o,~J., and Walker,~P.}{1999}
  {``Cactus: Thorny Problems in Numerical Relativity,''
   web page at {\tt http://cactus.aei-potsdam.mpg.de/}}
\RefPaper{McGuigan,~M.}{1991}
  {Gowdy Cosmology and Two-Dimensional Gravity}
  {Phys.\ Rev.~D}{43}{1199--1211}
\RefPaper{Misner,~C.~W.}{1968}
  {The Isotropy of the Universe}{Astrophys.~J.}{151}{431--457}
\RefPaper{Misner,~C.~W.}{1973}
  {A Minisuperspace Example: The Gowdy~$T^3$ Cosmology}
  {Phys.\ Rev.~D}{8}{3271--3285}
\RefBook{Misner,~C.~W., Thorne,~K.~S., and Wheeler,~J.~A.}{1973}
  {Gravitation}{W.~H.~Freeman and Company}
\Ref{Mitra,~S., Parashar,~M., and Browne,~J.}{1998}
  {``DAGH: User's Guide,'' at 
   {\tt http://www.cs.utexas .edu/users/dagh}}
\RefPaper{Moncrief,~V.}{1981}
  {Global Properties of Gowdy Spacetimes with $T^3 \times R$ Topology}
  {Ann.\ Phys.\ (N.Y.)}{132}{87--107}
\RefPaper{Moncrief,~V.}{1986}
  {Reduction of Einstein's Equations of Vacuum Space-Times with
   Spacelike~$U(1)$ Isometry Groups}
  {Ann.\ Phys.~(N.Y.)}{167}{118--142}
\RefPaper{Moncrief,~V.}{1987}
  {An Infinite-Dimensional Family of Curvature Singular Vacuum
   Spacetimes with $U(1)$ Isometry Groups}
  {Class.\ Quantum Grav.}{4}{1555--1563}
\RefPaper{Moncrief,~V.}{1989}
  {Reduction of the Einstein Equations in $2+1$ Dimensions to a 
   Hamiltonian System over Teichm{\" u}ller Space}
  {J.~Math.\ Phys.}{30}{2907--2914}
\RefPaper{Moncrief,~V.}{1990}
  {Reduction of the Einstein-Maxwell and Einstein-Maxwell-Higgs
   Equations for Cosmological Spacetimes with Spacelike~$U(1)$
   Isometry Groups}
  {Class.\ Quantum Grav.}{7}{329--352}
\Ref{Musgrave,~P., Pollney,~D., and Lake,~K.}{1996}
  {``GRTensorII,'' documentation available at 
   {\tt http:// www.astro.queensu.ca/{\Tilde}grtensor/}}
\RefPaper{New,~K.~C.~B., Watt,~K., Misner,~C.~W., and
   Centrella,~J.~M.}{1998}
  {Stable 3-Level Leapfrog Integration in Numerical Relativity}
  {Phys.\ Rev.~D}{58}{064022 (14~pages)}
\Ref{Ove,~R.}{1986}
  {``Numerical Investigations of Cosmic Censorship on $T^3 \times R$,'' 
   in {\it Dynamical Spacetimes and Numerical Relativity},
   edited by J.~M.~Centrella (Cambridge University Press)}
\RefPaper{Ove,~R.}{1989}
  {Symmetry and Singularity Formation in Expanding Spacetimes}
  {Phys.\ Rev.~D}{39}{3587--3595}
\RefPaper{Ove,~R.}{1990a}
  {Nonlinear Gravitational Effect}{Phys.\ Rev.\ Lett.}{64}{1200--1203}
\RefPaper{Ove,~R.}{1990b}
  {Analysis of Cosmological Gravitational Wave Residue}
  {Class.\ Quantum Grav.}{7}{2225--2236}
\RefPaper{Ove,~R.}{1990c}
  {Anomalous Foliations of Einstein Space-Times}
  {J.~Math.\ Phys.}{31}{2688--2693}
\RefPaper{Papadopoulos,~P., Seidel,~E., and Wild,~L.}{1998}
  {Adaptive Computation of Gravitational Waves from Black Hole Interactions}
  {Phys.\ Rev.~D}{58}{084002 (15~pages)}
\RefPaper{Petrich,~L.~I., Shapiro,~S.~L., and Teukolsky,~S.~A.}{1985}
  {Oppenheimer-Snyder Collapse with Maximal Time Slicing and 
   Isotropic Coordinates}{Phys.\ Rev.~D}{31}{2459--2469}
\RefPaper{Petrich,~L.~I., Shapiro,~S.~L., and Teukolsky,~S.~A.}{1986}
  {Oppenheimer-Snyder Collapse in Polar Time Slicing}
  {Phys.\ Rev.~D}{33}{2100--2110}
\RefPaper{Piran,~T.}{1981}
  {Problems and Solutions in Numerical Relativity}
  {Ann.\ N.~Y.~Acad.\ Sci.}{375}{1--14}
\Ref{Plewa,~T.}{1999}{``Adaptive Mesh Refinement for Structured Grids,''
   web page at {\tt http://www.camk .edu.pl/{\Tilde}tomek/AMRA/amr.html}}
\RefBook{Potter,~D.}{1973}{Computational Physics}{Wiley}
\RefBook{Press,~W.~H., Teukolsky,~S.~A., Vetterling,~W.~T., and
         Flannery,~B.~P.}{1992}
  {Numerical Recipes, Second Edition}
  {Cambridge University Press}
\Ref{Quirk,~J.~J., and Hanebutte,~U.~R.}{1993}
  {``A Parallel Adaptive Mesh Refinement Algorithm,''
   NASA Contractor Report 191530, ICASE Report No.~93-63}
\Ref{Reula,~O.~A.}{1998}
  {``Hyperbolic Methods for Einstein's Equations,''
   Living Review,
   {\tt http://www .livingreviews.org/Articles/Volume1/1998-3reula/}}
\RefPaper{Roe,~P.~L.}{1981}
  {Approximate Riemann Solvers, Parameter Vectors, and Difference Schemes}
  {J.~Comp.\ Phys.}{43}{357--372}
\RefPaper{Rupright,~M.~E., Abrahams,~A.~M., and Rezzolla,~L.}{1998}
  {Cauchy-Perturbative Matching and Outer Boundary Conditions.
   I.~Methods and Tests}
  {Phys.\ Rev.~D}{58}{044005 (9~pages)}
\RefBook{Ryan,~M.~P., and Shepley,~L.~C.}{1975}
  {Homogeneous Relativistic Cosmologies}
  {Princeton University Press}
\RefPaper{Scheel,~M.~A., Baumgarte,~T.~W., Cook,~G.~B., 
   Shapiro,~S.~L., and Teukolsky,~S.~A.}{1997}
  {Numerical Evolution of Black Holes with a Hyperbolic Formulation 
   of General Relativity}
  {Phys.\ Rev.~D}{56}{6320--6335}
\RefPaper{Scheel,~M.~A., Baumgarte,~T.~W., Cook,~G.~B., 
   Shapiro,~S.~L., and Teukolsky,~S.~A.}{1998}
  {Treating Instabilities in a Hyperbolic Formulation of Einstein's Equations}
  {Phys.\ Rev.~D}{58}{044020 (12~pages)}
\RefPaper{Seidel,~E.}{1996}
  {New Developments in Numerical Relativity}
  {Helv.\ Phys.\ Acta}{69}{454--471}
\Ref{Seidel,~E.}{1998}
  {``Numerical Relativity: Towards Simulations of 3D Black Hole 
   Coalescence,'' e-print {\tt gr-qc/9806088}, 
   to appear in GR15 proceedings}
\Ref{Shinkai,~H., and Maeda,~K.}{1993}
  {``Can Gravitational Waves Prevent Inflation?'' 
   {\it Phys.\ Rev.~D\/}, {\bf 48\/}, 3910--3913}
\RefPaper{Shinkai,~H., and Maeda,~K.}{1994}
  {Generality of Inflation in a Planar Universe}
  {Phys.\ Rev.~D}{49}{6367--6378}
\Ref{Shinkai,~H., and Yoneda,~G.}{1999}
  {``Asymptotically Constrained and Real-Values System Based On
     Ashtekar's Variables,'' 
   e-print {\tt gr-qc/9906062}}
\RefPaper{Smarr,~L., and York,~J.~W.}{1978}
  {Kinematical Conditions in the Construction of Spacetime}
  {Phys.\ Rev.~D}{17}{2529--2551}
\RefBook{Sod,~G.~A.}{1985}
  {Numerical Methods in Fluid Dynamics: 
   Initial and Initial Boundary-Value Problems}
  {Cambridge University Press}
\Ref{Stewart,~J.~M.}{1984}
  {``Numerical Relativity,'' 
   in {\it Classical General Relativity},
   edited by W.~B.~Bonnor, J.~N.~Islam and M.~A.~H.~MacCallum 
   (Cambridge University Press)}
\RefPaper{Stewart,~J.~M.}{1998}
  {The Cauchy Problem and the Initial Boundary Value Problem in 
   Numerical Relativity}
  {Class.\ Quantum Grav.}{15}{2865--2889}
\RefPaper{Strang,~G.}{1968}
  {On the Construction and Comparison of Difference Schemes}
  {SIAM J.~Numer.\ Anal.}{5}{506--517}
\RefBook{Stroustrup,~B.}{1997}{The C++ Programming Language, Third Edition}
  {Addison-Wesley Publishing Company}
\RefPaper{Tanimoto,~M.}{1998}{New Varieties of Gowdy Spacetimes}
  {J.~Math.\ Phys.}{39}{4891--4898}
\RefBook{Toro,~E.~F.}{1997}
  {Riemann Problems and Numerical Methods for Fluid Dynamics:
   A Practical Introduction}
  {Springer-Verlag}
\RefPaper{van Elst,~H., and Ellis,~G.~F.~R.}{1999}
  {Causal Propagation of Geometrical Fields in Relativistic Cosmology}
  {Phys.\ Rev.~D}{59}{024013 (22~pages)}
\RefPaper{van Putten,~M.~H.~P.~M.}{1997}
  {Numerical Integration of Nonlinear Wave Equations for General Relativity}
  {Phys.\ Rev.~D}{55}{4705--4711}
\RefPaper{van Putten,~M.~H.~P.~M., and Eardley,~D.~M.}{1996}
  {Nonlinear Wave Equations for Relativity}
  {Phys.\ Rev.~D}{53}{3056--3063}
\RefPaper{Varadarajan,~M.}{1995}
  {Gauge Fixing of One Killing Field Reductions of Canonical Gravity: 
   The Case of Asymptotically Flat Induced Two-Geometry}
  {Phys.\ Rev.~D}{52}{2020--2029}
\RefBook{Wald,~R.~M.}{1984}{General Relativity}
  {University of Chicago Press}
\Ref{Wild,~L.~A.}{1996}
  {``Adaptive Mesh Refinement in Numerical Relativity,'' 
   Ph.D.\ dissertation, University of Wales}
\RefPaper{Witten,~L.}{1959}
  {Invariants of General Relativity and the Classification of Spaces}
  {Phys.\ Rev.}{113}{357--362}
\RefPaper{Yoneda,~G., and Shinkai,~H.}{1999a}
  {Symmetric Hyperbolic System in the Ashtekar Formulation}
  {Phys.\ Rev.\ Lett.}{82}{263--266}
\Ref{Yoneda,~G., and Shinkai,~H.}{1999b}
  {``Constructing Hyperbolic Systems in the Ashtekar Formulation of 
     General Relativity,'' 
   e-print {\tt gr-qc/9901053}}
\Ref{York,~J.~W.}{1979}
  {``Kinematics and Dynamics of General Relativity,'' 
   in {\it Sources of Gravitational Radiation},
   edited by L.~Smarr (Cambridge University Press)}
\Ref{York,~J.~W.}{1989}
  {``Initial Data for Collisions of Black Holes and Other 
     Gravitational Miscellany,'' 
   in {\it Frontiers in Numerical Relativity},
   edited by C.~Evans, S.~Finn and D.~Hobill (Cambridge University Press)}

\end